\def\answ{b }
\language=1
\catcode`é\active
\defé{\'e}
\catcode`à\active
\defà{\`a}
\catcode`è\active
\defè{\`e}
\catcode`ê\active
\defê{\^e}
\catcode`î\active
\defî{\^\i{}}
\catcode`ô\active
\defô{\^o}
\catcode`û\active
\defû{\^u}
\catcode`â\active
\defâ{\^a}
\catcode`ù\active
\defù{\`u}
\catcode`ü\active
\defü{\"u}
\catcode`ç\active
\defç{\c c{}}
\catcode`ö\active
\defö{\"o}
\def\petitblanc{\kern .2em}
\def\exclam{!}
\catcode`!\active
\def!{\penalty10000\petitblanc\exclam}
\def\interrog{?}
\catcode`?\active
\def?{\penalty10000\petitblanc\interrog}
%

\input thesemac
\catcode`\#=12%
\input labeldef.tmp%
\hyperdiese
\writedefs
\overfullrule=0pt
\def\X{{\rm XXX}}
\def\XZ{{\rm XXZ}}
\def\NJL{{\rm NJL}}
\def\Z{{\bf T}}
\def\A{{\bf A}}
\def\B{{\bf B}}
\def\C{{\bf C}}
\def\D{{\bf D}}
\def\E{{\rm E}}
\def\F{{\rm F}}
\def\HH{{\rm H}}
\def\K{{\rm K}}
\def\H{{\bf H}}
\def\P{{\bf P}}
\def\s{{\rm s}}
\def\S{{\rm S}}
\def\Sh{\widehat{S}}

\def\verylongrightarrow{
\relbar\joinrel\relbar\joinrel\relbar\joinrel\relbar\joinrel
\longrightarrow}
\def\figdir#1{#1}
\def\artic0#1{%
\uppercase\expandafter{\romannumeral#1}%
}
\everypar={\looseness=-1}
\nopagenumbers\bigskip
\centerline{UNIVERSITE PIERRE ET MARIE CURIE}\bigskip
\centerline{LABORATOIRE DE PHYSIQUE THEORIQUE}
\centerline{DE L'ECOLE NORMALE SUPERIEURE}\bigskip
\centerline{\epsfbox{\figdir{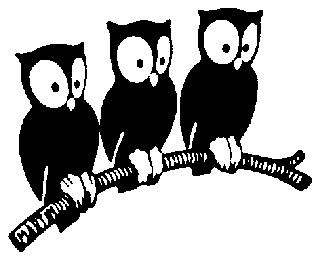}}}%
\medskip
\centerline{\bf THESE DE DOCTORAT DE L'UNIVERSITE PARIS 6}
\medskip
\centerline{Spécialité: {\bf PHYSIQUE THEORIQUE}}\medskip
\centerline{présentée par}\medskip
\centerline{\bf Paul ZINN-JUSTIN}\medskip
\centerline{pour obtenir le titre de}\medskip
\centerline{DOCTEUR DE L'UNIVERSITE PARIS 6}\bigskip\bigskip
\centerline{Sujet:}\medskip
\titlefont\centerline{\bf QUELQUES APPLICATIONS}\smallskip
\centerline{\bf DE L'ANSATZ DE BETHE}\vskip2.5cm
\tenpoint\centerline{Soutenance le 8 juillet 1998
devant le jury composé de:}\medskip
{\parindent=90pt
M.~ANDREI Natan\par
M.~BREZIN Edouard\par
M.~DE VEGA Hector (rapporteur)\par
M.~GERVAIS Jean-Loup\par
M.~KAZAKOV Vladimir (directeur)\par
M.~KOSTOV Ivan (rapporteur)\par
M.~ZUBER Jean-Bernard\par}
\vfill\supereject
\ \vfill\supereject
\long\def\rem#1{}\norefpages
\centerline{\bf Remerciements}
Je tiens à exprimer ma reconnaissance à tous ceux qui m'ont aidé à mener
à bien mon travail de thèse.

Je souhaite remercier tout d'abord les membres du jury:
les rapporteurs H.~De~Vega et I.~Kostov, pour avoir accepté cette tâche
ingrate, et pour un grand nombre de passionnantes discussions,
sur les modèles intégrables et sur les modèles de matrices respectivement;
N.~Andrei, pour avoir fait le déplacement depuis l'autre côté de
l'Atlantique, et pour m'avoir expliqué la physique
du modèle de Kondo lors de notre fructueuse collaboration; J-B.~Zuber
et E.~Brézin, avec lesquels j'ai eu des rapports
aussi amicaux qu'instructifs; J.L.~Gervais,
qui m'a conseillé dans mon travail à de nombreuses reprises;
et V.~Kazakov, mon directeur de thèse, qui a accepté de
diriger mes recherches, et qui m'a prodigué des conseils utiles.

Je remercie le Laboratoire de Physique Théorique de l'Ecole Normale
Supérieure qui m'a accueilli pendant ces trois ans: aussi bien les
chercheurs avec qui j'ai eu des contacts enrichissants, en particulier
A.~Georges, B.~Julia, N.~Sourlas et les 
thésards D.~Carpentier, P.~Chose, O.~Parcollet, S.~Silva, T.~Wynter,
que les secrétaires M-C.~Launay, M.~Leliepvre et N.~Ribet
pour leur aide inestimable.

Je remercie également les chercheurs du LPTHE (Jussieu)
et du SPhT (Saclay) qui
m'ont aidé dans mon travail de recherche, parmi lesquels
O.~Babelon, D.~Bernard, F.~Smirnov et bien sûr J.~Zinn-Justin;
ainsi que les nombreuses personnes
que j'ai rencontrées lors de mes déplacements, en particulier
les chercheurs du Yukawa Institute à Kyoto pour leur accueil, et S.~Hikami 
qui m'a invité à l'université de Tokyo.

Enfin, je remercie mon entourage qui m'a soutenu tout au long de mon
travail de thèse.
\vfill\eject
\ \vfill\supereject
\centerline{\bf RESUME: Quelques applications de l'Ansatz de Bethe}
L'Ansatz de Bethe est une méthode utilisée dans les modèles
quantiques intégrables pour une résolution explicite de ceux-ci.
Cette méthode est exposée ici dans un cadre général, valable
pour les chaînes de spin quantiques 1D, les modèles statistiques
sur réseau (du type modèles de vertex) 2D et les théories des champs
relativistes à 1 dimension d'espace et 1 dimension de temps. Le lien
avec les groupes quantiques est explicité. Plusieurs applications
sont alors présentées. Le calcul des corrections de taille finie
est effectué par deux méthodes:
les Equations Non-Linéaires Intégrales, que l'on applique à l'étude
des états du modèle de Toda affine en constante de couplage
imaginaire sur un espace compactifié, et leur interpolation
entre la région de haute énergie (ultra-violette) et de basse
énergie (infra-rouge); et les Equations d'Ansatz de Bethe
Thermodynamique, ainsi que les Equations de Fusion qui leur
sont associées, dont on se sert pour déterminer la
thermodynamique du modèle
de Kondo multi-canal généralisé. Ce dernier
est ensuite étudié plus en détail, toujours par l'Ansatz de Bethe
et les groupes quantiques, de façon à caractériser le spectre
des excitations de basse énergie.

\noindent$\star$ {\it mots-clés}: modèles intégrables,
diffusion factorisée, Ansatz de Bethe, groupes quantiques, effet Kondo.
\bigskip
\language=0
\centerline{\bf ABSTRACT: A few applications of the Bethe Ansatz}
The Bethe Ansatz is a method that is used in quantum integrable
models in order to solve them explicitly. This method is
explained here in a general framework, which applies to
1D quantum spin chains, 2D statistical lattice models (vertex models)
and relativistic field theories with 1 space dimension and
1 time dimension. The connection with quantum groups is expounded.
Several applications are then presented.
Finite size corrections are calculated via two methods:
The Non-Linear Integral Equations, which are applied to the
study of the states of the affine Toda model with imaginary coupling,
and their interpolation between the high energy (ultra-violet) and
low energy (infra-red) regions; and the Thermodynamic Bethe Ansatz
Equations, along with the associated Fusion Equations, which are used
to determine the thermodynamic properties
of the generalized multi-channel Kondo model. The latter
is then studied in more detail, still using the Bethe Ansatz and
quantum groups, so as to characterize the spectrum of
the low energy excitations.

\noindent$\star$ {\it keywords}: integrable models,
factorized scattering, Bethe Ansatz, quantum groups, Kondo effect.
\vfill\eject
\ \vfill\supereject
\centerline{\bf Table des matières}\nobreak\medskip{\baselineskip=12pt%
\parskip=0pt\catcode`\@=11%
\catcode`\#=12%
\input toc.tmp%
\hyperdiese%
\catcode`\@=12%
\bigbreak\bigskip}
\writetoc
\newnochap{Introduction}
\footline={\hss\tenrm\folio\hss}\pageno=1
Pendant les trois ans qu'a duré mon travail de thèse,
j'ai consacré un temps à peu près égal à deux sujets:
les modèles de matrice et les modèles intégrables.
Nous allons dans cette introduction expliquer les enjeux
de ces deux thèmes de recherche;
par la suite, il ne sera plus question que de modèles intégrables,
pour des raisons qui vont être exposées ci-dessous.
Dans tout ce qui suit, les chiffres romains désignent les
articles que j'ai écrits et qui sont placés en appendice;
ainsi, [\artic04.3.1] désigne la section 3.1 de l'article \artic04,
tandis que (\artic03.2.7) désigne l'équation (2.7) de l'article \artic03.
Nous ne donnerons
aucune référence dans cette introduction, celles-ci étant contenues
dans les articles et/ou dans le texte de la thèse.

Les modèles de matrices ont de multiples applications physiques:
gravitation quantique bidimensionnelle, statistique des
membranes, physique des systèmes désordonnés. Le point commun
de ces différents domaines est la possibillité de
se ramener au calcul de la fonction de partition -- ou
de fonctions de corrélations d'observables -- de modèles
dont les champs sont des matrices $N\times N$, où
la taille $N$ est généralement supposée grande.
Dans les cas les plus simples, le calcul est effectué en
se ramenant à un point de col sur les valeurs propres
des matrices.
Je me suis pour ma part intéressé à un certain nombre de méthodes
mathématiques pour traiter des cas plus compliqués:
développement en caractères, points de col non-triviaux,
problèmes d'inversion fonctionnelle, etc. J'ai appliqué
certaines de ces méthodes dans les articles [\artic01,\artic02,\artic05],
que je vais maintenant brièvement présenter. 

Le problème qui est à la base de [\artic01,\artic02] est le suivant:
en physique nucléaire et en physique des systèmes désordonnés,
on représente les Hamiltoniens
de systèmes complexes ou désordonnés
par des matrices aléatoires de grande taille.
On s'intéresse alors à la statistique des niveaux (valeurs
propres des matrices). Une donnée {\it a priori}
est la loi de probabilité des matrices: de celle-ci
dépendent évidemment les propriétés statistiques des niveaux.
Ainsi, si le Hamiltonien possède une symétrie, de telle sorte
qu'on peut le diagonaliser par blocs dans les différentes
représentations de cette symétrie, alors les valeurs propres
appartenant à des représentations distinctes seront totalement
incorrélées. A l'opposé, pour une mesure raisonnablement
régulière sur l'espace des matrices, des arguments élémentaires
(déterminant de Van Der Monde dans la mesure exprimée en
termes des valeurs propres) montrent que les différents
niveaux ont tendance à {\it se repousser}. Il est donc essentiel
de comprendre jusqu'à quel point les propriétés statistiques
des niveaux sont affectées par le choix de la mesure.
On recherche en particulier des {\it classes d'universalité}
de mesures pour lesquelles, dans la limite de taille infinie
des matrices, les corrélations entre valeurs propres acquièrent
la même forme.

Il existe une certaine classe de modèles pour lesquels
les fonctions de corrélation des valeurs propres se
factorisent sous forme de déterminant (Eq. (\artic01.1.2));
dans ce cas, l'analyse des corrélations est grandement
facilitée, et on a en particulier accès
à la fonction d'espacement des niveaux, qui est l'une des
principales quantités physiques recherchées. Dans [\artic01],
il est prouvé que pour une classe de modèles assez large,
une telle formule de déterminant existe. Dans [\artic02]
est développée une nouvelle méthode pour exprimer le noyau
qui apparaît dans la formule de déterminant,
à travers des équations aux dérivées partielles qu'il
satisfait. On parvient alors à démontrer {\it l'universalité
à courte distance} de l'espacement des niveaux (courte
distance voulant dire un espacement de l'ordre de $1/N$,
$N$ taille de la matrice). Les appendices de [\artic02] contiennent
quelques résultats complémentaires obtenus par des méthodes
similaires, dont une démonstration élémentaire de formules
de développements en caractères.
Les deux articles [\artic01,\artic02]
s'auto-suffisent, et c'est pourquoi j'ai choisi de ne plus parler
de cette partie de mon travail dans la suite de cette thèse.

J'ai également appliqué des techniques assez proches, en
particulier le développement en caractères déjà évoqué dans
[\artic02], dans mon dernier travail sur le modèle à deux
matrices avec terme en $ABAB$ [\artic05].
Ce modèle, de par son développement en diagrammes de Feynmann,
est en fait un modèle de surfaces discrétisées (soit encore
de gravité quantique bidimensionnelle), 
sur lesquelles on dessine des boucles
de deux couleurs (qui forment la matière couplée à la gravité).
L'article, tel qu'il est présenté dans l'appendice,
est encore au stade de brouillon,
mais il montre que ce nouveau modèle de 
possède des propriétés intéressantes (transition de phase,
point critique $c=1 \ldots$). Etant donné que le travail, au moment
où j'ai écrit ce mémoire de thèse, n'était pas encore
terminé,
il n'en sera plus question non plus dans ce qui suit.

Passons donc au deuxième sujet,
les modèles intégrables, que j'étudie depuis
environ deux ans. Je me suis plus précisément intéressé
à un certain nombre de théories des champs
à deux dimensions, qui s'avèrent être intégrables. Qu'entend-on par
là? La dé\-finition classique d'intégrabilité pour un système
qui possède un nombre fini de degrés de liberté suggère
qu'une telle théorie doit posséder une infinité de quantités
conservées qui commutent entre elles;
on impose généralement à celles-ci une condition
de localité appropriée (sur des états asymptotiques composés
de particules éloignées les unes des autres, ces quantités 
se décomposent en somme de termes correspondant à chaque particule).

Ces lois de conservation ont pour
conséquence que la cinématique des théories intégrables
est extrêmement contrainte; c'est là qu'intervient
de manière cruciale le fait que l'on se place à $2$ dimensions.
On trouve les propriétés suivantes: a) Il y a {\it
conservation
du nombre de particules}, c'est-à-dire plus précisément que
le nombre de particules d'une masse donnée est fixe, les particules
de même masse étant interchangeables\foot{Pour
une théorie non-intégrable, la conservation du nombre
de particules n'a
en général lieu que dans la limite non-relativiste ($c\to\infty$),
où la production/annihilation de particules est
supprimée, et qui nous ramène à un problème de mécanique
quantique non-relativiste.}.
b) Il y a {\it diffusion factorisée} des particules (sans réflexion
possible pour des particules de masses différentes):
la diffusion d'un nombre quelconque de particules se factorise
comme produit de diffusions de $2$ particules; la condition
de cohérence de cette factorisation est l'équation dite
de Yang--Baxter, sur laquelle nous reviendrons en détail.
Il existe des solutions non-triviales des équations de Yang--Baxter,
c'est-à-dire correspondant à des matrices $S$ de diffusion
non-diagonale; nous utiliserons pour notre part les
solutions dites rationnelles et trigonométriques.

L'existence d'une infinité de quantités conservées qui commutent
entre elles s'avère n'être qu'une partie
de la symétrie des modèles intégrables; en étudiant l'algèbre
de Yang--Baxter associée aux équations du même nom,
il s'avère que l'objet algébrique qu'il convient
de considérer est ce qu'on appelle les groupes quantiques
Chaque solution des équations de Yang--Baxter est naturellement
reliée à un groupe quantique\foot{Ceci
n'est strictement vrai que pour les solutions
rationnelles/trigonométriques. Pour les solutions elliptiques,
il faut considérer un objet mathématique un peu plus général
(groupe quantique elliptique).}, qui est représenté par une algèbre
d'opérateurs sur le Hilbert de la théorie.

Arrivons-en au sujet central de cette thèse: l'Ansatz de Bethe
est une méthode développée originellement pour diagonaliser
le Hamiltonien de chaînes de spins intégrables, et dont on
verra qu'il peut également conduire, dans une limite
d'échelle appropriée, à une description complète
de théories quantiques des champs intégrables
(spectre, matrices $S$, etc).
L'``Ansatz'', dans la formulation originelle
de Bethe, consiste à supposer une certaine forme de la fonction
d'onde, afin de diagonaliser le Hamiltonien. Dans cette thèse,
nous ferons appel à une version plus abstraite de l'Ansatz
de Bethe, nommée Ansatz de Bethe algébrique.

En comparant les deux paragraphes précédents, on se rend compte,
que, pour que la procédure de diagonalisation de l'Ansatz de Bethe
ait un sens, il faut que l'algèbre de symétrie du modèle
ne soit pas trop grosse (sinon, il y aurait une trop grande
dégénérescence du spectre). Or, les groupes quantiques associés
aux solutions de Yang--Baxter qui nous intéressent (solutions
avec paramètres spectraux) sont des groupes quantiques affines,
c'est-à-dire intuitivement des symétries de dimension infinie.
Ceci nous amène à l'une des ambiguités fondamentales de l'Ansatz de
Bethe: pour résoudre un modèle intégrable, l'Ansatz de Bethe
nous force à briser l'énorme symétrie sous-jacente! En pratique,
on diagonalise le Hamiltonien sur un espace compactifié de longueur
finie, alors que les groupes quantiques (affines)
ne sont typiquement des symétries que sur un espace
non-compactifié\foot{Notons que cette approche est strictement
non-historique, puisqu'on a d'abord r\'esolu des mod\`eles avec
des conditions p\'eriodiques aux bords (o\`u l'alg\`ebre de
Yang--Baxter suffit \`a assurer la solubilit\'e), puis on s'est
rendu compte que la limite thermodynamique faisait appara\^\i tre
une sym\'etrie plus grande $\ldots$}.
La symétrie complète ne réapparaît que dans
la limite de taille infinie de l'espace. On verra que l'on
peut tout de même restaurer un ``sous-groupe de dimension finie''
de la symétrie complète sur un espace de taille finie, à condition
de prendre des conditions aux bords particulières: celui-ci
joue alors le rôle habituel de groupe de symétrie globale (comme
$SU(2)$ dans le modèle de spins de Heisenberg) du modèle.
On verra également que cette subtilité rend difficile la comparaison
entre l'Ansatz de Bethe et la Théorie Conforme (deux domaines
dans lesquels la notion de groupe quantique joue un rôle-clé), du fait
qu'en théorie conforme on suppose généralement l'espace compactifié
(quantification radiale).

Une fois émise cette réserve, on verra que l'on peut tout de même
extraire de l'Ansatz de Bethe l'essentiel de la physique
des modèles intégrables connus. 
Nous reprendrons tout d'abord l'exemple historique
de la chaîne de spins XXX (chapitre \chapbase),
et nous verrons comment une légère
généralisation de ses équations d'Ansatz de Bethe décrit
également d'autres théories reliées. Nous introduirons sur
ce cas particulier l'une des notions-clé de cette thèse, celle
d'Equations d'Ansatz de Bethe physiques. Le contenu
du chapitre \chapbase\ est très classique, et nous
adopterons donc une démarche synthétique et moderne (mais
qui reste suivable pas-à-pas) pour le parcourir.

Le chapitre \chapstrucbae\ contient l'essentiel de la technique
nécessaire pour les applications de l'Ansatz de Bethe que l'on a en
vue. Tout d'abord, une brève introduction aux groupes quantiques, qui,
volontairement, ne rentre pas plus dans les détails mathématiques
qu'il n'est jugé utile, permet de replacer le cas particulier
étudié au chapitre \chapbase\ dans un contexte plus général;
ceci permet alors d'en étudier un certain nombre de généralisations,
et d'analyser les Equations d'Ansatz de Bethe associées.
Bien qu'une partie du chapitre \chapstrucbae\ soit constituée de
résultats bien connus, la manière dont ils sont présentés et les
commentaires qui les accompagnent me sont propres. En particulier,
les sections \secphyssup, \secfusmix\ et \secbaephysgen\ contiennent
des observations nouvelles. Egalement, nous nous appuierons de
plus en plus, au fur et à mesure des généralisations, sur les
diagrammes qui représentent les Equations d'Ansatz de Bethe:
bon nombre de phénomènes physiques se ``voient'' sur ces diagrammes.

Nous rentrerons dans le vif du sujet avec le chapitre \chaptba.
Celui-ci présente les Equations d'Ansatz de Bethe Thermodynamique,
qui gouvernent les systèmes à température finie, puis les
Equations Non-Linéaires Intégrales qui sont reliées aux systèmes
sur un espace de taille finie. Nous
tenterons en particulier de comprendre en détail la signification
des résultats de [\artic04], obtenus grâce aux Equations
Non-Linéaires Intégrales du modèle de Toda
affine en constante de couplage imaginaire.

Enfin, nous appliquerons les techniques des chapitres \chapstrucbae\
et \chaptba\ à l'étude du modèle $\sigma$ principal $SU(N)$ et de
sa généralisation, le modèle de Wess--Zumino--Witten $SU(N)$.
Nous obtiendrons pour ce dernier un certain nombre
de résultats non publiés, en particulier une description explicite
des excitations de basse énergie. Nous nous intéresserons
alors à un modèle intimement lié au modèle de Wess--Zumino--Witten
qu'est le modèle de Kondo. En plus des résultats physiques
nouveaux obtenus dans [\artic03], nous tenterons
d'élaborer une vision intuitive de la physique de basse énergie
du modèle de Kondo dans ses différents régimes, et ce en particulier
grâce aux Equations d'Ansatz de Bethe Physiques.\goodbreak

%
%
%
%
%
\rem{
encore quelques signes random dans les matrices $S$?

eventuellement a la fin faire un replace bourrin des ``:''

penser a remettre (POUR MA THESE SEULEMENT)
DDV dans l'abstract de artic04
}

\newchap\chapbase{L'Ansatz de Bethe: la chaîne XXX}
L'Ansatz de Bethe est une m\'ethode qui permet la diagonalisation explicite
du Hamiltonien de théories intégrables.
Les états
propres dépendent de nombres complexes appelés les paramètres spectraux;
ceux-ci vérifient un système d'équations algébriques couplées, les
Equations d'Ansatz de Bethe (BAE).
Dans une limite thermodynamique à définir,
ces équations se réduisent à des équations linéaires intégrales
pour les densités de ses racines.

Les paragraphes qui suivent rappellent brièvement
la solution de la chaîne de spins XXX (\secXXX)
et de modèles reliés (\secGN-\secTin) par l'Ansatz de Bethe 
(\secdiag-\secfond); puis les équations continues qui apparaissent dans
la limite thermodynamique (section \secbaeco). Nous
expliquerons en particulier la distinction entre Equations
d'Ansatz de Bethe Nues et Equations d'Ansatz de Bethe Physiques
(\secphysiso).

De manière générale,
nous insisterons plus sur la théorie des champs continue que
l'on obtient dans une limite d'échelle relativiste que sur
les modèles de départ (de type ``chaîne de spins''); à plusieurs
endroits, tout au long de cette thèse, nous supposerons que
nous avons affaire à une théorie des champs relativiste pour
faire un calcul, bien que ce ne soit souvent pas indispensable.

\newsec\secaba{L'Ansatz de Bethe Algébrique}
Nous allons maintenant passer en revue plusieurs modèles dont
la résolution passe par l'introduction d'un opérateur que nous
nommerons matrice de transfert, et dont la forme la plus générale sera
considérée au \secTin.
On procèdera alors à la diagonalisation de cette matrice de transfert,
grâce à l'{\it Ansatz de Bethe Algébrique}; celui-ci nous
conduira aux {\it Equations d'Ansatz de Bethe}.

\subsec\secXXX{La chaîne de spins $1/2$ XXX}
Considérons une chaîne de spins quantiques $1/2$ placés sur $M$
sites, avec des conditions de bord périodiques.
Soit $V={\Bbb C}^2$ la représentation de spin $1/2$ de $SU(2)$, et $V_k=V$
l'espace du \eme{k} spin; le Hamiltonien
$\H_\X$ de la chaîne de spin XXX, qui agit donc sur l'espace de Hilbert
${\cal H}=\bigotimes_{k=1}^M V_k=V^{\otimes M}$, est
$$\H_\X=2\sum_{k=1}^M {\vec\s}_k\cdot {\vec\s}_{k+1}\eqn\hamxxx$$
où les composantes  $\s_k^A$ ($A=1,2,3$) de ${\vec\s}_k$
sont les générateurs (hermitiens) de ${\goth su}(2)$,
agissant sur le \eme{k} site. On a la condition de
périodicité ${\vec\s}_{k+M}\equiv {\vec\s}_k$.

Le Hamiltonien $\H_\X$ possède une invariance évidente $SU(2)$:
$[\H_\X,{\vec\S}]=0$ où ${\vec\S}=\sum_{k=1}^M {\vec\s}_k$.
L'interaction entre spins est antiferromagnétique, donc même le vide
de la théorie n'est pas trivial.

Pour diagonaliser $\H_\X$, on introduit la
matrice $R$ agissant dans l'espace $V\otimes V$:
$$R(\lambda)={\lambda -i{\cal P}\over \lambda-i}\eqn\defR$$
où $\cal P$ permute les deux facteurs du produit tensoriel
$V\otimes V$, et $\lambda$ est un nombre complexe (le {\it
paramètre spectral}). 
Nous représenterons $R$ par une intersection (figure \R).
\fig\R{La matrice $R(\lambda)$. Le paramètre $\lambda$ doit être 
considéré comme la différence des paramètres spectraux des deux lignes
qui se croisent: $R(\lambda)\equiv 
R_{12}(\lambda_1-\lambda_2$).}{\figdir{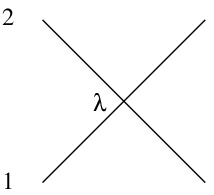}} 
On utilisera une notation courante dans les modèles intégrables,
qui consiste à placer en indice d'un opérateur
les espaces sur lequel il agit; ainsi, on notera $R_{12}$
pour indiquer que la matrice $R$ agit sur l'espace $V_1\otimes V_2$,
où $V_1$ et $V_2$ sont deux espaces quelconques isomorphes à $V$.
Sur un dessin, le premier indice correspond à la ligne qui est en-dessous
de l'autre à gauche de l'intersection.

Avec la convention de normalisation choisie\foot{Pour
son application à la chaîne XXX,
la normalisation de la matrice $R$
n'a aucune importance et l'on pourrait la multiplier par une fonction
(scalaire) arbitraire de $\lambda$. Pour le modèle NJL (cf section \secGN),
l'unitarité
est indispensable, sans qu'elle détermine pour autant la normalisation de
manière unique -- pour ce faire, on a besoin d'une autre condition
que l'on verra plus loin: la {\it symétrie de croisement}, qui donne
la ``bonne'' normalisation de $R$.}, $R$ vérifie la condition
dite {\it d'unitarité}, qui peut être représentée par
la figure \figunit.
\fig\figunit{La condition d'unitarité.}{\figdir{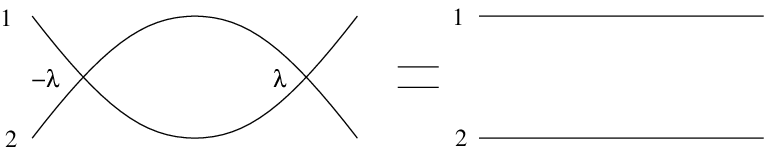}} 
On voit d'emblée un premier avantage de la notation indiciée: en effet,
on peut écrire grâce à elle la condition d'unitarité:
$$R_{12}(\lambda) R_{21}(-\lambda)=1\eqn\unit$$
alors qu'avec des notations
plus explicites il faudrait écrire:
$R(\lambda){\cal P}R(-\lambda){\cal P} =1$\foot{Comme
$R$ commute avec $\cal P$, on voit que l'on pourrait en fait
écrire plus simplement $R(\lambda)R(-\lambda)=1$. Mais
cette dernière propriété est moins générale.}.
Pour $\lambda$ réel, $R_{21}(-\lambda)=R_{12}(\lambda)^\dagger$ et
donc $R(\lambda)$ est unitaire.

$R$ possède aussi une propriété d'invariance $SU(2)$ en ce sens
que $R$ commute avec l'action de $SU(2)$ sur $V\otimes V$.
En termes de projecteurs $P_1$ et $P_0$
sur les sous-représentations irréductibles (de spins $1$ et $0$)
de $SU(2)$ dans $V\otimes V$, on a 
$R(\lambda)=P_1 + {\lambda+i\over \lambda-i} P_0$.

De plus, la matrice $R$ vérifie la célèbre équation de Yang--Baxter 
[\ref\YBEref{J.B.~Mc~Guire, {\it J.\ Math.\ Phys.} 5 (1964), 622\semi
E.~Brézin et J.~Zinn-Justin, {\it C.\ R.\ Acad.\ Sci.} 263 (1966), 670\semi
C.N.~Yang, {\it Phys.\ Rev.\ Lett.} 19 (1967), 1312\semi
R.J.~Baxter, {\it Ann.\ Phys.} 70 (1972), 192.}]:
$$R_{12}(\lambda_1-\lambda_2) 
R_{13}(\lambda_1-\lambda_3)
R_{23}(\lambda_2-\lambda_3)=
R_{23}(\lambda_2-\lambda_3)
R_{13}(\lambda_1-\lambda_3)
R_{12}(\lambda_1-\lambda_2)\eqn\YBE$$
que l'on peut représenter graphiquement par la figure \YBEfig.
\fig\YBEfig{L'équation de Yang--Baxter. On peut faire franchir
à une ligne l'intersection de deux autres.}{\figdir{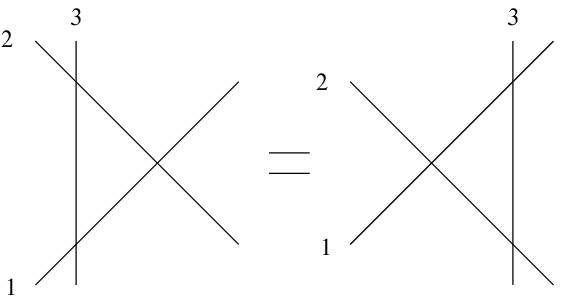}} 
Les matrices $R$
agissent dans \YBE\ sur deux de trois espaces $V_1$, $V_2$, $V_3$ 
quelconques isomorphes à $V$, les indices indiquant lesquels.
Notons que si on lit les dessins de gauche à droite (et de haut en bas
pour les lignes verticales), alors on obtient les expressions
correspondantes de droite à gauche.
Ce choix est conventionnel, et peut varier selon les auteurs.

Le fait que la matrice $R$ v\'erifie l'\'equation de Yang--Baxter est
essentiel,
car c'est cette relation qui assure qu'il existe une alg\`ebre
de Yang--Baxter sous-jacente [\ref\DVEM{H.J.~De~Vega, H.~Eichenherr
et J.M.~Maillet, {\it Comm. Math. Phys.} 92 (1984), 507; {\it
Nucl. Phys.} B240 (1984), 377.}], et donc au final, l'int\'egrabilit\'e.

Introduisons maintenant un espace auxiliaire $V_a=V$ en plus de nos espaces physiques de spins $V_i$,
et considérons la {\it matrice de monodromie} $T_a(\lambda)$
agissant sur $V_a\otimes {\cal H}$:~\foot{Pour
suivre rigoureusement la notation qui consiste à placer en indice
l'espace sur lequel les opérateurs agissent, il faudrait noter:
$T_{a{\cal H}}(\lambda)$, mais nous supprimerons systématiquement les
indices $\cal H$.}
$$T_a(\lambda)=R_{aM}(\lambda)R_{aM-1}(\lambda)\ldots
R_{a1}(\lambda)\eqn\defT$$
Les matrices $R_{ai}$ agissent sur le produit tensoriel de l'espace
auxiliaire $V_a$ et d'un espace physique $V_i$ (figure \T).
Rappelons que $a$ n'est pas un indice (qui pourrait prendre des valeurs),
mais bien le label désignant l'espace auxiliaire $V_a$.
\fig\T{La matrice de monodromie $T_a(\lambda)$.}{\figdir{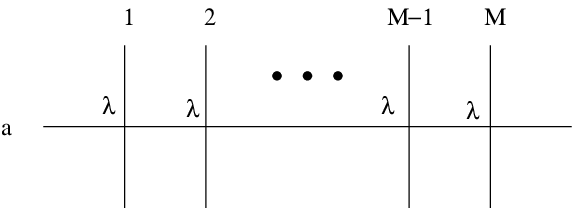}} 

En considérant {\it deux} espaces auxiliaires
$V_a$ et $V_b$, donc deux matrices de monodromie $T_a$ et $T_b$, et
en appliquant à répétition \YBE, on obtient les ``relations $RTT$''
(figure \RTTfig):
$$R_{ab}(\lambda-\mu)T_a(\lambda)T_b(\mu)=T_b(\mu)T_a(\lambda)R_{ab}(\lambda-\mu)\eqn\RTT$$
\fig\RTTfig{Les relations $RTT$.}{\figdir{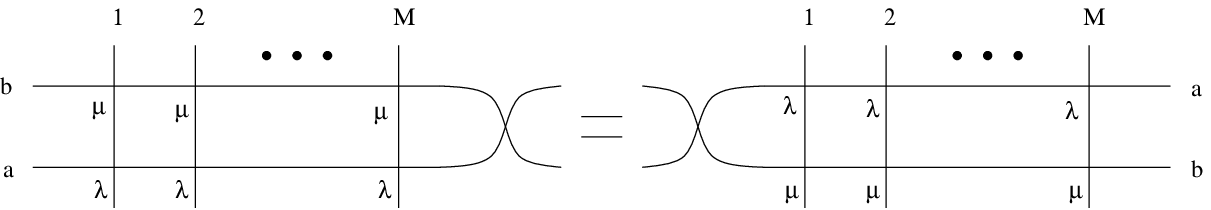}} 
On a pris dans \YBE\ l'un des $3$ paramètres spectraux égaux à zéro; 
il aurait pu être non nul si nous l'avions incorporé dans la définition \defT\ de
la matrice de monodromie (en fait, on peut même plus généralement considérer
une matrice de monodromie {\it inhomogène}, ce que nous ferons 
au \secTin).
Si l'on avait laissé le troisième paramètre spectral, l'équation \RTT\ serait
exactement de la même forme que \YBE; nous reviendrons sur ce point
lorsqu'il sera question de fusion (section \secfus).

Introduisons finalement {\it la matrice de transfert} $\Z(\lambda)$
$$\Z(\lambda)=\tr_a(T_a(\lambda))\eqn\defZ$$
qui est un opérateur sur $\cal H$ obtenu en prenant la trace de $T_a(\lambda)$
sur l'espace auxiliaire $V_a$.

Les relations $RTT$ \RTT\ impliquent, $R$ étant inversible, que les matrices
$\Z(\lambda)$ commutent entre elles
pour des valeurs différentes du paramètre spectral:
$$[\Z(\lambda),\Z(\mu)]=0\eqn\comZ$$
Elles peuvent donc être
diagonalisées simultanément. Comme $(\lambda+i)^M \Z(\lambda)$ 
est un polynôme de degré
$M$ en $\lambda$, on obtient ainsi $M$ quantités indépendantes qui commutent
(cf la définition classique d'intégrabilité). Calculons les premières en
développant $\Z(\lambda)$ autour de $\lambda=0$; on trouve que
$$\Z(\lambda=0)={\cal P}_{aM} {\cal P}_{aM-1}\ldots {\cal P}_{a1}\eqn\ZzP$$
est l'opérateur de translation discrète sur la chaîne périodique
(figure \Tzero).
\fig\Tzero{La matrice de transfert $\Z(\lambda=0)$. C'est
en fait la matrice de monodromie qui est représentée,
car pour prendre
la trace sur l'espace auxiliaire, il faudrait refermer le trait
horizontal sur lui-même.}{\figdir{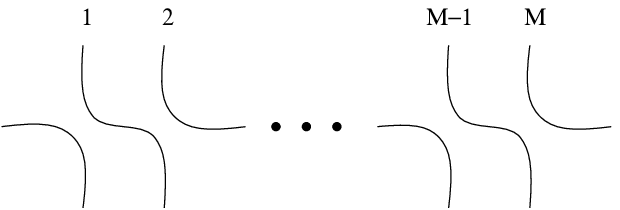}} 

On a donc $\Z(\lambda=0)=\e{-i\P a}$, où $\P$ est l'opérateur d'impulsion,
et $a$ est la distance entre $2$ sites
voisins. On peut aussi considérer l'impulsion par site:
$p=\P/M$ de sorte que $\Z(\lambda=0)=\e{-ipL}$ où $L=Ma$ est la longueur de
la chaîne.

Ensuite, on développe au premier ordre $\Z(\lambda)$
(figure \Tzerob) et on utilise la relation ${\cal P}_{k\,k+1}=
1/2+2\,{\vec\s}_k \cdot {\vec\s}_{k+1}$; on obtient
\fig\Tzerob{Un terme dans le d\'eveloppement au premier ordre
de $\Z(\lambda)$ en $\lambda=0$ (multiplié
par $\Z(0)^{-1}$).}{\figdir{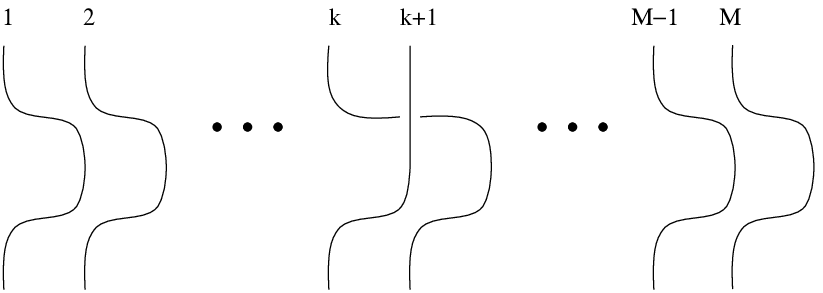}} 
$${1\over i} \Z(\lambda)^{-1} {\d\over\d\lambda}
\Z(\lambda)_{|\lambda=0}=-M/2+\H_\X\eqn\Zham$$
donc $\H_\X$ est l'un des $M$ Hamiltoniens commutant entre eux.
On a ainsi ramené la diagonalisation de $\H_\X$ à celle de $\Z(\lambda)$.

\subsec\secGN{Le modèle de Nambu--Jona-Lasinio (NJL) $SU(2)$}
Considérons à présent un autre modèle, le modèle de
Nambu--Jona-Lasinio [\ref\NaJL{Y.~Nambu et G.~Jona-Lasinio,
{\it Phys. Rev.} 122 (1961), 345.}] (parfois aussi nommé modèle
de Gross--Neveu chiral) à deux dimensions.
C'est une théorie des champs relativiste (une dimension
d'espace, une dimension de temps);
le Lagrangien est exprimé en termes de fermions de Dirac (il y a donc
$2$ chiralités) qui sont des doublets
de $U(2)$ et qui interagissent via une interaction à quatre fermions:
$$L=i\psib^a \dsl \psi^a - g {\vec\jmath}_\mu \cdot {\vec\jmath}^\mu\eqn\lagGN$$
$a=1,2$ est l'indice de doublets de $U(2)$ et
$\mu=0,1$ est l'indice d'espace-temps.
${\vec\jmath}^\mu$ est le courant $SU(2)$;
explicitement, $j^{\mu A}=\psib_a \S^A_{ab}\gamma^\mu \psi_b$,
où les $\S^A$ sont les générateurs (hermitiens) de ${\goth su}(2)$ 
dans la représentation doublet des fermions. Dans \lagGN, il y
a sommation sur les indices (tensoriels) répétés, comme
partout dorénavant.

Pour assurer la renormalisabilité du modèle, il faudrait ajouter
également un terme courant-courant $U(1)$, mais comme il va s'avérer
que le secteur $U(1)$ de la théorie est trivial et 
se découple, nous omettons ce terme.

Le modèle NJL est intéressant car il exhibe un certain nombre de
phénomènes physiques universels, tels que la génération dynamique de
masse, la liberté asymptotique [\ref\GN{D.~Gross et A.~Neveu,
{\it Phys. Rev.} D10 (1974), 3235.}], et d'autres plus spécifiques
à la dimension $2$, comme l'apparition d'un secteur
non-massif malgré l'impossibilité d'avoir des bosons de Goldstone,
du fait que la symétrie chirale ne peut pas être brisée.
Tout ceci apparaît dans sa solution par l'Ansatz de Bethe
[\ref\AL{N.~Andrei et J.H.~Lowenstein,
{\it Phys. Rev. Lett.} 43 (1979), 1698.}].
Précisons maintenant quelles sont les symétries du Lagrangien:

\item{$\diamond$} invariance sous $SU(2)$. En fait, on a bien sûr
une symétrie $U(2)$, mais, comme on l'a déjà signalé,
le secteur $U(1)$ de la théorie ne nous intéresse pas,
donc on ne considère qu'une symétrie
$SU(2)$\foot{Notons
que $U(2)$ n'est que localement isomorphe à $SU(2)\times U(1)$:
plus précisément $U(2)=(SU(2)\times U(1))/{\Bbb Z}_2$, donc
on ne peut pas ``complètement'' découpler $U(1)$ et $SU(2)$.
Nous reviendrons sur cette subtilité au chapitre 3.}.
L'invariance séparée des deux chiralités sous l'action de $SU(2)$
(symétrie $SU(2)_+\times SU(2)_-$) est, elle, brisée par l'interaction.

\item{$\diamond$} symétrie chirale $U(1)$\foot{Cette symétrie -- dont
le courant développe une anomalie, mais qui n'est pas brisée -- interdit
aux fermions d'avoir une masse:
nous verrons qu'il y a effectivement un secteur non-massif de la
théorie (le secteur $U(1)$), mais il est trivial et se découple
du secteur en interaction (le secteur $SU(2)$), qui, lui, échappe
à l'interdiction et possède une masse.},
donc conservation du nombre de fermions gauches et droits.
Cette dernière propriété est caractéristique d'une théorie
dont la diffusion est factorisée (cf introduction); elle
permet de se placer d'un point de vue dit de ``première
quantification'', c'est-à-dire de considérer le Hamiltonien
à nombre fixé de particules.

Fixons donc le nombre de particules $M$ du système. Il est important
de souligner que ces particules ``nues'', c'est-à-dire les fermions qui
apparaissent dans le Lagrangien, {\it ne sont pas} les excitations
physiques du modèle. En effet, le vrai vide de la théorie contient
un condensat de fermions, dont les excitations physiques
sont des excitations collectives de spin. 
Nous extrairons ces dernières
par une procédure de {\it limite d'échelle} dans laquelle nous
enverrons la densité linéique de fermions $M/L$
à l'infini tout en maintenant
une échelle de masse fixée, et recouvrerons ainsi
l'invariance relativiste de la théorie.

Commençons par considérer
le cas $M=2$. Si les deux particules sont de chiralités différentes,
le Hamiltonien s'écrit:
$$\H=i(\der_1-\der_2)+4g\,\delta(x_1-x_2) {\cal P}\eqn\hamGN$$
où $x_1$ et $x_2$ sont les positions des particules gauche et droite,
et $\cal P$ permute les deux particules.
Cherchons à résoudre l'équation de Schrödinger associée $\H|\Phi\rangle
=E|\Phi\rangle$, avec 
$$|\Phi\rangle=\int \d x_1\, \d x_2\,
\Phi^{ab}(x_1,x_2) \psi^a(x_1) \psi^b(x_2)|0\rangle$$
Dans chacune des régions $x_1<x_2$ et $x_1>x_2$, on peut choisir
$|\Phi\rangle$ sous forme d'onde plane:
$$\Phi^{ab}(x_1,x_2)= \e{i(p_1 x_1 + p_2x_2)}
\left\{ \eqalign{&\phi^{ab}_1\quad x_1<x_2\cr &\phi^{ab}_2\quad
x_1>x_2\cr}\right.
\eqn\schr$$
de façon à satisfaire l'équation de Schrödinger avec $E=p_2-p_1$.
Les composantes de spin de la fonction d'onde $\phi_1$ et $\phi_2$
appartiennent, pour reprendre les notations du \secXXX, à
$V\otimes V$.
L'Ansatz \schr\ sur la forme de la
fonction d'onde contient implicitement le fait
que l'énergie/impulsion est conservée séparément pour chaque
particule lors de la diffusion (ce qui est caractéristique
d'une théorie intégrable).

Il faut ensuite assurer le ``recollement'' des deux solutions à $x_1=x_2$;
on vérifie que si $\phi_2=S_{-+}\,\phi_1$,
où la matrice $S$ ``nue'' $S_{-+}$ entre particules gauches
et droites agit sur $V\otimes V$ et vaut
$$S_{-+}={1+ig{\cal P}\over 1-ig{\cal P}}\eqn\bareS$$
alors $|\Phi\rangle$ est vecteur propre de $\H$.

Si les deux particules sont de même chiralité, il n'y a pas
d'interaction entre elles, et on pourrait être tenté de prendre
les ondes planes comme base des fonctions propres de $\H$. Cependant, il
y a une grande dégénérescence dans le spectre, du fait que $E=\pm(p_1+p_2)$
ne dépend que de $p_1+p_2$ et pas de chacune des impulsions
séparément. On peut tirer parti de ce fait pour prendre
les vecteurs propres du même type que \schr,
avec une matrice $S_{++}$ ou $S_{--}$ reliant $\phi_1$ et $\phi_2$
{\it a priori} quelconque.
Chaque choix de matrice $S$ constitue un choix de base de fonctions
propres. Gardons pour l'instant $S_{++}$ et $S_{--}$ arbitraires;
nous les fixerons ultérieurement.

Considérons maintenant le système avec un nombre supérieur de
particules $M$; la \eme{i} particule est de chiralité $\epsilon_i$
($+$ pour les particules droites, $-$ pour les particules gauches).
On peut opérer de la même manière 
que pour $M=2$ et séparer l'espace des
positions $(x_1,\ldots,x_M)$ de nos particules en régions du type
$x_{\sigma(1)}<x_{\sigma(2)}<\ldots<x_{\sigma(N)}$, où $\sigma$
est une permutation
quelconque; dans chacune de ces régions, on a, avec des notations
évidentes
$$\Phi^{a_1\ldots a_M}(x_1,\ldots,x_M)=\phi^{a_1\ldots a_M}_\sigma \e{i(p_1
x_1+\cdots+p_M x_M)}\eqn\schrb$$
et l'énergie totale vaut $E=\epsilon_1 p_1+\cdots+\epsilon_M p_M$.

A nouveau, à la frontière entre deux domaines, soit quand $x_i=x_j$,
on doit utiliser la matrice $S_{ij}\equiv S_{\epsilon_i\epsilon_j}$
correspondant aux chiralités des
particules $i$ et $j$ et agissant sur les \eme{i} et \eme{j} indices
de $\phi^{a_1\ldots a_M}_\sigma$. Nous utilisons ici la factorisation
de la diffusion, en termes de diffusions successives entre $2$ particules
(lire à ce sujet [\ref\ZZb{A.B.~Zamolodchikov
et Al.B.~Zamolodchikov, {\it Annals Phys.} 120 (1979), 253.}]).

Evidemment, cette procédure n'est cohérente que si la
définition de la fonction d'onde ne dépend pas du chemin choisi
lors des franchissements de frontières successifs pour aller d'une
région à une autre; on peut vérifier que ceci impose pour
seule condition,
en dehors des relations triviales $S_{ij}S_{ji}=1$ (soit explicitement
$S_{+-}{\cal P}S_{+-}{\cal P}=S_{++}{\cal P}S_{++}{\cal P}
=S_{--}{\cal P}S_{--}{\cal P}=1$),
l'équation de Yang--Baxter pour les matrices $S$:
$$S_{ij} S_{ik} S_{jk}=S_{jk} S_{ik} S_{ij}\eqn\YBES$$
Nous allons maintenant choisir $S_{++}=S_{--}$ de telle façon que
\YBES\ soit satisfaite. Pour ce faire, on observe que si
$S_{ij}=R(\theta_i-\theta_j)$, où $R$ est la matrice qui 
a été définie dans \defR,
pour des paramètres $\theta_i$ bien choisis, alors d'après \YBE, \YBES\
sera automatiquement satisfaite (et $S_{ij}S_{ji}=1$ n'est autre
que la condition d'unitarité de $R(\lambda)$).

Prenons donc $\theta_i=\epsilon_i/c$, où $c$ est une nouvelle
paramétrisation de la constante de couplage: $c=4g/(1-g^2)$.
On vérifie que $S_{-+}$, donnée par \bareS, vaut bien $R(-2/c)$\foot{Ceci
n'est en fait vrai qu'à une phase près; mais le lecteur pourra vérifier
que la phase supplémentaire n'a qu'un effet trivial,
et qu'on peut donc l'omettre.}.
De plus, ce choix impose une matrice $S$ non-triviale:
$S_{++}=S_{--}={\cal P}$ entre particules de même chiralité.

Maintenant que l'on a trouvé les matrices $S$ nues, on utilise une
procédure standard pour parvenir à la matrice de transfert (et donc
aux équations d'Ansatz de Bethe): celle du déphasage {\it (phase shift)}.
Elle consiste à considérer un espace compactifié
de longueur $L$, ce qui revient à imposer des conditions au bord
périodiques à la fonction d'onde $\Phi^{a_1\ldots
a_M}(x_1,\ldots,x_M)$. Plaçons-nous par exemple dans la région
$x_1<x_2<\ldots<x_M$, choisissons un certain $x_k$
et augmentons-le de manière à lui faire faire
le tour des autres $x_i$; on voit facilement que la condition
$x_k\equiv x_k+L$ se récrit:
$$\e{-i p_k L}= S_{k\,k-1}\ldots S_{k1} S_{kM}\ldots S_{k\,k+1}\eqn\phsh$$
\phsh\ doit être interprétée comme une équation aux valeurs propres
de l'opérateur du second membre 
$\Z_k=S_{k\,k-1}\ldots S_{k1} S_{kM} S_{k\,k+1}$,
le vecteur propre correspondant étant 
$\phi^{a_1\ldots a_M}_1$. Cette équation affirme que le déphasage
de la \eme{k} particule -- tenant compte à la fois du déphasage
libre $\e{i p_k L}$ et de la diffusion sur les autres particules --
après un tour complet de l'espace compactifié vaut $1$.
La diagonalisation de $\Z_k$ permet donc de calculer $p_k$, et de là,
le spectre du Hamiltonien.

Finalement, on introduit comme au \secXXX\ un espace auxiliaire
$V_a\equiv V$, et la matrice de transfert $\Z(\lambda)$ définie
par:
$$\Z(\lambda)=\tr_a(R_{aM}(\lambda-\theta_M)R_{aM-1}(\lambda
-\theta_{M-1})\ldots
R_{a1}(\lambda-\theta_1))\eqn\defTb$$
(où, rappelons-le, $\theta_k=\pm 1/c$).
On a alors la relation (figure \Tcba):
$$\Z_k=S_{k\,k-1}\ldots S_{k1} S_{kM}\ldots
S_{k\,k+1}=\Z(\lambda=\theta_k)\eqn\ZkZ$$
\fig\Tcba{La matrice de transfert $\Z(\lambda=\theta_k)$. La ligne
horizontale doit se refermer sur elle-m\^eme.}{\figdir{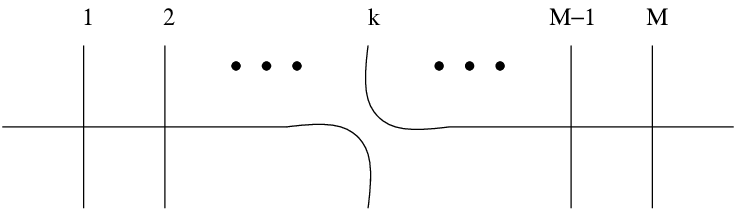}} 
Cette identification nous permet d'affirmer que les $\Z_k$ commutent
entre eux, ce qui est une condition de cohérence de la construction
que nous avons effectuée (sinon il serait impossible de les
diagonaliser simultanément, cf eq. \phsh); et que nous avons ramené
la diagonalisation du Hamiltonien à celle de la matrice de transfert
$\Z(\lambda)$ définie par \defTb.

\subsec\secTin{Modèles statistiques sur réseau: modèles de vertex}
Les ``matrices de transfert'' que nous avons introduites pour les deux
modèles précédents sont très similaires: nous allons présenter
maintenant un formalisme unificateur, qui justifiera en même temps la
dénomination de matrice de transfert.

Considérons un modèle défini sur un réseau rectangulaire 2D.
A chaque ligne verticale (respectivement horizontale),
on associe un paramètre spectral $\theta_i$ (resp. $\theta'_i$).
Les poids statistiques
sont assignés aux vertex du réseau: chaque arête peut être dans $2$ états,
souvent représentés par des flèches (figure \slm)
et le poids de Boltzmann
associé au vertex dépend des valeurs des arêtes qui en partent.
\fig\slm{Mod\`ele de vertex sur r\'eseau.
Si on suppose le réseau doublement périodique,
les arêtes sortantes doivent être identifiées deux
à deux.}{\figdir{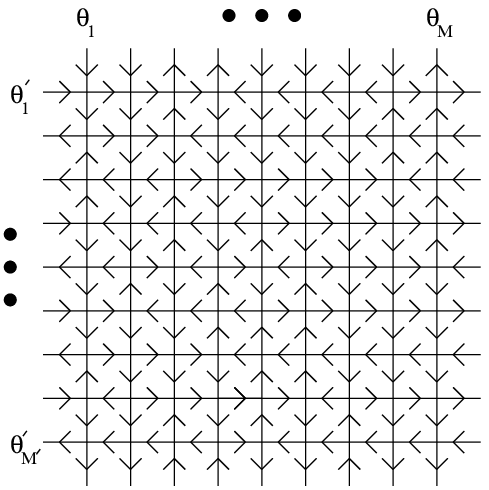}} 
Il y a dans notre cas $2^4=16$ configurations d'arêtes possibles
autour d'un vertex donné, mais seules
$6$ ont des poids non-nuls (notre modèle est un cas particulier du {\it modèle
à 6 vertex}). Pour faire le lien avec la matrice $R$, nous allons
placer les $16$ poids (nuls ou non-nuls) sur une matrice $4\times 4$:
la ligne correspond à la configuration des arêtes au-dessus et à gauche 
du vertex (dans l'ordre: $\rightarrow\downarrow$, $\rightarrow\uparrow$,
$\leftarrow\downarrow$, $\leftarrow\uparrow$), la colonne correspond
à celles au-dessous et à droite (dans le même ordre).
Alors, le poids correspondant au vertex à l'intersection de la 
\eme{j} ligne horizontale et de la 
\eme{k} ligne verticale est par définition
la matrice $R(\theta'_j-\theta_k)$ dans la base standard de $V={\Bbb C}^2$.
Le moment est venu d'écrire explicitement la matrice $R(\lambda)$:
$$R(\lambda)=\pmatrix{
1&0&0&0\cr
0&{\lambda\over\lambda-i}&{-i\over\lambda-i}&0\cr
0&{-i\over\lambda-i}&{\lambda\over\lambda-i}&0\cr
0&0&0&1\cr
}\eqn\explR$$
On vérifie qu'il y a bien $6$ poids non-nuls. Cette matrice $R$
est dite {\it rationnelle}, du fait de sa dépendance en $\lambda$.
Evidemment, pour que les poids soient réels, il faut que les
$\theta_j$ soient purement imaginaires: notre paramétrisation
de la matrice $R$ est adaptée à un espace-temps à 1+1 dimensions,
mais pour passer à la mécanique statistique à 2+0 dimensions, il
faut effectuer la ``rotation de Wick'' $\lambda\rightarrow i\lambda$.

Pour sommer sur les différentes
configurations du système, il est naturel d'introduire les matrices de
monodromie
$$T(\theta'_j|\theta_1,\ldots,\theta_M)=
R_{aM}(\theta'_j-\theta_M)R_{aM-1}(\theta'_j-\theta_{M-1})\ldots R_{a1}
(\theta'_j-\theta_1)\eqn\defTin$$
agissant comme au \secXXX\ sur le produit tensoriel de l'espace de Hilbert
``physique''
${\cal H}=\otimes_{k=1}^M V_k=V^{\otimes M}$, qui est ici associé aux valeurs
des arêtes verticales (plus précisément $V_k$ est associé à la 
\eme{k} ligne verticale), et d'un ``espace auxiliaire'' qui s'identifie ici
aux valeurs des arêtes horizontales de la \eme{j} ligne.
Ce sont les matrices de monodromie {\it inhomogènes} (les
inhomogénéités sont les $\theta_1,\ldots,\theta_M$).

Considérons maintenant
la fonction de partition du modèle sur un réseau doublement périodique, 
de périodes
$M$ et $M'$; pour tenir compte de la périodicité horizontale, on
doit prendre la trace sur le ``\eme{j} espace auxiliaire'', c'est-à-dire
la \eme{j} ligne horizontale:
$$\Z(\theta'_j|\theta_1,\ldots,\theta_M)=\tr_a(
T_a(\theta'_j|\theta_1,\ldots,\theta_M))
\eqn\defZin$$
On voit ainsi apparaître $\Z_j\equiv
\Z(\theta'_j|\theta_1,\ldots,\theta_M)$, les
matrices de transfert (inhomog\`enes). Elles méritent ce nom, puisque
la fonction de partition $Z$ s'écrit alors
$$Z=\tr_{\cal H} (\Z_{M'} \Z_{M'-1}
\ldots \Z_1)\eqn\ZZin$$
Dans la limite $M'\to\infty$, $Z$ est dominée
par les valeurs propres les plus grandes des $\Z_j$, et on est donc
amené à les diagonaliser. En vertu de l'équation de Yang--Baxter,
les relations $RTT$ \RTT\ sont toujours
valables pour les matrices de monodromie inhomogènes, et ainsi 
les $\Z_j$ commutent: on peut donc les diagonaliser simultanément.

Il est clair
que les matrices de transfert introduites aux paragraphes précédents
ne sont que des cas particuliers de la matrice de transfert
$\Z(\theta'|\theta_1,\ldots,\theta_M)$, pour des inhomogénéités $\theta_k$
particulières: $\theta_k=0$ pour la chaîne XXX, $\theta_k=\pm 1/c$
pour le modèle NJL.

\subsec\secdiag{Diagonalisation de la matrice de transfert inhomogène}
Nous allons maintenant diagonaliser la matrice de transfert 
$\Z(\lambda|\theta_1,\ldots,\theta_M)$, qui
est la trace sur l'espace auxiliaire de la matrice de monodromie 
$T_a(\lambda|\theta_1,\ldots,\theta_M)$ (figure \Tin).
\fig\Tin{La matrice de monodromie inhomogène.}{\figdir{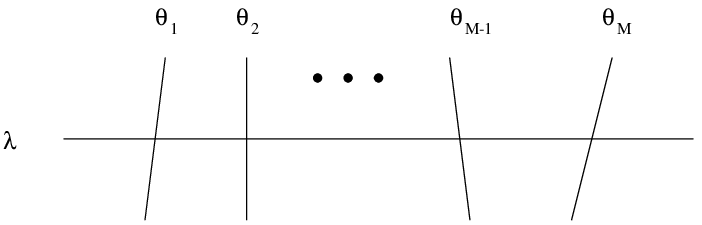}} 
Nous représentons les lignes verticales quelque peu inclinées (contrairement
à la figure \slm), pour signifier qu'elles ont des paramètres
spectraux associés différents (on verra en effet
que les paramètres spectraux s'interprètent comme des rapidités
dans le Minkowskien, ou des angles dans l'Euclidien).

La construction qui suit {\it (Ansatz de Bethe Algébrique)} 
est particulièrement classique [\ref\STF{L.D.~Faddeev,
E.K.~Sklyanin et L.A.~Takhtajan, {\it Theor. Math. Phys.} 40 (1979),
194\semi
L.D.~Faddeev, {\it Integrable models in $1+1$ dimensional quantum
field theory}, dans {\it Recent advances in field theory and
statistical mechanics}, Les Houches 1982 (Editeurs: J.B.~Zuber et
R.~Stora), North-Holland (1984).}]
et nous ne ferons donc que la décrire brièvement.

Rappelons que $SU(2)$ agit naturellement sur le Hilbert physique $\cal H$ et
sur l'espace auxiliaire $V_a$.
Si l'on décompose la matrice de monodromie $T_a$ sur la base usuelle
de l'espace auxiliaire $V_a$ (base de diagonalisation
de $s_3$, composante $z$ du spin), on obtient
$$T_a(\lambda|\theta_1,\ldots,\theta_M)=
\pmatrix{\A(\lambda)&\B(\lambda)\cr
\C(\lambda)&\D(\lambda)\cr}\eqn\ABCD$$
où les $\A(\lambda)$, $\B(\lambda)$, $\C(\lambda)$, $\D(\lambda)$ 
(ces notations sont standard) sont
des opérateurs sur le Hilbert $\cal H$. Ainsi
$\Z(\lambda)=\A(\lambda)+\D(\lambda)$.
Du fait de l'invariance $SU(2)$ de la matrice $R$, $\B(\lambda)$
diminue la composante $\S^3$ du spin total (dans $\cal H$) de $1$, tandis que
$\C(\lambda)$ l'augmente de $1$, et $\A(\lambda)$ et $\D(\lambda)$
commutent avec elle.

Commençons donc avec l'état de plus haut poids $|\Omega\rangle$ de $\cal H$
(vis-à-vis de la symétrie $SU(2)$); dans la base standard de $V$, on a
$$|\Omega\rangle = \left( 1\atop 0 \right) 
\otimes\left( 1\atop 0 \right) \otimes \cdots \otimes
\left( 1\atop 0 \right)\eqn\php$$
On vérifie aisément que
$$\eqalign{
\A(\lambda)|\Omega\rangle&=|\Omega\rangle\cr
\D(\lambda)|\Omega\rangle&=\prod_{k=1}^M {\lambda - \theta_k \over
\lambda-\theta_k-i} |\Omega\rangle}\eqn\ADOm$$
donc $|\Omega\rangle$ est un état propre de $\Z(\lambda)$. 

L'idée est alors d'utiliser $\B(\lambda)$ comme opérateur de création
d'une pseudo-particule (une excitation de spin
du système, parfois nommée {\it onde de spin}) au-dessus
du pseudo-vide que constitue $|\Omega\rangle$. Ainsi, on définit l'état
$|\Psi\rangle$ (figure \ABA)
$$|\Psi\rangle = \B(\lambda_1) \ldots \B(\lambda_m) |\Omega\rangle\eqn\BBBOm$$
où les $\lambda_\alpha$ sont des nombres complexes, qui doivent
être tels que $|\Psi\rangle$ soit un état propre de
$\A(\lambda)$ et $\D(\lambda)$. 
\fig\ABA{Construction de 
l'état $|\Psi\rangle$. Les flèches entourées de cercles
symbolisent les états: $\uparrow\equiv \left({1\atop 0}\right)$,
$\downarrow\equiv \left({0\atop 1}\right)$.}{\figdir{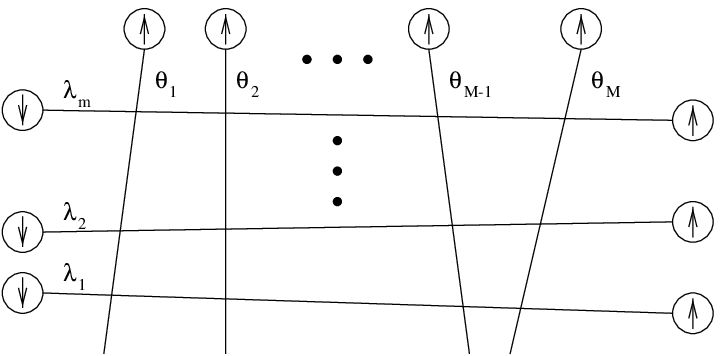}} 

Comme la matrice de monodromie $T(\lambda)$
vérifie les relations $RTT$ \RTT, on dispose de relations de commutation
pour ses composantes $\A(\lambda)$, $\B(\lambda)$, $\D(\lambda)$, qui
nous permettent de calculer l'action de $\A(\lambda)$ et $\D(\lambda)$
sur $|\Psi\rangle$. A la suite d'un calcul classique, on trouve que
$|\Psi\rangle$ est vecteur propre si et seulement si les $\lambda_\alpha$
vérifient le système d'équations algébriques suivant:
$$\prod_{\scriptstyle \beta=1\atop\scriptstyle \beta\ne \alpha}^m 
{\lambda_\alpha-\lambda_\beta+i\over \lambda_\alpha-\lambda_\beta-i}
=\prod_{k=1}^M {\lambda_\alpha-\theta_k\over \lambda_\alpha-\theta_k-i}\eqn\xxxbae$$
Ce sont les {\it Equations d'Ansatz de Bethe} (BAE).
La valeur propre correspondante est alors donnée par:
$$\Z(\lambda)|\Psi\rangle=\left[
\prod_{\alpha=1}^m {\lambda-\lambda_\alpha+i\over\lambda-\lambda_\alpha}
+\prod_{k=1}^M {\lambda-\theta_k\over\lambda-\theta_k-i}
\prod_{\alpha=1}^m {\lambda-\lambda_\alpha-i\over\lambda-\lambda_\alpha}
\right]|\Psi\rangle
\eqn\xxxbaeb$$

Nous admettrons également les résultats suivants: 
\item{$\bullet$} Les états $|\Psi\rangle$ ainsi construits (``états de Bethe'')
sont des {\it états de plus haut poids} vis-à-vis de $SU(2)$;
donc leur spin $\S^3=s=M/2-m$.
En fait, quand $\lambda\to\infty$, on a $\B(\lambda)\sim {1\over\lambda}
\S^-$, où $\S^-=\S^1-i\S^2$ est l'opérateur de descente.
Admettons alors que les $\lambda_\alpha$
puissent prendre la valeur $\infty$. On remarque que
si $(\lambda_1,\ldots,\lambda_m)$ est une solution des BAE \xxxbae,
alors $(\lambda_1,\ldots,\lambda_m,\infty,\ldots,\infty)$ en est une
autre, avec la même valeur propre \xxxbaeb: on peut insérer un nombre 
arbitraire de racines infinies, 
ce qui est bien sûr une manifestation de la symétrie $SU(2)$,
et donc atteindre ainsi les états de poids inférieur dans les multiplets.
\item{$\bullet$} En considérant comme cela est
expliqué ci-dessus les multiplets
associés aux vecteurs de plus haut poids $|\Psi\rangle$, on obtient
une base d'états propres de $\Z(\lambda)$. 
\item{$\bullet$} Les $\lambda_\alpha$
sont nécessairement tous distincts (principe d'exclusion pour les
excitations de spin).

Ceci conclut la diagonalisation de la matrice de transfert.

Pour interpréter les BAE \xxxbae-\xxxbaeb, on remarque
qu'aussi bien au \secXXX\ (chaîne XXX) qu'au \secGN\
(modèle NJL) on ne considère que la valeur particulière $\lambda=
\theta_k$ du paramètre spectral. Pour cette
valeur $\D(\lambda)|\Psi\rangle=0$, et, décalant
les $\lambda_\alpha$ de $i/2$, les
équations \xxxbae-\xxxbaeb\ se récrivent:
$$\eqnalign\xxxbaec{
\prod_{\scriptstyle \beta=1\atop\scriptstyle \beta\ne\alpha}^m 
{\lambda_\alpha-\lambda_\beta+i\over \lambda_\alpha-\lambda_\beta-i}
&=\prod_{k=1}^M
{\lambda_\alpha-\theta_k+i/2\over \lambda_\alpha-\theta_k-i/2} &\xxxbaec a\cr
\e{-ip_kL}&=
\prod_{\alpha=1}^m {\theta_k-\lambda_\alpha+i/2\over\theta_k-\lambda_\alpha-i/2}
&\xxxbaec b\cr
}$$
où, conformément aux notations du \secGN,
$\e{-ip_kL}$ est la valeur propre de $\Z(\lambda=\theta_k)$.

On remarque alors
que les équations \xxxbaec{} ont exactement la
même forme que les équations de déphasage \phsh:
en effet, on peut considérer que \xxxbaec a décrit l'interaction
de nos particules initiales de paramètres spectraux $\theta_k$
avec de nouvelles particules de paramètres spectraux
$\lambda_\alpha$, et que \xxxbaec b décrit
l'interaction de ces nouvelles particules entre elles. Ces
dernières sont elles-mêmes sujettes à la diffusion factorisée,
puisque seules des fonctions de $\lambda_\alpha-\lambda_\beta$ apparaissent
dans \xxxbaec b.
La différence essentielle entre \phsh\ et \xxxbaec{} tient 
au fait que dans \xxxbaec{}, la diffusion
est {\it diagonale}: la matrice de monodromie correspondante
n'a pas de structure matricielle. On peut encore dire que les
pseudo-particules soumises aux équations \xxxbaec{} n'ont pas de nombres
quantiques internes.
Ainsi, l'Ansatz de Bethe algébrique ramène un problème de déphasage
non diagonal à un problème de déphasage diagonal. 

%
Terminons l'analyse des BAE par une autre remarque importante:
les $\theta_k$, qui sont les paramètres spectraux des particules
``nues'' de notre modèle, sont fixes et ne dépendent pas des
impulsions $p_k$ de ces mêmes particules. De manière plus générale
(comme cela apparaît dans les modèles de Hubbard ou de Kondo
multi-canal, cf section \seckondo, mais aussi dans les ``Equations
d'Ansatz de Bethe physiques'' que nous introduirons plus loin), 
les $\theta_k$ peuvent être fonctions
des $p_k$. L'interprétation intuitive de l'absence
de dépendance des $\theta_k$ en les $p_k$ dans le modèle NJL
est le {\it découplage spin/charge}:
les valeurs des $p_k$, qui représentent le secteur de charge $U(1)$,
n'influent plus sur les valeurs des rapidités $\lambda_\alpha$,
qui représentent le secteur de spin $SU(2)$. 
A l'inverse, les $\lambda_\alpha$ ne déterminent les $p_k$ que modulo
$2\pi/L$ (eq. \xxxbaec b): on peut ajouter
à $p_k$ un terme $2\pi n_k/L$, et le choix des $n_k$
correspond au choix de l'état dans le secteur de charge.
Ce dernier
étant totalement trivial (fermion de Dirac libre découplé), nous
supposerons à l'avenir que nous sommes dans l'état fondamental du
secteur de charge $U(1)$ (ce qui correspond à un choix particulier des $n_k$),
et porterons notre attention sur le secteur $SU(2)$.

\subsec\secfond{L'état fondamental et les premiers états excités}
Pour simplifier la discussion, nous allons à présent nous concentrer
sur la chaîne XXX et le modèle NJL (et laisser
de côté, bien qu'ils puissent être traités de la même
manière, les modèles de vertex).
De plus, nous supposerons que le nombre de sites/particules $M$ est
pair, et, pour NJL, que l'état que nous considérons est
de chiralité totale nulle, i.e.\ qu'il y a autant de particules
gauches que droites: $M_+=M_-=M/2$ 
(en effet, le vide et les états de basse énergie du secteur de spin
vérifient cette condition). Nous reviendrons plus loin sur
l'hypothèse de parité. Enfin, pour le restant de cette section,
nous garderons la distance entre deux sites/inverse de la densité 
linéique de fermions fixée (selon le modèle),
de sorte que nous pouvons nous placer dans des unités
telles que: $a=1$, soit encore $M=L$.

Il s'agit désormais de chercher l'état fondamental, c'est-à-dire de 
minimiser l'énergie $E$. L'impulsion vaut $P=\sum_k p_k$ pour les deux modèles
considérés, où les $p_k$ sont donnés par \xxxbaec{}:
$$\eqalign{
P_\X&=-\sum_{\alpha=1}^m [2\arctan(2\lambda_\alpha)+\pi]\cr
P_\NJL&=-\sum_{\alpha=1}^m
\left[\arctan(2(\lambda_\alpha-1/c))+\arctan(2(\lambda_\alpha+1/c))+\pi\right]
\cr}$$
(modulo $2\pi$).
Par contre, l'énergie n'a pas la même expression; pour la chaîne XXX,
il faut dériver l'impulsion par rapport au paramètre spectral
(cf eq. \Zham), donc
$$E_\X=-\sum_{\alpha=1}^m {1\over \lambda_\alpha^2+1/4} \eqn\EXXX$$
Pour le modèle NJL, avec des notations
évidentes, on a: $E_\NJL=M/2\, (p_+ - p_-)$, donc
$$E_\NJL= \sum_{\alpha=1}^m
\left[\arctan(2(\lambda_\alpha+1/c))-\arctan(2(\lambda_\alpha-1/c))\right]
\eqn\EGN$$
Le choix de modulo pour $E_\NJL$ équivaut
à n'écrire que la contribution à l'énergie du secteur de spin.

La caractéristique commune aux deux expressions de
l'énergie \EXXX\ et \EGN\ est que la contribution
de chaque pseudo-particule de
rapidité $\lambda_\alpha$ est toujours négative. C'est d'ailleurs
la raison pour laquelle on les appelle pseudo-particules. Le vide (ou état
fondamental) n'est donc pas le pseudo-vide $|\Omega\rangle$ introduit
au \secdiag: il est obtenu au contraire en remplissant la ``mer 
de Fermi'' de pseudo-particules (qui, rappelons-le, vérifient
un principe d'exclusion).

La limite thermodynamique est obtenue en prenant $M\to\infty$.
Pour l'état du vide,
on peut montrer que les $\lambda_\alpha$ solutions des BAE \xxxbaec a
forment une densité continue $\rho_\gs(\lambda)$ de racines sur l'axe réel.
Pour éviter les redondances avec
les sections qui suivent, nous n'allons pas faire de calcul explicite ici.
La plupart des résultats que nous allons citer dans les lignes
qui suivent seront
redémontrés plus loin par des méthodes générales.

Les états excités sont obtenus en créant des {\it trous}
dans la mer de Fermi de racines (nous donnerons
plus loin une définition rigoureuse d'un trou):
si l'on place un trou au paramètre spectral
$\tilde{\lambda}$, on peut dire que l'on a créé une
excitation physique dont l'énergie-impulsion est paramétrisée par 
$\tilde{\lambda}$ (le tilde sert à distinguer toutes les quantités
associées aux trous).
Pour $\tilde{m}$ trous placés aux paramètres spectraux
$\tilde{\lambda}_1,\ldots,
\tilde{\lambda}_{\tilde{m}}$, on peut montrer que l'énergie-impulsion s'écrit
sous la forme:
$$\eqalign{
E&=E_\gs+\sum_{\alpha=1}^{\tilde{m}} \tilde{\epsilon}(\tilde{\lambda}_\alpha)\cr
P&=\sum_{\alpha=1}^{\tilde{m}} \tilde{p}(\tilde{\lambda}_\alpha)\cr
}\eqn\EPphys$$
où les fonctions $\tilde{\epsilon}(\lambda)$ et $\tilde{p}(\lambda)$ dépendent
du modèle considéré:
$$\left\{\eqalign{
\tilde{\epsilon}_\X(\lambda)&={\pi\over\cosh\pi\lambda}\cr
\tilde{p}_\X(\lambda)&=2\arctan(\e{\pi\lambda})\cr
}\right.\quad
\left\{\eqalign{
\tilde{\epsilon}_\NJL(\lambda)&=\arctan(\e{\pi(\lambda-1/c)})
+\arctan(\e{\pi(-\lambda-1/c)})\cr
\tilde{p}_\NJL(\lambda)&=\arctan(\e{\pi(\lambda-1/c)})
-\arctan(\e{\pi(-\lambda-1/c)})\cr
}\right.
\eqn\EPphysb$$
(notons que $\tilde{\epsilon}_\X={\d\over\d\lambda} \tilde{p}_\X$).
L'état à $\tilde{m}$ trous constitue donc
sans surprise
un état à $\tilde{m}$ particules physiques, de rapidités
$\tilde{\lambda}_1,\ldots,\tilde{\lambda}_{\tilde{m}}$.
\EPphysb\ nous donne également la {\it relation de dispersion} de
la chaîne XXX:
$\tilde{\epsilon}=\pi\sin\tilde{p}$. En particulier, il n'y a pas
de gap.
Nous n'écrivons rien d'analogue pour le modèle NJL,
puisque nous discuterons sa relation de dispersion
après avoir pris la limite d'échelle (section \secbaeco).
Remarquons simplement qu'au contraire, le modèle NJL possède
un gap ($\tilde{\epsilon}$ est minimum en $\lambda=0$).

Considérons maintenant le spin d'un état:
on peut montrer qu'il est simplement donné par $s=\tilde{m}/2$.
Les excitations physiques élémentaires sont donc de spin $1/2$. 
On les appelle des {\it spinons}.
L'état formé de $\tilde{m}$ trous est l'état à $\tilde{m}$ trous
``le plus symétrique'', c'est-à-dire de plus haut poids dans
la représentation 
produit tensoriel de $\tilde{m}$ représentations de spin $1/2$.
Cependant, comme cette représentation est 
réductible pour $\tilde{m}>1$, le multiplet du vecteur de plus haut poids
ne décrit pas tout l'espace des états à $\tilde{m}$ particules.

Prenons un exemple concret, celui de $\tilde{m}=2$, soit
de deux particules de rapidités $\tilde{\lambda}_1$ et $\tilde{\lambda}_2$.
On a l'état triplet ($s=1$) qui est le plus
symétrique, et qui est l'état à deux trous,
mais aussi l'état singlet ($s=0$). Comment obtenir
ce dernier? Il faut se rappeler pour cela que les racines des
BAE \xxxbaec{} ne sont pas nécessairement réelles: ici, pour obtenir
l'état singlet à partir de l'état triplet, on ajoute une {\it
$2$-corde} centrée en $(\tilde{\lambda}_1+\tilde{\lambda}_2)/2$, 
c'est-à-dire deux racines complexes conjuguées
$(\tilde{\lambda}_1+\tilde{\lambda}_2)/2\pm i/2$.
On peut montrer que, quand $L\to\infty$,
la contribution de la $2$-corde à l'énergie est nulle, de sorte
que $E=E_\gs+\tilde{\epsilon}(\lambda_1)+\tilde{\epsilon}(\lambda_2)$
(et de même pour l'impulsion).
Les états triplet et singlet sont donc dégénérés: c'est le signe
qu'à $L=\infty$ apparaît une symétrie plus grande que $SU(2)$. Nous
reviendrons sur ce point au \secphysisoq. 
En étudiant les corrections en $1/L$ à l'énergie/impulsion des
premiers états excités, on peut obtenir la matrice $S$ de diffusion
des excitations physiques\foot{La méthode en question
ne donne pas la matrice $S$ correcte pour des modèles plus
généraux: c'est pourquoi nous exposerons plus loin une autre méthode,
celle des Equations d'Ansatz de Bethe physiques (voir section \secphysiso),
qui est applicable à tous les modèles.}. On trouve que la matrice
de diffusion entre deux particules dont la différence des rapidités
vaut $\lambda$, est donnée par:
$$S(\lambda)=
{ \Gamma\left(1+i{\lambda\over 2}\right)
  \Gamma\left({1\over 2}-i{\lambda\over 2}\right)
\over
  \Gamma\left(1-i{\lambda\over 2}\right)
  \Gamma\left({1\over 2}+i{\lambda\over 2}\right)}
R(\lambda)\eqn\Sphys$$
où $R(\lambda)$ est la matrice $R$
habituelle (eq. \defR). La matrice $S$ est donc
une autre version de la matrice $R$, avec un préfacteur
scalaire supplémentaire; elle vérifie comme $R$ l'équation
de Yang--Baxter, l'unitarité, mais vérifie de surcroît
la {\it symétrie de croisement}: si $S(\lambda)=f(\lambda)+
g(\lambda){\cal P}$=$\hat{f}(\lambda)+\hat{g}(\lambda)E$ 
avec $E_{ij}^{kl}=\varepsilon_{ij} \varepsilon^{kl}$,
alors $f(i\pi-\lambda)=\hat{f}(\lambda)$ et 
$g(i\pi-\lambda)=\hat{g}(\lambda)$.

Il convient de remarquer que puisqu'on a supposé $M$ pair, le spin
du système est nécessairement entier et donc $\tilde{m}$ est pair: on ne
peut créer les particules élémentaires que par paires\foot{C'est
d'ailleurs la raison pour laquelle $\tilde{p}$ est défini
modulo $\pi$ et non $2\pi$.}.
En particulier, ceci pourrait donner l'illusion qu'on a un spectre
d'excitations de spin $1$, ce qui est une interprétation erronée.
On peut résoudre cette difficulté en considérant
``simultanément'' $M$ pair et $M$ impair dans la limite
$M\to\infty$, de fa\c con à construire un espace de Fock complet.

Après cette brève description des états de Bethe (vide,
états excités), nous allons passer à la limite thermodynamique.

\newsec\secbaeco{BAE continues}
On a vu au \secfond\ que le vide de la chaîne XXX/modèle NJL
est constitué d'une densité continue de racines des BAE. Nous allons
maintenant généraliser cette idée à des états thermodynamiques
quelconques. Par ``états thermodynamiques'', il faut entendre états
dans lesquels le nombre d'excitations physiques\foot{Cette quantité
est bien définie dans une théorie intégrable, car
on peut nécessairement parler d'états asymptotiques.}
est grand: la densité linéique de ces excitations
reste fixée dans la limite $L\to\infty$.
Il convient de distinguer ces états thermodynamiques des ``états
excités'' que sont les premiers états propres du Hamiltonien
au-dessus du vide, et qui contiennent un nombre fini d'excitations
physiques. Ce sont de tels états que nous avons considéré
au \secfond. Pour ces derniers, les équations que nous allons écrire
à présent ne sont pas valables; nous reviendrons à eux lorsque
nous parlerons d'Equation Non-Linéaire Intégrale (section \secddv).
Nous traiterons d'abord en détail (\secsimpbae\ et
\secdualsimp) un modèle simplifié qui met en
lumière les mécanismes essentiels de la limite continue des BAE;
puis nous indiquerons comment les généraliser aux BAE de
XXX/NJL, ce qui nécessitera l'introduction de l'hypothèse de corde.
Une des idées-clé que nous développerons
alors sera l'introduction de nouvelles équations,
les Equations d'Ansatz de Bethe physiques, qui s'obtiennent
directement du spectre des excitations physiques, et qui sont
équivalentes aux Equations d'Ansatz de Bethe nues, qui
sont celles que l'on a considérées jusqu'à présent.

\subsec\secsimpbae{Diffusion diagonale}
Pour montrer dans le cadre le plus simple comment écrire des équations
d'Ansatz de Bethe continues, nous allons
considérer un modèle de diffusion (factorisée) diagonale dans lequel
des particules d'un seul type diffusent entre elles
avec une certaine matrice $S$. L'énergie-impulsion $(\epsilon,p)$
est exprimée en termes d'un
paramètre spectral $\lambda$ {\it réel} (la {\it rapidité}),
de telle manière que la matrice $S$
de la diffusion de deux particules de rapidités
$\lambda_1$ et $\lambda_2$ ne dépende que de $\lambda_1-\lambda_2$.
Par exemple, 
dans le cas de particules relativistes de masse $m$ non nulle,
la paramétrisation de la couche de masse en termes de la rapidité
$\lambda$ s'écrit $(\epsilon,p)=(m \cosh\lambda,m\sinh\lambda)$, et
$\lambda_1-\lambda_2$ est l'unique invariant relativiste de la
diffusion (dans la paramétrisation traditionnelle,
$s=4m^2 \cosh^2{\lambda_1-\lambda_2\over 2}$).

On peut utiliser ici le même procédé de ``déphasage''
qui nous a conduit précédemment à écrire l'équation \phsh\ 
pour le modèle NJL. On considère un état à $m$ particules
de rapidités $\lambda_1,\ldots,\lambda_m$; on suppose que les
particules vérifient un principe d'exclusion, de sorte que les
$\lambda_\alpha$ sont tous distincts. On impose des conditions
p\'eriodiques au bord \`a la fonction d'onde\foot{Evidemment,
la notion de fonction d'onde n'a pas grand sens ici, \`a moins que les
interactions ne soient ponctuelles. Dans le cas g\'en\'eral,
il faut supposer que $mL\gg 1$, o\`u $1/m$ est l'\'echelle de longueur
du syst\`eme.}.
L'équation de déphasage pour une particule s'écrit:
$$\e{-i p(\lambda_\alpha)L}
=\prod_{\scriptstyle\beta=1\atop\scriptstyle\beta\ne\alpha}^m 
S(\lambda_\alpha-\lambda_\beta)
\eqn\phshdiag$$

Conformément aux remarques faites au \secdiag,
les équations \phshdiag\ sont du même type que \xxxbaec a 
(pour des $\theta_k$ fixés jouant le rôle du terme
d'impulsion dans \phshdiag), mais les $\lambda_\alpha$
sont par définition réels dans \phshdiag.

Comme la diffusion est diagonale, la matrice $S(\lambda)$ est une
simple phase: $S(\lambda)=\exp(-i\phi(\lambda))$. Nous supposerons
pour simplifier que $\phi(0)=0$.
Introduisons alors la {\it fonction de comptage} $I(\lambda)$:
$$I(\lambda)={1\over 2\pi} \bigg[
p(\lambda)L+\sum_{\beta=1}^m \phi(\lambda-\lambda_\beta) \bigg]
\eqn\defI$$
On voit que \phshdiag\ se récrit simplement: $\exp(2\pi i
I(\lambda_\alpha))=1$, soit encore $I(\lambda_\alpha)$ entier pour
tout $\alpha$. Avec des hypothèses appropriées sur $p(\lambda)$ et
$\phi(\lambda)$, la fonction $I(\lambda)$ est strictement croissante;
alors, comme les $\lambda_\alpha$ sont des réels distincts, les
$I(\lambda_\alpha)$ sont des entiers distincts.

Les $\lambda_\alpha$ n'épuisent pas nécessairement toutes les valeurs
entières de $I(\lambda)$; pour compléter, on introduit les {\it trous}
$\tilde{\lambda}_\alpha$, qui sont les réels distincts entre eux
et distincts des $\lambda_\alpha$, et tels que
$I(\tilde{\lambda}_\alpha)$ soit entier.

On considère maintenant
les équations \phshdiag\ dans la limite thermodynamique
$L\to\infty$, en nous
concentrant sur les états dits thermodynamiques.
Pour saisir intuitivement ce qui se passe dans
cette limite, et identifier lesdits états,
supposons un instant que $\phi=0$: on a alors affaire
à un système de fermions libres, et les impulsions
autorisées $p=2\pi I/L$ (avec $I$ entier) décrivent,
quand $L\to\infty$, un continuum
d'états accessibles aux particules du système. Certains de ces états sont
effectivement occupés: ce sont les $p(\lambda_\alpha)=2\pi
I(\lambda_\alpha)/L$; d'autres ne le sont pas: ce sont les
$p(\tilde{\lambda}_\alpha)=2\pi I(\tilde{\lambda}_\alpha)/L$.
Remarquons la dualité complète trou-particule 
qui apparaît dans cette description;
cette dualité est brisée par l'ajout de l'interaction ($\phi\ne 0$),
puisque ce sont les particules qui interagissent entre elles, et non
les trous (on verra cependant au paragraphe suivant comment rétablir
cette dualité en la généralisant).

Pour décrire un état donné, on introduit très classiquement des densités
$\rho(p)$ et $\rhot(p)$ de particules
et de trous tels que 
$$\rho(p)+\rhot(p)={1\over 2\pi}\eqn\baefree$$
Les états
thermodynamiques sont par définition les états pour lesquels on
peut définir de telles fonctions $\rho(p)$ et $\rhot(p)$ bien
régulières dans la limite $L\to\infty$.

Reprenons maintenant le cas général $\phi\ne 0$, ainsi que
notre paramétrisation $p=p(\lambda)$ qui est la plus naturelle
dans le cas en interaction. On définit les densités $\rho(\lambda)$
et $\rhot(\lambda)$ de la manière évidente: $\rho(\lambda)
d\lambda$ est le nombre de $\lambda_\alpha$ compris entre
$\lambda$ et $\lambda+d\lambda$ divisé par $L$, 
et de même pour $\rhot(\lambda)$.
La généralisation au cas en interaction de \baefree, qui
traduisait la quantification des impulsions, consiste à utiliser
la ``condition de quantification des impulsions généralisée'':
$I(\lambda_\alpha)$ entier, d'où l'on déduit immédiatement que,
quand $L\to\infty$,
$$\d I/\d\lambda=L(\rho(\lambda)+\rhot(\lambda))\eqn\prebaediag$$
On exprime maintenant $I(\lambda)$ en termes de $\rho(\lambda)$:
d'après \defI, $2\pi I(\lambda)/L=p(\lambda) + \int \d\mu\, \rho(\mu)
\phi(\lambda-\mu)$, et l'équation \prebaediag\ s'écrit explicitement:
$$2\pi(\rho(\lambda)+\rhot(\lambda))={\d p\over\d\lambda}+K\star\rho(\lambda)
\eqn\baediag$$
où l'on
a dénoté par $K\star\rho$ le produit de convolution de $K\equiv 
\d\phi/\d\lambda$ et de $\rho$. Remarquons que \baediag\ se réduit bien
à \baefree\ pour $K=0$ et $\rho(p) \d p=\rho(\lambda) \d\lambda$,
$\rhot(p) \d p=\rhot(\lambda) \d\lambda$.

L'équation \baediag\ est l'Equation d'Ansatz de Bethe continue
que nous recherchions. C'est une équation intégrale linéaire qui admet
une infinité de solutions (en effet, il y a deux inconnues,
$\rho(\lambda)$ et $\rhot(\lambda)$, pour une seule équation)
correspondant aux différents états thermodynamiques du système.

\subsec\secdualsimp{Dualité particule-trou}
Continuons notre étude du
modèle de diffusion diagonale introduit au paragraphe précédent,
caractérisé par un seul type de particules et une matrice $S$ de
diffusion. Le mécanisme que nous allons décrire est à la base
des dualités plus compliquées présentes dans les
modèles de diffusion non diagonale.

Nous avons montré que les BAE s'écrivaient dans la limite
thermodynamique pour le modèle de diffusion diagonale (Eq.\ \baediag):
$2\pi(\rho(\lambda)+\rhot(\lambda))=\d
p/\d\lambda+K\star\rho(\lambda)$,
où $K$ est la dérivée logarithmique de la matrice $S$. Comme nous
l'avons déjà signalé, cette équation n'est pas symétrique par échange
de $\rho$ et $\rhot$: en effet, le terme $K\star\rho$ n'a pas
d'équivalent $\tilde{K}\star\rhot$, ce qui revient à dire que ce sont
les particules qui interagissent, et non les trous.

Pour rétablir la dualité particule-trou, nous allons donc écrire
\baediag\ sous la forme
$$\rhot(\lambda)+A\star\rho(\lambda)={1\over 2\pi}{\d
p\over\d\lambda}\eqn\baediagb$$
où $A=1-K/(2\pi)$ est un noyau de Fredholm,
qui admet donc un inverse de la même forme:
$A^{-1}=1-\tilde{K}/(2\pi)$. Multiplions alors \baediagb\ par $A^{-1}$
et définissons $\tilde{p}=-A^{-1}\star p$; on obtient:
$$2\pi(\rho(\lambda)+\rhot(\lambda))=
-{\d\tilde{p}\over\d\lambda}+\tilde{K}\star\rhot(\lambda)\eqn\baediagdu$$
Comparant \baediag\ et \baediagdu, on voit que les rôles de $\rho$
et $\rhot$ ont été échangés (la seule différence tient
au signe de $\tilde{p}$, que nous avons choisi opposé à celui de $p$
pour des raisons qui vont apparaître clairement par la suite).

Avant d'interpréter intuitivement ces résultats, écrivons
aussi l'énergie/impulsion du système. Par définition de
$\rho(\lambda)$, on a:
$$\eqalign{
E/L=\int\d\lambda\, \rho(\lambda) \epsilon(\lambda)\cr
P/L=\int\d\lambda\, \rho(\lambda) p(\lambda)\cr
}\eqn\EPdiag$$
En utilisant \baediagb, on peut récrire $E$ et $P$ en fonction
de $\rhot$:
$$\eqalign{
E/L={\rm cst}+\int\d\lambda\, \rhot(\lambda) \tilde{\epsilon}(\lambda)\cr
P/L={\rm cst}+\int\d\lambda\, \rhot(\lambda) \tilde{p}(\lambda)\cr
}\eqn\EPdiagdu$$
où $\tilde{p}$ a déjà été défini précédemment, et
$\tilde\epsilon(\lambda)=-A^{-1}\star\epsilon(\lambda)$.

En général, $\epsilon(\lambda)$ et $\tilde\epsilon(\lambda)$ sont
de signes opposés: ainsi, si
les particules que nous avons considérées initialement
ont leur énergie positive,
alors créer un trou la diminue. Inversement,
supposons que les particules ont leur énergie négative (ce sont
alors des pseudo-particules, au même titre que les excitations
de spin que nous avons introduites pour la chaîne XXX):
le vrai vide de la théorie est alors caractérisé
par $\rhot(\lambda)=0$, et ce sont les trous qui tiennent lieu
d'excitations physiques. 

D'après \baediagdu, on peut alors
considérer que ce sont les trous qui interagissent
avec une matrice $\tilde{S}$ dont la dérivée logarithmique vaut
$\tilde{K}$: en effet, si l'on procédait à l'envers, partant
d'un système défini par la matrice de diffusion $\tilde{S}$ et
déterminant les BAE de ce système, on obtiendrait précisément
\baediagdu\ (avec les notations $\rho$ et $\rhot$ échangées).

Ainsi, partant de la matrice $S$ des pseudo-particules, on calcule
$K=i(\d/\d\lambda)\log S(\lambda)$, puis $\tilde{K}(\lambda)
=-A^{-1}\star K(\lambda)$, d'où finalement la vraie matrice de
diffusion $\tilde{S}$
de la théorie, définie (à une constante multiplicative près)
par $\tilde{K}=i(\d/\d\lambda)\log\tilde{S}(\lambda)$.

\subsec\secbaehypco{Chaîne XXX et hypothèse de corde}
Reprenons maintenant l'étude des BAE \xxxbaec{}, qui gouvernent en
particulier les modèles XXX et NJL. Afin d'appliquer
le procédé du paragraphe précédent 
à ces équations, il faut d'abord
se ramener à des paramètres spectraux réels: en effet,
les $\lambda_\alpha$ solutions de \xxxbaec{} sont {\it a priori}
complexes. Cependant il existe
un argument ``na\"\i f'' 
qui permet de justifier que dans la limite $M\to\infty$,
les $\lambda_\alpha$ ne
peuvent former que certaines configurations particuli\`eres
dans le plan complexe.

Supposons donc qu'une racine $\lambda_\alpha$ n'est pas r\'eelle,
par exemple que sa partie imaginaire est strictement positive.
Alors chaque facteur du second membre de l'\'equation \xxxbaec{a} est de module
strictement sup\'erieur \`a $1$, de sorte que le module du second membre
diverge exponentiellement avec $M$. Cette divergence doit donc \^etre
\'egalement pr\`esente dans le membre de gauche, de sorte qu'il doit
exister une autre racine $\lambda_\beta$ qui s'approche
exponentiellement vite dans la limite $M\to\infty$ d'un p\^ole:
$$\lambda_\beta=\lambda_\alpha-i+O(\e{-aM})$$

Si $\lambda_\beta$ est encore de partie imaginaire strictement
sup\'erieure \`a z\'ero, on peut recommencer le processus:
on obtient une nouvelle racine $\lambda_\gamma=\lambda_\alpha-2i$,
etc. Le processus s'arr\^ete quand la partie imaginaire
de la derni\`ere racine consid\'er\'ee devient n\'egative.
En effet, pour les racines $\lambda_\alpha$ de partie imaginaire
strictement n\'egative, il faut au contraire une autre racine
$\lambda_\beta=\lambda_\alpha+i$, ce dont on dispose par construction.

On peut de plus d\'emontrer que si $\lambda_\alpha$ est une racine
de \xxxbaec{}, alors $\bar{\lambda}_\alpha$ l'est aussi.
Ceci nous am\`ene naturellement
\`a l'{\it hypothèse de corde} [\ref\BET{H.~Bethe,
{\it Z. Phys.} 71 (1931), 205.}]: celle-ci stipule
que les solutions de \xxxbaec{} forment des ``cordes'', c'est-à-dire
des groupes composés de nombres complexes de même partie réelle et de
parties imaginaires équidistantes. On définit ainsi une
$j$-corde ($j\ge 1$) comme un ensemble de $j$ racines de la forme
$$\left\{ \lambda_{j;\alpha}+i\left({j+1\over 2}-k\right)
,\ 1\le k\le j\right\}\eqn\strhyp$$
où le {\it centre de la corde} $\lambda_{j;\alpha}$ est réel.
On peut imaginer les cordes comme des {\it états liés}
de pseudo-particules, du fait de la forme de \strhyp\
(bien sûr, les cordes ne sont pas plus des
excitations physiques que les pseudo-particules elles-mêmes). 
Ceci sugg\`ere qu'il existe une interpr\'etation plus profonde
de l'hypoth\`ese de corde, qui est li\'ee \`a la proc\'edure de
fusion, ce que nous verrons au \secfusb.

Pourquoi l'hypoth\`ese de corde n'est-elle qu'une hypoth\`ese,
c'est-\`a-dire pourquoi l'argument na\"\i f n'en est-il pas
une preuve? Le probl\`eme vient du fait que pour appliquer cet
argument, il faut supposer que le nombre de racines $m$ reste
fini quand $M\to\infty$, ce qui n'est le cas que pour des \'etats
non-physiques tr\`es loin du vrai vide de la th\'eorie. Si l'on
consid\`ere au contraire des \'etats excit\'es au-dessus
du vrai vide, on peut d\'emontrer [\ref\DL{C.~Destri
et J.H.~Lowenstein, {\it Nucl. Phys.} B205 (1982), 369.}] que
la divergence du membre de droite de \xxxbaec{a} quand $M\to\infty$
est exactement compens\'ee par la divergence du membre
de gauche due au fait que $m\sim M$, sans qu'il y ait besoin
d'avoir de cordes\foot{En simplifiant quelque peu.
Voir [\DL] et la g\'en\'eralisation XXZ [\ref\BDVV{O.~Babelon,
H.J.~De~Vega et C.M.~Viallet, {\it Nucl. Phys.} B220 (1983), 13.}]
pour plus de d\'etails.
Notons que le fait que l'hypoth\`ese de corde soit alors viol\'ee
s'interpr\`ete tout naturellement comme le fait que les
racines qui d\'efinissent ces \'etats excit\'es satisfont
aux BAE physiques, qui par d\'efinition restent finies
quand le nombre d'excitations physiques est fini,
m\^eme si $M$ a \'et\'e envoy\'e \`a l'infini (voir \secphysiso).}.
Lorsque nous reviendrons à ces états excités (section
\secddv), nous ne ferons donc pas usage de l'hypoth\`ese de corde.

Pour ce qui est des états
thermodynamiques, au sens défini précédemment,
il est raisonnable de supposer (bien que cela n'aie jamais
\'et\'e prouv\'e rigoureusement) que la compensation de la divergence n'est que
partielle, de sorte que l'hypoth\`ese de corde appliqu\'ee \`a ces
\'etats est valide.

Notons enfin que l'hypoth\`ese de corde a souvent \'et\'e utilis\'ee
en dehors de son domaine de validit\'e, car elle reste
``qualitativement'' vraie; ainsi,
on sait faire un comptage
des états correct en supposant l'hypothèse de corde, même pour
une chaîne finie [\ref\KIb{A.N.~Kirillov,
{\it Zap. Nauch. Semin. LOMI} 131 (1984), 88.}].

\subsec\secbaecoxxx{BAE continues de la cha\^\i ne XXX}
Admettons donc l'hypoth\`ese de corde telle qu'elle a \'et\'e
d\'efinie au paragraphe pr\'c\'edent.
On peut maintenant multiplier \xxxbaec a pour les différentes racines
d'une même corde et on obtient des équations pour les centres des
cordes; ainsi,
pour le centre $\lambda_{j;\alpha}$ (où $\alpha$ parcourt les
différentes $j$-cordes, $1\le\alpha\le m_j$) d'une $j$-corde, on obtient
$$\prod_{\scriptstyle j'\ge 1,\, 1\le \beta\le m_{j'}\atop\scriptstyle (j',\beta)\ne (j,\alpha)}
S_{jj'}(\lambda_{j;\alpha}-\lambda_{j';\beta})
=\prod_{k=1}^M
{\lambda_\alpha-\theta_k+i\,j/2\over \lambda_\alpha-\theta_k-i\,j/2}
\eqn\xxxbaed$$
où $S_{jj'}(\lambda)$ est
la matrice $S$ entre les $j$-cordes
et les $j'$-cordes: elle est obtenue par définition
en faisant le produit des
facteurs $(\lambda+i)/(\lambda-i)$ sur toutes les paires de racines
d'une corde et de l'autre. Le second membre est obtenu de la même
manière. On peut faire subir une transformation similaire à \xxxbaec b.

Les équations \xxxbaed\ ne font intervenir que les centres des cordes
qui sont par définition réels; de plus, elles
constituent manifestement une généralisation à plusieurs
types de particules de la situation du paragraphe précédent.
On peut donc introduire des densités
de $j$-cordes $\rho_j(\lambda)$ et de trous de $j$-cordes
$\rhot_j(\lambda)$, et opérer comme au \secsimpbae\foot{Notons que cette
procédure est explicitée dans un cadre à peine plus compliqué dans
[\artic03.1.2]; nous adopterons
ici des notations et conventions proches
de celles de cet article; 
la normalisation des rapidités, elle, diffère, la correspondance étant:
$\lambda=\Lambda/c$.}. On a: $\int \d\lambda\,\rho_j(\lambda)=m_j/L$,
où $m_j$ est le nombre de $j$-cordes, et de même on peut définir
$\tilde{m}_j$ par $\int \d\lambda\, \rhot_j(\lambda)=\tilde{m}_j/L$.
Dans la limite thermodynamique $L\to\infty$,
$m_j/L$ et $\tilde{m}_j/L$
restent fixés de l'ordre de la seule échelle d'énergie que l'on
ait: $M/L$. Nous serons plus tard amenés à considérer
deux échelles d'énergie divergentes l'une par rapport à l'autre,
mais pour l'instant, elles sont confondues.

Le système d'équations linéaires
intégrales couplées ainsi obtenu peut être écrit de manière
particulièrement esthétique si l'on introduit la
``matrice de Cartan avec paramètre spectral'' $C_{jk}(\lambda)$
définie par:
$$C_{jk}(\lambda)=\delta_{jk}\delta(\lambda)-(\delta_{jk-1}
+\delta_{jk+1})s(\lambda)\quad j,k\ge 1
\eqn\defC$$
où $s(\lambda)=1/2\cosh(\pi\lambda)$.
La matrice de Cartan agit par convolution sur
les fonctions de $\lambda$; nous noterons par $\star$ le produit
de convolution: $f\star g(\lambda)\equiv\int \d\mu\, f(\lambda-\mu)
g(\mu)$. Dans \defC, les indices vont de $1$ à $\infty$,
ce qui signifie que $C_{jk}$ est la matrice de Cartan du diagramme de Dynkin
$A_\infty$. La normalisation est telle que
$\int\d\lambda\, C_{jk}(\lambda)$ vaut {\it la moitié} de
la matrice de Cartan au sens usuel, c'est-à-dire sans
paramètre spectral.

Si l'on définit la transformée de Fourier d'une fonction
$\phi(\lambda)$ par 
$$\phi(\kappa)\equiv \int \phi(\lambda)
\exp(2i\kappa\lambda) \d\lambda$$
alors $s(\kappa)=1/(2\cosh(\kappa))$, et la matrice de Cartan inverse
$C^{-1}$ est donnée par
$$C^{-1}_{jk}(\kappa)=C^{-1}_{kj}(\kappa)=2 \coth(\kappa) \exp(-j|\kappa|)
\sinh(k\kappa)
\qquad j\ge k$$

Les équations d'Ansatz de Bethe peuvent alors
être mises sous la forme:
$$\rhot_j+ C^{-1}_{jk}\star\rho_k=\sigma\star K_j(\lambda)\eqn\xxxcoobae$$
c'est-à-dire que la dérivée logarithmique des
matrices $S$ entre les différents types de cordes s'exprime naturellement
en termes de la matrice de Cartan inverse $C^{-1}$.
On a introduit dans \xxxcoobae\ le noyau $\sigma(\lambda)
={1\over L}\sum_{k=1}^M \delta(\lambda-\theta_k)$,
et les fonctions
$K_j(\lambda)\equiv (1/2\pi) j/(\lambda^2+j^2/4)$ 
($K_j(\kappa)=\e{-j|\kappa|}$), afin
de récrire simplement le second membre. Rappelons
que nous utilisons la convention de sommation
sur les indices répétés.

Multiplions \xxxcoobae\ par la matrice $C_{jk}$ (qui
agit par produit de convolution); on obtient la forme finale
des BAE:
$$C_{jk}\star \rhot_k + \rho_j=\delta_{j1} \sigma\star s(\lambda)
\eqn\xxxcobae$$
où $\sigma(\lambda)={M\over L}\delta(\lambda)$ pour la chaîne XXX,
$\sigma(\lambda)={M\over 2L}(\delta(\lambda-1/c)+\delta(\lambda+1/c))$
pour le modèle NJL.

Nous représenterons le système d'équations \xxxcobae\ par la figure \bae.
\fig\bae{Le système de BAE de la chaîne XXX/modèle NJL
$SU(2)$.}{\figdir{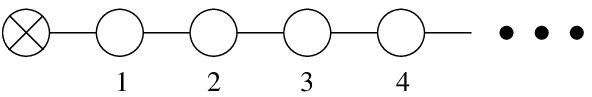}} 
Sur ce diagramme, chaque noeud (sauf le noeud coché)
correspond à une équation; un trait entre $2$ noeuds $j$ et
$k$ (resp. entre le noeud $j$ et le noeud coché)
signifie qu'un terme $-s\star\rhot_k$ (resp. $-s\star\sigma$)
apparaît dans l'équation $j$. Si l'on oublie le noeud coché,
ceci équivaut simplement à représenter le diagramme
de Dynkin dont $C_{jk}$ est la matrice de Cartan.

Ecrivons également l'énergie/impulsion du système: elles s'expriment
en termes de racines des BAE (eqs. \EXXX\ et \EGN), mais
en utilisant \xxxcobae, on peut les récrire en fonction des $\rhot_j$;
on obtient:
$$\eqalign{
E/L&=E_\gs/L+\int\d\lambda\,\rhot_1(\lambda) \tilde{\epsilon}(\lambda)\cr
P/L&=\int\d\lambda\,\rhot_1(\lambda) \tilde{p}(\lambda)\cr
}\eqn\EPphysc$$
où $\tilde{\epsilon}$ et $\tilde{p}$ sont donnés par \EPphysb. Ceci
est donc une démonstration (pour les états vérifiant l'hypothèse de
corde, et bien qu'elle soit en l'occurrence inutile) des formules
\EPphys-\EPphysb.

Le vide est obtenu pour $\rhot_1(\lambda)=0$, c'est-à-dire quand
il n'y a aucun trou dans la mer de racines. La densité $\rho_1(\lambda)$
vaut alors: $\rho_1(\lambda)\equiv
\rho_\gs(\lambda)=\sigma\star s(\lambda)$, tandis que
$\rho_j(\lambda)=\rhot_j(\lambda)=0$ pour $j>1$. Les $j$-cordes ($j>1$)
n'ont pas d'effet sur l'énergie \EPphysc: c'est un phénomène
que nous avions déjà observé au \secfond\ (dégénérescence de
l'état triplet et de l'état singlet). Leur rôle est élucidé en
calculant le spin de l'état correspondant: partant
de $2s=M-\sum_{j\ge 1} j\, m_j$, on arrive à
$$2s=\tilde{m}_1 - \sum_{j\ge 2} (j-1) m_j\eqn\spinphys$$
Les $j$-cordes ($j>1$) permettent donc de diminuer le spin à nombre fixé
de trous: ainsi, elles permettent de sélectionner dans quelle
sous-représentation irréductible de $V^{\otimes\tilde{m}_1}$ se trouve
le système.

Prenons enfin la limite d'échelle du modèle NJL. Pour cela, il faut
réintroduire la densité linéique de fermions $D=M/L=a^{-1}$, que nous
avions pour simplifier fixée à $1$ dans la définition \EPphysb\
de $\tilde{\epsilon}$ et $\tilde{p}$.
Ainsi $\tilde{\epsilon}_\NJL(\lambda)=D[\arctan(\e{\pi(\lambda-1/c)})
+\arctan(\e{\pi(-\lambda-1/c)})]$. Prenons alors les limites
simultanées $D\to\infty$, $c\to 0$ de sorte que l'on garde
fixée l'échelle d'énergie $m\equiv 2D \e{-\pi/c}$ (cette limite
est caractéristique de la {\it génération dynamique de masse} dans
le modèle NJL): on a alors $\tilde{\epsilon}_\NJL(\lambda)=m
\cosh(\pi\lambda)$ et de même $\tilde{p}_\NJL(\lambda)=m\sinh(\pi\lambda)$:
c'est la paramétrisation usuelle (à un facteur $\pi$ près)
d'une particule {\it relativiste
massive} de masse $m$. Egalement, les BAE \xxxcobae\ se simplifient
dans la limite d'échelle et deviennent:
$$C_{jk}\star \rhot_k + \rho_j=\delta_{j1}{m\over 2}\cosh(\pi\lambda)
\eqn\xxxcobaeb$$

\noindent{\it Remarques:}
\item{$\star$} Supposons que nous ayons considéré un modèle NJL avec une
seule chiralité (par exemple, uniquement des particules ``droites'');
évidemment, l'interaction (qui est entre particules de chiralités
différentes) n'existe plus, et on obtient une théorie de doublets
$SU(2)$ de fermions de Weyl libres non-massifs.
L'Ansatz de Bethe, le découplage
charge/spin, la limite d'échelle, etc.\ constituent une manière
compliquée de résoudre un modèle simple, mais qui n'en est pas moins
intéressante. Reprenant les calculs, on voit que les équations sont
inchangées, à part la relation de dispersion qui s'écrit 
$\tilde{\epsilon}(\lambda)=\tilde{p}(\lambda)=E \e{\pi\lambda}$ (où
$E$ est une échelle de masse arbitraire, puisqu'elle peut
être modifiée par décalage de $\lambda$), ce qui est comme il se doit
la relation de dispersion d'une particule droite non-massive. De même,
les BAE s'écrivent:
$$C_{jk}\star \rhot_k + \rho_j=\delta_{j1} {E\over 2} e^{\pi\lambda}
\eqn\xxxcobaec$$
Remarquons que dans l'Ansatz de Bethe (tel qu'on l'a formulé ici),
une masse n'apparaît que lorsque l'on met deux chiralités différentes
de particules nues (non-massives)
ensemble, ce qui est assez compréhensible cinématiquement.

\item{$\star$} Nous aurions également pu prendre une limite
relativiste dans la chaîne XXX. En effet, dans la limite {\it infra-rouge},
on peut développer la relation de dispersion au voisinage de
$\tilde{\epsilon}\sim 0$, et on voit facilement qu'on obtient
une théorie conforme qui est le point fixe infra-rouge
(du groupe de renormalisation) de la théorie.
A part une subtile différence dans la définition des deux
chiralités (sur laquelle nous reviendrons au \secddv),
ce point fixe infra-rouge est identique au point fixe {\it
ultra-violet} du modèle NJL.
La ressemblance entre un point fixe infra-rouge (de la chaîne de spins)
et un point fixe ultra-violet (du modèle NJL) ne doit pas surprendre:
en effet, c'est réellement la même région d'énergies que décrivent
les deux théories conformes. Dans le premier cas,
la limite infra-rouge consiste à regarder des énergies $\epsilon$
très petites comparées à la seule échelle d'énergie existante,
c'est-à-dire le cutoff ultra-violet $D$: $\epsilon\ll D$. 
Dans le second cas, la limite ultra-violette consiste à
regarder des énergies beaucoup plus grandes que la masse $m$,
mais après la limite d'échelle qui envoie $D$ à l'infini,
de sorte que: $m\ll \epsilon \ll D$. On a simplement
dans le second cas une condition supplémentaire
$m\ll \epsilon$, du fait qu'on a affaire à une théorie
massive (ce qui provient
des inhomogénéités $\theta_k=\pm 1/c$).

\subsec\secphysiso{BAE nues/BAE physiques: les excitations isotopiques $SU(2)$}
Le passage des BAE nues aux BAE physiques est un concept un 
peu difficile à expliquer; nous allons donc utiliser
les diagrammes décrivant la structure des BAE afin de mieux le
visualiser. Les BAE physiques sont des équations obtenues
{\it en supposant connus} le spectre et les matrices $S$ des
excitations physiques de la théorie. On leur applique alors
la même procédure qu'aux particules nues dans le modèle NJL
(déphasage sur un espace compactifié), ce qui nous amène
à de nouvelles BAE, les BAE physiques, qui doivent être
équivalentes par une transformation appropriée aux BAE que
nous avons déjà écrites, c'est-à-dire les BAE nues.

Evidemment, les excitations physiques du modèle NJL
étant des ``trous'', le passage
des BAE nues aux BAE physiques est intimement lié à la
dualité particule-trou. Cependant,
du fait de la présence des excitations isotopiques (ou excitations de
spin) qui reflètent la symétrie $SU(2)$, l'équivalence
comprend une étape supplémentaire
par rapport à ce qui a été fait au \secdualsimp:
$$\matrix{
{\hbox{particules nues}\atop\hbox{(multiplets de $SU(2)$)}}&
\buildrel {\rm ABA}\over\verylongrightarrow & \hbox{ondes de spin\hskip2cm}\cr
&&\Bigg\downarrow {\hbox{dualité}\atop\hbox{particule-trou}}\cr
{\hbox{particules physiques}\atop\hbox{(multiplets de $SU(2)$)}} &
\buildrel {\rm ABA}\over\verylongrightarrow &
\hbox{spinons} + \hbox{leurs excitations isotopiques}\cr
}$$
(ABA=Ansatz de Bethe Algébrique)

Tout le principe des BAE physiques consiste à ``remonter''
la flèche du bas qui va vers la droite, en devinant un
spectre et des matrices $S$ (ou éventuellement en les obtenant par
bootstrap) qui donnent des BAE équivalentes aux BAE nues.

Montrons comment ce procédé fonctionne pour le modèle NJL (avec des
modifications mineures, on pourrait l'appliquer à la chaîne XXX).
Supposons que les excitations physiques sont des doublets $SU(2)$,
avec une matrice $S$ donnée par \Sphys. Considérons un état
constitué de $M^{ph}$ particules de rapidités $\theta_1,\ldots,\theta_{M^{ph}}$
(c'est-à-dire d'énergie/impulsion $\epsilon_k=m \cosh(\pi\theta_k)$
et $p_k=m\sinh(\pi\theta_k)$). Consid\'erons le syst\`eme sur un
espace compactifi\'e de taille $L$, avec $mL\gg 1$, de sorte que l'on
puisse supposer que les diff\'erentes particules sont s\'epar\'ees par
des distances beaucoup plus grandes que $1/m$.
On peut alors utiliser les matrices $S$ pour \'ecrire
la condition de
déphasage d'une particule qui fait le tour de l'espace
compactifi\'e:
$$\e{-i p_k L}= S(\theta_k-\theta_{k-1})\ldots 
S(\theta_k-\theta_1) S(\theta_k-\theta_{M^{ph}})\ldots 
S(\theta_k-\theta_{k+1})\eqn\phshphys$$
Cette équation est tout à fait analogue à l'équation \phsh, 
bien que son interprétation soit différente: dans \phsh, les rapidités
$\theta_k$ (des particules nues) sont fixées, tandis que dans
\phshphys, elles ne le sont pas (en fait, à $L<\infty$,
toutes les valeurs des $\theta_k$ ne sont pas pour autant
admises puisqu'elles
doivent être telles que l'équation aux valeurs propres \phshphys\
ait une solution, ce qui implique une quantification des $\theta_k$).

Pour résoudre \phshphys, on remarque qu'ici encore le second
membre de \phshphys\ est la trace de la matrice de monodromie
inhomogène $T(\lambda=\theta_k|\theta_1,\ldots,\theta_{M^{ph}})$, et on
aboutit aux BAE (analogues de \xxxbaec{}):
$$\eqnalign\xxxbaecphys{1&=
\prod_{\scriptstyle \beta=1\atop\scriptstyle \beta\ne\alpha}^m 
{\lambda_\alpha-\lambda_\beta-i\over \lambda_\alpha-\lambda_\beta+i}
\prod_{k=1}^{M^{ph}}
{\lambda_\alpha-\theta_k+i/2\over\lambda_\alpha-\theta_k-i/2}&\xxxbaecphys a\cr
\e{-ip_kL}&=\prod_{\scriptstyle l=1\atop\scriptstyle
l\ne k}^{M^{ph}} \Sh(\theta_k-\theta_l)
\prod_{\alpha=1}^m{\theta_k-\lambda_\alpha+i/2\over\theta_k-\lambda_\alpha-i/2}
&\xxxbaecphys b\cr
}$$
Remarquons la présence des
facteurs $\Sh(\theta_k-\theta_l)$,
où $\Sh(\lambda)={\Gamma(1+i\lambda/2)\over\Gamma(1-i\lambda/2)}
{\Gamma((1-i\lambda)/2)\over\Gamma((1+i\lambda)/2)}$,
dus au fait que
la matrice $S$ physique est de la forme $S(\lambda)=\Sh(\lambda)R(\lambda)$.
Ces facteurs assurent que, même dans la représentation de plus
haut poids (spin $1$), c'est-à-dire en l'absence d'excitations de
spin, les particules physiques ont une interaction.

Prenons enfin la limite thermodynamique des équations \xxxbaecphys{}.
L'équation
\xxxbaecphys{a} est identique à \xxxbaec{a}, et, en appliquant
l'hypothèse de corde, on obtient immédiatement les BAE (cf \xxxcobae)
$$C_{jk}\star \rhot_k^{ph} +\rho_j^{ph}=\delta_{j1} s\star \sigma^{ph}
\eqn\xxxcobaephys$$
où $\rho_j^{ph}$ et $\rhot_j^{ph}$ sont les densités
de $j$-cordes/trous de $j$-cordes des BAE \xxxbaecphys{},
et $\sigma^{ph}$ est la densité
d'excitations physiques: $\sigma^{ph}(\lambda)={1\over L}\sum_{k=1}^{M^{ph}}
\delta(\lambda-\theta_k)$.

Comme les $\theta_k$ sont des inconnues dans \xxxbaecphys{},
il faut également récrire \xxxbaecphys{b} dans la limite
thermodynamique; on prend la dérivée logarithmique et on obtient
l'Equation d'Ansatz de Bethe supplémentaire:
$$2\pi(\sigma^{ph}(\lambda)
+\tilde{\sigma}^{ph}(\lambda))
=m \pi\cosh (\pi\lambda)
+2\pi K_j\star\rho_j^{ph} +
\left({1\over i}{\d\over\d\lambda} \log \Sh\right)
\star\sigma^{ph}\eqn\xxxcobaebphys$$
Divisant par $2\pi$ et identifiant les noyaux, on obtient:
$$\tilde{\sigma}^{ph} + (C_{11}^{-1})^{-1}\star\sigma^{ph}
+(C_{11}^{-1})^{-1}\star C_{1\,j+1}^{-1}\star\rho_j^{ph}={m\over 2}
\cosh(\pi\lambda)\eqn\xxxcobaecphys$$
En multipliant \xxxcobaecphys\ par $C_{11}^{-1}$,
on peut finalement identifier \xxxcobaephys\ et \xxxcobaecphys\
avec \xxxcobaeb; l'identification consiste à poser: $\sigma^{ph}=
\rhot_1$ et $\tilde{\sigma}^{ph}=\rho_1$, ce qui est la confirmation
que les excitations physiques sont des trous de $1$-cordes;
mais également $\rho^{ph}_j=\rho_{j+1}$ et $\rhot^{ph}_j=\rhot_{j+1}$,
c'est-à-dire que les $j$-cordes physiques sont les $(j+1)$-cordes
nues! On peut représenter la transformation des BAE nues aux
BAE physiques par la figure \baeb.
\fig\baeb{Passage des BAE nues aux BAE physiques
dans le modèle NJL. Les numéros sont les longueurs
des cordes {\it nues}.}{\figdir{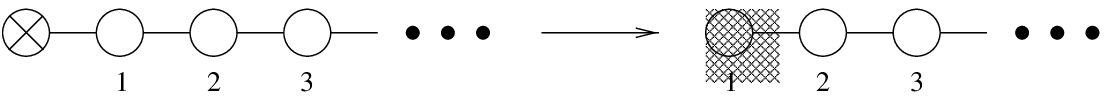}} 
Dans les BAE nues, ce sont les particules nues (représentées par
le noeud coché) qui servent de ``source'' (de second membre) aux
excitations de spin. De même, dans les BAE physiques,
ce sont les particules physiques, maintenant représentées
par le noeud quadrillé, qui servent de source à leurs excitations de
spin. Le noeud est quadrillé pour les BAE physiques, et non
coché comme pour les BAE nues, pour indiquer que les particules
physiques ont des rapidités qui sont elles-mêmes des paramètres
physiques qui peuvent prendre des valeurs arbitraires, ce qui
n'est pas le cas des particules nues.

Remarquons enfin que le spin est donné par $2s=M^{ph}-\sum_{j\ge 1}
j\,m_j^{ph}$, ce qui concorde avec l'expression \spinphys\ obtenue
à partir des BAE nues, du fait que $M^{ph}\equiv \tilde{m}_1$,
$m_j^{ph}\equiv m_{j+1}$.

\newchap\chapstrucbae{Structure des Equations d'Ansatz de Bethe}
On peut généraliser de nombreuses manières les modèles considérés précédemment,
et donc les BAE associées. Nous allons à présent considérer
celles qui nous seront utiles pour les modèles physiques
que nous avons en vue.
Ceci nous contraint à un petit interlude mathématique; en effet, la
notion algébrique sous-jacente aux modèles intégrables est celle de
groupe quantique, et elle nous est indispensable pour aller plus loin
dans nos constructions.

\newsec\secqg{Généralités sur les groupes quantiques}
En théorie quantique des champs, les symétries d'un modèle jouent
un rôle essentiel dans son étude. Par symétrie, on entend généralement
groupe de symétrie, bien que naïvement, toute observable
qui commute avec le Hamiltonien puisse faire office de symétrie.
Une remarque-clé est que la
structure de groupe provient de l'exigence suivante:
on demande que lorsqu'on
a l'action de la symétrie sur un état à $1$ particule, alors on sait
aussi la faire agir sur un état à plusieurs particules. C'est ce qui
nous amènera à la notion de co-multiplication pour les groupes
quantiques.

Dans le cadre de la théorie des groupes, le théorème de
Coleman--Mandula restreint énormément les symétries possibles
d'une théorie quantique des champs. Un premier pas vers les groupes
quantiques consiste à remarquer qu'en incluant les super-algèbres
de Lie comme symétries possibles, on obtient des algèbres de
symétrie plus grosses (supersymétrie). A quatre dimensions,
on ne peut aller plus loin, car le théorème spin-statistique
nous limite aux symétries bosonique/fermionique. Par contre,
à deux dimensions, du fait qu'on peut avoir des statistiques
arbitraires, on peut imaginer des symétries plus générales
que les groupes ou les super-groupes: ce sont les groupes quantiques,
qui sont obtenus en relaxant la condition de co-commutativité de
la co-multiplication. Effectivement, les groupes quantiques
s'avèrent être la symétrie naturelle des modèles intégrables
à deux dimensions; les générateurs
de groupes quantiques y apparaissent comme des charges non-locales
[\ref\BF{D.~Bernard et G.~Felder, {\it Nucl. Phys.} B365 (1991), 98.}].
Celles-ci forment en général des groupes quantiques qui
sont des déformations d'algèbres de Lie de dimension infinie:
ce sont les groupes quantiques affines, par opposition
aux groupes quantiques usuels (de dimension finie).

Nous allons rappeler brièvement la définition d'un groupe
quantique (\secdefqg), et donner l'exemple de $U_q({\goth sl}(2))$
(\secalglie);
puis nous passerons aux groupes quantiques affines et aux
Yangiens (\secqgaff), et développerons le cas
de $U_q(\widehat{{\goth sl}(2)})$ (\secqgaffsl).

\subsec\secdefqg{Quelques définitions}
Il n'est pas question ici de traiter de manière détaillée la théorie
des groupes quantiques. Nous renvoyons le lecteur aux
références [\ref\CP{V.~Chari et A.~Pressley,
{\it A guide to quantum groups}, Cambridge University Press (1994).},
\ref\FUC{J.~Fuchs, {\it Affine Lie algebras and quantum groups},
Cambridge monographs in mathematical physics,
Cambridge University Press (1992).},
\ref\GRS{C.~G\'omez, M.~Ruiz-Altaba et G.~Sierra, {\it Quantum groups
in two-dimensional physics},
Cambridge monographs in mathematical physics,
Cambridge University Press (1996).}] pour plus de détails.

Rappelons simplement que pour nous, un groupe
quantique est essentiellement une algèbre de Hopf quasi-triangulaire,
c'est-à-dire, pour expliciter nos notations:
\item{a)} une algèbre $\CA$ associative.
\item{b)} la donnée d'un co-produit
$\Delta\colon \CA\to \CA\otimes\CA$ (homomorphisme d'algèbre co-associatif),
d'une antipode $\gamma\colon \CA\to\CA$ (anti-homomorphisme d'algèbre)
et d'une co-unité $\eps\colon \CA\to {\Bbb C}$, qui vérifient des
conditions de compatibilité appropriées.
\item{c)} la donnée de la matrice $R$ universelle: ${\bf
R}\in\CA\otimes\CA$. Celle-ci assure que l'algèbre de Hopf $\CA$ est
``presque co-commutative'':
$$\sigma(\Delta(a))={\bf R} \Delta(a) {\bf R}^{-1}\quad\forall a\in\CA
\eqn\prcocom$$
où $\sigma$ permute les facteurs dans $\CA\otimes\CA$:
$\sigma(a\otimes b)=b\otimes a$.
La matrice ${\bf R}$ vérifie de plus les conditions de quasi-triangularité:
$$\eqalign{
(1 \otimes\Delta)({\bf R})&={\bf R}_{13} {\bf R}_{12}\cr
(\Delta\otimes 1)({\bf R})&={\bf R}_{13} {\bf R}_{23}\cr
}\eqn\quatri$$
où par exemple ${\bf R}_{12}={\bf R}\otimes 1\in
\CA\otimes\CA\otimes\CA$. Pour mémoire, on a aussi
la condition $(\gamma\otimes 1)({\bf R})={\bf R}^{-1}$.

Deux remarques élémentaires:
\item{$\bullet$} En combinant \prcocom\ et \quatri, on obtient l'équation de
Yang--Baxter sans paramètres spectraux:
$${\bf R}_{12} {\bf R}_{13} {\bf R}_{23} = {\bf R}_{23}
{\bf R}_{13} {\bf R}_{12}\eqn\YBEsans$$
en tant que relation dans $\CA\otimes\CA\otimes\CA$.

\item{$\bullet$} Posons $\overline{\bf R}=\sigma({\bf R}^{-1})$. Alors
\prcocom\ implique que $\overline{\bf R}$ vérifie aussi la relation
$\sigma(\Delta(a))=\overline{\bf R} \Delta(a) \overline{\bf R}^{-1}$.
Si ${\bf R}=\overline{\bf R}$ alors on dit que l'algèbre de Hopf $\cal A$
est triangulaire.

Nous nous intéresserons essentiellement aux représentations de groupes
quantiques. Traduisons-donc nos définitions en termes de
représentations:
\item{a)} une représentation de $\CA$ sur un Hilbert $V$ est un
morphisme d'algèbre de $\CA$ dans les endomorphismes de $V$; c'est
donc donner à $V$ une structure de $\CA$-module à gauche.
Nous abrègerons dorénavant représentation en ``\rep''.
\item{b)} Le co-produit $\Delta$ permet de définir le produit tensoriel
de deux \rep\ de $\CA$: si $\rho_1$ est une \rep\ sur $V_1$ et
$\rho_2$ sur $V_2$, alors $(\rho_1\otimes\rho_2)\Delta$ est une
\rep\ sur $V_1\otimes V_2$. 
L'antipode permet de même de définir la \rep\ contragrédiente\foot{D'où,
en combinant co-produit et antipode, la représentation sur les opérateurs d'un
Hilbert sur lequel $\CA$ agit,
ce qui est important en physique.}: si $\rho$ est une \rep\ sur $V$,
alors $(\rho\gamma)^\star$ est une \rep\ sur l'espace dual
$V^\star$.
Enfin, la co-unité constitue la \rep\ triviale de $\CA$.
\item{c)} Le produit tensoriel de \rep\ n'est pas commutatif
du fait de la non co-commutativité de $\Delta$, c'est-à-dire
que la simple permutation des facteurs ${\cal P}_{21\leftarrow 12}\colon
V_1\otimes V_2\to V_2\otimes V_1$ n'entrelace pas
les deux \rep\ $\rho_{12}\equiv(\rho_1\otimes\rho_2)\Delta$ et
$\rho_{21}\equiv
(\rho_2\otimes\rho_1)\Delta$ sur $V_1\otimes V_2$ et $V_2\otimes V_1$.
Cependant, la matrice
${\bf R}$, prise dans les deux \rep\ $\rho_1$ et
$\rho_2$: $R_{12}\equiv(\rho_1\otimes\rho_2)({\bf R})$
permet justement de construire un tel entrelacement: en effet,
la relation \prcocom\ s'écrit en appliquant $\rho_1\otimes\rho_2$:
$${\cal P}_{12\leftarrow 21} \rho_{21}(a) {\cal P}_{21\leftarrow 12} 
= R_{12} \rho_{12}(a)
R_{12}^{-1}\eqn\prcocomrep$$
(où ${\cal P}_{12\leftarrow 21}={\cal P}_{21\leftarrow 12}^{-1}$).
On voit donc que $\check{R}_{12}\equiv{\cal P}_{21\leftarrow 12} R_{12}$
est un opérateur d'entrelacement entre $\rho_{12}$ et
$\rho_{21}$\foot{On dit souvent que $\check R_{12}$ ``commute''
avec l'action du groupe quantique. Ceci n'est vrai
au sens littéral du terme
que si $\rho_1=\rho_2$, de sorte que $\rho_{12}=\rho_{21}$.}.
Remarquons qu'avec nos notations qui consistent à placer en indice 
les espaces sur lesquels les matrices $R$ agissent, il est inutile
d'écrire explicitement les permutations ``${\cal P}_{ba\leftarrow
ab}$''\foot{C'est ce que nous avons fait implicitement au
chapitre \chapbase; cf l'équation d'unitarité \unit.}.
Nous utiliserons donc $R_{12}$ (et non $\check R_{12}$) autant que possible.

\item{$\bullet$} De plus, pour trois \rep\ $\rho_1$, $\rho_2$,
$\rho_3$ données, on peut définir $3$ matrices $R$: $R_{12}$,
$R_{13}$, $R_{23}$ agissant sur $V_1\otimes V_2$, $V_1\otimes V_3$,
$V_2\otimes V_3$; elles vérifient elles aussi d'après \YBEsans\
l'équation de Yang--Baxter sans paramètres spectraux: $R_{12} R_{13}
R_{23}=R_{23} R_{13} R_{12}$.

\item{$\bullet$} Enfin, si l'on a deux espaces de \rep\
$V_1$ et $V_2$, alors on a des opérateurs
d'entrelacement: $\check R_{12}$ qui va
de $V_1\otimes V_2$ vers $V_2\otimes V_1$,
et $\check R_{21}$ qui va de $V_2\otimes V_1$ vers $V_1\otimes V_2$, mais
ils ne sont inverses l'un de l'autre que si l'algèbre est triangulaire
(${\bf R}=\overline{\bf R}$).

L'ensemble des représentations d'une algèbre de Hopf $\cal A$
(ou plus exactement, l'espace formé par leurs combinaisons
linéaires), muni de la somme directe et du produit tensoriel,
forme une algèbre associative $\cal R$. 
La ``presque co-commutativité'' de $\cal A$
implique de plus que $\cal R$ est
commutative.

\subsec\secalglie{Algèbres de Lie simples et leur déformation; exemple
de ${\goth sl}(2)$}
Le prototype de l'algèbre de Hopf co-commutative est l'algèbre
enveloppante universelle $U({\goth g})$ d'une algèbre de Lie ${\goth g}$.
Le co-produit et l'antipode sont alors engendrés par leur définition sur
l'algèbre de Lie ${\goth g}$ (et sur $1$):
$$\matrix{\Delta(a)=1\otimes a+a\otimes 1\hfill& \gamma(a)=-a\hfill&
\eps(a)=0\hfill&\forall a\in{\goth g} \cr
\Delta(1)=1\otimes 1\hfill&\gamma(1)=1\hfill&\eps(1)=1\hfill&\cr}
\eqn\cocomhopf$$
Dans ce cas, la matrice ${\bf R}$ est triviale: ${\bf R}=1\otimes 1$.

Si $\goth g$ est l'algèbre de Lie d'un groupe de Lie compact 
simplement connexe $G$, alors l'algèbre des représentations $\cal R$
s'identifie à l'algèbre des fonctions de classe
(i.e. invariantes par conjugaison) sur $G$: ${\cal R}=L^2_{class}(G)$.
La base naturelle de $\cal R$ constituée par les représentations
irréductibles (que nous abrégerons en ``\irr'') 
s'identifie à la base des {\it caractères des \irr}
de $L^2_{class}(G)$. On rappelle qu'à une représentation
$R$ et un élément $g\in G$ on peut associer le
caractère $\chi_R(g)$ qui est par définition la trace
de $g$ pris dans la \rep\ $R$. Pour $R$ fixé, $g\to\chi_R(g)$ est
une fonction de classe; inversement, pour $g$ fixé, $R\to\chi_R(g)$
est une représentation (uni-dimensionnelle) de $\cal R$. 
L'algèbre $\cal R$ possède une
représentation naturelle qui est son action sur elle-même
par multiplication; explicitement, à chaque représentation $R$
on peut associer une matrice $A$ (de taille infinie) à coefficients
positifs qui donne
son action par produit tensoriel dans la base des irrep:
$$R\otimes R_1=\bigoplus_{R_2} A^{R_2}_{R_1} R_2\qquad R_1,R_2\hbox{ irrep}$$
Du fait de l'interprétation physique que nous lui donnerons,
nous appellerons $A$ la {\it matrice d'adjacence} associée à $R$.
Comme $\cal R$ est commutative, la représentation $R\rightarrow A(R)$
est décomposable en représentations uni-dimensionnelles, ce qui
veut dire que les $A(R)$ sont simultanément diagonalisables,
de valeurs propres les caractères;
on montre qu'on obtient ainsi tous les caractères exactement une fois,
grâce à l'identité (qui nous donne aussi
les vecteurs propres de $A^t$):
$$\chi_R(g)\chi_{R_1}(g)=\sum_{R_2} A^{R_2}_{R_1} \chi_{R_2}(g)$$
En particulier, la dimension d'une représentation $R$ constitue
la valeur propre de Perron--Frobenius de sa matrice d'adjacence.

De plus, si l'on veut éviter d'avoir à passer
par le groupe $G$, on peut définir un caractère en restant
au niveau de l'algèbre $\goth g$ de la manière suivante: on considère
une sous-algèbre de Cartan ${\goth h}\subset{\goth g}$, et alors
pour $x\in{\goth h}$, $\chi_R(x)\equiv\tr_R(\exp(x))$
(où l'on considère $x$ dans la représentation $R$, puis
on exponentie).

Les exemples connus de groupes quantiques non co-commutatifs sont
obtenus par déformation des algèbres enveloppantes des algèbres de Lie
usuelles: on les note $U_q({\goth g})$ [\ref\JIMb{M.~Jimbo,
{\it Lett. Math. Phys.} 10 (1985), 63; 11 (1986), 247.}],
où $q$ est le paramètre de
déformation. Nous allons maintenant donner la définition
précise de $U_q({\goth sl}(2))$.
Une construction similaire peut être effectuée
pour toutes les algèbres de Lie ${\goth g}$ simples. 

$U_q({\goth sl}(2))$ est
l'algèbre engendrée par les générateurs 
$\K$ (inversible d'inverse $\K^{-1}$), $\E$, $\F$ et les
relations [\ref\KRb{P.P.~Kulish et N.Yu.~Reshetikhin,
{\it J. Sov. Math.} 23 (1983), 2435.},\JIMb]:
$$\eqalign{
\K\E\K^{-1}&=q^2 \E\cr
\K\F\K^{-1}&=q^{-2} \F\cr
\E \F - \F \E &= {\K - \K^{-1}\over q-q^{-1}} \cr
}\eqn\uqsl$$
Il faut considérer que $\K=q^\HH$ où $\HH\equiv 2\S^3$ est l'équivalent 
du générateur de la sous-algèbre de Cartan
de ${\goth sl}(2)$\foot{$\HH$ n'est pas défini
en tant qu'élément abstrait de $U_q({\goth sl}(2))$; cependant,
dans les représentations que nous considérerons,
$\HH$ sera bien défini en tant qu'opérateur.}. Si l'on utilise la notation
$\q{x}\equiv (q^x-q^{-x})/(q-q^{-1})$, alors $[\E,\F]=\q{\HH}$.
De même, on pose $\E=\S^+ \K^{1/2}$ et $\F=\K^{-1/2} \S^-$
(on a introduit $\K^{1/2}=q^{\HH/2}$), de telle
sorte qu'on ait encore $[\S^+,\S^-]=[\E,\F]=\q{\HH}$; $\S^\pm$ est alors analogue
à l'opérateur  $\S^1\pm i \S^2$ de ${\goth sl}(2)$.

On voit que dans la limite $q\to 1$,
\uqsl\ se réduit aux relations:
$$\eqalign{
[\S^3,\S^\pm]&=\pm \S^\pm\cr
[\S^+,\S^-]&=2\S^3\cr
}$$
qui ne sont autres que les relations de commutation de ${\goth sl}(2)$.

Définissons maintenant le co-produit [\ref\SKLb{E.K.~Sklyanin,
{\it Uspehi Mat. Nauki} 40 (1985), 214.}], l'antipode et la co-unité:
$$\matrix{\Delta(\K)=\K\otimes \K&\gamma(\K)=\K^{-1}&\eps(\K)=1\cr
\Delta(\E)=\E\otimes \K+1\otimes \E&\gamma(\E)=-\E \K^{-1}&\eps(\E)=0\cr
\Delta(\F)=\F\otimes 1+\K^{-1}\otimes \F&\gamma(\F)=-\K\F &\eps(\F)=0\cr
\Delta(1)=1\otimes 1&\gamma(1)=1&\eps(1)=1\cr
}\eqn\couqsl$$

En termes de $\S^\pm$, le co-produit se met sous la forme
$\Delta(\S^\pm)=\S^\pm \otimes \K^{1/2}
+\K^{-1/2} \otimes \S^\pm$. Cette reformulation a le mérite de montrer
que la transformation $q\to q^{-1}$ (et donc $\K\to \K^{-1}$),
qui est une symétrie de \uqsl, échange $\Delta$ et $\sigma\circ\Delta$
(et $\gamma$ et $\gamma^{-1}$).

La théorie de la représentation de $U_q({\goth sl}(2))$, pour
des valeurs génériques de $q$ (c'est-à-dire pour $q$ non racine
de l'unité) ressemble à celle de ${\goth sl}(2)$:
pour chaque demi-entier $s$, on a une \irr\
de dimension $2s+1$, l'action des générateurs dans la base de
diagonalisation de $\S^3$ étant donnée par: ($m=-s,-s+1,\ldots,s$)
$$\eqalign{
\S^\pm |s,m\rangle&=\sqrt{\q{s\mp m}\q{s\pm m+1}} |s,m\pm 1\rangle\cr
\S^3 |s,m\rangle&=m |s,m\rangle\cr}\eqn\stdrep$$
On voit que $\HH=2\S^3$ est bien défini en tant qu'opérateur.
Les \irr\ sont classifiées par la valeur
du $q$-Casimir quadratique ${\rm C}$,
générateur du centre de $U_q({\goth sl}(2))$:
$${\rm C}\equiv \S_- \S_+ + \q{\S^3+1/2}^2=\q{s+1/2}^2$$
On appellera les \irr\ données par \stdrep\ les \rep\ standard.

L'algèbre des représentations $\cal R$ est donc isomorphe à celle de ${\goth sl}(2)$.
En d'autres termes, les coefficients de Littlewood-Richardson
(les constantes de
structure de $\cal R$) sont les mêmes; par contre, les symboles $3j$ 
(coefficients de Clebsch--Gordan)
et $6j$ (Wigner--Racah) sont $q$-déformés [\ref\KRc{A.N.~Kirillov
et N.Yu.~Reshetikhin, {\it Zap. Nauch. Semin. LOMI} 168 (1988),
68.}].
On peut également définir des caractères $\chi_R(x)=\tr_R(\exp(x\HH))$,
qui sont les mêmes que les caractères usuels de ${\goth sl}(2)$.

La situation est plus compliquée pour $q$ racine de l'unité. Posons
$f$ le plus petit entier tel que $q^{f+2}=\pm 1$. Alors, on remarque
que le centre de $U_q({\goth sl}(2))$ est aggrandi, puisque
$(\S^\pm)^{f+2}$ et $\K^{f+2}$ commutent avec tous les opérateurs
de $U_q({\goth sl}(2))$. Les représentations irréductibles de
$U_q({\goth sl}(2))$ sont maintenant également classifiées par
les valeurs propres de ces trois opérateurs, si on laisse
ces dernières arbitraires: c'est ce qu'on appelle
la {\it spécialisation non-restreinte} de $U_q({\goth sl}(2))$.

Nous ne considérerons pas cette situation
ici: nous n'aurons affaire qu'à des \rep\ de $U_q({\goth
sl}(2))$ telles que $\K^{f+2}=\pm 1$ et $(\S^\pm)^{f+2}=0$
(ce qui implique en particulier que $\HH$ est bien défini comme
opérateur dans ces \rep). 
La théorie de la représentation n'en reste
pas moins non-triviale: en effet, en imaginant $q$ comme un paramètre
amené à varier continûment, il est naturel de considérer
que les opérateurs $(\S^\pm)^{f+2}/\q{f+2}\exclam$ restent bien définis
dans la limite où $\q{f+2}\exclam\equiv\prod_{i=1}^{f+2}\q{i}\to 0$
et $(\S^\pm)^{f+2}\to 0$. L'algèbre $U_q({\goth sl}(2))$ munie
de ces nouveaux éléments constituent
la {\it spécialisation restreinte} de $U_q({\goth sl}(2))$.
Les opérateurs correspondants 
aggrandissent les modules de $U_q({\goth sl}(2))$. 
En particulier, certaines \rep\ deviennent réductibles
mais indécomposables. Qu'advient-il des représentations standard
quand $q$ devient une racine de l'unité? On trouve la classification
suivante:
\item{a)} les représentations standard de spin $s$ tel que $0\le 2s\le f$
restent irréductibles.
\item{b)} les représentations standard de dimension $2s+1=n(f+2)$
restent également irréductibles. 
\item{c)} les autres se retrouvent groupées ensemble
en \rep\ réductibles mais
indécomposables. 

Appelons ``bonnes'' (ou de type II)
les \rep\ a), et ``mauvaises'' (ou de type I) les \rep\ b) et c).
Il existe un critère simple pour distinguer les bonnes
\rep\ des mauvaises:
on introduit pour cela la notion de $q$-dimension, qui n'est autre
qu'un caractère particulier. Par définition,
$$\qdim(R)=\tr_R(\K)\eqn\defqdim$$
Pour $q=1$, on retrouve la notion de dimension usuelle.
On calcule aisément pour la \rep\ standard de spin $s$
que $\qdim(R)=\q{2s+1}$.
En particulier,
pour $q=\pm\exp(\pm i\pi/(f+2))$, $\qdim(R)>0$ pour $0\le 2s\le f$.
Par contre, pour les mauvaises \rep, on a: $\qdim(R)=0$ (c'est
évident pour les \rep\ de type b), et prouvable pour
celles de type c)).

Ceci suggère de tronquer l'algèbre des
représentations en ne conservant
que les bonnes \rep, c'est-à-dire de $q$-dimension non nulle,
et en jetant les mauvaises [\ref\AGGS{L.~Alvarez-Gaumé,
C.~G\'omez et G.~Sierra, {\it Nucl. Phys.} B330 (1990), 347.}].
Pour cela, il est nécessaire
de définir un nouveau produit tensoriel tronqué:
évidemment, il ne s'agit pas seulement d'éliminer
les \rep\ de $q$-dimension nulle, car ceci briserait
l'associativité de l'algèbre $\cal R$ (des bonnes \rep\
peuvent apparaître par produit tensoriel de mauvaises).
Dans le cas de $U_q({\goth sl}(2))$, une méthode simple
pour construire les matrices d'adjacence tronquées
(ce qui équivaut à donner le produit tensoriel tronqué)
consiste à partir de la matrice $A$ associée à la \rep\ fondamentale
(spin $1/2$) et à la tronquer au sens littéral du terme: on obtient
une matrice $(f+1)\times (f+1)$ (dans la base des irreps classées
par ordre de dimension croissante):
$$A=\pmatrix{0&1& & &0\cr
            1&0&1& &\cr
             &\ddots&\ddots&\ddots&\cr
                             &&1&0&1\cr
                            0&& &1&0\cr      
}$$
Ensuite, on obtient les autres matrices d'adjcacence dans
la décomposition du produit tensoriel de \rep\ de spin $1/2$.
Ainsi, la matrice d'adjacence de spin $1$ vérifie $A_1=(A_\ha)^2 - 1$.
Nous donnerons une vision intuitive de ces matrices d'adjacence
tronquées au \secanistronc.
Donnons la formule finale du produit tensoriel de deux bonnes \irr\
de spins $s$ et $s'$ ($s'\le s$):
$$s\otimes s'=\bigoplus_{s''=s-s'}^{\min(s+s',f-(s+s'))} s''\eqn\troncprod$$
L'algèbre des représentations ${\cal R}_{tronc}$ ainsi
définie reste associative et commutative; elle est de dimension
finie, et n'admet donc plus qu'un nombre fini d'\irr\ (caractères),
la $q$-dimension étant l'un d'entre eux (valeur
propre de Perron-Frobenius des matrices d'adjacence). Ceci
revient à dire que les règles d'addition et de multiplication
des caractères ne sont plus valables pour des caractères quelconques,
du fait de la troncation (ainsi, la dimension du produit
tensoriel tronqué n'est pas le produit des dimensions; par contre,
la $q$-dimension du produit tensoriel tronqué est le produit
des $q$-dimensions).

Finalement, on peut montrer que $U_q({\goth sl}(2))$ 
est quasi-triangulaire\foot{Pour $q$ générique.
Pour $q$ racine de l'unité, et dans la spécialisation restreinte,
on peut tout de même trouver des matrices $R$ pour toutes
les représentations.}
et possède donc une matrice ${\bf R}$
universelle. Nous ne donnerons pas la formule abstraite de ${\bf R}$
(cf [\FUC]), mais seulement sa matrice 
(et celle de $\overline{\bf R}$) dans deux représentations de spin
$1/2$:
$$R=\pmatrix{q&0&0&0\cr 0&1&q-q^{-1}&0\cr 0&0&1&0\cr 0&0&0&q\cr}
\qquad\overline{R}=\pmatrix{q^{-1}&0&0&0\cr 0&1&0&0\cr 0&q^{-1}-q&1&0\cr 
0&0&0&q^{-1}\cr}\eqn\Ruqsl$$

\subsec\secqgaff{Yangiens, groupes quantiques affines et matrices $R(\lambda)$}
Nous avons vu au \secdefqg\ des expressions pour les
matrices $R$ semblables à celles qui avaient été utilisées dans
l'Ansatz de Bethe. Cependant, les matrices $R$ des groupes quantiques
ne dépendent {\it a priori} pas d'un paramètre
spectral; en particulier, elles vérifient l'équation de Yang--Baxter,
mais sans paramètres spectraux. Pour y remédier, il suffit
de considérer des groupes quantiques dont les représentations
sont indexées par un paramètre spectral $\lambda$: c'est ce qui nous
conduit naturellement aux Yangiens et aux algèbres de boucles quantiques.

On sait qu'à une algèbre de Lie simple ${\goth g}$ on peut associer une
algèbre affine (non tordue) $\widehat{\goth g}$
qui peut être considérée comme une extension centrale de l'algèbre de boucles
$\widehat{\goth g}_0$ de 
${\goth g}$. Si l'on oublie l'extension centrale, on voit que l'on a
précisément introduit un nouveau paramètre $x$ (de sorte
que l'algèbre de boucles soit: $\widehat{\goth g}_0={\goth g}\otimes {\Bbb
C}[x,x^{-1}]$). Notons que dans l'application usuelle des algèbres
affines à la physique (pour les théories conformes), $x$ joue
le rôle de variable d'espace-temps, alors qu'ici il constitue le paramètre
spectral (on verra qu'il
est relié à notre paramètre spectral habituel $\lambda$ par 
$x=q^{i\lambda}$).

Pour des raisons qui vont apparaître plus loin, il est plus naturel
de commencer par considérer le cas déformé.
De la même manière que l'on peut déformer l'algèbre de Lie
simple ${\goth g}$ en $U_q({\goth g})$, on peut déformer $\widehat{\goth g}$
en un {\it groupe quantique affine} $U_q(\widehat{\goth g})$.
En fait, nous ne nous intéresserons qu'aux représentations de niveau $0$
de $U_q(\widehat{\goth g})$\foot{Les représentations de niveau non nul
peuvent quand même apparaître elles aussi, cf [\ref\JM{M.~Jimbo
et T.~Miwa, {\it Algebraic analysis of solvable lattice models},
Regional Conference Series in Mathematics number 85,
Conference Board of the Mathematical Sciences (1995).}].},
c'est-à-dire aux représentations de
l'algèbre de boucles quantique $U_q(\widehat{\goth g}_0)$: en effet,
celles-ci sont, comme dans la limite non-déformée $q=1$, des
{\it représentations d'évaluation} du paramètre $x$.
On peut montrer que toute représentation de $U_q(\widehat{\goth g}_0)$
s'obtient par produit tensoriel de représentations d'évaluation.

On peut ensuite prendre la limite non-déformée $q\to 1$
(ou $q\to -1$, voir \secanis); dans cette limite,
et pour les modèles physiques qui nous intéressent,
l'algèbre de boucles quantique $U_q(\widehat{\goth g}_0)$ ne
redonne pas l'algèbre de boucles usuelle, mais ``dégénère'' en un
{\it Yangien} [\ref\DRI{V.G.~Drinfel'd,
{\it Soviet. Math. Dokl.} 36 (1988), 212.},\ref\BER{D.~Bernard,
{\it Commun. Math. Phys.} 137 (1991), 191.}], noté $Y({\goth g})$.
$Y({\goth g})$ est encore une algèbre 
de Hopf non co-commutative, qui contient
comme sous-algèbre l'algèbre universelle enveloppante
$U({\goth g})$. La théorie de la représentation de $Y({\goth g})$ est
complètement analogue à celle de $U_q(\widehat{\goth g}_0)$:
ses représentations peuvent
être obtenues par produit tensoriel de représentations dites d'évaluation.
Nous n'expliciterons pas ici $Y({\goth g})$,
et le lecteur est donc renvoyé aux références [\CP,\ref\BERb{D.~Bernard, 
{\it Int. J. Mod. Phys.} B7 (1993), 3517 \pre{hep-th/9211133}.}]
pour plus de détails.

Pour deux \rep\ d'évaluation $V_1$ et $V_2$ de paramètres
spectraux $\lambda_1$ et $\lambda_2$ d'un groupe quantique
affine/Yangien, la matrice $R(\lambda)$ entre
les deux \rep\ est une fonction de $\lambda=\lambda_1-\lambda_2$.
Ainsi, l'équation de Yang--Baxter acquiert la forme \YBE\ avec
paramètres spectraux. De plus, remarquons que l'équation d'unitarité \unit\
s'interprète alors comme la triangularité
du groupe quantique affine/Yangien: ${\bf R}=\overline{\bf R}$
prise dans $V_1\otimes V_2$.

Il convient de remarquer que les Yangiens et les groupes quantiques
affines ne sont pas {\it stricto sensu} des algèbres de Hopf 
quasi-triangulaires; en effet, ils ne sont pas 
tout à fait ``presque co-commutatifs'',
fait qui s'avèrera important pour nous puisqu'il
est lié à la procédure de fusion (section \secfus).
Cependant, on peut tout de même définir
pour toute paire de \rep\ $V_1$ et $V_2$
des matrices $R$ qui vérifient l'équation de Yang--Baxter;
et pour des valeurs {\it génériques} des
paramètres spectraux caractérisant $V_1$ et $V_2$, 
$\check R_{12}$ entrelace bien $V_1\otimes V_2$ et $V_2\otimes V_1$.

Explicitons maintenant $U_q(\widehat{\goth g})$ 
dans le cas ${\goth g}={\goth sl}(2)$.

\subsec\secqgaffsl{Le groupe quantique affine
$U_q(\widehat{{\goth sl}(2)})$ et ses représentations de niveau $0$}
L'algèbre $U_q(\widehat{{\goth sl}(2)})$ est engendrée par les
éléments $\K_i$ (inversibles d'inverses $K_i^{-1}$), $\E_i$, $\F_i$
($i=0,1$) et les relations:
$$\eqalign{
\K_i \K_j &= \K_j \K_i \cr
\K_i \E_j \K_i^{-1} &= q^{a_{ij}} \E_j \cr
\K_i \F_j \K_i^{-1} &= q^{-a_{ij}} \F_j \cr
\E_i \F_j - \E_j \F_i &= \delta_{ij} {\K_i-\K_i^{-1}\over q-q^{-1}}\cr
({\rm ad}_{\E_i})^{1-a_{ji}}(\E_j)&=0\quad (i\ne j)\cr
({\rm ad}_{\F_i})^{1-a_{ji}}(\F_j)&=0\quad (i\ne j)\cr
}\eqn\uslaff$$
où $(a_{ij})$ est la matrice de Cartan de $A_1^{(1)}$: 
$(a_{ij})=\bigl({2\atop -2}{-2\atop 2}\bigr)$. Nous n'écrirons pas plus
explicitement les deux dernières conditions (voir [\CP] pour
une définition de l'action adjointe $q$-déformée ${\rm ad}$). 
On remarque que $\K_1 \K_0$ appartient
au centre de l'algèbre: dans une représentation irréductible,
$\K_1 \K_0 = q^n$, où $n$ est le {\it niveau} de la représentation.

$U_q(\widehat{{\goth sl}(2)})$ contient deux 
sous-algèbres $U_q({\goth sl}(2))$ (engendrées par $\K_i$, $\E_i$, $\F_i$
à $i$ fixé), qui jouent {\it a priori} des rôles symétriques.
Evidemment, ceci est à contraster avec la vision usuelle de
$U_q(\widehat{{\goth sl}(2)}_0)$ comme algèbre de boucles (déformée),
dans laquelle on a une sous-algèbre privilégiée, la sous-algèbre
{\it horizontale}: les générateurs de cette dernière
seront identifiés avec $\K_1$, $\E_1$, $\F_1$.
Le co-produit, l'antipode et la co-unité sont donnés pour chaque
sous-algèbre $U_q({\goth sl}(2))$ par les relations
\couqsl.

Définissons à présent les représentations d'évaluation: elles
sont obtenues par {\it affinisation} des représentations de
$U_q({\goth sl}(2))$. Ainsi, si l'on a défini une représentation
de $U_q({\goth sl}(2))$ par l'action de ses générateurs $\E$, $\F$, $\K$,
alors l'action des générateurs de $U_q(\widehat{{\goth sl}(2)})$ est donnée
par
$$\matrix{
\E_1=x\E \hfill& \F_1=x^{-1} \F \hfill& \K_1=\K \hfill\cr
\E_0=x\F \hfill& \F_0=x^{-1} \E \hfill& \K_0=\K^{-1} \hfill\cr
}\eqn\affrep$$
On a $\K_0 \K_1=1$, donc c'est une représentation de niveau $0$, caractérisée
par son spin (celui de la représentation de $U_q({\goth sl}(2))$
dont on est parti) et un paramètre spectral $x$.
Remarquons également qu'à la représentation \affrep\ est naturellement
associée une {\it gradation} de $U_q(\widehat{{\goth sl}(2)})$,
c'est-à-dire que l'on associe aux générateurs $\E_1$, $\E_0$ un degré
$+1$ (car ils multiplient par $x$), à $\F_1$, $\F_0$ un degré $-1$,
et à $\K_1$, $\K_0$ un degré $0$. C'est la {\it gradation principale},
la plus naturelle pour nous.

Nous serons cependant amenés à considérer une autre gradation.
Pour cela, effectuons un changement de base {\it dépendant de $x$},
de façon que $\E$ acquière un degré égal à $-1$, et $\F$ à $+1$. Alors, on
obtient une nouvelle gradation, la {\it gradation homogène}, dans
laquelle la sous-algèbre horizontale $\E_1$, $\F_1$, $\K_1$ est de degré
$0$, tandis que $\E_0$ est de degré $2$ et $\F_0$ de degré $-2$.

Prenons un exemple concret. Dans la représentation de spin $1/2$,
les matrices des générateurs \affrep\ sont
$$\matrix{
\E_1=\pmatrix{0&x\cr 0&0}& \F_1=\pmatrix{0&0\cr x^{-1}&0}&
\K_1=\pmatrix{q&0\cr 0&q^{-1}}\cr
\E_0=\pmatrix{0&0\cr x&0}&\F_0=\pmatrix{0&x^{-1}\cr 0&0}& 
\K_0=\pmatrix{q^{-1}&0\cr 0&q}\cr}
\eqn\affrepb$$
La matrice $R(x)$ entre deux représentations de spin $1/2$,
de paramètres spectraux $x_1$ et $x_2$ ($x=x_1/x_2$) vaut
$$R(x)=\pmatrix{1&0&0&0\cr
0& {x-x^{-1}\over xq-x^{-1}q^{-1}}&{q-q^{-1}\over xq-x^{-1}q^{-1}}&0\cr
0&{q-q^{-1}\over xq-x^{-1}q^{-1}}&{x-x^{-1}\over xq-x^{-1}q^{-1}}&0\cr
0&0&0&1}\eqn\affR$$
Effectuons maintenant le changement de base 
$|\tilde{\pm}\rangle=x^{\pm 1/2}|\tilde{\pm}\rangle$; on obtient les
nouvelles matrices
$$\matrix{
\tilde{\E}_1=\pmatrix{0&1\cr 0&0}& \tilde{\F}_1=\pmatrix{0&0\cr 1&0}&
\tilde{\K}_1=\pmatrix{q&0\cr 0&q^{-1}}\cr
\tilde{\E}_0=\pmatrix{0&0\cr x^2&0}&\tilde{\F}_0=\pmatrix{0&x^{-2}\cr 0&0}& 
\tilde{\K}_0=\pmatrix{q^{-1}&0\cr 0&q}\cr}
\eqn\affrepc$$
L'action de la sous-algèbre horizontale s'identifie à l'action usuelle
de $U_q({\goth sl}(2))$. Quant à la matrice $\tilde{R}(x)$, elle s'écrit
$$\tilde{R}(x)=\pmatrix{1&0&0&0\cr
0& {x-x^{-1}\over xq-x^{-1}q^{-1}}&x{q-q^{-1}\over xq-x^{-1}q^{-1}}&0\cr
0&x^{-1}{q-q^{-1}\over xq-x^{-1}q^{-1}}&{x-x^{-1}\over xq-x^{-1}q^{-1}}&0\cr
0&0&0&1}={xR-x^{-1}\overline{R}\over xq-x^{-1}q^{-1}}\eqn\affRb$$
où $R$ et $\overline{R}$ sont les matrices $R$ de sa sous-algèbre
horizontale (cf \Ruqsl).

Qu'est-ce qui motive physiquement le choix d'une base ou de l'autre, et donc
d'une gradation ou de l'autre? Pour se faire
une idée, supposons que le groupe quantique affine
agisse sur le Hilbert d'une théorie invariante relativiste,
de telle sorte que $\lambda$ ($x=q^{-i\lambda}$)
s'identifie à la rapidité des états asymptotiques.
Un choix naturel consiste
à prendre des vecteurs de base $|\pm,x\rangle$ (où l'on a spécifié
explicitement le paramètre spectral $x$ de la représentation) tels que
$|\pm,xx_1\rangle=\tau_x |\pm,x_1\rangle$ où $\tau_x$ est l'opérateur
de {\it boost} dans l'espace de Hilbert.
Ainsi, le degré d'un opérateur ${\cal O}$ dans la gradation correspondante
s'identifie au {\it comportement sous le groupe de Lorentz} de
cet opérateur (son ``spin''): $\tau_x^{-1} {\cal O} \tau_x = x^{\deg
{\cal O}} {\cal O}$.
Le choix de la gradation est donc lié au ``spin''
des générateurs du groupe quantique.

Remarquons enfin que, dans la limite $q\to 1$,
la matrice $R(x)$ \affR\ exprimée en termes de $\lambda$,
avec $x=q^{-i\lambda}$ 
(on parle alors de matrice $R$ {\it trigonométrique}),
tend vers la matrice $R(\lambda)$ rationnelle \explR.
Cette dernière s'interprète dans
ce contexte comme la matrice $R(\lambda)$ associée au Yangien
$Y({\goth sl}(2))$.

Présentons maintenant la théorie de la représentation
de $U_q(\widehat{{\goth sl}(2)}_0)$ [\ref\CPb{V.~Chari et
A.N.~Pressley, {\it Commun. Math. Phys.} 142 (1991), 261.}].
Nous supposerons $q$ générique.
On montre que toute représentation irréductible de dimension finie
est caractérisée par un polynôme $P(x)=\prod_k (x-x_k)$; les $x_k$
sont les paramètres spectraux associés à la représentation.
Pour deux \irr\ de polynômes $P(x)$ et $Q(x)$ et des valeurs
génériques de leurs racines, leur produit tensoriel est irréductible
de polynôme $P(x) Q(x)$. Voyons précisément ce qu'on entend ici
par génériques.

Partons de la \rep\ d'évaluation de spin $1/2$, de paramètre
spectral $x_0$. Le polynôme associé est $P(x)=x-x_0$. Pour deux telles
\rep\ sur $V_1={\Bbb C}^2$ et $V_2={\Bbb C}^2$,
de polynômes $P(x)=x-x_1$ et $Q(x)=x-x_2$, le
produit tensoriel est irréductible sauf si $\lambda_2-\lambda_1=\pm i$
(on a repris notre paramétrisation $x=q^{-i\lambda}$). Dans ce
cas,la représentation est réductible
(mais indécomposable), et on a un sous-espace stable. Pour
$\lambda_2-\lambda_1=+i$, ce sous-espace constitue la \rep\ d'évaluation
de spin $1$, de polynôme $P(x)Q(x)$\foot{Ce sous-espace
s'identifie avec la sous-représentation de spin $1$ dans
la décomposition $\ha\otimes\ha=0\oplus 1$ pour une {\it quelconque}
des deux sous algèbres $U_q({\goth sl}(2))$.};
pour $\lambda_1-\lambda_2=-i$,
c'est la représentation de spin $0$ (triviale), de polynôme $1$.
Remarquons la dissymétrie manifeste entre les deux cas: dans le
premier cas, le polynôme est comme dans le cas générique le produit
des polynômes, et non dans le second. Ceci est dû au fait que
le vecteur de plus haut poids du produit tensoriel se trouve
dans la sous-représentation dans le premier cas, et non dans le
second. La sous-représentation du vecteur de plus haut poids joue
donc un rôle spécial, ce que nous aurons
l'occasion de revoir plus loin, lorsque nous mettrons
en pratique ces résultats pour la procédure de fusion.

De même, la représentation d'évaluation de spin $s$ admet pour
polynôme $P(x)=\prod_{k=1}^{2s} (x-x_k)$ avec $x_k=q^{-i\lambda_k}$
et $\lambda_k=\lambda+i(s+1/2-k)$, c'est-à-dire que les $\lambda_k$ forment
une $2s$-corde. Nous justifierons d'ailleurs au \secfusb\ le lien
avec l'hypothèse de corde. La représentation d'évaluation de spin $s$
s'obtient donc comme sous-représentation du vecteur de plus haut
poids de la représentation $V^{\otimes 2s}$ ($V$ \rep\ de spin $1/2$)
avec les paramètres spectraux $\lambda_1$, $\ldots$, $\lambda_{2s}$.

De manière générale, pour deux représentations de polynômes $P(x)$
et $Q(x)$, leur produit tensoriel est irréductible si et seulement si
les racines de $P(x)$ et de $Q(x)$ mises ensemble ne forment pas de
nouvelles cordes (cas ``générique''). Pour une analyse complète du
cas non-générique, nous renvoyons le lecteur à [\CP].

\newsec\secanis{Anisotropie}
Nous avons maintenant les outils nécessaires pour
généraliser la chaîne XXX ou le modèle NJL
en introduisant une anisotropie dans l'interaction spin-spin.
Remarquons que nous ne considèrerons pas l'anisotropie la plus générale
(les trois constantes de couplages différentes), car
elle serait associée aux solutions {\it elliptiques} de
l'Equation de Yang--Baxter (et aux groupes quantiques elliptiques)
dont nous n'avons pas parlé; cependant, ce cas de figure
ne nous intéresse pas directement car il est lié à des modèles
sur réseau non-critiques.

\subsec\secanisb{La chaîne XXZ}
Considérons donc un Hamiltonien de la chaîne XXZ, qui
agit sur le même Hilbert ${\cal H}=V^{\otimes M}$ avec $V={\Bbb C}^2$,
et qui est donné par
$$\H_\XZ=\H_\XZ(\Delta)=
-2\sum_{i=1}^M (\s_k^1 \s_{k+1}^1+\s_k^2 \s_{k+1}^2+\Delta \s_k^3 \s_{k+1}^3)
\eqn\hamxxz$$
où $\Delta$ est un réel fixé. Ce Hamiltonien ne généralise pas
de manière évidente le Hamiltonien XXX \hamxxx, du fait du signe $-$
dans \hamxxz. Cependant, si l'on effectue la transformation
$T=\prod_{k\ impair} \s_k^3$,
alors le nouveau Hamiltonien $\tilde{\H}_\XZ(\Delta)\equiv
T \H_\XZ(\Delta)T^{-1}$ vaut
$$\tilde{\H}_\XZ(\Delta)=
2\sum_{i=1}^M (\s_k^1 \s_{k+1}^1+\s_k^2 \s_{k+1}^2-\Delta \s_k^3
\s_{k+1}^3)$$
soit encore: $\H_\XZ(\Delta)=-\tilde{\H}_\XZ(-\Delta)$.
En particulier, $\H_\X=\tilde{\H}_\XZ(\Delta=-1)$. Evidemment,
les deux Hamiltoniens $\H_\XZ(\Delta)$ 
et $\tilde{\H}_\XZ(\Delta)$ n'exhibent pas les mêmes
symétries: par exemple, $\tilde{\H}_\XZ(\Delta=-1)=\H_\X$
a une invariance $SU(2)$, alors que cette dernière n'est pas manifeste
pour $\H_\XZ(\Delta=-1)$. De manière générale, on verra que
$\H_\XZ(\Delta)$ 
est naturellement associé à un groupe quantique avec un paramètre
de déformation $q$, où $\Delta=(q+q^{-1})/2$; 
alors, $\tilde{\H}_\XZ(\Delta)=-\H_\XZ(-\Delta)$ est
associé à un paramètre de déformation $-q$. Cependant,
ces deux groupes quantiques sont reliés de manière triviale
l'un à l'autre.
Dorénavant, de façon à nous
conformer aux conventions de la littérature
sur le sujet, nous ne considérerons plus le Hamiltonien
de la chaîne XXZ que sous la forme $\H_\XZ$ (et non
$\tilde{\H}_\XZ$); en particulier, la limite isotrope correspond
pour nous à $q\to -1$, et non à $q\to 1$.

Le Hamiltonien \hamxxz\ n'est pas invariant sous $SU(2)$, mais seulement
sous l'action de sa sous-algèbre de Cartan $U(1)$: $[\H_\XZ,\S^3]=0$. Y a-t-il
une symétrie naturelle de la théorie,
qui puisse généraliser $SU(2)$?

Une première suggestion de réponse apparaît en remarquant
que la chaîne XXZ, {\it avec des
conditions aux bords ouvertes particulières} [\ref\PS{V.~Pasquier
et H.~Saleur, {\it Nucl.\ Phys.} B330 (1990), 523.}], 
possède une symétrie $U_q({\goth sl}(2))$ où
$$\Delta={q+q^{-1}\over 2}\eqn\valq$$

Cependant, cette symétrie n'est ``pas suffisante'' pour nos besoins;
en effet, on a vu aux paragraphes précédents que pour construire
une matrice $R(\lambda)$ avec paramètre spectral, il nous faut
une symétrie beaucoup plus grosse qu'un simple groupe quantique.
Il est donc naturel de considérer le groupe quantique affine
$U_q(\widehat{{\goth sl}(2)})$ (pour $q$ donné par la même valeur
\valq), et de le faire agir sur le Hilbert ${\cal H}=V^{\otimes M}$
de la chaîne XXZ.
$U_q(\widehat{{\goth sl}(2)})$ agit sur $V$ par la
représentation de spin $1/2$ et de paramètre spectral $x=1$,
et donc sur $\cal H$ par co-produit. 
On introduit ensuite
un espace auxiliaire $V_a$, de rapidité $x$,
et la matrice de monodromie
$$T_a(x)=R_{aM}(x)R_{aM-1}(x)\ldots
R_{a1}(x)\eqn\defTani$$
où $R(x)$ est la matrice $R$ donnée par \affR.
Choisissons maintenant la paramétrisation suivante: posons
$q=-\e{-i\gamma}$, soit $\Delta=-\cos\gamma$,
et $x=\e{\gamma\lambda}$. La matrice $R$ est alors donnée par
$$R(\lambda)=\pmatrix{
1&0&0&0\cr
0&{\sinh(\gamma\lambda)\over\sinh(\gamma(i-\lambda))}&
{\sinh(i\gamma)\over\sinh(\gamma(i-\lambda))}&0\cr
0&{\sinh(i\gamma)\over\sinh(\gamma(i-\lambda))}&
{\sinh(\gamma\lambda)\over\sinh(\gamma(i-\lambda))}&0\cr
0&0&0&1\cr
}\eqn\explRq$$

Pour avoir des conditions aux bords périodiques,
on pose $\Z(x)=\tr_a(T_a(x))$. On a alors la relation:
$$\H_\XZ=  i{\sin\gamma\over \gamma} \Z(\lambda)^{-1}{\d\over\d\lambda}
\Z(\lambda)_{|\lambda=0} + {\rm cst}\eqn\Zhamb$$
Il est pratique de redéfinir $\H_\XZ \to {\gamma\over\sin\gamma} \H_\XZ$
pour se débarrasser de la constante multiplicative dans \Zhamb.
C'est ce que nous ferons dorénavant.

Par définition de la matrice $R(\lambda)$, on a
$$T_a(\lambda)g_{a\cal H}=g_{{\cal H}a} T_a(\lambda)$$
où $g_{a{\cal H}}$ (resp. $g_{{\cal H}a}$)
est l'action d'un élément $g$ de $U_q({\widehat{\goth sl}(2)})$
sur $V_a\otimes{\cal H}$ (resp. ${\cal H}\otimes V_a$);
donc ``modulo les termes de bord'', c'est-à-dire
en oubliant l'espace auxiliaire, $T_a(\lambda)$ commute avec
l'action de $U_q(\widehat{{\goth sl}(2)})$ sur $\cal H$.
Cependant, ces termes de bord ne disparaissent
pas quand on prend des conditions aux bords périodiques, et donc
ni l'algèbre entière $U_q(\widehat{{\goth sl}(2)})$ ni même
une quelconque sous-algèbre $U_q({\goth sl}(2))$ ne commutent avec
le Hamiltonien $\H_\XZ$; on peut seulement considérer 
que le groupe quantique affine constitue en un sens vague une symétrie du
Hamiltonien de la chaîne XXZ dans la limite $M\to\infty$ (puisque
le manque de commutation entre $\H_\XZ$ et l'action de
$U_q(\widehat{{\goth sl}(2)})$ est lié à des termes de 
bord).
Ce point de vue est celui qui est la base des travaux [\JM], mais 
leur optique est suffisamment différente de la nôtre pour
que nous ne nous attardions pas sur ce point.

Signalons immédiatement que,
de la même manière que l'on passe de la
chaîne XXX à la chaîne XXZ, on peut considérer un modèle NJL déformé,
dont le Lagrangien est [\ref\JNW{G.I.~Japaridze,
A.A.~Nersesyan et P.B.~Wiegmann, {\it Nucl. Phys.} B230
(1984), 511.}]:
$$L=i\psib^a \dsl \psi^a - g j^3_\mu j^3{}^\mu
-f (j^1_\mu j^1{}^\mu+j^2_\mu j^2{}^\mu)\eqn\lagGNdef$$
Aussi bien la chaîne XXZ que le modèle NJL déformé
nous conduisent à des cas particuliers de la matrice de monodromie inhomogène
$$T_a(x|y_1,\ldots,y_M)=
R_{aM}(x y_M^{-1})R_{aM-1}(x y_{M-1}^{-1})\ldots R_{a1}
(x y_1^{-1})\eqn\defTinb$$
Pour le modèle NJL déformé, $\gamma$ est donné
par $\cos\gamma=\cos g/\cos f$
et les inhomogénéités $y_k=\e{\pm A}$
où $\tanh A=\cos f\, \sin\gamma/\sin g$ (de sorte que $A\to\infty$
quand $f\to 0$).

Comme au \secTin, on peut définir un modèle statistique
sur réseau à partir de la matrice de monodromie \defTinb: on obtient
ainsi le {\it modèle à 6 vertex général}. En effet,
la matrice la plus générale qui conserve la
charge $U(1)$ et qui est ${\Bbb Z}_2$-invariante par
retournement des flèches a les mêmes $6$ poids non-nuls
que $R(\lambda)$ et dépend de trois paramètres indépendants,
qui sont pour nous la normalisation globale de la matrice $R$, 
le paramètre de déformation $q$ et le paramètre spectral $x$.

Une fois que l'on a compris que l'anisotropie nous forçait simplement
à remplacer le Yangien $Y({\goth sl}(2))$ par le groupe
quantique affine $U_q(\widehat{{\goth sl}(2)})$, on peut recommencer
l'Ansatz de Bethe algébrique pour
diagonaliser la matrice de transfert $\Z(\lambda)=\tr_a(T_a(\lambda))$. 
La procédure est inchangée, mais elle nous conduit
à de nouvelles équations d'Ansatz de Bethe:
$$\prod_{\scriptstyle \beta=1\atop\scriptstyle \beta\ne \alpha}^m 
{x_\alpha x_\beta^{-1} q^{-1}-x_\alpha^{-1} x_\beta q
\over x_\alpha x_\beta^{-1} q-x_\alpha^{-1} x_\beta q^{-1}}
= \prod_{k=1}^M
{x_\alpha y_k^{-1}-x_\alpha^{-1} y_k
\over x_\alpha y_k^{-1}q-x_\alpha^{-1} y_k q^{-1}}\eqn\xxzbae$$

Nous nous limiterons dorénavant à la phase dite non-massive,
c'est-à-dire $-1<\Delta<1$, soit encore $|q|=1$. Dans la
paramétrisation $q=-\e{-i\gamma}$, $\gamma$ est réel, $0<\gamma<\pi$.

Les BAE \xxzbae\
se récrivent alors, après le décalage des $\lambda_\alpha$ de $i/2$:
$$\prod_{\scriptstyle \beta=1\atop\scriptstyle \beta\ne \alpha}^m 
{\sinh(\gamma(\lambda_\alpha-\lambda_\beta+i))
\over \sinh(\gamma(\lambda_\alpha-\lambda_\beta-i))}
=\prod_{k=1}^M 
{\sinh(\gamma(\lambda_\alpha-\theta_k+i/2))
\over \sinh(\gamma(\lambda_\alpha-\theta_k-i/2))}\eqn\xxzbaeb$$
où $y_k\equiv \e{\gamma\theta_k}$.
De même, la valeur propre $\e{-ip_kL}$ de $\Z(\theta_k)$ correspondant
à l'état de Bethe caractérisé par les $\lambda_\alpha$ vaut:
$$\e{-ip_kL}=
\prod_{\alpha=1}^m {\sinh(\gamma(\theta_k-\lambda_\alpha+i/2))
\over\sinh(\gamma(\theta_k-\lambda_\alpha-i/2))}
\eqn\xxzbaec$$
Une différence essentielle avec le cas isotrope
consiste en ce que, le système n'ayant plus
l'invariance ${\goth sl}(2)$ (ni l'invariance de groupe
quantique du fait des conditions de bord), on ne peut plus
rajouter de racines $\lambda_\alpha=\infty$ à une solution donnée;
ainsi, on obtient {\it tous} les vecteurs propres de la matrice
de transfert (et non pas seulement des vecteurs de plus haut poids)
en ne considérant que des racines $\lambda_\alpha$ finies
[\ref\KIL{A.N.~Kirillov et N.A.~Liskova, {\it J.\ Phys.} A30 (1997),
1209 \pre{hep-th/9403107}\semi
A.N.~Kirillov et N.A.~Liskova, \pre{hep-th/9607012}.}].

Dans la région $-1<\Delta<1$ que nous considérons, le vide de la
théorie, comme dans le cas isotrope, est non-trivial: il est obtenu
en remplissant de racines des BAE \xxzbaeb\ l'axe réel. 

Dans la limite thermodynamique, il faut trouver une hypothèse de corde
appropriée: celle-ci existe [\ref\TS{M.~Suzuki et M.~Takahashi,
{\it Prog.\ Theor.\ Phys.} 48 (1972), 2187.}], mais
elle est relativement compliquée pour une valeur quelconque
de $\gamma$ (elle dépend du développement en fraction continue de
$\gamma/\pi$). Nous nous limiterons donc ici à la valeur
$\gamma=\pi/(p+1)$, $p>1$ entier (nous
reviendrons à une valeur arbitraire de $\gamma$ quand
nous parlerons de NLIE, section \secddv), qui a un intérêt tout particulier
pour nous. L'hypothèse de corde s'exprime alors ainsi:
les racines des BAE \xxzbaeb\ forment soit des $j$-cordes (centrées
sur l'axe réel) avec $1\le j\le p$, soit des ``$1^-$-cordes'',
c'est-à-dire des racines de partie
imaginaire ${\pi\over 2\gamma}$. En utilisant des
techniques similaires à celles employées au \secbaehypco,
on obtient le nouveau système d'Equations d'Ansatz de Bethe continues:
$$\eqalign{
C_{jk}\star \rhot_k + \rho_j-\delta_{j\,p-1} s\star\rho_{1^-}
&=\delta_{j1} \sigma\star s(\lambda)\quad 1\le j\le p\cr
\rhot_{1^-}-s\star\rhot_{p-1}+\rho_{1^-}&=0\cr}
\eqn\xxzcobae$$
$C_{jk}$ est la matrice de Cartan avec paramètre spectral
de $A_p$, c'est-à-dire qu'elle est donnée par \defC,
mais avec les conditions de bord $1\le j,k\le p$. Comme nous
nous resservirons de la matrice de Cartan de $A_p$ plus
tard, donnons son inverse en transformée de Fourier:
$$C_{jk}^{-1}(\kappa)=C_{kj}^{-1}(\kappa)=2 \coth(\kappa) 
{\sinh((p+1-k)\kappa) \sinh(j\kappa)
\over \sinh((p+1)\kappa)}\qquad k\ge j\eqn\invCp$$
Les BAE \xxzcobae\
peuvent être décrites par la figure \baeq.
\fig\baeq{BAE de la chaîne XXZ à $\gamma=\pi/(p+1)$.}{\figdir{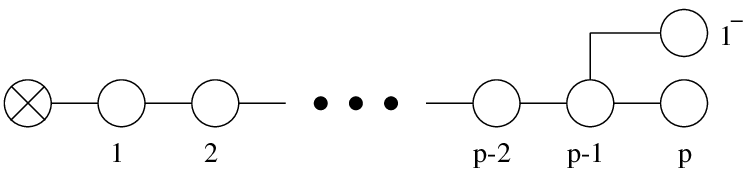}} 
Le noeud supplémentaire lié au noeud $p-1$ est celui qui correspond
aux $1^-$-cordes: le lien entre les deux noeuds signifie qu'il
y a un terme $-s\star\rhot_{p-1}$ dans l'équation du noeud $1^-$,
et qu'il a un terme $-s\star\rho_{1^-}$ dans l'équation du noeud
$p-1$.

L'énergie-impulsion a précisément la même forme \EPphys-\EPphysb\ que
dans le cas isotrope (rappelons que nous avons redéfini la
normalisation de $\H_\XZ$ à cet effet). Les excitations physiques
sont donc des trous dans la mer de Fermi de racines réelles des BAE,
et elles ont les relations de dispersion
$\tilde{\epsilon}=\pi\sin\tilde{p}$ pour la chaîne XXZ,
$\tilde{\epsilon}^2=\tilde{p}^2+m^2$ pour le modèle NJL anisotrope
(après avoir pris la limite d'échelle $D\to\infty$, $m\equiv
2D\e{-\pi/c}$ fixée).

Bien sûr, le manque de symétrie du Hamiltonien à $L<\infty$
n'empêche pas
les excitations physiques d'avoir une structure ``isotopique'',
liée à la possibilité d'insérer des $j$-cordes, $j\ge 2$, et
des $1^-$-cordes. Nous reviendrons sur ce point au \secphysisoq;
nous déterminerons alors également la matrice $S$ des excitations
physiques.

\subsec\secanistronc{Troncation de groupe quantique}
Comme nous l'avons dit, le système que nous venons de décrire n'est
pas invariant par l'action d'un groupe quantique du fait
des conditions de bord. Il y a deux manières de rétablir
cette invariance: 
\item{$\star$} soit, comme nous l'avons expliqué précédemment,
en prenant des conditions de bord ouvertes particulières [\PS];
ceci a le désavantage de modifier l'aspect des BAE,
du fait qu'elles doivent maintenant incorporer les
effets de réflection aux bord [\ref\SKL{E.K.~Sklyanin, {\it J.\ Phys.} 
A21 (1988), 2375.}].
\item{$\star$} soit, pour $q$ racine de l'unité (donc en particulier
pour $\gamma=\pi/(p+1)$), on peut modifier la matrice de transfert
(tout en conservant les conditions de bord périodiques)
[\ref\KZ{M.~Karowski et A.~Zapletal, {\it Nucl.\ Phys.} B419
(1994), 567 \pre{hep-th/9312008}.}] de façon à restaurer la symétrie
$U_q({\goth sl}(2))$. Il est alors naturel de changer
la gradation du groupe quantique affine
en la gradation homogène, de façon à faire de la symétrie
$U_q({\goth sl}(2))$ une sous-algèbre horizontale\foot{En fait,
le choix de la gradation nous est imposé par le fait
que, dans la limite conforme,
il existe une nouvelle action de l'algèbre de
Virasoro qui modifie le spin
des générateurs du groupe quantique -- voir chapitre \chaptba.}.
Il est par contre impossible de restaurer
la symétrie du groupe quantique affine complet.

Nous choisirons donc la deuxième solution: elle conduit à des
BAE quasiment identiques à \xxzbaeb\ (il y a simplement des
``twists'' supplémentaires\foot{Ceux-ci compensent exactement
les ``twists'' qui apparaissent par l'action des générateurs
du groupe quantique, assurant ainsi l'invariance des BAE.}); en
particulier dans la limite thermodynamique, les BAE continues
ont la même forme \xxzcobae.

On peut alors implémenter la ``troncation de groupe
quantique''. Rappelons que celle-ci consiste à tronquer le produit
tensoriel $V^{\otimes M}$ qui constitue le Hilbert physique de façon
à retirer les mauvaises \rep\ au fur et à mesure
qu'on effectue les produits tensoriels succesifs par $V$. Comme
nous l'avons déjà signalé, cette
procédure n'équivaut pas à simplement effectuer
le produit tensoriel complet $V^{\otimes M}$, {\it puis} retirer
les mauvaises \rep: en effet, il existe des bonnes
\rep\ qui ont été obtenues par produit tensoriel
de mauvaises \rep\ et qui sont ainsi conservées
alors qu'elles ne devraient pas appartenir au Hilbert tronqué 
${\cal H}_{tronc}$.
Une bonne visualisation de la procédure de troncation se fait
grâce aux {\it diagrammes de Bratteli} (figure \brat). 
\fig\brat{Un diagramme de Bratteli. Le spin est
placé en ordonnée.}{\figdir{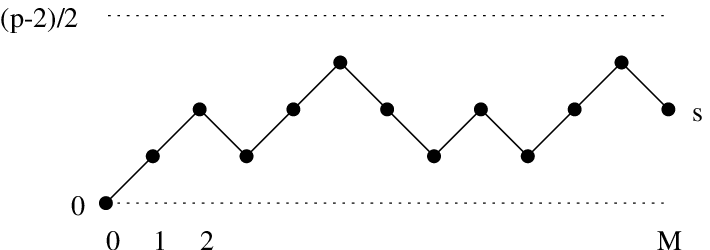}} 
On part de la
représentation triviale, de spin $s=0$,
puis on effectue le produit tensoriel par $V$ une première
fois, et on obtient $s=1/2$, puis une seconde fois et on
obtient $s=0$ ou $s=1$, etc,
jusqu'au cran $M$ où l'on obtient le spin de l'état. 
A chaque diagramme allant du spin $0$ au spin $s$ correspond
exactement une sous-rep de spin $s$ de $V^{\otimes M}$, et on
a ainsi construit une décomposition de $V^{\otimes M}$ en
sous-rep irréductibles.
La troncation consiste alors simplement à ne retenir que les
diagrammes qui sont entièrement en-dessous de la valeur
maximale $s=(p-1)/2$.
Les diagrammes de Bratteli
sont directement liés à la formulation IRF (``Interaction Round
a Face'') [\ref\ABF{G.E.~Andrews, R.J.~Baxter et P.J.~Forrester,
{\it J. Stat. Phys.} 35 (1984), 193.}] des modèles qui nous intéressent (en
l'occurrence les modèles IRF équivalents sont les
modèles SOS/RSOS), mais il n'est
pas possible, par manque de place, de discuter de cette formulation ici.

Au niveau de l'Ansatz de Bethe, la troncation implique de ne
conserver que les états de Bethe qui sont les vecteurs
de plus haut poids de bonnes représentations (de spin
$s\le (p-1)/2$) de $U_q({\goth sl}(2))$,
et qui sont dans ${\cal H}_{tronc}$. Il y a plusieurs manières
équivalentes de caractériser ces états (voir par exemple
[\ref\JK{G.~Jüttner et M.~Karowski, {\it Nucl.\ Phys.} B430 (1994), 615
 \pre{hep-th/9406183}.},
\ref\DDVQG{C.~Destri and H.J.~De~Vega, {\it Nucl.\ Phys.}
B385 (1992), 361 \pre{hep-th/9203065}\semi
H.J.~De~Vega, Lectures given at the Vth. Nankai Workshop (Tianjin,
P. R. of China, June 1992) \pre{hep-th/9308008}.}]
pour les conditions de bord ouvertes);
dans la limite thermodynamique,
on peut les identifier par les propriétés suivantes\foot{Cette
caractérisation a été
trouvée dans [\ref\BR{V.V.~Bazhanov et Yu.N.~Reshetikhin,
{\it Int.\ J.\ Mod.\ Phys.} A4 (1989), 115.}],
bien qu'elle ait été obtenue en supposant
le spin $U_q({\goth sl}(2))$ nul, ce qui est inutile -- voir note suivante.}:
$\rho_p=\rhot_p=\rho_{1^-}=\rhot_{1^-}=0$,
et donc, d'après les BAE \xxzcobae, $\rhot_{p-1}=0$. La densité
de $(p-1)$-cordes $\rho_{p-1}$ n'est, elle, pas nécessairement nulle,
(ce qui jouera un rôle par la suite),
mais comme les différentes équations
ne sont couplées entre elles que par les $\rhot_j$, on peut
supprimer le noeud $p-1$ sur le diagramme, d'où la figure \truncbae.
\fig\truncbae{Troncation de groupe quantique
dans les BAE.}{\figdir{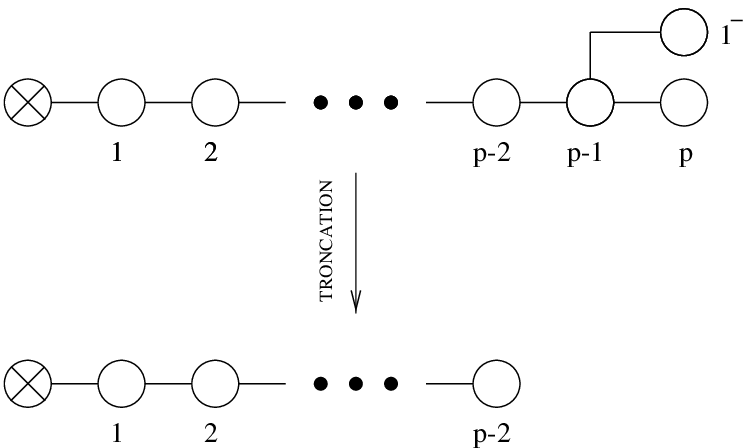}} 
Il y a encore correspondance entre noeuds des BAE tronquées
et bonnes \rep\ (à condition d'exclure les \rep\ ``triviales''
de spins $0$ et $(p-1)/2$).
Nous ne donnerons pas la valeur du spin pour les états de Bethe
de ${\cal H}_{tronc}$, car la situation est plus compliquée que
dans le cas isotrope: ceci est dû au fait que le groupe quantique
``nu'' (celui dont on a parlé jusqu'à présent)
et le groupe quantique ``physique'' (qui sera introduit
au \secphysisoq) sont distincts. Bornons nous à constater que,
dans la limite thermodynamique, $s/L$ (le spin par unité de longueur)
est nécessairement nul puisque $s\le (p-1)/2$ tandis que
$L\to\infty$\foot{En particulier,
en remplaçant l'hypothèse $s=0$ par l'hypothèse plus
faible $s/L\to 0$ dans [\BR], on rend le raisonnement de cet article
applicable à des états de spin quelconque.}.

\subsec\secphysisoq{BAE nues/BAE physiques: cas anisotrope}
On peut également obtenir des BAE physiques pour le cas anisotrope.
Pour ce faire, dessinons les diagrammes (non tronqués)
correspondant aux BAE physiques
et aux BAE nues pour $\gamma=\pi/(p+1)$ (figure \baec).
\fig\baec{Passage des BAE nues aux BAE physiques
dans le modèle NJL anisotrope. Les numéros sont les longueurs
des cordes {\it nues}.}{\figdir{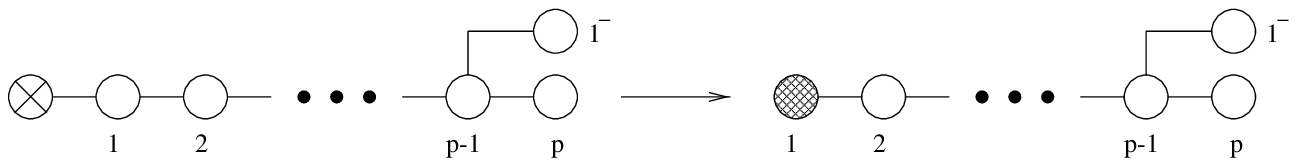}} 
La situation est essentiellement la même que dans le cas isotrope,
à une différence cruciale près: le diagramme auquel on a retiré
un noeud n'est pas identique au diagramme complet; il correspond
maintenant à une nouvelle valeur $\tilde{\Delta}=-\cos\tilde{\gamma}$,
avec $\tilde{\gamma}=\pi/p$, de l'anisotropie.

Une fois cette différence observée, on conjecture que les excitations
physiques sont des doublets du ``groupe quantique affine
physique'' $U_{\tilde{q}}(\widehat{{\goth sl}(2)})$
($\tilde{q}=-\e{-i\tilde{\gamma}}$), et que leur matrice $S$ est donnée
par $S(\lambda)=\Sh(\lambda)R(\lambda)$, où $\Sh$ est un facteur
scalaire et $R$ est donnée par \affR\ 
pour le paramètre de déformation $\tilde{q}$
(et où $x=\e{\tilde{\gamma}\lambda}$).
Un calcul analogue à celui qui a été fait dans le cas
isotrope nous permet de calculer $\Sh(\lambda)$ (à une phase globale
près):
$$\eqalign{
\Sh(\lambda)&=\exp\left(i \int_0^{+\infty} \d\kappa\, 
{\sin(2\kappa\lambda)\over \kappa}
{\sinh((p-1)\kappa) \over \sinh(p\kappa)\cosh(\kappa)}
\right)\cr
&=
{ \Gamma\left(1+i{\lambda\over 2}\right)
  \Gamma\left({1\over 2}-i{\lambda\over 2}\right)
\over
  \Gamma\left(1-i{\lambda\over 2}\right)
  \Gamma\left({1\over 2}+i{\lambda\over 2}\right)}
\prod_{n=1}^\infty
{\Gamma\left({np\over 2}+1+i{\lambda\over 2}\right)
\Gamma\left({np+1\over 2}-i{\lambda\over 2}\right)^2
\Gamma\left({np\over 2}+i{\lambda\over 2}\right)
\over
\Gamma\left({np\over 2}+1-i{\lambda\over 2}\right)
\Gamma\left({np+1\over 2}+i{\lambda\over 2}\right)^2
\Gamma\left({np\over 2}-i{\lambda\over 2}\right)}
\cr}
\eqn\sGS$$
Cette matrice $S$ s'identifie à la matrice $S$ soliton-soliton de
Sine--Gordon pour une constante de couplage $\beta^2/8\pi=p/(p+1)$,
ce qui est un signe de l'équivalence des modèles NJL anisotrope
et de Sine--Gordon. Le lien exact avec Sine--Gordon sera discuté
au \secddvuv.
Notons aussi que la nature des excitations physiques,
et l'expression de leur matrice $S$, que nous avons
obtenu pour $\gamma=\pi/(p+1)$, $p$ entier, sont encore
valables pour $\gamma$ arbitraire, $0<\gamma<\pi/2$, soit
$p+1=\pi/\gamma$ non nécessairement entier.

Dans la limite thermodynamique $L\to\infty$,
on voit d'après l'expression de l'énergie, qui ne dépend
que de la position des trous, et pas de leur spin, que
le Hamiltonien commute avec l'action du groupe quantique affine
$U_{\tilde{q}}(\widehat{{\goth sl}(2)})$: 
c'est cette symétrie
qui explique la dégénérescence supplémentaire du spectre
à $L=\infty$. De même, dans la limite isotrope
$\gamma\to 0$, on voit que la symétrie $SU(2)$ est étendue en une
symétrie $Y({\goth sl}(2))$ à $L=\infty$.

La situation reste inchangée quand on opère la troncation: comme
on le voit diagrammatiquement sur la figure \baed\ (et comme on peut
en effet le vérifier explicitement [\DDVQG]), la troncation vis-à-vis
du groupe quantique nu $U_q({\goth sl}(2))\subset
U_q(\widehat{{\goth sl}(2)})$
est {\it équivalente} à la troncation vis-à-vis
d'un groupe quantique physique $U_{\tilde{q}}({\goth sl}(2))\subset
U_{\tilde{q}}(\widehat{{\goth sl}(2)})$. Ainsi le Hilbert des excitations
de basse énergie a la forme d'un espace de Fock {\it tronqué}.
\fig\baed{Passage des BAE nues aux BAE physiques
dans le modèle NJL anisotrope tronqué.}{\figdir{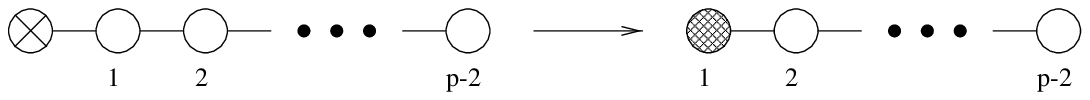}} 
Une manière agréable de visualiser la troncation consiste
à considérer les excitations physiques comme
des {\it solitons} (pour une revue des solitons
dans les modèles intégrables,
voir [\ref\KM{T.~Klassen et E.~Melzer, {\it Nucl.\ Phys.} B382 (1992),
441 \pre{hep-th/9202034}.}]).
Pour cela, assignons à chaque nombre quantique $s$ 
(spin de $U_{\tilde{q}}({\goth sl}(2))$, $0\le 2s\le f\equiv p-2$),
un vide\foot{Ceci veut-il dire
que la théorie que nous considérons admet réellement plusieurs vides?
En fait, la réponse à cette question dépend des conditions de
bord. Pour celles que nous avons prises (voir plus loin),
le vide correspond nécessairement à la représentation triviale $s=0$.}.
D'après la loi de composition des spins $s\otimes\ha=(s-\ha)\oplus(s+\ha)$
(avec la restriction $0\le 2s\le f$),
une particule de spin $1/2$ doit être considérée
comme un soliton qui connecte deux vides adjacents (figure \soli).
\fig\soli{Vision intuitive des différents vides.}{\figdir{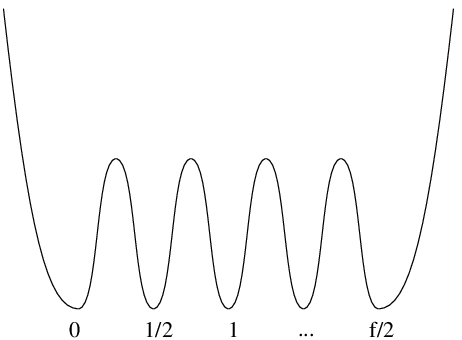}} 
On part donc d'un vide à $x=-\infty$ qui est nécessairement pour
nous $s=0$; puis on place des solitons qui produisent des
transitions vers des vides $s>0$; la troncation est naturellement
implémentée par le fait qu'on ne peut aller plus loin que
le vide $s=f/2$.

Dans le langage solitonique, un état à deux solitons est spécifié
par la donnée de trois vides, tandis que
la diffusion de deux solitons est donnée par {\it quatre} vides:
en effet, seul le vide intermédiare peut changer entre avant
et après la collision. La matrice $S$ est donnée dans ce langage
dans par exemple [\ref\FSZ{P.~Fendley, H.~Saleur et Al.B.~Zamolodchikov,
{\it Int.\ J.\ Mod.\ Phys.} A8 (1993), 5751 \pre{hep-th/9304051}.}]; 
retrouvons-la explicitement à partir de notre matrice $S(\lambda)
=\Sh(\lambda)\tilde{R}(\lambda)$, où l'on utilise
la matrice $\tilde{R}$ donnée par \affRb\ 
(gradation homogène) de $U_{\tilde{q}}(\widehat{{\goth sl}(2)})$.
Du fait que $\check{S}(\lambda)\equiv {\cal P} S(\lambda)$ commute
avec l'action du groupe quantique $U_{\tilde{q}}({\goth sl}(2))$\foot{mais
ne commute pas avec l'action du
groupe quantique affine tout entier! En effet,
$\check{S}$ ne fait qu'{\it entrelacer} les représentations
sur $V_1\otimes V_2$ et $V_2\otimes V_1$, qui sont distinctes
à cause des paramètres spectraux. Par contre, grâce au changement
de gradation, l'action du sous-groupe horizontal
$U_{\tilde{q}}({\goth sl}(2))$ ne dépend pas du paramètre spectral.},
il peut se décomposer
en projecteurs sur les différentes sous-irrep: on trouve
$$\check{S}(\lambda)=\Sh(\lambda)\left(P_1-{\sinh(\tilde{\gamma}(\lambda+i))
\over\sinh(\tilde{\gamma}(\lambda-i))}P_0\right)\eqn\decS$$
Il faut maintenant convertir
ce résultat dans la base naturelle du langage solitonique:
celle des chemins sur un diagramme de Bratteli (figure \solibrat).
\fig\solibrat{Les quatre chemins associés à deux excitations
de spin $1/2$.}{\figdir{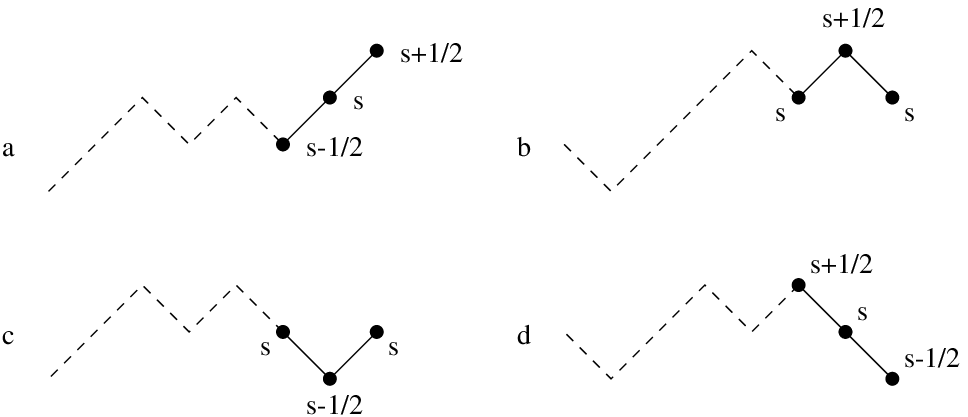}} 
Ceci revient à décomposer la représentation $s\otimes\ha\otimes\ha$
de deux manières différentes: $s\otimes(1\oplus 0)$ pour \decS,
$(s+\ha\oplus s-\ha)\otimes\ha$ pour les diagrammes de la figure
\solibrat. Les diagrammes a et d sont clairement contenus
dans $s\otimes 1$, et donc l'élément de matrice $S$
associé n'est autre que
$\Sh(\lambda)$. Par contre, les diagrammes b et c mélangent
$s\otimes 1$ et $s\otimes 0$, et la matrice $\check{S}$ n'est plus
diagonale dans la base (b,c): il peut y avoir des transitions
b$\leftrightarrow$c. Le calcul des coefficients $6j$ qui
donnent le changement de base nous conduit aux formules suivantes:
$$\eqalign{
S(a\rightarrow a)=S(d\rightarrow d)&=\Sh(\lambda)\cr
S(b\rightarrow c)=S(c\rightarrow b)&={\q{2s+2}^{1/2} \q{2s}^{1/2}\over\q{2s+1}}
{\sinh(\tilde{\gamma}\lambda)\over\sinh(\tilde{\gamma}(\lambda-i))}\Sh(\lambda)\cr
S(b\rightarrow b)&=-{1\over\q{2s+1}}
{\sinh(\tilde{\gamma}(\lambda+i(2s+1)))\over\sinh(\tilde{\gamma}(\lambda-i))}
\Sh(\lambda)\cr
S(c\rightarrow c)&={1\over\q{2s+1}}
{\sinh(\tilde{\gamma}(\lambda-i(2s+1)))\over\sinh(\tilde{\gamma}(\lambda-i))}
\Sh(\lambda)\cr
}\eqn\kinkS$$
où $\q{x}\equiv \sin(\tilde{\gamma}x)/\sin(\tilde{\gamma})$.
La formulation solitonique de la matrice $\check{S}$ \kinkS\ est
plus compliquée que sa formulation usuelle \decS, du fait
qu'intervient explicitement $s$, le nombre quantique qui décrit
dans quels vides se situe l'interaction.

Notons également que la transition $c\to b$ est interdite
pour $s=f/2$ (car $\q{f+2}=0$), ce qui est une justification
{\it a posteriori} de la stabilité du Hilbert tronqué ${\cal
H}_{tronc}$ (cf [\ABF] pour une remarque similaire).

\newsec\secfus{Spin plus élevé et fusion}
Jusqu'à présent, nous avons uniquement utilisé la représentation de
spin $1/2$ de $SU(2)$ (ou sa déformation dans le cas anisotrope). 
Pourquoi ne pas utiliser les représentations
de spin supérieur? Pour cela, il faut construire les matrices $R(\lambda)$
correspondantes. Ce problème est intimement lié à la théorie
de la représentation de $Y({\goth sl}(2))$ (ou $U_q(\widehat{{\goth sl}(2)})$),
et en fournit une bonne illustration.
Pour simplifier la discussion, nous nous replaçons tout
d'abord dans le cas isotrope.

\subsec\secfusb{La procédure de fusion}
Nous avons vu que les matrices $R(\lambda)$ sont simplement
les matrices dans une représentation donnée de la 
matrice $R$ universelle de $Y({\goth sl}(2))$. 
Mais ceci ne nous fournit pas de manière
explicite de calculer $R(\lambda)$.
Le plus simple, pour y parvenir,
est d'utiliser la procédure de fusion. Pour le restant de cette
section, nous redéfinirons la matrice $R$: $R(\lambda)=\lambda-i{\cal P}$,
supprimant le dénominateur $\lambda-i$ qui est un simple facteur scalaire.
Evidemment, cette redéfinition brise l'unitarité: on a
la condition plus faible $R_{12}(\lambda)R_{21}(-\lambda)=\rho(\lambda)1$.

La fusion est implémentée par l'équation de quasi-triangularité:
pour $3$ \rep\ $V_1$, $V_2$, $V_3$ de paramètres spectraux
$\lambda_1$, $\lambda_2$, $\lambda_3$, on a, d'après \quatri:
$$R_{1(23)}(\lambda_1|\lambda_2,\lambda_3)=R_{13}(\lambda_1-\lambda_3)
R_{12}(\lambda_1-\lambda_2)\eqn\quatrirep$$
où $R_{1(23)}$ est la matrice $R$ entre $V_1$ et $V_2\otimes V_3$\foot{Du
fait de l'inhomogénéité de la condition
de quasi-triangularité par rapport à $R$, la normalisation de $R$
importe; si l'on multiplie $R$ par un facteur scalaire,
alors \quatrirep\ doit être modifiée de la même
façon que l'équation d'unitarité.}.
Ainsi, la quasi-triangularité permet de construire la matrice $R$
dans des représentations produit tensoriel à partir de celles des
composantes du produit tensoriel. La matrice 
de monodromie inhomogène $T_a$ introduite au \secaba, en
particulier, n'est autre
que la matrice $R$ entre l'espace auxiliaire $V_a$ et l'espace de
Hilbert entier $\cal H$.
\fig\bs{Représentation imagée de la quasi-triangularité,
ou du bootstrap.}{\figdir{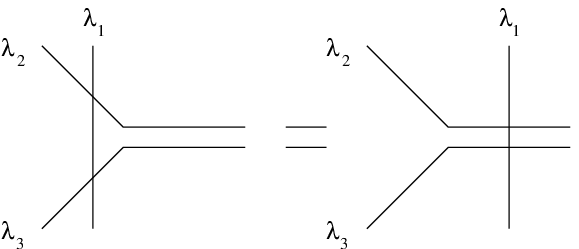}} 

Le lecteur avisé aura reconnu en \quatrirep\ une version algébrisée de la
procédure de bootstrap (figure \bs) pour les matrices $S$. Evidemment,
il n'y a plus de pôles, car on a supprimé tous les facteurs scalaires dans
les matrices $R$: les ``états liés'' potentiels
seront donc déterminés par la
condition que la matrice $R$ laisse stable un sous-espace propre
de $V_2\otimes V_3$. L'idée est dès lors que,
toute représentation de $Y({\goth sl}(2))$ pouvant être extraite
du produit tensoriel des \rep\ d'évaluation de spin $1/2$,
on obtiendra la matrice $R$ dans toutes les \rep.

Nous avons vu au \secqgaffsl\ que, pour des valeurs génériques des
paramètres spectraux, le produit tensoriel de deux \rep\
d'évaluation est irréductible. Par contre, si les racines des deux
polynômes associés forment une nouvelle corde, on a réductibilité.
Ici, le cas se présente pour $\lambda_2-\lambda_3=\pm i$.
Et en effet, l'équation de Yang--Baxter
$$
R_{1(23)}(\lambda_1|\lambda_2,\lambda_3)
R_{32}(\lambda_3-\lambda_2)
=
R_{32}(\lambda_3-\lambda_2)
R_{12}(\lambda_1-\lambda_2) R_{13}(\lambda_1-\lambda_3) 
\eqn\ybfus$$
pour $\lambda_3-\lambda_2=-i$ (resp. $\lambda_3-\lambda_2=+i$)
prouve, puisque $R_{32}$ vaut alors le projecteur $P_1$
sur la sous-\rep\ de spin $1$ (resp. $P_0$, de spin $0$),
que $R_{1(23)}$ laisse stable cette sous-\rep.
Posons donc $\lambda_2=\lambda\pm i/2$, $\lambda_3=\lambda\mp i/2$,
et projetons $R_{1(23)}$ sur la sous-\rep\ stable:
on obtient la matrice $R$ entre la \rep\ de spin $1/2$
et de paramètre spectral $\lambda_1$ et la \rep\ de spin $1$
ou $0$ et de paramètre spectral $\lambda$. Explicitement, pour
la \rep\ de spin $1$ (le cas le plus intéressant, puisque
la \rep\ de spin $0$ est triviale), \ybfus\ implique la décomposition
en blocs suivante:
$$\eqalign{
R_{1(23)}(\lambda_1|\lambda+i/2,\lambda-i/2)=
R_{13}(\lambda_1-\lambda+i/2)
R_{12}(\lambda_1-\lambda-i/2)&=\pmatrix{R_{1\{ 23\}}(\lambda_1-\lambda)&\star\cr 0&1}\cr
R_{12}(\lambda_1-\lambda-i/2) R_{13}(\lambda_1-\lambda+i/2)
&=\pmatrix{R_{1\{ 23\}}(\lambda_1-\lambda)&0\cr \star&1}}\eqn\bloc$$
où les deux blocs correspondent aux deux sous-\rep\ de spin $1$ et $0$
dans $V_2\otimes V_3$;
$R_{1\{ 23\}}(\lambda)$ est donc
la matrice $R$ recherchée entre les \rep\ de spin $1/2$ et $1$.
Après avoir remis les facteurs scalaires, et changé
de notation: $R_{1\{ 23\}}\equiv R_{\ha,1}$ (on met
en indice le {\it spin} des représentations), on obtient:
$$R_{\ha,1}(\lambda)={\lambda-i/2-2 {\vec\s}_\ha
\cdot {\vec\s}_1\over\lambda-3i/2}$$
où ${\vec\s}_\ha$ et ${\vec\s}_1$ sont les générateurs de ${\goth sl}(2)$ dans
les \rep\ de spin $1/2$ et $1$.

On peut recommencer la procédure de fusion pour obtenir les matrices
$R$ pour des spins plus élevés. En particulier, d'après les résultats
du \secqgaffsl, on sait que la matrice $R_{\ha,s}$ peut être extraite
de la matrice $R$ agissant sur $V\otimes V^{\otimes 2s}$:
$$R_{1\,2s+1}(\lambda+i(s-1/2)) R_{13}(\lambda+i(s-3/2))\ldots R_{12}
(\lambda-i(s-1/2))\eqn\fuss$$
De plus, la sous-représentation de $V^{\otimes 2s}$ de spin $2s$ joue
un rôle spécial: c'est la sous-représentation de plus haut poids
(i.e. elle contient le vecteur de plus haut poids de $V^{\otimes 2s}$).
Ceci nous permettra d'utiliser la construction \fuss\ pour généraliser
les BAE à des spins supérieurs.

Appliquons donc la procédure de fusion à la matrice de monodromie.
On voit qu'il y a deux possibilités: on peut augmenter soit le spin 
de l'espace auxiliaire $V_a$, soit celui des espaces physiques $V_i$.

$\star$ Si l'on fusionne l'espace auxiliaire de la matrice de monodromie:
on obtient alors une nouvelle matrice de monodromie $T_a(\lambda)$.
On peut encore écrire des relations $RTT$ du type de \RTT\ pour
deux matrices $T_a$ et $T_b$ dont les espaces auxiliaires sont de spins
quelconques; les matrices de transfert associées
commutent donc entre elles. Ainsi, nous avons
apparemment aggrandi l'ensemble des quantités qui commutent entre elles
(et par exemple avec
le Hamiltonien de la chaîne XXX); en fait, les nouvelles quantités
conservées obtenues par fusion ne sont pas indépendantes des
précédentes. Par exemple, en utilisant une
décomposition du type \bloc, on a l'équation
$$\Z_\ha(\lambda+i/2)\Z_\ha(\lambda-i/2)=\Z_1(\lambda)+\Z_0(\lambda)\eqn\ffus$$
où l'indice correspond au spin de l'espace auxiliaire; en particulier,
$\Z_0(\lambda)$ est une fonction scalaire calculable. Comme tous
ces opérateurs commutent, \ffus\ est une équation pour leurs valeurs
propres: c'est un cas particulier
des {\it équations de fusion} satisfaites par les matrices
de transfert, que nous étudierons à la section \secfuseq.

De plus, puisque les nouvelles matrices de transfert fusionnées commutent
avec la matrice de transfert non-fusionnée, la procédure de
diagonalisation (\secdiag) n'est pas modifiée; 
seules les valeurs propres correspondantes
sont différentes. 

Remarquons qu'il existe une autre manière d'utiliser la fusion
dans l'espace auxiliaire: bien que nous ne nous en servions pas par la suite,
cette remarque est intéressante car elle donne une interprétation
intuitive de l'hypothèse de corde. Considérons donc les opérateurs
$\B(\lambda)$ utilisés dans la diagonalisation de la matrice de transfert:
on peut eux aussi les fusionner. Il suffit de fusionner la matrice
de monodromie correspondante, puis de prendre l'élément de matrice
entre le vecteur de plus bas poids et le vecteur de plus haut poids du
produit tensoriel des espaces auxiliaires.

Ainsi, l'opérateur $\B(\lambda_1)\ldots \B(\lambda_m)$ peut être considéré
comme un seul opérateur $\B(\lambda_1,\ldots,\lambda_m)$! La matrice
de monodromie correspondante a pour espace auxiliaire $V_1\otimes\cdots
\otimes V_m$. Ceci n'a pas d'intérêt tel quel. On peut faire mieux:
supposons que les $\lambda_\alpha$ vérifient l'hypothèse
de corde et fusionnons ensemble les différents $\lambda_\alpha$ appartenant
à une même corde (figure \ABAb).
\fig\ABAb{Fusion dans l'espace auxiliaire. Les flèches entourées
d'un cercle sont les états de plus haut/bas poids.}{\figdir{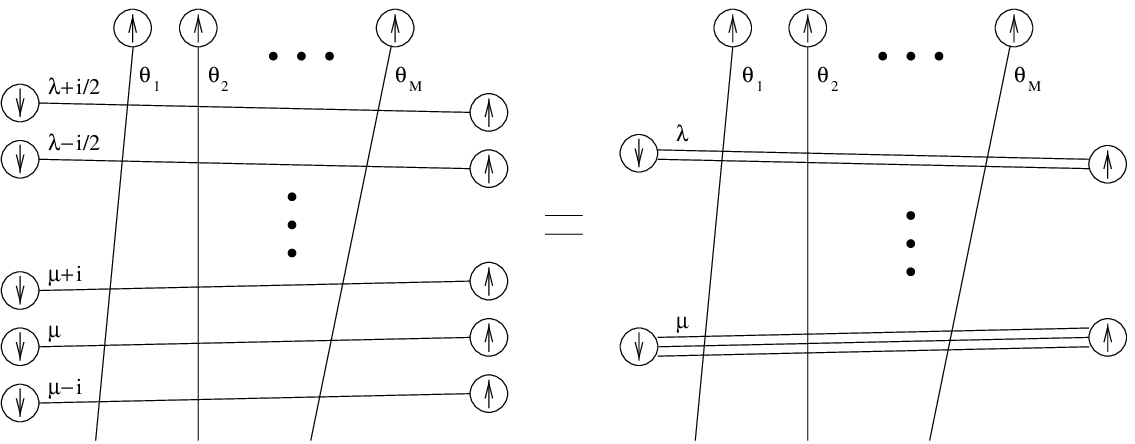}} 
Pour une corde de longueur $j$, on obtient précisément\foot{A condition
de placer les racines dans l'ordre approprié. Les placer dans
un autre ordre ne changerait évidemment rien, mais la situation serait
plus confuse car on serait amené à considérer une partie
indécomposable de la représentation du Yangien.}
la représentation réductible du Yangien de sorte que
le sous-espace irréductible du vecteur de plus haut poids (et de celui
de plus bas poids) soit la \rep\ de spin $j/2$ et de
paramètre spectral le centre de la corde. La matrice $R$ laisse stable
ce sous-espace, et on peut donc remplacer les opérateurs
$\B(\lambda_{\alpha_1})\B(\lambda_{\alpha_2})\ldots \B(\lambda_{\alpha_j})$,
où les $\lambda_{\alpha_k}$ sont les différents membres d'une $j$-corde,
par un seul opérateur $\B(\lambda_{j;\alpha})$
où $\B$ est un élément de matrice de la matrice de monodromie
dont l'espace auxiliaire est de spin $j/2$ et de paramètre spectral
le centre de la corde $\lambda_{j;\alpha}$.
Ceci nous donne une correspondance simple entre les différentes
cordes, et donc les différents noeuds du diagramme \bae, avec les
représentations irréductibles (non triviales) de $SU(2)$: on retrouvera cette
correspondance lorsqu'il sera question d'équations de fusion.

$\star$ Si l'on fusionne les espaces physiques: 
la nouvelle matrice de transfert agit sur un
espace de Hilbert constitué de spins dans des représentations plus élevées
de $SU(2)$. Toute représentation de spin $s$ pouvant être considérée,
grâce à la procédure de fusion (figure \ABAc), 
comme sous-représentation {\it de plus
haut poids} (parmi les différentes sous-représentations) du produit
de $2s$ spins $1/2$, il est inutile de recommencer la procédure de
diagonalisation.
\fig\ABAc{Fusion dans l'espace physique.}{\figdir{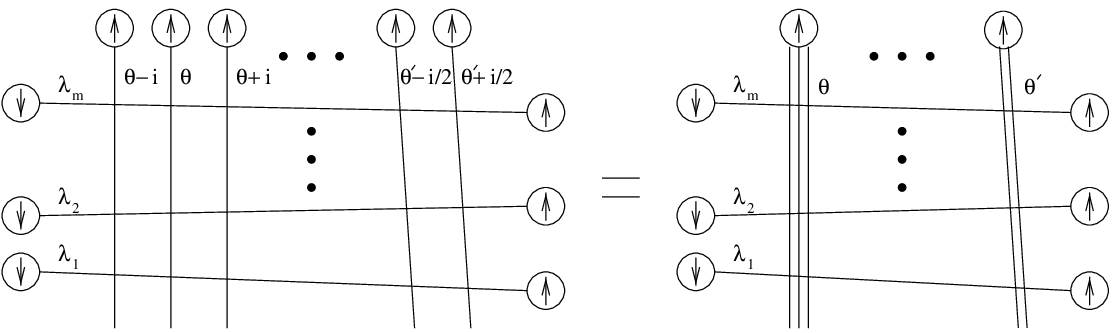}} 
On obtient immédiatement les nouvelles Equations d'Ansatz de Bethe:
$$\prod_{\scriptstyle \beta=1\atop\scriptstyle \beta\ne\alpha}^m 
{\lambda_\alpha-\lambda_\beta+i\over \lambda_\alpha-\lambda_\beta-i}
=\prod_{k=1}^M
{\lambda_\alpha-\theta_k+is\over \lambda_\alpha-\theta_k-is} \eqn\xxxbaee$$
L'équation \xxxbaee\ est obtenue à partie de l'équation non-fusionnée
\xxxbaec a en appliquant \fuss\ (ce sont les $\theta_k$
que l'on fusionne).

\subsec\secfusc{Chaîne de spins XXX et modèle NJL fusionnés}
On cherche maintenant à généraliser la chaîne de spins XXX (ou le modèle
NJL) aux représentations de spin
supérieur: on suppose que tous les spins (ou toutes les particules nues)
sont dans la même représentation de spin $s$. Il faut alors
combiner la fusion dans l'espace auxiliaire et dans l'espace
physique, de sorte que l'on ait encore $V_a=V_k$: en effet, pour
définir le Hamiltonien (ou pour relier la matrice
de monodromie au déphasage d'une particule nue), on utilise le
fait que $R(\lambda=0)={\cal P}$, ce qui n'a de sens que pour $V_a=V_k$.
A nouveau, la chaîne XXX correspond au choix d'inhomogénéités $\theta_k=0$.
On peut alors développer au voisinage de $\lambda=0$ $\Z(\lambda)$,
qui est une fonction génératrice de Hamiltoniens locaux. Ainsi, au
premier ordre, on obtient un Hamiltonien de plus proches voisins,
le Hamiltonien de la chaîne de spins $s$ XXX;
mais il est essentiel de noter que l'on n'obtient ainsi qu'un certain
Hamiltonien de plus proches voisins invariant $SU(2)$: d'autres sont
possibles, qui ne sont pas intégrables, et qui peuvent
avoir des propriétés physiques très différentes. Par exemple,
tout Hamiltonien de plus proches voisins invariant $SU(2)$ 
d'une chaîne de spins $s=1$ est une combinaison
linéaire de deux termes indépendants\foot{De manière générale,
des arguments élémentaires de théorie des groupes montrent
qu'il y a $2s$ termes indépendants dans un Hamiltonien
de plus proches voisins de spin $s$.}:
$$\H=\sum_{i=1}^M\left[\alpha\, {\vec\s}_k\cdot {\vec\s}_{k+1}
+\beta\, ({\vec\s}_k\cdot {\vec\s}_{k+1})^2\right]$$
Pour des valeurs génériques
des deux constantes de couplage $\alpha$ et $\beta$, 
le modèle a un gap, et seulement pour $\alpha=-\beta$ 
obtient-on la chaîne intégrable (qui, 
comme on va le voir, n'a pas de gap pour $\alpha<0$).

De la même manière, on peut considérer des modèles NJL fusionnés.
Notons que ceci n'a pas été fait
explicitement pour toutes les valeurs de $s$ (en effet,
nous verrons au chapitre \chappcf\ 
qu'il existe une autre manière d'arriver naturellement
aux BAE fusionnées); le cas $s=1$ (modèle fermionique
invariant $O(3)$), par exemple, a été considéré dans
[\ref\AD{N.~Andrei et C.~Destri, {\it Nucl.\ Phys.} B231 (1984), 445.}].
Par extrapolation du cas non-fusionné, il est clair que l'on doit
considérer la matrice de transfert inhomogène avec des inhomogénéités
$\theta_k=\pm 1/c$ correspondant aux deux chiralités de fermions,
$c$ étant envoyé à $0$ dans la limite d'échelle.

Pour éviter les confusions entre le spin des états de Bethe
et le spin des constituants de la chaîne de spin (ou
des particules nues), nous noterons ce dernier: $s\equiv f/2$,
où $f$ est un entier.
Récapitulons alors les équations d'Ansatz de Bethe que nous obtenons
pour les modèles fusionnés:
$$\eqnalign\xxxbaef{
\prod_{\scriptstyle \beta=1\atop\scriptstyle \beta\ne\alpha}^m 
{\lambda_\alpha-\lambda_\beta+i\over \lambda_\alpha-\lambda_\beta-i}
&=\prod_{k=1}^M
{\lambda_\alpha-\theta_k+if/2\over \lambda_\alpha-\theta_k-if/2} &\xxxbaef a\cr
\e{-ip_kL}&=
\prod_{\alpha=1}^m {\theta_k-\lambda_\alpha+if/2\over\theta_k-\lambda_\alpha-if/2}
&\xxxbaef b\cr
}$$
La seconde équation a été obtenue en appliquant
la procédure de fusion dans l'espace auxiliaire (en plus de la fusion
des $\theta_k$);
remarquons qu'on utilise ici le fait que
$\D(\lambda=\theta_k)|\Psi\rangle=0$, ce qui implique
que la trace dans l'espace auxiliaire ne se fait que
dans la sous-\rep\ de plus haut poids.

On peut maintenant appliquer l'hypothèse de corde aux
équations \xxxbaef{}, et
prendre la limite thermodynamique $L\to\infty$; on obtient la forme
finale des BAE
$$C_{jk}\star \rhot_k + \rho_j=\delta_{jf} \sigma\star s(\lambda)
\eqn\xxxcobaef$$
que l'on peut représenter graphiquement, avec les mêmes conventions
de dessin qu'au \secbaehypco, par la figure \baefus.
\fig\baefus{Le système de BAE de la chaîne de spins $s$ intégrable
(ici, $s=3/2$).}{\figdir{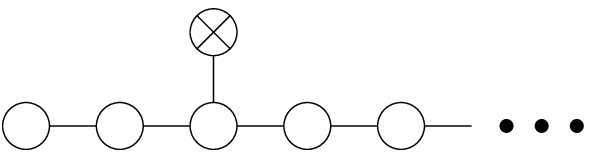}} 
On remarque la confirmation de la correspondance
noeuds/représentations: pour une chaîne de spins $s$, le second
membre agit sur le noeud $f=2s$.

L'énergie/impulsion s'écrit
$$\eqalign{
E/L&=E_\gs/L+\int\d\lambda\,\rhot_f(\lambda) \tilde{\epsilon}(\lambda)\cr
P/L&=\int\d\lambda\,\rhot_f(\lambda) \tilde{p}(\lambda)\cr
}\eqn\EPphysf$$
où les fonctions $\tilde{\epsilon}$ et $\tilde{p}$ sont
encore données par \EPphysb, 
c'est-à-dire que l'énergie est minimisée quand il n'y a pas de trous
de $f$-cordes: le vide de la théorie est un état constitué d'une
densité continue (et sans trous) de $f$-cordes. Les excitations
physiques sont alors créées en insérant des trous dans la mer de
$f$-cordes. Le spin est maintenant donné par
$$2s=\tilde{m}_f - \sum_{j>f} (j-f) m_j\eqn\spinphysb$$
donc les excitations physiques sont encore de spin $1/2$. Cependant,
on remarque que les $j$-cordes, pour $j<f$, n'influent plus sur le spin.
Elles doivent avoir un autre rôle, qui sera élucidé au prochain
paragraphe. En particulier,
nous pourrons alors donner la matrice $S$ des excitations physiques.

\subsec\secphyssup{BAE nues/BAE physiques: spin supérieur}
Commençons par dessiner le diagramme des BAE physiques
pour le modèle NJL fusionné: on se
convainc facilement qu'il a l'aspect du diagramme de
droite de la figure \baefusb.
\fig\baefusb{BAE nues et BAE physiques
du modèle NJL fusionné.}{\figdir{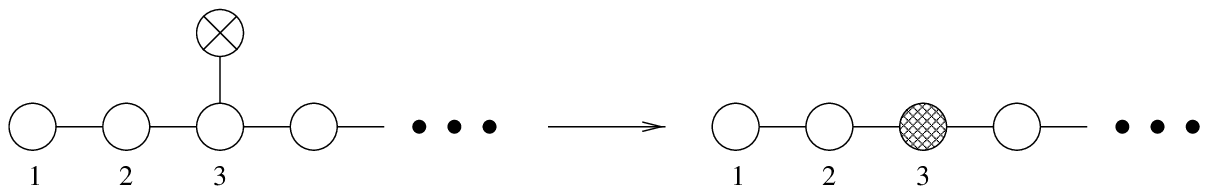}} 
Naïvement, ce diagramme suggère qu'il y a maintenant {\it deux}
symétries qui classifient les excitations physiques:
\item{$\diamond$} La symétrie $SU(2)$ habituelle, qui,
dans la vision ``physique'' des BAE, est liée aux noeuds $j>f$
(ce qui est en accord avec l'expression du spin \spinphysb). Cette symétrie
s'étend en une symétrie $Y({\goth sl}(2))$.
\item{$\diamond$} Une nouvelle symétrie $U_q({\goth sl}(2))$, avec
$q=-\e{-i\pi/(f+2)}$, qui est liée aux noeuds $j<f$. Cette
nouvelle symétrie apparaît de la manière suivante:
on considère d'abord un espace de Fock complet avec une
symétrie $U_q(\widehat{{\goth sl}(2)})$, tel que
les particules élémentaires en soient des doublets (en plus
de leur structure de doublet de $Y({\goth sl}(2))$). Puis
on tronque l'espace de Fock vis-à-vis de
la sous-algèbre horizontale $U_q({\goth sl}(2))$.

Partons alors d'un modèle massif
relativiste dans lequel les particules
élémentaires sont des doublets de $SU(2)$ {\it et} de $U_q({\goth
sl}(2))$, et dont la matrice $S$ est donnée par
$$S(\lambda)=\Sh(\lambda)\, R(\lambda)
\otimes R'(\lambda)\eqn\Sphysdbl$$
où $R(\lambda)$ est la matrice $R$ rationnelle
\explR, et $R'(\lambda)$ est
la matrice $R$ trigonométrique \explRq\ ($\gamma=\pi/(f+2)$).
On considère pour l'instant le modèle non tronqué (on pourrait
aussi partir directement de la matrice $S$ solitonique,
comme cela est fait dans [\ref\RESH{N.~Reshetikhin, {\it J.\ Phys.}
A24 (1991), 3299.}], mais celle-ci est plus
lourde à manipuler).

La procédure habituelle de déphasage d'une particule physique
quand elle fait le tour de l'espace compactifié nous conduit
à diagonaliser deux matrices de transfert qui correspondent
aux deux nombres quantiques. Leur diagonalisation
par l'Ansatz de Bethe algébrique nécessite deux
jeux de paramètres spectraux $\{\lambda_\alpha\}$ et $\{\mu_\alpha\}$
qui vérifient les équations d'Ansatz de Bethe suivantes:
$$\eqnalign\supbaephys{
1&=\prod_{\scriptstyle \beta=1\atop\scriptstyle \beta\ne\alpha}^m 
{\lambda_\alpha-\lambda_\beta-i\over \lambda_\alpha-\lambda_\beta+i}
\prod_{k=1}^{M^{ph}}
{\lambda_\alpha-\theta_k+i/2\over \lambda_\alpha-\theta_k-i/2}
&\supbaephys a\cr
1&=
\prod_{\scriptstyle \beta=1\atop\scriptstyle \beta\ne\alpha}^{m'}
{\sinh(\gamma(\mu_\alpha-\mu_\beta-i))\over
\sinh(\gamma(\mu_\alpha-\mu_\beta+i))}
\prod_{k=1}^{M^{ph}}
{\sinh(\gamma(\mu_\alpha-\theta_k+i/2))
\over \sinh(\gamma(\mu_\alpha-\theta_k-i/2))} &\supbaephys b\cr
\e{-ip_kL}&=\prod_{\scriptstyle l=1\atop\scriptstyle
l\ne k}^{M^{ph}} \Sh(\theta_k-\theta_l)
\prod_{\alpha=1}^m {\theta_k-\lambda_\alpha+i/2
\over\theta_k-\lambda_\alpha-i/2}
\prod_{\alpha=1}^{m'} {\sinh(\gamma(\theta_k-\mu_\alpha+i/2))
\over\sinh(\gamma(\theta_k-\mu_\alpha-i/2))}
&\supbaephys c\cr
}$$
On prend ensuite
la limite thermodynamique $L\to\infty$
et on applique l'hypothèse de corde aux $\lambda_\alpha$
et aux $\mu_\alpha$: les $\rho_j^{ph}$ et $\rhot_j^{ph}$
sont les densités associées aux $j$-cordes de $\lambda_\alpha$
($j\ge 1$),
les $\rho_j^{ph'}$ et $\rhot_j^{ph'}$ sont associées
aux $j$-cordes de $\mu_\alpha$ ($1\le j\le f+1$ ou $j=1^-$),
et $\sigma^{ph}$ est la densité des
rapidités des particules physiques. Les BAE sont:
$$\eqnalign\supcobaephys{
C_{jk}\star \rhot_k^{ph} +\rho_j^{ph}&=\delta_{j1} s\star
\sigma^{ph}\quad j\ge 1&\supcobaephys a\cr
C_{jk}\star \rhot_k^{ph'} + \rho^{ph'}_j-\delta_{j\,p-2} s\star\rho^{ph'}_{1^-}
&=\delta_{j1} s\star\sigma^{ph}\quad 1\le j\le f+1&\supcobaephys b\cr
\rhot^{ph'}_{1^-}-s\star\rhot^{ph'}_{p-2}
+\rho^{ph'}_{1^-}&=0&\supcobaephys{b'}\cr
\sigma^{ph}(\lambda)
+\tilde{\sigma}^{ph}(\lambda)
-{m\over 2}\cosh (\pi\lambda)
+\sum_{j\ge 1} K_j\star\rho_j^{ph}&+\sum_{j=1}^{f+1} K'_j\star\rho_j^{ph'}&\cr
+K'_{1^-}\star\rho_{1^-}^{ph'}
+\left({1\over 2\pi i}{\d\over\d\lambda} \log \Sh\right)
\star\sigma^{ph}&=0&\supcobaephys c}$$
où on a utilisé les fonctions
$$\eqalign{
K_j(\lambda)={1\over 2\pi} {j\over\lambda^2+j^2/4}
&\qquad K_j(\kappa)=\e{-j|\kappa|}\cr
K'_j(\lambda)={1\over 2\pi} {\gamma\sin(j\gamma)
\over\sinh(\gamma(\lambda-ij/2))\sinh(\gamma(\lambda+ij/2))}
&\qquad K'_j(\kappa)={\sinh((f+2-j)\kappa)\over\sinh((f+2)\kappa)}\cr
K'_{1^-}(\lambda)={1\over 2\pi} {\gamma\sin(\gamma)
\over\cosh(\gamma(\lambda-i/2))\cosh(\gamma(\lambda+i/2))}
&=-K'_{p-1}(\lambda)\cr}$$
On voit que les équations \supcobaephys a et \supcobaephys b vont
reproduire (après la troncation) les BAE \xxxcobaef\ pour $j<f$ et
$j>f$, avec les identifications $\rho^{ph}_j=\rho_{j+f}$,
$\rhot^{ph}_j=\rhot_{j+f}$ ($j\ge 1$), et $\rho^{ph'}_j=\rho_{f-j}$,
$\rhot^{ph'}_j=\rhot_{f-j}$ ($1\le j\le f+1$), $\rho^{ph'}_{1^-}=\rho_{1^-}$,
$\rhot^{ph'}_{1^-}=\rhot_{1^-}$.
Ce faisant, nous avons introduit trois noeuds
en excédent: $j=-1$, $j=0$ et $j=1^-$, que la troncation
doit supprimer. Cependant, avant
de pouvoir opérer la troncation, il faut transformer
\supcobaephys c, car $\rho_0=\rho^{ph'}_f$ y apparaît explicitement,
et $\rho_0\ne 0$ après la troncation,
bien qu'il soit ``invisible'' dans le diagramme des BAE
tronquées\foot{Si l'on oublie ce point, on
aboutit à une formule incorrecte pour $\Sh(\lambda)$,
comme cela est expliqué dans [\RESH].}.
Considérons donc la matrice de Cartan $C$ agissant
sur les indices $\{ -1,0,1,\ldots\}$ (et non sur seulement
$\{1,2,\ldots\}$) et 
identifions les noyaux dans \supcobaephys c:
$$\eqalign{
&\left(1-{1\over 2\pi i}{\d\over\d\lambda}\log \Sh\right)\star
\sigma^{ph}+\tilde{\sigma}^{ph}
+\sum_{j\ge 1} (C^{-1}_{ff})^{-1}\star C^{-1}_{f\,j+f}\star\rho_j^{ph}\cr
&+\sum_{j=1}^{f+1} (C^{-1}_{ff})^{-1}\star
C^{-1}_{f\,f-j}\star\rho_j^{ph'}
-(C^{-1}_{ff})^{-1}\star C^{-1}_{f0}\star\rho_{1^-}^{ph'}
={m\over 2} \cosh(\pi\lambda)\cr}
\eqn\supcobaephysc$$
On voit qu'en multipliant \supcobaephysc\ par $C^{-1}_{ff}$, 
puis en recombinant \supcobaephysc, \supcobaephys a et
\supcobaephys b, et finalement en tronquant, on obtient
les équations \xxxcobaef\ avec les identifications
supplémentaires: $\sigma^{ph}=\rhot_f$, $\tilde{\sigma}^{ph}=\rho_f$
et $\delta(\lambda)-
{1\over 2\pi i}{\d\over\d\lambda}\log \Sh=(C^{-1}_{ff})^{-1}$.
On en déduit
$$\eqalign{
\Sh(\lambda)&=\exp\left[-2i \int_0^{+\infty} \d\kappa\, 
{\sin(2\kappa\lambda)\over \kappa}
\left( {1\over 2\cosh\kappa} {\sinh\kappa\over\sinh((f+2)\kappa)}
\e{(f+2)|\kappa|}-1\right)
\right]\cr
&=\exp\left[i \int_0^{+\infty} \d\kappa\, 
{\sin(2\kappa\lambda)\over \kappa}
{1\over\cosh\kappa}
\left(\e{-|\kappa|}+{\sinh((f+1)\kappa)\over\sinh((f+2)\kappa)}
\right)\right]\cr}
\eqn\Sfus$$
On observe (seconde ligne de \Sfus) une factorisation ``miraculeuse''
(jusque dans les facteurs scalaires)
de la matrice $S$ de notre modèle fusionné en la matrice $S$
de Gross--Neveu isotrope et celle de Gross--Neveu anisotrope
à $\tilde{\gamma}=\pi/(f+2)$. Nous reviendrons sur ce phénomène
dans la section \secbaephysgen.

Ecrivons donc, après troncation, la matrice $S$ dans
le langage solitonique:
$$S(\lambda)=
{ \Gamma\left(1+i{\lambda\over 2}\right)
  \Gamma\left({1\over 2}-i{\lambda\over 2}\right)
\over
  \Gamma\left(1-i{\lambda\over 2}\right)
  \Gamma\left({1\over 2}+i{\lambda\over 2}\right)}
{\lambda-i{\cal P}\over\lambda-i}\otimes S_{\rm soli}(\lambda)
\eqn\kinkSb$$
où le premier facteur agit sur le nombre quantique $SU(2)$,
et la seconde partie est la matrice $S$ solitonique donnée par
\kinkS.

Notons enfin 
que $\Sh(\lambda)$ n'a de sens physique que pour $f\ge 2$.
En effet, pour $f=1$, le spin $U_q({\goth sl}(2))$ est $0$ ou $1/2$,
donc entièrement déterminé par la parité du nombre
de particules. En particulier, deux particules diffusent
nécessairement avec un spin $U_q({\goth sl}(2))$ nul;
on vérifie que les facteurs
${\sinh({\pi\over 3}(\lambda+i))\over
\sinh({\pi\over 3}(\lambda-i))}$ dans \kinkSb\ 
se compensent (soit encore
$S_{\rm soli}(b\to b)=1$ pour $f=1$, $s=0$)
de sorte que \kinkSb\
reproduit alors la matrice $S$ de Gross--Neveu
isotrope (Eq.\ \Sphys), comme il se doit.

Nous avons donc trouvé que les BAE physiques tronquées
sont équivalentes aux BAE nues. Doit-on en conclure
que le Hilbert du modèle NJL fusionné est exactement
isomorphe à l'espace de Fock tronqué dont on a donné la construction
au \secphysisoq?
Pour répondre à cette question, considérons plutôt la chaîne
XXX, qui correpond au cas de figure le plus simple (la matrice
de transfert est homogène), et supposons que le nombre de sites
$M$ est fixé et pair.

On peut alors calculer la dimension $D$ de l'espace des états à $\tilde{m}_f$
trous ($\tilde{m}_f$ pair),
soit par des simultations numériques, soit par un calcul
analytique mais en supposant l'hypothèse de corde valide (même
si ce n'est pas le cas): tenant compte de la dimension
du multiplet $SU(2)$ associé aux états de Bethe, on obtient le résultat
exact (par un calcul classique): 
$D=2^{\tilde{m}_f}\,(A^{\tilde{m}_f})_0^0$,
où $(A^{\tilde{m}_f})_0^0$ est par définition
la multiplicité de la \rep\ triviale dans
le produit tensoriel tronqué de $\tilde{m}_f$ \rep\ de spin
$1/2$. Ce résultat s'interprète très simplement: le facteur
$2^{\tilde{m}_f}$ est lié à la structure doublet $SU(2)$, tandis
que le facteur $(A^{\tilde{m}_f})_{00}$ suggère
que l'on n'obtient que les états qui sont
dans la représentation triviale de $U_q({\goth sl}(2))$.
Ceci se vérifie explicitement grâce à la correspondance
BAE nues/BAE physiques: en effet, on vérifie que
les états solutions des BAE nues, retraduits
dans le langage des BAE physiques, sont les états
de spin $U_q({\goth sl}(2))$ nul, {\it à condition}
de rajouter les $f$-cordes physiques ``cachées''.

On se trouve donc dans la situation suivante: bien que l'espace physique
soit invariant par $U_q({\goth sl}(2))$, cette
symétrie est tout de même
nécessaire pour expliquer le facteur
$(A^{\tilde{m}_f})_0^0$ dans $D$: on parle alors
de {\it symétrie cachée}.

Y a-t-il un moyen de récupérer les représentations non-triviales
de $U_q({\goth sl}(2))$? Comme premier pas, supposons que l'on
ait fixé au contraire le nombre de sites $M$ à une valeur impaire.
On peut refaire le calcul de la dimension de l'espace des
états à $\tilde{m}_f$ trous, et on obtient\foot{Je n'ai
pas tenté de démontrer
ce résultat, mais je l'ai vérifié numériquement dans de
très nombreux cas. 
J'ai en particulier vérifié, comme test non-trivial
de ces simulations numériques, qu'en sommant
sur les rapidités possibles, puis sur $\tilde{m}_f$,
on obtient exactement la dimension du Hilbert complet.}:
$D=2^{\tilde{m}_f}\,(A^{\tilde{m}_f})_0^{f\over 2}$,
où cette fois $(A^{\tilde{m}_f})_0^{f\over 2}$ est la multiplicité
de la \rep\ de spin $f/2$ dans le produit tensoriel tronqué
de $\tilde{m}_f$ \rep\ de spin $1/2$. Comme on le vérifie
explicitement, on atteint ainsi l'équivalent des états de
plus haut poids de $U_q({\goth sl}(2))$ dans la \rep\ de spin $f/2$.
L'interprétation en termes de ``symétrie cachée'' est la suivante:
on prend l'espace de Fock tronqué total ${\cal H}_{tronc}$, on le tensorise
par la \rep\ de spin $f/2$ de $U_q({\goth sl}(2))$ et on prend
le sous-espace invariant: $({\cal H}_{tronc}\otimes V^{f/2})_{inv}$.
Celui-ci est précisément isomorphe à l'espace des états de
la chaîne XXX fusionnée avec un nombre impair de sites.

Il ne reste plus qu'à obtenir les \rep\ de spin $s$, où $0<s<f/2$.
Ceci est possible à condition de recourir à un artifice:
considérons une chaîne de spins modifiée qui contient $M$
sites de spin $f/2$ ($M$ pair) et $1$ site de spin $s$. Les cas
étudiés précédemment sont les cas limites $s=0$ et $s=f/2$.
Le spin supplémentaire doit être considéré comme une condition
de bord spéciale de la chaîne de spins (figure \bcspin).
\fig\bcspin{L'ajout du spin $s$ modifie
les conditions de bord dans la vision
solitonique.}{\figdir{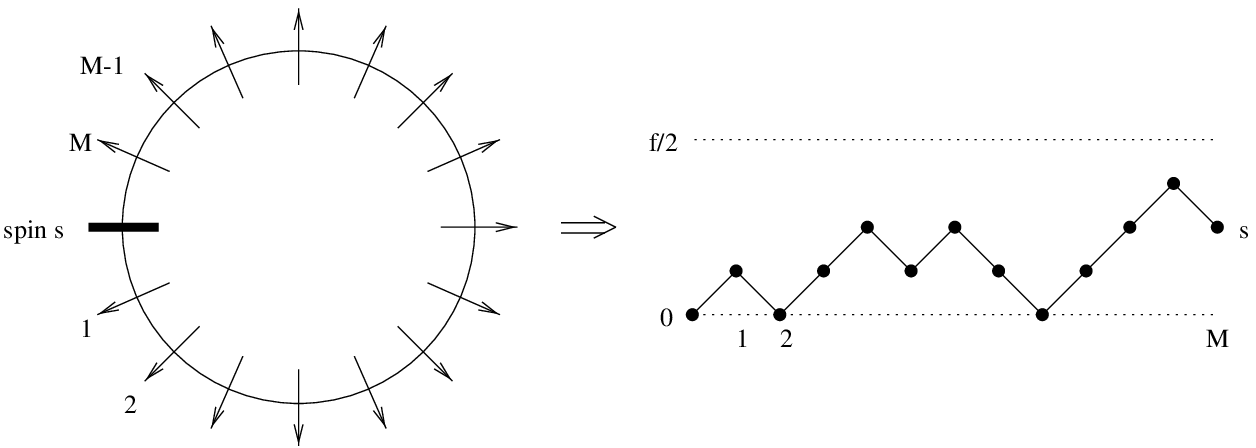}} 
On associe à cette chaîne le Hamiltonien intégrable obtenu
comme la dérivée logarithmique de la matrice de transfert homogène,
c'est-à-dire le Hamiltonien usuel plus un terme de bord.

Dans la limite thermodynamique, les excitations physiques
sont encore des trous dans la mer de $f$-cordes, et on peut
une fois de plus calculer la dimension de l'espace des états
à $\tilde{m}_f$ trous: on obtient, comme prévu\foot{A nouveau,
il ne s'agit que d'une conjecture (probablement facile à démontrer)
basée sur des simulations numériques.}:
$D=2^{\tilde{m}_f}\,(A^{\tilde{m}_f})_0^s$.
Ainsi, la bonne
interprétation de l'espace des états est: 
$({\cal H}_{tronc}\otimes V^s)_{inv}$, où $V^s$ est la \rep\ de
spin $s$ de $U_q({\goth sl}(2))$.

Résumons la construction qui vient d'être faite: nommons ${\cal H}^s$
le Hilbert ``limite'' quand $M\to\infty$ de la chaîne de spins
fusionnée avec un spin $s$ supplémentaire. Alors le Hilbert complet
s'écrit:
$${\cal H}_{tronc}=\bigoplus_{s=0}^{f/2} {\cal H}^s \otimes V^s\eqn\refrac$$
Dans le langage de [\AD], on peut dire que la resommation \refrac\
du Hilbert
permet de ``refractionniser'' le nombre quantique solitonique.

La construction que nous avons faite peut paraître un
peu artificielle, mais nous verrons plusieurs situations
physiques dans lesquelles elle apparaît naturellement:
pour les modèles fermioniques invariants $U(Nf)$ (couleur $\times$
saveur) dans lesquels il y a fusion dynamique des spins (le
spin supplémentaire constitue alors les fermions restants
qui n'ont que partiellement fusionné), et surtout
pour le modèle de Kondo dans le régime surécranté
(le spin supplémentaire s'identifie à l'impureté).
Nous reviendrons donc en détail sur l'interprétation
de cette construction.

Terminons par un exemple concret qui montre comment
la correspondance BAE nues/BAE physiques fonctionne: classifions
les états à deux trous (de $f$-cordes)
dans la chaîne de spins $f/2$ ($f\ge 2$).
Supposons d'abord qu'il y a $M$ spins $f/2$ (et pas de spin supplémentaire);
on trouve alors deux états à deux trous:
\item{$\bullet$} L'état constitué de deux trous $\lambda_1$,
$\lambda_2$, et d'une $(f-1)$-corde de centre $(\lambda_1+\lambda_2)/2$.
Dans la vision des BAE physiques, la $(f-1)$-corde est une $1$-corde
pour la partie $U_q({\goth sl}(2))$ 
de la matrice de transfert; chaque
racine faisant diminuer de $1$ le spin, et chaque trou
correspondant à un spin $1/2$,
le spin $U_q({\goth sl}(2))$ vaut donc $s=0$.
Par contre, le spin ${\goth sl}(2)$ vaut $s=1$.
\item{$\bullet$} L'état constitué de deux trous $\lambda_1$,
$\lambda_2$, d'une $(f-1)$-corde et d'une $(f+1)$-corde toutes deux
de centres $(\lambda_1+\lambda_2)/2$.
Dans la vision des BAE physiques, la $(f+1)$-corde est une $1$-corde
pour la partie ${\goth sl}(2)$ 
de la matrice de transfert; donc le spin ${\goth sl}(2)$,
aussi bien que le spin $U_q({\goth sl}(2))$, est nul.

Si l'on suppose que l'on a rajouté dans la chaîne un spin $1$,
on trouve de même deux états:
\item{$\bullet$} L'état constitué de deux trous,
dont les spins ${\goth sl}(2)$ et $U_q({\goth sl}(2))$ valent
tous deux $1$.
\item{$\bullet$} L'état constitué de deux trous et d'une $(f+1)$-corde,
de spin ${\goth sl}(2)$ nul et de spin $U_q({\goth sl}(2))$ égal à $1$.

\subsec\secfusmix{Fusion mixte}
Nous avons jusqu'à présent considéré le cas le plus simple
où toutes les particules nues (ou tous les spins) appartiennent
à la même représentation. Rien ne nous empêche de considérer
des matrices de transfert mixtes, comprenant différentes représentations.
Du point de vue de la chaîne de spins,
le seul problème consiste à définir un Hamiltonien raisonnable
physiquement. Ainsi, dans
[\ref\DVMN{H.J.~De~Vega et F.~Woynarovich, 
{\it J.\ Phys.} A25 (1992), 4499\semi
H.J.~De~Vega, L.~Mezincescu et R.I.~Nepomechie, 
{\it Int.\ J.\ Mod.\ Phys.} B8 (1994), 3473
\pre{hep-th/9402053}.}] est introduite une chaîne de spins alternés
$1/2$ et $1$, mais le Hamiltonien contient des couplages
entre seconds plus proches voisins.
De manière générale, si l'on considère un modèle NJL fusionné
avec des particules nues dans les représentations $f_1/2,f_2/2,\ldots,f_k/2$,
de sorte que l'énergie d'une $j$-corde ($j=f_1,\ldots f_k$) soit
négative, alors on aboutit par un raisonnement identique
à ce qui a été fait précédemment au diagramme de BAE physiques
donné par la figure \baemix.
\fig\baemix{BAE physiques d'un modèle NJL mixte.}{\figdir{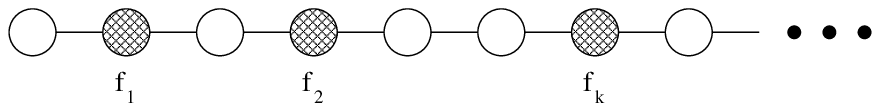}} 
On a alors $k$ types différents d'excitations physiques;
la \eme{j} particule est obtenue en créant un trou de $f_j$-corde,
et possède deux nombres quantiques: $U_{q_j}({\goth sl}(2))$
et $U_{q_{j+1}}({\goth sl}(2))$ ($q_j=-\e{-i\pi/(f_j-f_{j-1}+2)}$
pour $1\le j\le k$, $f_0\equiv 0$, $f_{k+1}\equiv\infty$). 
On vérifie que cette vision
est compatible avec les résultats trouvés dans [\DVMN,\ref\AM{S.R.~Aladim
et M.J.~Martins, {\it J.\ Phys.} A26 (1993), L529.}]
pour les chaînes de spins mixtes (dans le langage
de [\AM], le nombre quantique de groupe quantique
constitue le comportement ``parafermionique'' des excitations physiques).

Nous aurons nous-mêmes à considérer de tels modèles mixtes,
par exemple
pour le modèle de Wess--Zumino--Witten (chapitre \chappcf). D'autres
exemples de modèles relativistes avec fusion mixte ont été
récemment étudiés (voir par exemple
[\ref\ADJ{N.~Andrei, M.~Douglas et A.~Jerez,
\pre{cond-mat/9502082}, \pre{cond-mat/9803134}.},\ref\SaSi{H.~Saleur
et P.~Simonetti, \pre{hep-th/9804080}.}]).

\subsec\secfusanis{Fusion et anisotropie}
La procédure de fusion s'applique également au cas anisotrope;
elle permet de la même manière de déplacer le second membre
des BAE vers le noeud de son choix. Nous aurons l'occasion
d'étudier un exemple concret de ce cas de figure à propos
du régime surécranté de Kondo (section \chappcf).

\newsec\secrang{Rang plus élevé}
Une dernière généralisation consiste à remplacer l'algèbre
${\goth sl}(2)$, de rang $1$, par une algèbre de rang plus élevé.
On peut définir des matrices $R(\lambda)$
pour toute algèbre de Lie ${\goth g}$
simple\foot{Comme nous l'avons vu dans la section précédente, ces matrices
$R(\lambda)$ sont plus précisément associées aux Yangiens $Y({\goth g})$ basés
sur ${\goth g}$.}
et pour des représentations appropriées $V$ de
${\goth g}$, et de là la matrice de transfert
inhomogène, la chaîne de spins XXX et le modèle 
NJL ${\goth g}$-symétriques.

Nous ne parlerons ici que du cas ${\goth g}={\goth sl}(N)$, où
toute \irr\ de ${\goth g}$ s'étend en une représentation
d'évaluation de $Y({\goth g})$ [\ref\KRS{P.P.~Kulish,
N.Yu.~Reshetikhin et E.K.~Sklyanin, {\it Lett. Math. Phys.} 5 (1981), 393.}].

\subsec\secrangN{Modèles invariants $SU(N)$}
Pour la représentation fondamentale $V={\Bbb C}^N$ de plus basse dimension
de ${\goth sl}(N)$, la matrice $R$ de $V\otimes V$ vaut alors:
$$R(\lambda)={\lambda-i{\cal P} \over\lambda-i}=
P_\squares + {\lambda+i\over\lambda-i} P_\squarea\eqn\RN$$
soit la même expression que dans le cas $SU(2)$.

On définit la matrice de monodromie inhomogène de la manière habituelle
(eq. \defTin). En prenant des inhomogénéités nulles,
et en développant par rapport au paramètre spectral,
on obtient le Hamiltonien $\H_\X$ $SU(N)$:
$$\H_\X=-i \Z(\lambda)^{-1} {\d\over\d\lambda}
\Z(\lambda)_{|\lambda=0}={\rm cst}+\sum_{k=1}^M
{\vec\s}_k\cdot {\vec\s}_{k+1}\eqn\hamxxxN$$
où les $\s^A_k$ ($A=1,\ldots,N^2-1$) sont les générateurs
de ${\goth sl}(N)$. En prenant des inhomogénéités
$\theta_k=\pm 1/c$, on obtient le modèle NJL $SU(N)$.

La diagonalisation de la matrice de transfert se fait grâce à
l'Ansatz de Bethe emboîté [\ref\KR{P.P.~Kulish
et N.Yu.~Reshetikhin, {\it J. Phys.} A16 (1983), L591.}]. 
Nous ne décrirons pas ici cette
procédure qui généralise l'Ansatz de Bethe usuel. La diagonalisation
se fait par étapes successives, chaque étape faisant diminuer le
rang de l'algèbre de $1$ et faisant apparaître un nouveau jeu de
paramètres spectraux. La structure des Equations d'Ansatz de Bethe
qui en résulte reproduit le diagramme de Dynkin de l'algèbre
correspondante (ici, $A_{N-1}$):
$$\prod_{\scriptstyle\beta=1\atop\scriptstyle\beta\ne\alpha}^{m^r} 
{\lambda_\alpha^r-\lambda_\beta^r+i
\over\lambda_\alpha^r-\lambda_\beta^r-i}
\,
\prod_{\scriptstyle t=1\atop\scriptstyle t=r\pm 1}^{N-1} \prod_{\beta=1}^{m^t}
{\lambda_\alpha^r-\lambda_\beta^t-i/2
\over\lambda_\alpha^r-\lambda_\beta^t+i/2}
=\prod_{k=1}^M
\left({\lambda_\alpha^1-\theta_k+i/2\over\lambda_\alpha^1-\theta_k-i/2}\right)
\eqn\Nbae$$
où $m_r$ est le nombre de racines $\lambda_\alpha^r$ 
au rang $r$, $1\le r\le N-1$. Les $\lambda_\alpha^r$ ont déjà été
décalés dans \Nbae.

Les propriétés énoncées pour le cas $SU(2)$ s'étendent au cas $SU(N)$;
en particulier, les états de Bethe sont de plus haut poids, et ces derniers
valent:
$$R=R_0-\sum_{r=1}^{N-1} m^r \alpha^r$$
où $\alpha^r$ est la \eme{r} racine simple de
${\goth sl}(N)$, et
$R_0$ est le plus haut poids du Hilbert complet: il correspond à la
représentation du pseudo-vide $|\Omega\rangle$. Pour une chaîne à $M$
sites dans la représentation fondamentale $V={\Bbb C}^N$, on a:
$R_0=M\omega_1$, où $\omega_1$ est le poids fondamental correspondant.

Dans la limite thermodynamique $M\to\infty$, on peut encore appliquer
l'hypothèse de corde pour les racines de chaque rang; on doit donc
introduire les densités de $j$-cordes de rang $r$ $\rho^r_j(\lambda)$
et les densités de trous $\rhot^r_j(\lambda)$. Les
Equations d'Ansatz de Bethe continues s'écrivent sous la forme:
$$C_{jk}\star\rhot^r_k+C^{qr}\star\rho^q_j=\delta_{j1}\delta^{r1} 
\sigma\star s(\lambda)\eqn\cobaeN$$
La matrice $C^{qr}$ est la matrice de Cartan
avec paramètre spectral (donnée par \defC) de $A_{N-1}$.
Nous pouvons représenter les BAE \cobaeN\ par la figure \baeN.
\fig\baeN{Le système de BAE de la chaîne XXX $SU(N)$/modèle
NJL $SU(N)$ (ici $N=4$).}{\figdir{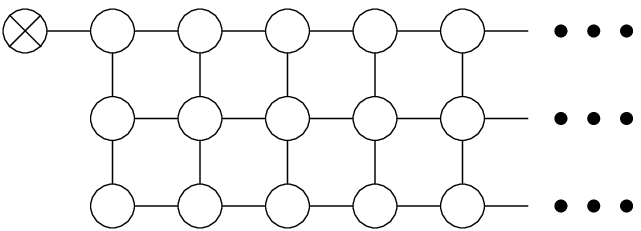}} 
Le noeud à la ligne $r$ et à la colonne $j$ correspond
à l'équation faisant intervenir $\rho^r_j+\rhot^r_j$.
En plus des traits horizontaux qui représentent des termes
$-s\star \rhot^r_{j\pm 1}$ (ou $-s\star \sigma$ pour le noeud coché), on
a dessiné des traits verticaux qui correspondent aux
termes $-s\star\rho^{r\pm 1}_j$.

L'énergie-impulsion est donnée par
$$\eqalign{
E/L&=E_\gs/L+\sum_{r=1}^{N-1} \int d\lambda\,
\rhot^r_j(\lambda) \tilde{\epsilon}^r_1(\lambda)\cr
P/L&=\sum_{r=1}^{N-1} \int d\lambda\,
\rhot^r_j(\lambda) \tilde{p}^r_1(\lambda)\cr}\eqn\EPphysN$$
où
$$\eqalign{
&\left\{\eqalign{
\tilde{\epsilon}^r_\NJL(\lambda)&=g^r(\lambda-1/c)+g^r(\lambda+1/c)\cr
\tilde{p}^r_\NJL(\lambda)&=g^r(\lambda-1/c)-g^r(\lambda+1/c)\cr}\right.
\quad
\left\{\eqalign{
\tilde{\epsilon}^r_\X(\lambda)&={\d\over\d\lambda} g^r(\lambda)\cr
\tilde{p}^r_\X(\lambda)&=g^r(\lambda)\cr}\right.\cr
&g^r(\lambda)=
2\arctan\left(
\tan\left({\pi\over 2}{N-r\over N}\right) 
\tanh\left({\pi\over N}\lambda\right)\right)
+\pi{N-r\over N}\cr}\eqn\EPphysbN$$
Le vide est donc rempli de racines réelles ($1$-cordes)
de tous les rangs, et
il y a $N-1$ types d'excitations physiques, créées en
pla\c cant des trous de rang $r$ ($r=1,\ldots,N-1$).
La représentation d'un état est donnée par:
$$R=\sum_{r=1}^{N-1} \tilde{m}^r_1 \omega^r
-\sum_{j=2}^\infty (j-1) \sum_{r=1}^{N-1} m^r_j \alpha^r$$
où $\omega^r$ est le \eme{r} poids fondamental de ${\goth sl}(N)$,
et $\tilde{m}^r_1$ est le nombre de trous de rang $r$.
L'excitation de rang $r$ appartient donc à la \eme{r} représentation
fondamentale de ${\goth sl}(N)$ (tableau de Young constitué d'une
colonne de taille $r$).
Elles diffusent bien sûr avec une matrice $S$ proportionnelle
à la matrice $R$ entre les \rep\ des deux particules;
par exemple,
$$\eqalign{
S^{11}(\lambda)&=\Sh^{11}(\lambda)
\left(P_\squares + {\lambda+i\over\lambda-i} P_\squarea\right)\cr
S^{1\bar{1}}(\lambda)&=\Sh^{1\bar{1}}(\lambda)
\left(P_{\rm ad}+{\lambda+iN\pi/2\over\lambda-iN\pi/2} P_\emptyset\right)\cr
}$$
où $\bar{1}\equiv N-1$.
Nous calculerons les préfacteurs scalaires dans le paragraphe sur les BAE
physiques (\secphysisoN).

Nous n'irons pas plus loin dans
l'analyse de ce modèle, car il est préférable de le généraliser
tout de suite aux modèles fusionnés. En effet,
on peut à nouveau appliquer la procédure de fusion au cas $SU(N)$,
ce qui permet d'obtenir la matrice $R$ dans
deux représentations arbitraires de ${\goth sl}(N)$.
Si l'on place au \eme{k} site un spin dans la représentation
$R_k=\sum_{r=1}^{N-1} \mu_k^r e^r$
(avec les notations du [\artic03.A], c'est-à-dire
dont la \eme{r} ligne du tableau de Young a $\mu^r_k$ boîtes),
et de paramètre spectral
$\theta_k$, alors les équations d'Ansatz de Bethe s'écrivent:
$$\prod_{\scriptstyle\beta=1\atop\scriptstyle\beta\ne\alpha}^{m^r} 
{\lambda_\alpha^r-\lambda_\beta^r+i
\over\lambda_\alpha^r-\lambda_\beta^r-i}
\,
\prod_{\beta=1}^{m^{r-1}}
{\lambda_\alpha^r-\lambda_\beta^{r-1}
\over\lambda_\alpha^r-\lambda_\beta^{r-1}+i}
\prod_{\beta=1}^{m^{r+1}}
{\lambda_\alpha^r-\lambda_\beta^{r+1}-i
\over\lambda_\alpha^r-\lambda_\beta^{r+1}}
=\prod_{k=1}^M
\left({\lambda_\alpha^1-\theta_k-i\mu_k^{r+1}
\over\lambda_\alpha^1-\theta_k-i\mu_k^r}\right)
\eqn\Nbaefus$$
(les $\lambda_\alpha$ n'ont pas encore été décalés;
par contre, les $\theta_k$ ont été décalés
d'une constante imaginaire qui dépend de la \rep\ $R_k$,
pour simplifier les équations). L'expression
de la valeur propre correpondante est un peu compliquée,
et nous ne la donnerons pas dans le cas général.

Nous nous limiterons désormais
au cas où les spins physiques vivent dans
des représentations $V_k$ qui correspondent à des tableaux de Young
{\it rectangulaires}: en effet, c'est seulement dans ce cas
que l'hypothèse de corde s'applique encore.
Supposons donc que tous les spins appartiennent à la représentation
$n\times f$; après décalage des $\lambda_\alpha$, et passage à la
limite thermodynamique, on obtient les BAE représentées
par la figure \baeNfus, c'est-à-dire la généralisation
évidente de la figure \baeN.
\fig\baeNfus{BAE $SU(N)$ fusionnées.}{\figdir{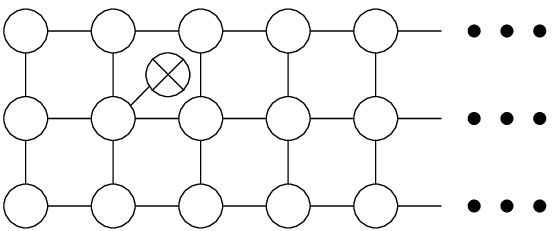}} 
On voit qu'il y a correspondance entre les noeuds de la figure
\baeNfus\ et les représentations rectangulaires (et elles seulement).
L'énergie-impulsion est maintenant donnée par une expression
similaire à \EPphysN, mais où ce sont les trous de $f$-cordes
qui contribuent à l'énergie:
le vide est donc obtenu en remplissant la mer de Fermi
de $f$-cordes de tous les rangs, et il y a à nouveau
$N-1$ types d'excitations physiques (trous de $f$-cordes de
type $r$, $r=1\ldots N-1$). Dans la limite d'échelle du modèle
NJL $SU(N)$ fusionné ($c\to 0$, $D\to\infty$), l'énergie
$$\eqalign{
\tilde{\epsilon}^r_\NJL(\lambda)&=m \sin(\pi r/N) \cosh(2\pi\lambda/N)\cr
\tilde{p}^r_\NJL(\lambda)&=m \sin(\pi r/N) \sinh(2\pi\lambda/N)
\cr}
$$
exhibe le spectre de masse: $m^r=m \sin(\pi r/N)$,
où $m=2D{\sin(\pi n/N)\over\sin(\pi/N)} \e{-2\pi/Nc}$ 
est l'échelle de masse engendrée
dynamiquement. On voit que la taille verticale du tableau de Young
des spins physiques peut être entièrement cachée dans l'échelle
de masse.

Enfin, les BAE représentés par la figure \baeNfus\ (qui
sont la généralisation fusionnée des équations
\cobaeN), pour être
écrites dans la limite d'échelle, doivent être multipliées 
par la matrice de Cartan inverse
de $A_{N-1}$ (qui agit
sur les indices placés en exposant)\foot{Sans cela, le second membre s'annule
dans la limite d'échelle pour $N>2$, et les BAE ne
contiendraient plus toute l'information nécessaire.}; on trouve alors:
$$C^{-1}{}^{qr}\star C_{jk}\star\rhot^q_k+\rho^r_j=\delta_{jf}
{m\over N}\sin(\pi r/N)\cosh(2\pi\lambda/N)\eqn\cobaeNb$$

\subsec\secranganis{Rang plus élevé et anisotropie}
On peut aussi combiner rang plus élevé et anisotropie. On sait
définir la matrice $R$ de $U_q(\widehat{{\goth sl}(N)})$ 
[\ref\JIM{M.~Jimbo, {\it Int. J. Mod. Phys.} A4
(1989), 3759.}]
(le rang de l'algèbre augmente mais le nombre de paramètres
de déformation disponibles reste égal à $1$):
dans la gradation homogène, on a encore la relation
$\tilde{R}(x)=(xR-x^{-1}\bar{R})/(xq-x^{-1}q^{-1})$, où
$R$ et $\bar{R}$ sont les matrices $R$ de $U_q({\goth sl}(N))$,
que l'on peut trouver dans [\JIM].
On peut alors définir
la chaîne XXZ $SU(N)$ et le modèle NJL déformé $SU(N)$.
Cependant, il faut signaler que le Hamiltonien XXZ $SU(N)$
n'est pas hermitien.\rem{et NJL, il est unitaire ou pas?}
Nous ne nous apesantirons donc pas sur ce cas de figure, et passerons
directement au modèle tronqué à $\gamma=\pi/(p+1)$
($q=-\e{-i\gamma}$), puisqu'à ces valeurs, la troncation permet
de restaurer l'unitarité [\ref\RS{N.Yu.~Reshetikhin
et F.~Smirnov, {\it Commun. Math. Phys.} 131 (1990), 157.}].

La procédure de troncation pour
$U_q({\goth sl}(N))$ avec $q$ racine de l'unité,
est similaire au cas $N=2$. On a une notion de $q$-dimension:
$\qdim R=\tr_R(\K)$ avec $\K=q^{\sum_{\alpha>0} \HH^\alpha}$ ($\HH^\alpha$
élément de la sous-algèbre de Cartan correspondant à la racine
positive $\alpha$), qui permet de distinguer les
``bonnes'' représentations des mauvaises.
Si $q^{f+N}=\pm 1$,
alors les ``bonnes'' représentations $R=\sum_{r=1}^{N-1} n^r \omega^r$
sont celles dont
le tableau de Young a moins de $f$ colonnes ($n^1+\cdots+n^{N-1}\le f$),
et on peut définir des matrices d'adjacences tronquées pour ces
représentations (cf [\artic03.5.3]).

Au niveau de l'Ansatz de Bethe, et bien que cela n'ait pas été prouvé
rigoureusement, on a de bonnes raisons de penser 
[\ref\KUN{A.~Kuniba, {\it Nucl.\ Phys.} B389 (1993), 209.}]\foot{Notons
que dans [\KUN], comme dans [\BR], le spin quantique est supposé
nul {\it a priori}, ce qui est inutile.} que la chaîne
XXZ $SU(N)$ tronquée (ou le
modèle NJL correspondant) admet les BAE décrites par
la figure \baeNtronc: seules les $j$-cordes, $1\le j\le f$,
sont autorisées, et les $f$-cordes
sont invisibles sur le diagramme car $\rhot_f^r=0$.
\fig\baeNtronc{BAE $SU(N)$ tronquées au niveau 
$f$.}{\figdir{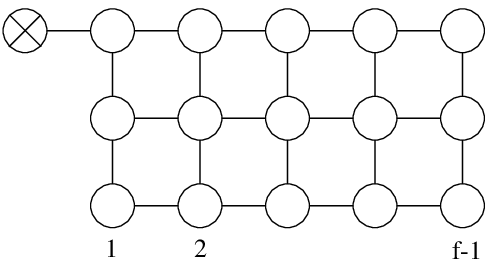}} 

Le reste des calculs est très similaire à ceux du \secrangN,
et nous ne reviendrons pas dessus. Pour nous, l'intérêt du
cas anisotrope est essentiellement de permettre d'écrire des
BAE physiques pour les modèles fusionnés, ce que nous allons
faire maintenant.

\subsec\secphysisoN{BAE nues/BAE physiques: rang supérieur}
Pour éviter les répétitions, nous traiterons directement
le cas (isotrope) fusionné. Les BAE physiques
sont représentées sur la figure \baeNphys.
\fig\baeNphys{BAE physiques $SU(N)$.}{\figdir{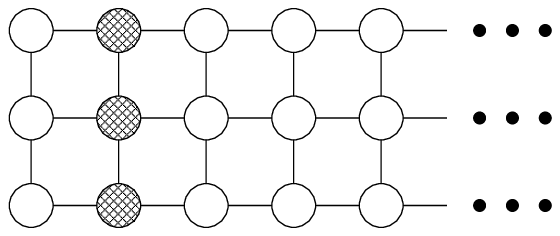}}
Comme dans le cas $SU(2)$, on conjecture que les excitations
physiques ont deux nombres quantiques: la symétrie $SU(N)$ dont
on est parti, et une nouvelle symétrie cachée $U_q({\goth sl}(N))$,
$q=-\exp(-i\pi/(f+N))$,
tronquée. La particule de type $r$ ($1\le r\le N-1$)
appartient à la même représentation fondamentale à
$r$ boîtes pour les deux symétries.
Les particules de différents types diffusent
sans réflexion (ils ne peuvent que modifier leurs nombres
quantiques lors de la collision), avec une matrice $S$ entre
particules de types $q$ et $r$ donnée par:
$$S^{qr}(\lambda)=\Sh^{qr}(\lambda)R^{qr}(\lambda)\otimes R'{}^{qr}(\lambda)$$
où $R^{qr}(\lambda)$ et $R'{}^{qr}(\lambda)$ sont les matrices
$R$ de $Y({\goth sl}(N))$ et de $U_q(\widehat{{\goth sl}(N)})$
dans les \eme{q} et \eme{r} représentations fondamentales,
et $\Sh^{qr}(\lambda)$ est le facteur scalaire qui donne la diffusion
dans le plus haut poids. En écrivant les Equations d'Ansatz
de Bethe physiques correspondantes et en les identifiant avec
les Equations d'Ansatz de Bethe nues, on arrive comme
au \secphyssup\ à calculer $\Sh^{qr}$, qui vaut:
$$\Sh^{qr}(\lambda)=
\exp\left[-2i \int_0^{+\infty} \d\kappa\, 
{\sin(2\kappa\lambda)\over \kappa}
\left(
{\sinh((N-q)\kappa)\sinh(r\kappa)\over\sinh(N\kappa)}
{\e{(f+N)|\kappa|}\over \sinh((f+N)\kappa)}-\delta^{qr}
\right)\right]\eqn\Ssup$$
($q\ge r$).

Ces matrices $S$ ont la particularité que toute particule
peut s'obtenir comme état lié de deux autres particules:
en effet, les particules $q$ et $r$ possèdent un état
lié dans le canal antisymétrique, qui est la particule $q+r\mod N$.
Les matrices $S$ physiques peuvent donc elles aussi être obtenues
les unes des autres par la procédure de fusion.

Donnons enfin l'interprétation solitonique du nombre
quantique $U_q({\goth sl}(N))$: à chaque tableau
de Young de moins de $f$ colonnes, on associe un vide. 
Dans le cas que nous avons considéré, les
excitations physiques sont dans les représentations fondamentales,
et il existe alors une règle particulièrement simple pour
déterminer les transitions entre vides autorisées pour
chaque type de particules (cf [\artic03.5.3]). Cependant,
il faut noter que pour des représentations plus élevées,
une différence avec le cas $N=2$ apparaît: certains
coefficients des matrices d'adjacence deviennent supérieurs
à $1$, ce qui indique des transitions multiples entre deux vides.
Pour décrire
un état du système comportant de telles transitions, il faut
donc spécifier, en plus de la donnée du vide
initial et du vide final, quelle est la transition choisie.
\rem{matrices $S$ solitoniques pour toutes les
rep fondamentales?}

\subsec\secrangniv{Dualité rang-niveau}
Pour les modèles les plus simples (diffusion diagonale),
la dualité particule-trou échange $\rho$ et $\rhot$ dans
les équations d'Ansatz de Bethe\foot{On verra
plus loin qu'elle échange aussi $\eta$ et $\eta^{-1}$ dans les TBA.}.
Cependant, on a vu au
\secphysiso\ à propos des BAE physiques que pour le modèle
NJL $SU(2)$, il y avait bien à nouveau une dualité particule-trou,
mais qu'elle n'affecte que le noeud $1$ des équations:
en effet, ce sont les trous de $1$-cordes qui sont les excitations
physiques, les $j$-cordes étant des ``excitations isotopiques''
et non des vraies particules. La même remarque vaut pour les modèles
fusionnés, où la dualité ne s'applique qu'au noeud $f$.

Quand on passe à $SU(N)$, il est pourtant particulièrement
tentant d'effectuer l'échange $\rho\leftrightarrow\rhot$. En effet,
si l'on oublie le second membre, la dualité $\rho\leftrightarrow\rhot$
a pour effet d'échanger dans les BAE \cobaeN\ les indices
du haut et du bas: c'est ce qu'on appelle la {\it dualité rang-niveau}.
Si l'on remet le second membre, naïvement il n'y a pas de problème,
et on obtient la transformation sur les BAE nues décrite
par la figure \dualbae. On voit que la dualité rang-niveau
est une transformation mathématique des BAE, sans signification
physique directe, qui par exemple échange un modèle tronqué
relié au groupe quantique nu $U_q({\goth sl}(N)$, $q=\exp(i\pi/(f+N))$,
avec un autre de groupe quantique nu $U_q({\goth sl}(f))$.
\fig\dualbae{Dualité rang-niveau et BAE nues.}{\figdir{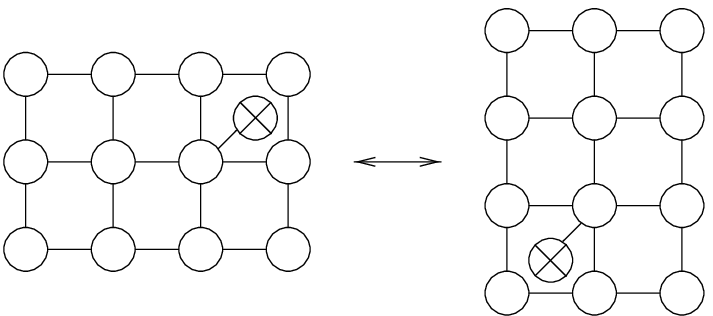}}

Cependant, les BAE ne déterminent pas entièrement la physique
du modèle correspondant: une autre donnée essentielle
est l'expression de l'énergie en fonction
des $\rho$ (ou des $\rhot$); en particulier, c'est le signe
de l'énergie des différents types de (pseudo-)particules
qui détermine quel est le vrai vide de la théorie.
Evidemment, c'est précisément la forme de l'énergie
qui brise la dualité rang-niveau: on le voit par exemple
sur la figure \dualbaeb, qui montre la même transformation
au niveau des BAE physiques. 
\fig\dualbaeb{Dualité rang-niveau et BAE 
physiques.}{\figdir{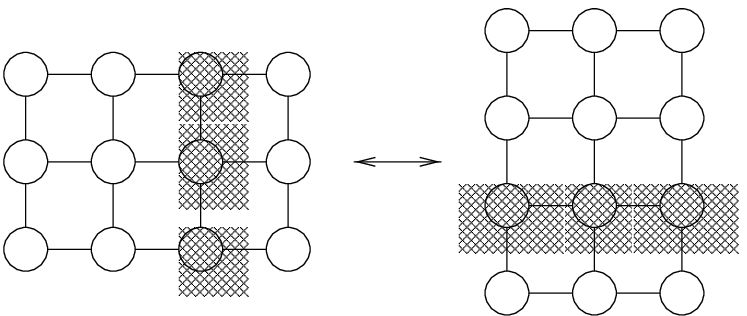}}
Cela veut-il dire que le modèle
décrit par le diagramme de droite de la figure \dualbaeb\ ne
correspond à rien de connu? En fait, il correspond tout simplement
à un cas que nous n'avons pas examiné, et qui
est le régime ferromagnétique, obtenu à partir
du régime antiferromagnétique en changeant
de signe l'énergie [\ref\BRb{V.V.~Bazhanov et N.~Reshetikhin,
{\it J. Phys.} A23 (1990), 1477.}]. La dualité rang-niveau
permet donc de relier, dans la limite thermodynamique,
les régimes ferromagnétique et
antiferromagnétique.

Plus généralement, en se rappelant que les noeuds des diagrammes
de BAE sont liés aux représentations rectangulaires,
on voit que la dualité rang-niveau correspond à la {\it transposition}
des tableaux de Young. Dans cette optique, la dualité rang-niveau
est basée sur le fait
que la transposition respecte le produit tensoriel,
c'est-à-dire que $R^T\otimes R'{}^T=(R\otimes R')^T$. En particulier,
les équations de fusion sont stables par transposition, ce qui
nous donne des relations entre quantités thermodynamiques
de différents modèles (cf [\artic03.4.2]).

\newsec\secbaephysgen{BAE physiques générales: comparaison $SU(2)$/$SU(N)$}
Cette dernière section a pour but de donner une méthode générale
pour passer des BAE nues aux BAE physiques dans
le cas de symétries $SU(2)$ ou $U_q({\goth sl}(2))$,
et de faire
quelques remarques sur la généralisation à $SU(N)$.
Les calculs qui suivent sont un peu formels, mais
les conclusions que l'on va en tirer seront, elles, bien concrètes.

Essayons donc de déduire de manière synthétique les BAE physiques
à partir des BAE nues pour un modèle général. Rappelons
l'idée générale: certains
noeuds du diagramme nu vont devenir les ``sources'' des BAE physiques.
On suppose que les particules physiques sont des multiplets
de $U_{q^{(a)}}({\goth sl}(2))$, $a=1\ldots b$, et que les matrices
$S$ de diffusion se factorisent en un produit des matrices $R$
correspondantes, fois un facteur scalaire $\Sh$. Le seul problème
consiste à calculer $\Sh$. Nous avons remarqué au \secphyssup\ que $\Sh$
se factorisait naturellement: nous allons maintenant montrer
ce fait.

Supposons plus précisément
que la particule de type $n$ ($n=1\ldots n_{\rm max}$)
appartienne à la représentation de spin $f_n^{(a)}/2$ du
groupe quantique $U_{q^{(a)}}({\goth sl}(2))$. 
Pour chaque symétrie $U_{q^{(a)}}({\goth sl}(2))$, on a une
matrice de transfert à diagonaliser qui nous donne des
équations du type:
$$C_{jk} \star \rhot^{(a)}_k + \rho^{(a)}_j=
s\star \sum_{n=1\atop f_n^{(a)}\ne 0}
^{n_{\rm max}} \delta_{j f_n^{(a)}}\sigma_n\eqn\genphysa$$
où $\sigma_n$ est la densité de particules de type $n$,
et $\rho^{(a)}_j$ est la densité de $j$-cordes. Nous n'incluons
pas les $1^-$-cordes, car du fait que seul $\rho_{1^-}$ apparaît
dans les équations, et non $\rhot_{1^-}$, pour les manipulations
formelles auxquelles nous procédens, il ne joue aucun rôle.

Ensuite, on repart des équations
de BAE nues au noeud coché lié à la particule $n$:
$$\sigma_n + \tilde{\sigma}_n - s\star \sum_{a=1\atop f^{(a)}_{n}\ne 0}^b
\rhot^{(a)}_{f^{(a)}_n}={m_n\over 2\pi} p'(\lambda)\eqn\genphysb$$
Le second membre $p'(\lambda)$ ne nous intéresse pas, donc
nous ne l'écrivons pas explicitement. Les équations \genphysb\ ne
sont pas encore des BAE physiques, car y apparaissent des
$\rhot^{(a)}$, alors que seuls les $\rho^{(a)}$ doivent y apparaître.
On exprime donc $\rhot^{(a)}_{f^{(a)}_n}$ en fonction des $\rho^{(a)}$
grâce à \genphysa:
$$\rhot^{(a)}_{f^{(a)}_n}=-C^{-1}_{jk}\star\rho^{(a)}_k
+s\star\sum_{m=1\atop f^{(a)}_m\ne 0}
^{n_{\rm max}} C^{-1}_{f_n^{(a)}f_m^{(a)}}\star\sigma_m
\eqn\genphysc$$
et on remplace dans \genphysb:
$$\sigma_n + \tilde{\sigma}_n =
{m_n\over 2\pi} p'(\lambda)
+ s^2\star \sum_{m=1}^{n_{\rm max}}
\sum_{a=1\atop f^{(a)}_{n}\ne 0}^b 
C^{-1}_{f_n^{(a)}f_m^{(a)}}\star\sigma_m
+\cdots\eqn\genphysd$$
où $s^2\equiv s\star s$, et les $\cdots$ sont les termes qui sont
fonctions des $\rho^{(a)}$, qui ne nous intéressent pas, puisque
dans la diffusion de plus haut poids $\rho^{(a)}=0$. On déduit
finalement de \genphysd\ l'expression générale
des déphasages:
$$\Sh_{nm}(\lambda)=\prod_{a=1\atop f_n^{(a)},\,f_n^{(b)}\ne 0}^b
\Sh_{nm}^{(a)}(\lambda)$$
c'est-à-dire que $\Sh_{mn}$ se factorise sous
forme de facteurs $\Sh_{nm}^{(a)}$ liés à la symétrie
$U_{q^{(a)}}({\goth sl}(2))$,
et donnés par
$${1\over 2\pi i} {\d\over\d\lambda} \Sh^{(a)}_{mn}(\lambda)
= s^2\star C^{-1}_{f_n^{(a)}f_m^{(a)}}$$
$C^{-1}$ est la matrice de Cartan inverse du diagramme
de Dynkin $A_{p-1}$ si $q^{(a)}=-\exp(-i\pi/p)$.
Ceci concorde avec les expressions trouvées précédemment
(Eq. \Sphys, \sGS, \Sfus).

Ce raisonnement se généralise-t-il à $SU(N)$? Pas directement,
car il faut, pour des rangs plus élevés,
tenir compte de la structure ``verticale'' des
diagrammes de BAE. Pour voir d'où vient le problème, supposons
que les particules physiques sont maintenant classifiées par deux
nombres $n$ et $r$, de sorte qu'elles appartiennent
à des tableaux de Young rectangulaires $f^{(a)}_n\times r$
pour une symétrie $U_{q^{(a)}}({\goth sl}(N))$.
On peut alors montrer que la généralisation de \genphysd\ est:
$$(C^{-1})^{qr}\sigma_n^q + \tilde{\sigma}_n^r =
{m^r_n\over 2\pi} p'{}^r(\lambda)
+ s^2\star \sum_{m=1}^{n_{\rm max}}
\sum_{a=1\atop f^{(a)}_{n}\ne 0}^b 
C^{-1}_{f_n^{(a)}\, f_m^{(a)}}\star(C^{-1})^{qr}\star\sigma_m^q
+\cdots\eqn\genphyse$$
On doit tenir compte du $(C^{-1})^{qr}$ du membre de gauche
dans le déphasage, si bien que l'on a:
$$\Sh_{nm}^{qr}(\lambda)=\left\{\matrix{
\displaystyle
X^{qr}(\lambda)
\prod_{a=1\atop f_n^{(a)}\ne 0}^b
\Sh_{nn}^{(a)}{}^{qr}(\lambda)&n=m\cr
\displaystyle
\prod_{a=1\atop f_n^{(a)},\,f_n^{(b)}\ne 0}^b
\Sh_{nm}^{(a)}{}^{qr}(\lambda)& n\ne m\cr}\right.$$
avec les facteurs habituels
$${1\over 2\pi i} {\d\over\d\lambda} \Sh^{(a)}_{mn}{}^{qr}(\lambda)
= s^2\star (C^{-1})^{qr}\star C^{-1}_{f_n^{(a)}f_m^{(a)}}$$
et les {\it facteurs CDD} supplémentaires $X^{qr}$ donnés par
$${1\over 2\pi i} {\d\over\d\lambda}
X^{qr}(\lambda)=\delta^{qr}-(C^{-1})^{qr}$$
Par exemple, on calcule que
$$X^{11}(\lambda)={\sinh{\pi\over N}(\lambda+i)
\over\sinh{\pi\over N}(\lambda-i)}$$
On vérifie que la matrice $S$ \Ssup\ admet une telle décomposition.

Remarquons finalement que le fait
que la matrice $S$ se factorise entièrement pour $SU(2)$,
et modulo un facteur CDD seulement pour $SU(N)$,
avait été remarqué empiriquement dans le modèle $\sigma$
principal (voir chapitre \chappcf) dans [\ref\WIEG{P.~Wiegmann,
{\it Phys. Lett.} B142 (1984), 173.}].

\newchap\chaptba{Corrections de taille finie et TBA}
Nous allons maintenant nous servir des Equations d'Ansatz de Bethe
pour calculer des quantités physiquement intéressantes. 
Une idée naturelle est d'étudier des systèmes
à température finie, ce qui nécessitera 
les Equations d'Ansatz de
Bethe Thermodynamiques (TBA, section \sectba) qui sont, elles, non-linéaires. 
Nous nous restreindrons ensuite aux théories quantiques des
champs relativistes. 
Pour celles-ci, la fonction de partition à température finie
n'est autre que la fonction de partition avec une direction de temps
euclidienne compactifiée, et on voit que l'Ansatz de Bethe
nous permet de faire des calculs de {\it corrections de taille finie},
c'est-à-dire des corrections aux quantités physiques dues à la
géométrie finie de l'espace-temps.
Nous ferons alors des calculs explicites de
charges centrales grâce aux TBA et aux sommes dilogarithmiques
(\secdilog). Puis nous montrerons
le lien entre ces équations et les {\it équations de fusion}
(section \secfuseq). Ceci nous conduira tout naturellement
aux Equations Non-Linéaires Intégrales (NLIE),
qui, comme les TBA, déterminent les corrections
de taille finie, mais cette fois dans la direction spatiale.
Les NLIE constituent une méthode alternative
aux TBA, et nous étudierons de telles équations
dans la section \secddv. Nous montrerons en particulier
qu'elles fournissent plus
d'informations que les TBA (charge centrale mais aussi poids
conformes, etc).

\newsec\sectba{Equations d'Ansatz de Bethe Thermodynamiques (TBA)}
\subsec\secbaediag{Principe}
Nous allons exposer la méthode d'obtention des TBA dans le cadre le
plus simple: le modèle de diffusion diagonale introduit au
\secsimpbae. On se place à température finie $T$,
et à potentiel chimique $\mu$, de sorte que l'énergie libre s'écrit
$F=E-TS-\mu N$. Pour un état constitué de $m$ particules
et de $\tilde{m}$ trous, on a un facteur entropique associé qui vaut:
$\exp S=(m+\tilde{m})\exclam/(m\exclam\,\tilde{m}\exclam)$.
Dans la limite thermodynamique, on trouve donc
$$\eqalign{F&/L=\int \d\lambda\, \rho(\lambda) \epsilon(\lambda)
-\mu\int\d\lambda\,\rho(\lambda)\cr
&-T\int\d\lambda\,
\left[
(\rho(\lambda)+\rhot(\lambda))\log(\rho(\lambda)+\rhot(\lambda))
-\rho(\lambda)\log\rho(\lambda)
-\rhot(\lambda)\log\rhot(\lambda)\right]
}\eqn\Fdiag$$
Quand $L\to\infty$, la fonction de partition est dominée par un point
de col par rapport aux densités $\rho(\lambda)$ et $\rhot(\lambda)$
[\ref\YY{C.N.~Yang et C.P.~Yang, {\it J. Math. Phys.} 10 (1969), 1115.}];
on différentie \Fdiag\ et on utilise l'Equation d'Ansatz de Bethe \baediag\
qui relie $\delta\rho(\lambda)$ et $\delta\rhot(\lambda)$;
on obtient alors l'équation de point de col:
$$0=\epsilon(\lambda)-\mu-T\left[\log{\rhot(\lambda)\over\rho(\lambda)}
+K\star\log\left(1+{\rho(\lambda)\over\rhot(\lambda)}\right)\right]$$
On voit que l'équation ne dépend que de $\eta\equiv\rhot/\rho$:
$$\log\eta(\lambda)+K\star\log(1+(\eta(\lambda))^{-1})=
{\epsilon(\lambda)-\mu\over T}\eqn\tbadiag$$
L'équation \tbadiag\ constitue l'Equation d'Ansatz de Bethe Thermodynamique
(TBA) de notre modèle: c'est une équation non-linéaire intégrale.

On peut ensuite calculer la valeur de $F$ au point de col; on trouve:
$$\eqalign{
F/L&=-T\int{\d\lambda\over 2\pi}{\d p\over\d\lambda}
\log(1+(\eta(\lambda))^{-1})\cr
&=-{T\over 2\pi} \int \d p\,\log(1+(\eta(p))^{-1})\cr}\eqn\Fdiagb$$
L'interprétation des équations \tbadiag-\Fdiagb\ est relativement
simple si l'on se rappelle l'analogie avec un système de fermions
sans interaction: pour cela, posons
$$\eta(\lambda)=\e{(\epsilon_{\rm hab}(\lambda)-\mu)/T}$$
Alors, \tbadiag\ se récrit
$$\epsilon_{\rm hab}(\lambda)+{T\over 2\pi} K\star
\log\left(1+\e{(\epsilon_{\rm hab}(\lambda)-\mu)/T}\right)=\epsilon(\lambda)$$
En l'absence d'interaction entre les particules, $\epsilon_{\rm hab}$ vaut
simplement $\epsilon$. Quand on rajoute l'interaction, on
continue à interpréter, d'après \Fdiagb, $\epsilon_{\rm hab}$
comme la fonction énergie des particules, mais ``habillée''
par l'interaction.

\subsec\secbaegen{Application aux modèles à diffusion non-diagonale}
Considérons maintenant la chaîne XXX (ou le modèle NJL) à température
finie $T$; il n'y a pas de notion évidente de
potentiel chimique, mais en revanche, on peut placer un champ magnétique $B$.
Celui-ci joue dans la limite thermodynamique le rôle de potentiel
chimique pour les racines des BAE. En effet, lorsque $L\to\infty$,
et pour $B=\left({-b\atop 0}{0\atop b}\right)$ ($b\ge0$),
on peut se contenter de la contribution du vecteur de plus haut poids
à la fonction de partition (voir [\artic03] pour une discussion
détaillée de ce point dans le cas plus général $SU(N)$), de sorte
que $B$ est couplé au spin: $\vec B\cdot\vec S=-bs={\rm cst}+bm$,
c'est-à-dire que $-b$ joue le rôle de potentiel chimique
des pseudo-particules\foot{Le signe
est important car $-b$ est toujours négatif.}.
On obtient finalement les TBA suivantes: ($\eta_j=\rhot_j/\rho_j$)
$$\log(1+\eta_j(\lambda))-C^{-1}_{jk}\star\log(1+(\eta_k(\lambda))^{-1})
={g_j(\lambda)\over T}\eqn\tbaxxx$$
où $g_j(\lambda)$ est la contribution à l'énergie d'une $j$-corde:
$g_j(\lambda)=-2\pi\,K_j(\lambda)+jb$ pour la chaîne XXX. La
multiplication par la matrice de Cartan fait disparaître la
dépendance en le champ magnétique $B$:
$$\log(1+(\eta_j(\lambda))^{-1})
-C_{jk}\star\log(1+\eta_k(\lambda))
=\delta_{j1}{\tilde{\epsilon}(\lambda)\over T}\eqn\tbaxxxb$$
qui se trouve cachée dans l'asymptotique de $\eta_j$ pour $j\to\infty$
(voir [\artic03.3]). L'énergie libre s'écrit quant à elle
$$\eqalign{
F&=-T\int\d\lambda\,\sigma\star K_j(\lambda) 
\log(1+(\eta_j(\lambda))^{-1})\cr
&=E_\gs-T\int\d\lambda\,\sigma\star s(\lambda) \log(1+\eta_1(\lambda))
\cr}\eqn\Fxxx$$

On voit que les TBA ont essentiellement la même structure
que les BAE physiques, de sorte que, au moins pour
les modèles relativistes, il existe des règles assez évidentes
qui permettent d'obtenir les équations de TBA \tbaxxxb\ et
l'expression de l'énergie libre (ici, Eq.\ \tbaxxxb\ et \Fxxx) à
partir des diagrammes de BAE physiques (ici,
le diagramme de droite de la figure \baeb). Nous ne
donnerons pas ces règles explicitement, car
nous en verrons suffisamment d'exemples dans les paragraphes qui suivent.

\subsec\secdilog{Théories conformes et calculs de dilogarithme}
Le cas de figure le plus simple est celui des TBA des théories
conformes, pour lesquelles l'énergie libre $F$ doit
être directement reliée à la charge centrale par la formule
[\ref\BCNI{H.W.J.~Blöte, J.L.~Cardy et M.P.~Nightingale,
{\it Phys. Rev. Lett.} 56 (1986), 742\semi
I.~Affleck, {\it Phys. Rev. Lett.} 56 (1986), 746.}]:
$$F/L=-{\pi\over 12} (c+\bar c)  T^2\eqn\Fc$$

Reprenons encore une fois le modèle de diffusion diagonale,
et supposons (condition d'invariance conforme)
que $p(\lambda)=\epsilon(\lambda)$. 
On suppose aussi que $\epsilon(\lambda)$ est une fonction
croissante de $\lambda$.
On remarque alors que si l'on pose:
$$\eqalign{
\rho(\lambda)&=-{T\over 2\pi} 
{\d\over\d\lambda} \log(1+(\eta(\lambda))^{-1})\cr
\rhot(\lambda)&=+{T\over 2\pi} 
{\d\over\d\lambda} \log(1+\eta(\lambda))\cr
}\eqn\baetba$$
(de sorte que $\rhot/\rho=\eta$ comme il se doit, et $\rho$, $\rhot>0$),
on résoud les BAE \baediag\ puisque celles-ci sont une simple conséquence
des TBA \tbadiag\ par dérivation par rapport à $\lambda$!
Calculons alors $S=-\der F/\der T$:
$$\eqalign{
S/L&= \int\d\lambda\,\left[\rho(\lambda)\log(1+\eta(\lambda))
+\rhot(\lambda)\log(1+(\eta(\lambda))^{-1})\right]\cr
&={T\over 2\pi} \int\left[
-\d(\log(1+\eta^{-1})) \log(1+\eta)+ \d(\log(1+\eta))\log(1+\eta^{-1})
\right]\cr
&={T\over \pi} L(f) \bigg|
^{\displaystyle f=1/(1+\eta(-\infty))}
_{\displaystyle f=1/(1+\eta(+\infty))}\cr
}\eqn\dilog$$
où la fonction dilogarithme 
[\ref\KRKM{A.N.~Kirillov et N.~Reshetikhin, {\it J. Phys.} A20 (1987), 1587\semi
T.R.~Klassen et E.~Melzer, {\it Nucl. Phys.} B338 (1990), 485.}]
est définie par
$L(f)=-{1\over 2}\int_0^f\big[{\d z\over z} \log(1-z)+{\d z\over 1-z} \log
z\big]$.

Les valeurs limites $\eta(+\infty)$ et $\eta(-\infty)$ sont
faciles à calculer: soit elles sont infinies,
soit elles vérifient d'après \tbadiag
$$\log\eta(\pm\infty)+K(\kappa=0) \log(1+(\eta(\pm\infty))^{-1})=
{\epsilon(\pm\infty)-\mu\over T}\eqn\tbalim$$

La généralisation aux modèles dont les TBA sont un système
d'équations couplées est élémentaire:
$$S/L={T\over\pi} \sum_j L(f_j)\big|
^{f=f_{\rm max}}_{f=f_{\rm min}}\eqn\dilogb$$
où $j$ parcourt l'ensemble des équations, c'est-à-dire {\it tous}
les noeuds du diagramme de BAE.
Pour rendre l'équation \dilogb\ indépendante de
la convention de signe choisie pour $\eta$,
la prescription est que $f$ va dans \dilogb\ de sa
valeur minimale à sa valeur maximale
pour l'équation qui contient le second membre.

Prenons maintenant un exemple concret de calcul
de charge centrale. Considérons le modèle décrit par le
système de BAE physiques de la figure \coset\ 
[\ref\Zam{Al.B.~Zamolodchikov, {\it Nucl. Phys.} B358 (1991), 497;
{\it Nucl.\ Phys.} B366 (1991), 122.}].
\fig\coset{TBA des modèles coset.}{\figdir{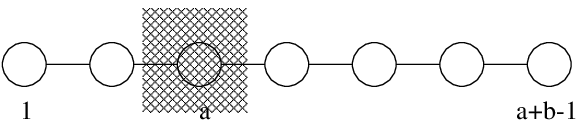}} 
Les excitations physiques sont des particules relativistes
sans masse ``droites'', avec des nombres quantiques
$U_{\e{i\pi/(a+2)}}({\goth sl}(2))$ et 
$U_{\e{i\pi/(b+2)}}({\goth sl}(2))$ tronqués. Il n'y
a pas de champ magnétique. Les TBA associées
sont:
$$\log(1+(\eta_j(\zeta))^{-1})
-C_{jk}\star\log(1+\eta_k(\zeta))
=\delta_{ja}{\e{\zeta}\over T}\qquad 1\le j\le a+b-1
\eqn\tbacoset$$
où l'on a utilisé la normalisation usuelle des rapidités:
$\zeta\equiv \pi\lambda$ (les noyaux changent
donc aussi de normalisation, de façon à ce que
$f(\lambda)\d\lambda=f(\zeta)\d\zeta$). Remarquons que la dépendance des TBA
(ainsi que de l'énergie libre $F$, que nous n'avons pas écrite
explicitement)
en la température $T$ est triviale du fait
que l'on peut décaler $\zeta\to\zeta+{\rm cst}$, ce qui est normal pour une théorie conforme.

Les valeurs limites sont:
$$\matrix{
\displaystyle
\eta_j(-\infty)={\sin(\pi j/(a+b+2))\sin(\pi(j+2)/(a+b+2))
\over \sin^2(\pi/(a+b+2))}\hfill & 1\le j\le a+b-1\cr
\displaystyle
\eta_j(+\infty)={\sin(\pi j/(a+2))\sin(\pi(j+2)/(a+2))
\over \sin^2(\pi/(a+2))}\hfill & 1\le j\le a\cr
\displaystyle
\eta_j(+\infty)={\sin(\pi (j-a)/(b+2))\sin(\pi(j-a+2)/(b+2))
\over \sin^2(\pi/(b+2))}\hfill & a\le j\le a+b-1\cr
}
\eqn\vallim$$
Nous donnerons ultérieurement une justification de ces valeurs (section
\secfuseq). En utilisant des formules connues de sommes
dilogarithmiques [\ref\KI{A.N.~Kirillov,
{\it Zap. Nauch. Semin. LOMI} 164 (1987), 121.}], on obtient (on
pose $\bar c=0$ dans \Fc):
$$c={6\over\pi^2}\sum_j L(f_j)\big|
^{+\infty}_{-\infty}
={3ab(a+b+4)\over (a+2)(b+2)(a+b+2)}\eqn\cosetc$$
qui est la charge centrale des modèles coset $SU(2)_a\times SU(2)_b
/SU(2)_{a+b}$ [\ref\GKO{P.~Goddard, A.~Kent and D.~Olive,
{\it Phys. Lett.} B152 (1985), 88.}].

La règle générale pour déduire la charge centrale du diagramme
des BAE physiques d'une théorie conforme à une
chiralité est la suivante:
\item{$\bullet$} Ajouter $1$ par noeud hachuré.
\item{$\bullet$} Ajouter la contribution du diagramme ``physique'',
c'est-à-dire en retirant les noeuds hachurés.
\item{$\bullet$} Retrancher la contribution du diagramme ``nu'',
c'est-à-dire au contraire en considérant les noeuds hachurés comme
des noeuds normaux.

La contribution des diagrammes qui nous
intéressent est donnée par la figure \dilogval.
\figg\dilogval{Table de sommes dilogarithmiques.}{%
$$\matrix{
a)\hfill&\hskip4.7mm\vcenter{\epsfbox{\figdir{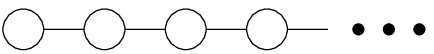}}}\hfill&1\cr
b)\hfill&\hskip4.7mm\vcenter{\epsfbox{\figdir{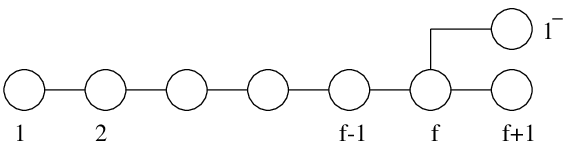}}\vskip1mm}\hfill&1\cr
c)\hfill&\hskip4.7mm\vcenter{\vskip1mm\epsfbox{\figdir{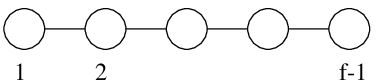}}}\hfill&{2(f-1)\over f+2}\cr
d)\hfill&\vcenter{\vskip1mm\epsfbox{\figdir{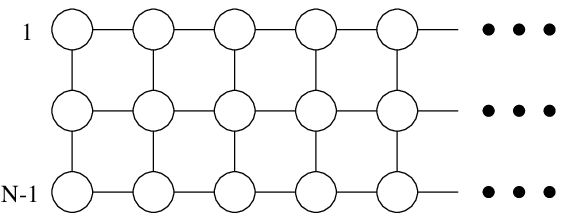}}}\hfill&{N(N-1)\over 2}\cr
e)\hfill&\vcenter{\vskip1mm\epsfbox{\figdir{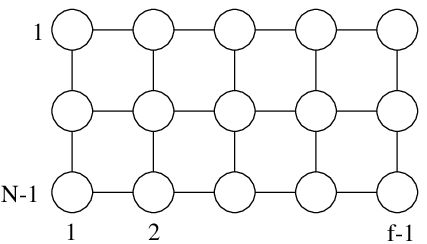}}}\hfill&{N(N-1)(f-1)\over f+N}\cr
}$$} 
Notons que a) n'est pas la limite de c) quand $f\to\infty$
(ou d) de e)),
car, pour $f$ grand, la somme dilogarithmique c) est
dominée par ses premiers termes ($j\ll f$) {\it et} par ses
derniers termes ($f-j\ll f$), et la somme a) n'est la limite
que de la première moitié. Cependant, du fait des compensations
entre les contributions du diagramme physique et du diagramme nu,
cela n'a pas de conséquence physique.
Montrons par exemple comment obtenir la formule b)
de manière élémentaire: on calcule tout d'abord les valeurs
limites des $\eta_j$ correspondant au diagramme: on
trouve $\eta_j=j(j+2)$ pour $1\le j\le f$,
$\eta_{f+1}=\eta_{1^-}=f+1$. On cherche donc à calculer
$$c={6\over\pi^2}
\sum_{j=1}^f L\left({1\over (j+1)^2}\right)+2L\left({1\over
f+2}\right)$$
On utilise la relation
$L(xy)=L(x)+L(y)-L(x(1-y)/(1-xy))-L(y(1-x)/(1-xy))$
satisfaite par $L$, avec $x=y=1/j$; on obtient alors
$$\eqalign{
c&={12\over\pi^2}\sum_{j=1}^f \left[ L\left({1\over j+1}\right)-L\left({1\over
j+2}\right)\right] + 2 L\left({1\over f+2}\right)\cr
&={12\over\pi^2} L\left({1\over 2}\right)\cr
&=1\cr}$$

Nous admettrons la valeur des autres sommes
dilogarithmiques de la figure \dilogval\ (voir
par exemple [\ref\FrSz{E.~Frenkel
et A.~Szenes, \pre{hep-th/9506215}.}] et références incluses).
Prenons encore un exemple d'application de ces formules
à un calcul de charge centrale qui
nous resservira par la suite: considérons la version
à une seule chiralité du modèle fusionné décrit
par les BAE physiques \baeNphys\foot{Ce modèle n'est bien sûr
pas un modèle de fermions chiraux libres dans des \rep\
supérieures: la première
remarque de la fin du \secbaehypco\ ne s'applique pas au cas fusionné,
cf l'observation similaire du paragraphe suivant
pour le cas anisotrope.}, et que nous reproduisons ici (figure
\tbaNwz).
\fig\tbaNwz{Diagramme de TBA de la CFT WZW chirale.}{\figdir{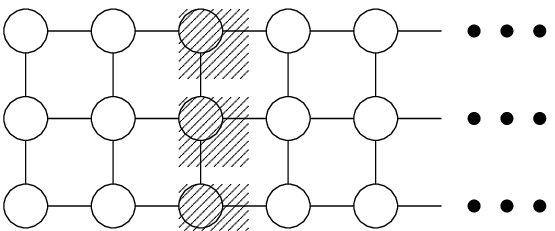}}
On calcule aisément la charge centrale, qui vaut: $c=(N^2-1)f/(f+N)$,
c'est-à-dire celle de la théorie conforme de Wess--Zumino--Witten
$SU(N)$ au niveau $f$ [\ref\WI{E.~Witten, {\it Commun. Math. Phys.} 92
(1984), 455.}].
Nous prouverons au chapitre \chappcf\ que si l'on considère une
chiralité de la CFT WZW,
alors elle admet effectivement le diagramme de BAE physiques (ou de
TBA) de la figure \tbaNwz.

Terminons sur quelques remarques générales:
\item{$\star$} Une fois que l'on a calculé l'énergie libre à champ magnétique
nul, on peut facilement réintroduire le champ magnétique, et calculer
la nouvelle somme dilogarithmique, par exemple en dérivant deux fois
par rapport au champ magnétique. Nous ferons un tel calcul au
paragraphe suivant.

\item{$\star$} Il existe une autre méthode,
plus traditionnelle, pour arriver aux formules
de dilogarithmes, qui n'utilise que les TBA (et pas les BAE).
C'est celle qui est utilisée dans les articles [\artic03,\artic04].
Notons que dans [\artic03.B], le modèle
n'est pas à strictement parler conforme, mais
cette méthode fonctionne tout de même; de plus, l'énergie libre est calculée
directement en présence d'un champ magnétique.

\item{$\star$}
Les structures ``nue'' et ``physique''
ne sont pas directement liées au groupe de renormalisation,
puisque même pour une théorie conforme, les deux structures coexistent
et donnent chacune leur contribution à la charge centrale\foot{Nous
verrons cependant un cas où le passage de la structure nue
à la structure physique est lié au flot de groupe de
renormalisation: le modèle de Kondo (section \seckondo).}.

\subsec\secdiloguv{Limite ultraviolette des théories massives et
découplage des chiralités}
Les calculs faits précédemment ne s'appliquent pas
au cas des théories massives. Cependant, on peut
espérer, dans la limite $T\to\infty$, explorer
la région ultra-violette et en particulier
retrouver la théorie conforme qui constitue le point
fixe ultra-violet de la théorie.

La limite $T\to\infty$ des TBA massives fait apparaître un
phénomène bien connu que nous nommerons découplage des chiralités.
Celui-ci est caractérisé par le fait que les fonctions
$\eta(\zeta)$ (on rappelle que l'on prend
maintenant la paramétrisation usuelle
de la rapidité; pour un modèle $SU(N)$,
on a la correspondance: $\zeta=2\pi\lambda/N$ avec les
notations du chapitre \chapbase)
se mettent à développer un plateau
dans la région $[-\log(T/m),+\log(T/m)]$, où $m$ est l'échelle
de masse du système. On peut alors définir des fonctions
$\eta^\pm(\zeta)=\eta(\zeta\pm\log(2T/m))$ qui admettent
(généralement) une limite finie quand $T\to\infty$: elles décrivent
séparément les deux chiralités de la théorie conforme ultra-violette. En effet les particules qui ont une énergie de l'ordre
de $T$ ont une rapidité $\zeta\sim \pm\log(T/m)$
et vérifient $\epsilon\sim |p|$ dès que $T\gg m$\foot{Remarquons
à ce propos que les particules d'une même chiralité 
dans la limite UV ont la même matrice $S$ de diffusion -- exprimée
en termes de la rapidité décalée -- que celle de la théorie massive
originale, puisque la matrice $S$ ne dépend que
de la différence des rapidités.}.

Prenons comme exemple le modèle NJL standard.
Diagrammatiquement, la limite UV est
représentée par la figure \uvbae.
\fig\uvbae{Découplage des chiralités dans le modèle NJL $SU(2)$.}{\figdir{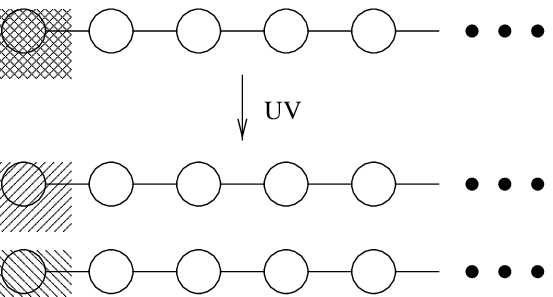}}
Les fonctions $\eta^\pm$ vérifient des équations séparées qui sont
les limites de \tbaxxxb\ 
(avec $\tilde{\epsilon}(\zeta)=m\cosh\zeta$):
$$\log(1+(\eta^\pm_j(\zeta))^{-1})
-C_{jk}\star\log(1+\eta^\pm_k(\zeta))
=\delta_{j1}{\e{\pm\zeta}\over T}\eqn\tbaxxxc$$
En appliquant les méthodes du \secdilog\ à \tbaxxxc, on calcule
les charges centrales UV $c$ et $\bar c$: on trouve sans difficulté\foot{Le
calcul est en fait trivial, du fait de la particularité du cas
isotrope: les symétries nue et physique sont identiques, d'où
compensation évidente de leur contribution à la charge centrale.}
$c=\bar c=1$, ce qui pouvait se deviner, puisque le modèle NJL est
asymptotiquement libre: dans l'UV, restent donc
deux fermions de Dirac libres ($c=2$) auquels on a retiré un secteur $c=1$
découplé.

On peut étendre le calcul de la charge centrale UV au modèle
NJL déformé, et on trouve qu'indépendamment de l'anisotropie $\gamma$,
$c=\bar c=1$. Ceci veut-il dire que le modèle NJL déformé est
encore asymptotiquement libre? Non, car la
limite UV du modèle NJL, quand on fait varier $\gamma$,
parcourt en fait la ligne de points critiques $c=1$ (du côté
massif de la transition de Kosterlitz--Thouless). Nous renvoyons
le lecteur aux références [\ref\JZJ{J.~Zinn-Justin,
{\it Quantum field theory and critical phenomena}, 3rd edition,
Oxford University Press (1996).},\ref\PG{P.~Ginsparg,
{\it Applied Conformal Field Theory}, dans {\it Fields,
strings and critical phenomena}, Les Houches 1988 (éditeurs:
E.~Brézin et J.~Zinn-Justin), North Holland.}]
pour une discussion générale de la transition de Kosterlitz--Thouless
et des mo\-dèles $c=1$. L'essentiel pour nous est que
la théorie UV {\it ne coïncide pas} avec la théorie non-perturbée;
en effet, si l'on regarde le flot de groupe de renormalisation
des deux constantes $f$ et $g$ du Lagrangien \lagGNdef,
on voit que $f$ tend vers $0$ (le terme
couplé à $f$ constitue la perturbation), 
mais que $g$ tend vers $\gamma$
(l'anisotropie $\gamma$ est un invariant du flot).
En particulier,
bien qu'il y ait génération dynamique de masse (par rapport
au modèle non-perturbé), il n'y a pas liberté asymptotique
puisque l'opérateur perturbant acquiert une dimension plus
petite que $1$ (donc devient relevant) quand on s'éloigne du
point isotrope. Une autre remarque reliée est que
les deux chiralités de la théorie UV {\it ne sont pas} les fermions chiraux de la
théorie non-perturbée, et donc la première remarque
de la fin du \secbaehypco\ ne s'applique pas au cas
anisotrope. Nous reviendrons sur la limite UV du modèle
NJL quand nous lui écrirons une NLIE (section \secddv).

Plaçons ensuite notre modèle NJL déformé à $\gamma=\pi/(p+1)$ dans
un champ magnétique $B=\left({-b\atop 0}{0\atop b}\right)$;
au lieu d'être lié à la limite $j\to\infty$ comme dans le cas
isotrope, il apparaît explicitement dans les deux derniers noeuds
des TBA [\ref\FS{P.~Fendley et H.~Saleur, {\it Nucl. Phys.} B388
(1992), 609 \pre{hep-th/9204094}.}].
On trouve facilement par les formules de dilogarithmes que,
dans la limite ultra-violette,
$$F\sim -{\pi\over 6}T^2\left(1+{6b^2/T^2\over \pi(\pi-\gamma)}\right)$$
On observe alors le fait suivant [\ref\DVK{H.J.~De~Vega et M.~Karowski,
{\it Nucl.\ Phys.} B285 (1987), 619.}]:
pour un champ magnétique purement imaginaire
$b/T=i\gamma=i\pi/(p+1)$, on a $F\sim -(\pi/6)T^2\,\hat c$ avec
la nouvelle charge centrale
$$\hat c=1-{6\over p(p+1)}$$
qui est celle du modèle minimal $M_p$ [\ref\BPZ{A.A.~Belavin,
A.M.~Polyakov et A.B.~Zamolodchikov, {\it Nucl. Phys.} B241
(1984), 333.}].
L'interprétation de ce résultat est la suivante: au
point fixe UV, il est possible
de définir une nouvelle action de l'algèbre de Virasoro,
de charge centrale inférieure à $1$.
Ceci est essentiellement la construction
de Dotsenko--Fateev [\ref\DF{V.S.~Dotsenko et V.A.~Fateev,
{\it Nucl. Phys.} B240 (1984), 312.}] (ou de Feigin--Fuks) du gaz de Coulomb
avec une charge de fond à l'infini. Il est 
bien connu que dans cette construction, on peut se ramener à un
sous-ensemble fini d'opérateurs Virasoro-primaires,
de sorte que le Hilbert peut être tronqué
pour donner le Hilbert des modèles minimaux. Qu'en est-il ici?

Le lecteur aura sans doute deviné que c'est précisément la troncation
de groupe quantique qui va jouer ce rôle [\PS]. En effet, on vérifie qu'en
imposant le champ magnétique $b/T=i\gamma$, la fonction
$\eta_{p-1}$ devient constante et égale à $0$, de sorte que
les noeuds $p-1$, $p$, $1^-$ se découplent et que l'on est
ramené au diagramme tronqué (cf figure \truncbae)!
Le champ magnétique
imaginaire constitue donc une manière alternative d'effectuer
la troncation de groupe quantique. Son avantage est qu'en modifiant
le Hamiltonien (par ajout du champ magnétique), on change
explicitement la gradation du groupe
quantique (``spin'' dans la limite relativiste). \rem{c'est vrai ça?}
Du point de vue mathématique, l'introduction du champ magnétique
imaginaire correspond à modifier la fonction de partition
$Z=\tr(\e{-\H/T})$ en $Z=\tr(\K\e{-H/T})$, où
$\K$ est l'élément de $U_q({\goth sl}(2))$ pris
dans sa représentation sur le Hilbert complet.
Ce sont les propriétés dites
de Markov [\ref\JON{V.~Jones, {\it Invent. Math.} 72 (1983), 1.}]
de cette trace modifiée $x\to \tr(\K x)$ 
(où $x$ appartient au commutant de $U_q({\goth sl}(2))$, soit
dans les cas qui nous occupent à l'algèbre de Temperley--Lieb
[\ref\TL{H.N.V.~Temperley et E.H.~Lieb, {\it Proc. Roy.
Soc.} A322 (1971), 251.}])
qui permettent de retrouver la troncation de groupe quantique.
Une construction similaire existe pour $U_q({\goth sl}(N))$ (cf \artic04.9).

Finalement, en calculant les corrections des $\eta(\zeta)$
à leur comportement dominant $\eta^\pm(\zeta\mp\log(2T/m)$,
on peut également déterminer le voisinage du
point fixe UV, et en particulier quelle est la dimension
de l'opérateur perturbant. Dans les cas les plus simples,
on peut systématiser le calcul de ces 
corrections [\ref\ZKM{Al.B.~Zamolodchikov,
{\it Nucl. Phys.} B342 (1990), 695\semi
T.R.~Klassen et E.~Melzer, 
{\it Nucl. Phys.} B350 (1991), 635.}]. Nous ferons nous-mêmes
de tels calculs pour le modèle de Kondo (section \seckondo).

\subsec\secdilogir{Les flots non-massifs}
Jusqu'ici, toutes les théories à deux chiralités que nous
avons considérées étaient massives. Rien n'empêche cependant d'avoir
des théories sans masse, mais non-conformes. On a alors un flot
de groupe de renormalisation d'un point fixe infra-rouge à
un point fixe ultra-violet. Les TBA doivent de nouveau exhiber
un découplage des chiralités. On peut aisément visualiser
ce phénomène au niveau diagrammatique: là où la limite UV faisait
apparaître deux TBA chirales essentiellement identiques
aux TBA de départ (à part pour le second membre), la limite IR
produit des TBA chirales telles que les noeuds hachurés d'une chiralité
``tuent'' le noeud correspondant de l'autre chiralité. Voyons
comment cette règle s'applique sur un exemple simple:
considérons le système de TBA donné par le diagramme du milieu
de la figure \minflow.
\fig\minflow{TBA d'un flot non-massif (le flot est de bas en haut). Les noeuds
hachurés sont différents pour les deux chiralités,
donc il n'y a pas apparition de masse.}{\figdir{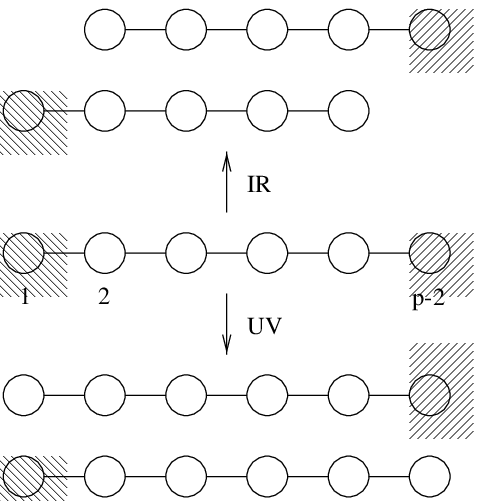}}
On a représenté les TBA dans la limite infra-rouge et dans la limite
ultra-violette. Les charge centrales $c_{\rm UV}=\bar{c}_{\rm UV}=1-6/(p(p+1))$
et $c_{\rm IR}=\bar{c}_{\rm IR}=1-6/(p(p-1))$, et
on peut se convaincre
[\FSZ] que les TBA de la figure \minflow\ représentent le flot
du modèle minimal $M_p$ vers $M_{p-1}$ 
par perturbation de $M_p$ par l'operateur $\Phi_{1,3}$.

Nous examinerons nous-mêmes un cas de flot non-massif:
le modèle de Wess--Zumino--Witten (chapitre \chappcf). Nous 
reviendrons alors sur la notion de diffusion des excitations
non-massives.

\newsec\secfuseq{$Y\!$-système et équations de fusion}
Il existe un lien profond entre les TBA et les ``équations
de fusion'' (nous avons déjà donné un exemple de ces dernières,
cf Eq.\ \ffus); celui-ci peut être deviné à partir
du lien déjà évoqué
entre la structure des TBA $SU(N)$ et les tableaux
de Young rectangulaires, suggérant une interprétation de
théorie de la représentation des TBA. Nous nous limiterons
dans cette section au cas isotrope (pour les TBA anisotropes,
il faut des équations de fusion plus générales
que celles que nous considérons, cf [\ref\KSS{A.~Kuniba,
K.~Sakai et J.~Suzuki, \pre{math/9803056}.}]).
Ce qui suit n'est qu'un survol d'un sujet qui est
encore en pleine activité à l'heure actuelle;
il nous permettra de mieux situer
un certain nombre de concepts introduits et utilisés
dans [\artic03,\artic04] (équation non-linéaire intégrale,
équations de fusion modifiées, etc).

Le point de départ des équations de fusion est la procédure
de fusion dans l'espace auxiliaire des matrices de monodromie.
La fusion exprime alors des relations dans l'algèbre
des représentations du Yangien correspondant $Y({\goth sl}(N))$.
En prenant la trace sur l'espace auxiliaire, on obtient
les matrices de transfert que l'on peut appeler
légèrement abusivement les caractères du Yangien $Y({\goth sl}(N))$.
Il s'agit d'un abus de langage car les matrices de transfert
sont encore des opérateurs sur l'espace physique, mais
qui se justifie du fait que, les matrices de transfert
commutant entre elles, on peut les identifier
à leurs valeurs propres; en particulier,
toute équation vérifiée par
les matrices de transfert est aussi une équation pour leurs
valeurs propres.

La plupart des relations qui existent pour l'algèbre
des représentations de ${\goth sl}(N)$ (et donc
leurs caractères) peuvent être
étendues en des relations pour $Y({\goth sl}(N))$ (et donc
les matrices de transferts associées),
à condition d'ajouter des décalages appropriés des
paramètres spectraux. Nous avons déjà vu que pour $SU(2)$,
l'identité $\ha\otimes\ha=1\oplus 0$ se traduit pour
les matrices de transfert par l'équation \ffus:
$$\Z_\ha(\lambda+i/2)\Z_\ha(\lambda-i/2)=\Z_1(\lambda)+\Z_0(\lambda)$$
On remarque que les équations de fusion ne dépendent
pas de l'espace physique, puisque la fusion n'a lieu que
dans l'espace auxiliaire. La simple donnée des
équations de fusion satisfaites par les matrices
de transfert ne peut donc évidemment pas contenir
toute la physique du modèle considéré. Les équations
de fusion ne dépendent pas non plus de la valeur
propre particulière considérée (état fondamental,
états excités). 

Pour $SU(N)$ et pour une représentation quelconque $R=\sum_r \mu^r e^r$
(avec les notations du [\artic03.A], c'est-à-dire
dont la \eme{r} ligne du tableau de Young a $\mu^r$ boîtes),
on a un analogue de la formule de Weyl
pour les matrices de transfert [\BRb]:
$$T_{\{ \mu^r \}}(\lambda)=\det\left[ T_{\mu_a+b-a}(\lambda+i(\mu_b-b))\right]
_{a,b=1\ldots N}$$
(avec des rapidités légèrement décalées par rapport à nos conventions).
Cependant, la structure des TBA suggère qu'il est nécessaire
de trouver des équations de fusion qui ne font appel qu'à
des tableaux de Young rectangulaires. Et en effet,
la figure \fuseqs\ montre qu'il existe de telles relations.
\fig\fuseqs{Egalité formelle entre des produits
tensoriels de représentations de $SU(N)$.}{\figdir{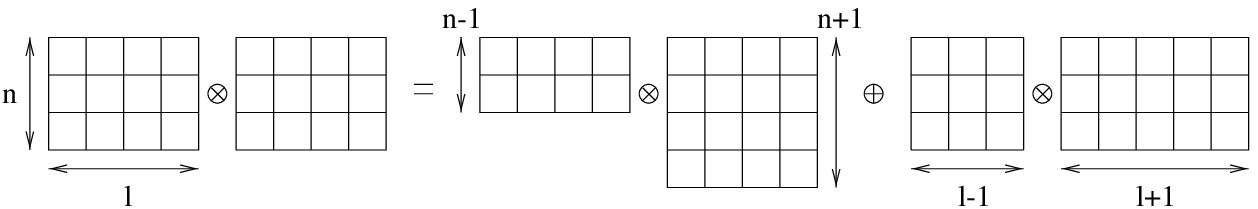}} 
La relation correspondante pour les matrices de transfert s'écrit
[\ref\KN{A.~Kuniba et T.~Nakanishi,
{\it Int. J. Mod. Phys. (Proc. Suppl.)} A3 (1993), 419.}]:
$$\Z^r_j(\lambda+i/2)\Z^r_j(\lambda-i/2)=\Z^{r+1}_j(\lambda)
\Z^{r-1}_j(\lambda)+\Z^r_{j+1}(\lambda) \Z^r_{j-1}(\lambda)\eqn\fusZ$$
$\Z^r_j$ est la matrice de transfert dont l'espace auxiliaire
est un tableau de Young à $j$ lignes et $r$ colonnes.
Considérons alors la combinaison [\ref\ZP{Y.K.~Zhou et P.~Pearce,
{\it Int. J. Mod. Phys.} B8 (1994), 3531 \pre{hep-th/9405019}.}]:
$$Y^r_j(\lambda)={\Z^r_{j+1}(\lambda) \Z^r_{j-1}(\lambda)
\over \Z^{r+1}_j(\lambda) \Z^{r-1}_j(\lambda)}\eqn\defY$$
On montre facilement à partir des équations de fusion \fusZ\
que les $Y^r_j$ vérifient le {\it $Y\!$-système}
[\ref\Zamb{Al.~B.~Zamolodchikov, {\it Phys. Lett.} B253
(1991), 391.}]:
$$Y^r_j(\lambda+i/2) Y^r_j(\lambda-i/2)=
{\left[1+Y^r_{j+1}(\lambda)\right] \left[1+Y^r_{j-1}(\lambda)\right]
\over \left[1+(Y^{r+1}_j(\lambda))^{-1}\right]
\left[1+(Y^{r-1}_j(\lambda))^{-1}\right]}.\eqn\Ysys$$
Le $Y\!$-système a été introduit
originellement en liaison avec les TBA; en effet,
si l'on pose $Y^r_j(\lambda)=\eta^r_j(\zeta)$ avec
$\zeta=2\pi\lambda/N$, alors
les équations \Ysys\ peuvent être déduites des
TBA $SU(N)$:
$$\log(1+(\eta^r_j(\zeta))^{-1})
-(C^{-1})^{qr}\star C_{jk}\star \log(1+\eta^q_k(\zeta)) 
= {\tilde{g}^r_j(\zeta)\over T}\qquad 1\le r\le N-1,\ 1\le j<\infty\eqn\tbaN$$
où les $\tilde{g}^r_j$ sont des fonctions
analytiques dans la bande $\Im\zeta\in[-\pi/N-\epsilon,
+\pi/N+\epsilon]$ et qui satisfont
$\tilde{g}^r_j(\zeta+i\pi/N)+\tilde{g}^r_j(\zeta-i\pi/N)
-\tilde{g}^{r+1}_j(\zeta)-\tilde{g}^{r-1}_j(\zeta)=0$ pour tout $j$.
Le $Y\!$-système permet alors de démontrer un certain nombre
de propriétés formelles des solutions des TBA \tbaN. Inversement,
la donnée des équations de fusions \fusZ\ et des comportements
asymptotiques des $\Z^r_j$ permet de ``remonter'' du $Y\!$-système
à des équations du type de \tbaN. Ceci a été fait
dans le cas $SU(2)$ dans
[\ref\KP{A.~Klümper et P.~Pearce, {\it Physica} A183 (1992),304; A194
(1993), 397.}],
et généralisé au cas $SU(N)$ (mais avec des erreurs
dans la dérivation -- misprints?) dans [\ZP].
Evidemment, les équations
ainsi obtenues ne sont pas réellement des TBA, puisque le
système considéré n'est pas à température finie mais dans un des états excités
au-dessus du vide. On les appelle
équations ``TBA-like''. Pour des modèles relativistes\foot{L'invariance
relativiste n'est en fait pas indispensable,
mais nous nous limiterons à ce cas.}, pour lesquels
$\tilde{g}^r_j=\sin(\pi r/N)(\alpha_j^+ \e{+\zeta}
+\alpha_j^- \e{-\zeta})$, le lien avec les vraies TBA
est le suivant: dans une limite d'échelle
appropriée, les TBA-like décrivent
le modèle correspondant sur un espace compactifié
de longueur finie; grâce à l'invariance
modulaire, on voit donc que les TBA-like pour l'état du vide
sont en fait les TBA usuelles après transformation
modulaire $\tau\to-1/\tau$; quant aux TBA-like pour
les états excités, elles correpondraient à d'éventuelles
``TBA pour les états excités'', qui n'ont pas d'interprétation
physique directe. De telles TBA ont effectivement été
introduites récemment [\ref\DT{P.~Dorey 
et R.~Tateo, \pre{hep-th/9706140}.}].

Remarquons que la dérivation
des TBA-like n'utilise jamais l'hypothèse de corde; et pourtant
on obtient la même structure des équations à deux indices que
dans les TBA. Ceci est bien sûr directement
lié à la remarque faite au \secfusb\ sur l'interprétation
de l'hypothèse de corde en termes de fusion.
En un certain sens, les TBA-like justifient donc
{\it a posteriori} l'usage de l'hypothèse de corde
dans les TBA.

Passons maintenant aux équations de fusion modifiées.
Celles-ci proviennent de la remarque suivante:
ni le $Y\!$-système ni les équations de fusion
usuelles ne peuvent contenir toute la physique d'un modèle,
puisque par rapport aux TBA, on a perdu le second membre.
Cependant il existe une manière de définir des
caractères modifiés, qui vérifient des
équations de fusion modifiées: celles-ci n'ont
pas de signification algèbrique profonde, mais
par contre elles incorporent le second membre des TBA.

Les équations de fusion modifiées apparaissent
naturellement dans le modèle de Kondo (voir [\artic03.3.3]),
puisqu'elles correspondent à l'idée intuitive
que l'on peut obtenir des impuretés de spin
supérieur en fusionnant plusieurs impuretés\foot{Cette idée
intuitive est déjà exprimée dans [\ref\FSb{P.~Fendley et H.~Saleur,
{\it Phys. Rev. Lett.} 75 (1995), 4492 \pre{cond-mat/9506104}.}],
mais l'analyse perturbative de [\FSb] ne leur
permet pas d'obtenir les équations de fusion modifiées 
qui contiennent des informations non-perturbatives.}.
Les caractères modifiés sont définis par l'équation
(\artic03.3.13), qui, dans le cas $SU(2)$, diffère de
la relation entre matrices de transfert (ou caractères non-modifiés)
et fonctions $Y$ utilisée dans [\KP] par des facteurs scalaires.
En tenant compte du fait que toutes les fonctions $\log(1+\eta(\zeta))$
sont {\it bornées}, on peut prolonger les caractères sur
une bande du plan complexe, et inverser (\artic03.3.13), ce qui
aboutit à l'équation (\artic03.3.15). On peut alors montrer
que les caractères $\chi^r_j$ vérifient des équations
de fusion modifiées du type:
$$\eqalign{
\chi^r_j(\zeta+i\pi/N)\chi^r_j(\zeta-i\pi/N)
&=\chi^{r+1}_j(\zeta)\chi^{r-1}_j(\zeta)\cr
&+\chi^r_{j+1}(\zeta)\chi^r_{j-1}(\zeta)
\exp\left(\sin(\pi r/N)(\alpha^+_j \e{+\zeta}+\alpha^-_j \e{-\zeta})\right)
\cr}\eqn\modfus$$
c'est-à-dire que le second membre des TBA apparaît de telle
sorte que l'on peut représenter les équations de fusion
modifiées par les mêmes diagrammes de BAE physiques que
les TBA. Les liens entre noeuds correspondent aux liens entre
caractères dans les équations de fusion, et les noeuds cochés
voient le terme qui les connecte à leurs voisins horizontaux
modifié.

Les équations de fusion modifiées, à la différence des TBA,
ne sont pas des équations ne faisant intervenir que des fonctions
réelles, et elles ne se prêtent donc pas bien aux simulations
numériques ou à certains types de calculs analytiques. Par contre,
elles permettent des calculs à un niveau formel qui peuvent
s'avérer intéressants: ainsi, dans [\artic03.4], de nombreuses
quantités thermodynamiques sont obtenues de manière élémentaire
grâce aux équations de fusion.

Les équations de fusion modifiées
permettent également d'interpréter les valeurs limites dans
les TBA (cf fin du [\artic03.3]). 
En effet, on voit que dans la limite $\zeta\to\pm\infty$,
on peut négliger
les décalages du paramètre spectral, de sorte que
les équations de fusion modifiées se résument à de
simples équations pour les caractères de $SL(N)$.
Cependant, le facteur 
$\exp(\sin(\pi r/N)\alpha_j \e{\zeta})$ (nous considérons
un modèle à une chiralité),
qui tend soit vers $1$ soit vers $0$,
restreint l'intervalle des indices des équations de fusion.
Dans les cas les plus simples, ces restrictions correspondent
simplement à remplacer les relations de caractères de $SU(N)$ par
les relations de caractères de l'algèbre
des représentations tronquée de $U_q({\goth sl}(N))$
avec $q=-\e{-i\pi/(f+N)}$. De manière générale,
on peut affirmer que la modificiation des équations de
fusion fait passer des équations de fusion
du groupe quantique nu ($\zeta=-\infty$) aux équations
de fusion du groupe quantique physique ($\zeta=+\infty$).

Dans les TBA, la condition de positivité des $\eta$ impose
le choix de la valeur propre de Perron--Frobenius des matrices
d'adjacence, c'est-à-dire la dimension pour $SU(N)$ et la
$q$-dimension pour $U_q({\goth sl}(N))$.
On reconstitue finalement les valeurs limites $\eta^r_j(\pm\infty)$
grâce aux équations du type \defY. 

Prenons un exemple: les valeurs limites \vallim\ s'obtiennent
facilement à partir des $q$-dimensions des trois
groupes quantiques
$U_{\e{i\pi/(a+2)}}({\goth sl}(2))$,
$U_{\e{i\pi/(b+2)}}({\goth sl}(2))$ et 
$U_{\e{i\pi/(a+b+2)}}({\goth sl}(2))$ (on a $\eta_j=d_{j+1} d_{j-1}$
où $d_j$ est la $q$-dimension de la représentation
de spin $j/2$: $d_j=\lfloor j+1\rfloor$).
On remarque alors que les sommes dilogarithmiques qui
apparaissent dans \cosetc\ sont
encore calculables si l'on remplace la $q$-dimension $d_j$ par
un autre caractère [\ref\KIc{A.N.~Kirillov, \pre{hep-th/9211137}.}], et 
font apparaître les poids conformes du modèle, indiquant
que pour les ``TBA pour les états excités'', les valeurs
limites sont données par les différents
caractères et non la seule $q$-dimension.

\newsec\secddv{Une alternative aux TBA: les Equation Non-Linéaires
Intégrales (NLIE)}
Nous avons vu dans la section précédente qu'il est possible
de considérer, par invariance modulaire, des théories
sur un espace compactifié de longueur finie,
et d'obtenir des équations similaires aux TBA.
Nous appellerons génériquement les équations correspondant
à un espace fini des Equations Non-Linéaires Intégrales (NLIE),
du fait de leur forme. Ici, nous ne intéresserons
pas aux NLIE ``TBA-like'' comme dans la section précédente,
mais au contraire à des NLIE
qui ont une structure plus simple que les TBA: nous
verrons que nos NLIE ne reproduisent que la structure
du diagramme de Dynkin de l'algèbre de symétrie correspondante,
sans qu'il y ait de second indice de corde. Des équations
de ce type ont été écrites pour les modèles à 6 vertex
et à 19 vertex (spin $1$) dans [\ref\BKP{M.~Batchelor,
A.~Kl\"umper et P.~Pearce, {\it J. Phys.} A24 (1991), 3111.}]
et ont été obtenues indépendamment dans le cadre
de la théorie des champs relativiste (Sine--Gordon) dans 
[\ref\DDV{C.~Destri and H.J.~De~Vega, {\it Phys.\ Rev.\ Lett.} 69
(1992), 2313}]. Des références plus complètes sont données
dans [\artic04], qui traite de la généralisation de ces NLIE
à une algèbre de Lie simplement lacée quelconque, et qui
utilise les méthodes de [\ref\DDVexc{C.~Destri and H.J.~De~Vega,
{\it Nucl. Phys.} B504 (1997), 621 \pre{hep-th/9701107}.}].

Les NLIE que nous allons voir maintenant se démarquent
notablement des TBA: tout d'abord, l'hypothèse de corde n'y est
pas applicable, les états excités pouvant contenir
des racines situées n'importe où dans le plan complexe;
par contre, l'étude d'un seul état (l'état fondamental) donne
autant d'information que les TBA (et en particulier, dans l'UV,
la charge centrale), tandis que l'étude des états excités
fournit des informations inaccessibles par les TBA (dans l'UV,
les poids conformes).

\subsec\secddvform{Les NLIE de Toda affine en couplage imaginaire}
Le modèle de Toda affine est présenté brièvement dans
[\artic04.1,9.1]. Reprenons cette présentation.
Si $\phi$ est un champ bosonique appartenant à la sous-algèbre
de Cartan d'une algèbre de Lie $\goth g$ de rang $n$,
$\alpha_s$, $1\le s\le n$ sont les racines simples
de $\goth g$, et $-\alpha_0$ est la racine la plus haute
de $\goth g$, alors l'action de Toda affine avec une
constante de couplage imaginaire est donnée par
$${\cal S}={1\over 4\pi}\int \d^2 x \left[
(\der_\mu \Phi)^2 + M^2 \sum_{s=0}^n
\e{-i \left< \alpha^s,\Phi\right>/R}\right]
\eqn\lagtoda
$$
Cette action pose plusieurs problèmes: tout d'abord,
elle n'est pas réelle, ce qui est lié au fait
que la théorie n'est pas unitaire.
Ensuite, elle n'est pas renormalisable telle quelle, puisque la
perturbation engendre tous les opérateurs du
type $\e{-i\langle \alpha,\phi\rangle}$ où $\alpha$
parcourt le réseau des racines. Ces opérateurs
apparaissent d'ailleurs quand on bosonise le modèle NJL
$SU(N)$ [\ref\BHN{T.~Banks,
D.~Horn et H.~Neuberger, {\it Nucl. Phys.} B108 (1976), 119.}]:
en effet, celui-ci est équivalent à la
théorie bosonique
$${\cal S}={1\over 4\pi}\int \d^2 x \left[
(\der_\mu \Phi)^2 + M^2 \sum_{a,b=1}^N
\cos((\Phi^a-\Phi^b)/R)\right]
\eqn\lagtodab$$
avec $R=1/\sqrt{2}$. \lagtodab\ se distingue de \lagtoda\ par
le fait que l'action est réelle et contient des opérateurs
perturbants supplémentaires qui rendent la théorie
renormalisable. Pour $R>1/\sqrt{2}$,
la perturbation ne produit pas d'opérateurs plus pertinents
que ceux de \lagtoda,
et il est concevable, par extension du cas isotrope,
que le modèle de Toda affine en couplage imaginaire
soit équivalent au modèle NJL anisotrope associé
à la même algèbre $\goth g$,
dont la bosonisation produit les opérateurs manquants
qui assurent la renormalisabilité\rem{ben faut
le faire mon gars}. Nous allons
donc écrire des Equations Non-Linéaires Intégrales
non pas pour Toda affine, mais pour le modèle NJL
anisotrope.

Nous supposerons dorénavant
que l'algèbre de Lie $\goth g$ est simplement lacée.
Le point de départ des NLIE est la donnée des Equations d'Ansatz
de Bethe qui sont la généralisation à $U_q(\widehat{\goth g})$
des équations \xxxbaeb\ pour ${\goth g}={\goth sl}(2)$ (Eq.
(\artic04.2.4)).
Les inhomogénéités sont alternées $\theta_k=\pm\theta$,
$\theta\to\infty$:
ce sont celles du modèle NJL anisotrope. On peut donner
une autre interprétation à ces inhomogénéités alternées 
en termes de régularisation sur réseau de la théorie des champs
que l'on obtient dans la limite d'échelle: c'est l'approche
dite du cône de lumière [\ref\DDVb{C.~Destri
et H.J.~De~Vega, {\it Nucl. Phys.} B290 (1987), 363.}]. Le réseau
est tourné de $45$ degrés par rapport
à l'ordinaire, c'est-à-dire que l'on considère
une matrice de transfert {\it diagonale à diagonale}.
Les $\e{-iE^\pm}=
\e{-i(E\pm P)}$ donnés par l'Eq.\ (\artic04.2.5)\foot{Notons
qu'ils diffèrent par des signes des expressions (Eq.\ \xxxbaec) que nous
avons utilisés jusqu'ici. Ce signe peut
être absorbé dans le choix de conditions aux bords
(périodiques ou anti-périodiques) pour les fermions nus.},
en particulier, sont les
translations des deux pas élémentaires du réseau, qui
sont dans les deux directions du cône de lumière. Dans cette approche,
on peut, dans le cas ${\goth g}={\goth sl}(2)$,
démontrer que certains opérateurs, obtenus par une transformation
du type Jordan--Wigner, vérifient les équations du mouvement
du modèle de Thirring massif, ce qui justifie le lien avec ce modèle.
Cependant, comme il est expliqué dans [\ref\KMb{T.R.~Klassen
et E.~Melzer, {\it Int. J. Mod. Phys.} A8 (1993), 4131 \pre{hep-th/9206114}.}],
ceci ne permet pas de distinguer entre les différents modèles ``équivalents''
à Sine--Gordon. Nous éluciderons ce point lors du calcul des
poids conformes UV (\secddvuv).

Partons donc des fonctions de comptage $Z_s$ (Eq.\ (\artic04.3.1)) pour
un état excité au-dessus du vide (qui contient un nombre
fini d'excitations physiques), et
distinguons deux types de racines des BAE: les racines réelles qui
constituent le vide de la théorie, et les racines complexes --
on ne parle plus de cordes puisque l'hypothèse de corde
ne s'applique pas. On utilise alors une intégrale de contour
pour récrire la contribution aux fonctions de comptage de
racines réelles. Notons que cette astuce ne fonctionne
que pour les modèles non-fusionnés, pour lesquels le vide
n'est constitué que de racines réelles. On parvient alors,
après une série de transformations, et le
passage à la limite d'échelle, aux NLIE (\artic04.3.17),
que nous récrivons ici dans les notations de cette thèse:
$$Z^s(\zeta)= M^s L \sinh \zeta
+X^{st} \star Q^t(\zeta)
+\sum_{\alpha=1}^{M_H^t} \chi^{st}(\zeta-\eta^t_\alpha)
-\sum_{\alpha=1}^{M_C^t} \chi^{st}(\zeta-\xi^t_\alpha) \eqn\ddv$$
où nous avons supposé pour simplifier qu'il n'y a pas de racines/trous
spéciaux. La fonction réelle $Q^s$ est définie
par $Q^s(\zeta)\equiv Z^s(\zeta)\mod 2\pi$ (là aussi,
nous simplifions en supposant $\delta^s=0$), et $|Q^s(\zeta)|<\pi$.
Les $X^{st}$ sont des noyaux donnés dans (\artic04.3.15-16),
et $\chi^{st}$ sont les primitives de $2\pi X^{st}$.
Enfin les $M^s$ sont les masses des différentes
excitations physiques.

Pour interpréter les NLIE \ddv, et en particulier les $X^{st}$,
regardons tout de suite la limite $L\to\infty$. On voit
facilement que les termes non-linéaires $X^{st}\star Q^t$
sont exponentiellement petits (en $\e{-ML}$, où
$M$ est une échelle de masse de l'ordre des $M_s$,
typiquement la plus petite masse). Une
fois ces termes retirés, les NLIE \ddv\
peuvent alors être considérées tout simplement comme le fait
que les $Z^s(\zeta)$ sont des fonctions de comptage ``physiques''
pour les trous et racines complexes. En effet, on a bien:
$Z^s(\eta^s_k)$ et $Z_s(\xi^s_k)$ égaux à $2\pi$ fois des
demi-entiers, ce qui est la définition d'une fonction de comptage.
En particulier, en considérant des configurations
sans racines complexes, les $\chi^{st}$
s'identifient au déphasage des trous
lors de la diffusion de plus haut poids. 
Ainsi, dans le cas ${\goth g}={\goth sl}(N)$,
la matrice $S^{11}$ (Eq. (\artic04.6.4)) se récrit explicitement:
$$\eqalign{
\Sh^{11}(\zeta)&=\exp\left(i \int_0^{+\infty} \d\kappa\, 
{\sin(N\kappa\zeta/\pi)\over \kappa}
{\sinh((p+1-N)\kappa)\sinh\kappa \over \sinh(p\kappa)\sinh(N\kappa)}
\right)\cr
&=
{ \Gamma\left(1+i{\zeta\over 2\pi}\right)
  \Gamma\left(1-{1\over N}-i{\zeta\over 2\pi}\right)
\over
  \Gamma\left(1-i{\zeta\over 2\pi}\right)
  \Gamma\left(1-{1\over N}+i{\zeta\over 2\pi}\right)}\cr
&\ \prod_{n=1}^\infty
{\Gamma\left({np\over N}+1+i{\zeta\over 2\pi}\right)
\Gamma\left({np+1\over N}-i{\zeta\over 2\pi}\right)
\Gamma\left({np-1\over N}+1-i{\zeta\over 2\pi}\right)
\Gamma\left({np\over N}+i{\zeta\over 2\pi}\right)
\over
\Gamma\left({np\over N}+1-i{\zeta\over 2\pi}\right)
\Gamma\left({np+1\over N}+i{\zeta\over 2\pi}\right)
\Gamma\left({np-1\over N}+1+i{\zeta\over 2\pi}\right)
\Gamma\left({np\over N}-i{\zeta\over 2\pi}\right)}
\cr}
\eqn\aTS$$
On conclut
donc que le passage de (\artic04.3.1) à (\artic04.3.17),
dans la limite $L\to\infty$, s'identifie à la
dualité particule-trou entre ondes de spin et spinons.

Cette interprétation montre aussi la différence essentielle
entre les fonctions de comptage des pseudo-particules, qui
sont celles dont nous sommes partis (Eq.\ (\artic04.3.1)), et les fonctions
de comptage physiques données par \ddv\ sans la non-linéarité.
Pour ce qui est de la première expression,
elle est exacte à toute longueur $L$,
ce qui provient en dernière analyse du fait que les interactions
entre pseudo-particules sont ponctuelles\foot{comme celles entre
particules nues; on peut le voir en utilisant l'Ansatz
de Bethe en Coordonnées, et non l'Ansatz de Bethe Algébrique,
pour la diagonalisation de la matrice de transfert.};
la seconde expression, en revanche, n'est valide qu'à $L\to\infty$,
puisqu'il y a des corrections en $\exp(-ML)$, qui
correspondent au fait que, lorsque le rayon $L$ devient trop
faible, de sorte que les particules deviennent
trop rapprochées (par rapport à l'échelle de distance du
système, qui est $1/M$), la notion d'état asymptotique
(composé de particules libres) n'est plus valable.
La non-linéarité modifie alors la quantification des rapidités,
et permet ainsi d'avoir une limite ultra-violette
non-triviale, que nous allons maintenant examiner.

\subsec\secddvuv{Limite ultra-violette et poids conformes}
Dans la limite $L\to 0$, les NLIE \ddv\ exhibent un comportement
similaire aux TBA, c'est-à-dire le découplage des chiralités
[\artic04.7-8].
Cependant, le découplage n'est total que pour l'état fondamental:
pour les états excités, il reste un {\it couplage résiduel} que
constitue le {\it ``twist'' des équations chirales l'une sur l'autre.}
Ce ``twist'' apparaît explicitement dans les fonctions de
comptage $Z_s$ comme des constantes qui traduisent
l'interaction entre particules de chiralités différentes
(et donc de différence de rapidités divergentes). Il conduit
à une valeur non-triviale de $Z^{s\pm}(\mp\infty)$; génériquement,
on a (Eq.\ (\artic04.7.8)):
$$Z^{s+}(-\infty)=Z^{s-}(+\infty)\equiv \gamma
\Delta r^s \mod 2\pi$$
où $\Delta r^s=r^{s+} - r^{s-}$ est la différence
des nombres quantiques $\goth g$ des particules droites
et gauches.

On peut alors calculer le terme en $1/L$ de l'énergie
d'un état excité pour $L\to 0$, ce qui par définition nous fournit
le poids conforme UV correspondant (ainsi que
la charge centrale, bien sûr, puisque les poids
conformes sont décalés de $c/24$ sur le cylindre).
Le calcul [\artic04.8] est assez
complexe, mais le résultat, lui, est simple:
la charge centrale vaut le rang de l'algèbre, et,
génériquement, on trouve que le poids conforme est
donné par:
$$\Delta^\pm={\sum_{s,t}(C^{-1})^{st} 
\left[ r^s \pm (1-\gamma/\pi) \Delta r^s \right]
\left[ r^t \pm (1-\gamma/\pi) \Delta r^t \right]
\over 8(1-\gamma/\pi)}+p^\pm\eqn\cw$$
où $(C^{-1})^{st}$ est l'inverse de la matrice
de Cartan {\it avec la normalisation usuelle},
c'est-à-dire $C^{st}=2\delta^{st}-A^{st}$
($A^{st}$ est la matrice d'adjacence
du diagramme de Dynkin de $\goth g$). $p^\pm$ sont des
entiers positifs. Les nombres quantiques $r^s$ sont
liés au groupe $U(1)^{N-1}$ (charges topologiques pour le
champ $\phi$ de Toda affine) qui est une symétrie de la théorie
tout le long du flot de groupe de renormalisation, tandis
que les $\Delta r^s$ n'ont de sens qu'au point fixe UV
(la notion de particules gauche et droite n'a de sens
qu'au point fixe UV), et correspond à une nouvelle
symétrie chirale $U(1)^{N-1}$ qui est brisée par
la perturbation hors du point fixe UV.

On trouve dans [\artic04.A]
une analyse générale de ces poids conformes. 
Nous allons nous limiter ici à quelques
remarques sur le cas ${\goth g}={\goth sl}(N)$,
avant d'examiner plus en détail le cas ${\goth g}={\goth sl}(2)$.
Rappelons qu'en posant $p^\pm=0$ dans \cw, on obtient
les poids conformes {\it primaires pour l'algèbre de Kac--Moody}
$\widehat{{\goth u}(1)}^{N-1}\!\!\oplus \widehat{{\goth u}(1)}^{N-1}\!\!$.

Supposons d'abord $\gamma=0$. On a alors la formule
particulièrement simple:
$$\Delta^\pm={1\over 2}\sum_{s,t} (C^{-1})^{st}
r^{s\pm} r^{t\pm}\eqn\cwiso$$
Les deux chiralités sont donc {\it totalement découplées}. Vérifions
que ces résultats décrivent
bien le point fixe UV du modèle NJL $SU(N)$, qui n'est
autre (d'après la liberté asymptotique)
que $N$ fermions de Dirac libres. Dans
une théorie de $N$ fermions de Dirac libres, les deux chiralités
sont effectivement complètement découplées, et pour
une chiralité donnée ($N$ fermions de Weyl), on sait
que les poids conformes primaires
pour l'algèbre de Kac--Moody $\widehat{{\goth u}(1)}^N$ sont donnés par
$$\Delta={1\over 2} \sum_{a=1}^N (\mu^a)^2$$
où $\mu^a$ est le poids du \eme{a} générateur de $U(1)^N$ (matrice
avec un $1$ à la ligne $a$, colonne $a$). En utilisant
la relation $r^s=\mu^s-\mu^{s+1}$, on obtient:
$$\Delta={1\over 2}\sum_{s,t} (C^{-1})^{st} r^s r^t + {1\over 2N}
\Big(\sum_a \mu^a\Big)^2\eqn\cwisob$$
$\sum_a \mu^a$ est la charge $U(1)$ globale qui est liée au
secteur $U(1)$ découplé; si l'on retire le terme correspondant dans
\cwisob, on obtient \cwiso. On en déduit que dans l'UV,
la NLIE \ddv, à $\gamma=0$, décrit $N$ fermions de Dirac
libres (soit la {\it fermionisation} [\ref\GO{P.~Goddard
et D.~Olive, {\it Int. J. Mod. Phys.} A1 (1986), 303.}] de
la théorie conforme de WZW $U(N)$ au niveau $1$), auxquels
on a retiré un secteur $U(1)$. Bien sûr, l'opération
qui consiste à retirer le secteur $U(1)$ est quelque
peu artificielle, car du fait que $U(N)=(SU(N)\times U(1))/{\Bbb
Z}_N$, les secteurs $SU(N)$ et $U(1)$ restent couplés
par des relations modulo $N$; explicitement,
dans \cwisob, on voit que la charge $U(1)$ $\sum_a \mu^a$
est congrue à $\sum_s s r^s$ modulo $N$. En particulier,
les propriétés modulaires de la fonction de partition
sur le tore,
dans laquelle a été retiré le secteur $U(1)$, sont détruites.

Pour $\gamma>0$, au contraire,
du fait des twists, les deux chiralités restent couplées 
(dans \cw, $r^+$ apparaît dans $\Delta^-$ et vice versa)
par ce qu'on interprète comme les {\it modes zéros}
de la théorie. 
Nous ne discuterons pas plus le cas anisotrope,
car il correspond à une théorie non-unitaire (sauf pour
${\goth g}={\goth sl}(2)$, que nous traiterons à part). 
Nous allons à présent procéder à la troncation à $\gamma=\pi/(p+1)$,
qui rétablit l'unitarité [\RS].

On a vu au \secdiloguv\ qu'en introduisant
un champ magnétique imaginaire approprié dans les TBA, on reproduisait
la troncation de groupe quantique. Quelle est la procédure
analogue pour la NLIE? C'est la question qui est éclaircie
au [\artic04.9.1]. Reprenons ce raisonnement sous une forme
légèrement différente.
Pour cela, partons de la fonction de partition
(sur un espace infini) à température finie $T=1/\beta$
en présence du champ magnétique
imaginaire: $Z(\beta)=\tr(\e{-\beta{\bf H}}\Omega)$, où ${\bf H}$ est le
Hamiltonien de la théorie fermionique (modèle NJL anisotrope),
et $\Omega=\e{-B/T}=q^{\sum_{\alpha>0}H_\alpha}$ 
est l'effet du champ magnétique.
$Z(\beta)$ est essentiellement la quantité que l'on calcule
par les TBA.
En termes d'intégrale fonctionnelle,
$$Z(\beta)=\int_{\psi(x,t=\beta)=-\Omega\psi(x,t=0)}
\hskip-2.5cm [\d\psi\d\psi^\dagger]\,
\e{-{\cal S}[\psi,\psi^\dagger]}\eqn\funcint$$
où ${\cal S}$ est l'action, dont nous n'avons pas besoin explicitement
car nous travaillons à un niveau formel.

Effectuons alors la transformation modulaire $x\leftrightarrow t$
dans \funcint: on trouve que $Z(\beta)$ est maintenant
la fonction de partition sur un espace compactifié de longueur
$\beta$, avec des {\it conditions aux bords twistées} par $\Omega$.
Ceci nous indique donc la marche à suivre:
il suffit d'introduire un ``twist'' dans la NLIE \ddv, qui
jouera le même rôle que le champ magnétique imaginaire.

On refait alors le calcul des corrections de taille finie en tenant compte
du twist [\artic04.9.2-9.3]; si celui-ci est suffisamment petit,
il a pour seul effet de rajouter un terme inhomogène dans les formules
de corrections de taille finie, d'où la charge centrale 
et les poids conformes (\artic04.9.13-9.14) qui sont,
dans le secteur pair de la théorie ($r^s$ et $\Delta r^s$ pairs) ceux
des Théories Conformes Rationnelles avec la symétrie conforme
étendue $W({\goth g})$.

\noindent{\it Remarque:} Les poids conformes UV \cw\ sont très
voisins des poids conformes IR de la chaîne de spins XXZ $SU(N)$
correspondante. Cependant,
ils sont distincts, puisque
ces derniers sont exactement donnés
par \cw, mais après échange de $r^s\leftrightarrow\Delta r^s$,
où les $r^s$ sont toujours les nombres quantiques
$U(1)^{N-1}$, et les $\Delta r^s$ sont les nombres quantiques d'un
$U(1)^{N-1}$ chiral qui apparaît au point fixe IR. Cette différence
est bien sûr liée à la manière différente dont
sont définies les deux chiralités dans les deux modèles,
ce qui au final est dû aux inhomogénéités différentes
($\theta_k\to\pm\infty$ dans un cas, $\theta_k=0$).

Une autre manière d'expliquer cette dualité est la suivante:
appelons $r^s$ les charges magnétiques de Toda affine,
et $\Delta r^s$ les charges électriques.
On voit que la dualité {\it électro-magnétique}
$r^s\leftrightarrow \Delta r^s$
revient à échanger $1-\gamma/\pi$ et $1/(1-\gamma/\pi)$.
Dans le cas de Toda affine, pour que la charge magnétique
soit conservée tout le long du flot, il faut que
les opérateurs perturbants soient purement électriques.
De plus, comme on est au point fixe UV, ils doivent être pertinents.
Au contraire, au point fixe IR de la chaîne de spins,
les opérateurs doivent être non-pertinents, et il est donc
naturel de supposer (intégrabilité du flot)
que ceux-ci sont précisément les duaux de ceux de Toda affine,
qui sont eux non-pertinents (puisque $\gamma=0$ est le point
de transition de type Kosterlitz--Thouless, et
si $1-\gamma/\pi<1$, alors $1/(1-\gamma/\pi)>1$).
Ils sont alors purement magnétiques, donc les charges conservées
tout le long du flot sont les charges électriques, ce sont
donc elles qu'on appelle $r^s$ dans la chaîne de spins,
d'où finalement l'échange $r^s\leftrightarrow \Delta r^s$.

\subsec\secddvuvb{Retour sur le cas ${\goth g}={\goth sl}(2)$}
Pour discuter des subtilités qui existent entre
les différents modèles bosoniques et fermioniques équivalents,
il est plus simple de revenir au cas ${\goth g}={\goth sl}(2)$.
Celui-ci présente l'originalité de posséder trois modèles
équivalents: en plus du modèle NJL anisotrope
et du modèle de Sine--Gordon (cas particulier
de Toda affine complexe), on a le modèle de Thirring massif, que
nous allons définir ci-dessous.

Ces trois modèles sont équivalents au sens où leurs
fonctions de partition sur un cylindre (mais pas sur un tore)
sont égales, ou encore au sens où on peut établir
des identifications entre certains opérateurs des différents
modèles. Cependant, ceci ne suffit pas à assurer
que les Hilberts des différents modèles sont identiques,
et, de fait, il est montré dans [\KMb] que ceux de Sine--Gordon
et Thirring massif sont distincts. L'idée de cet article,
que nous allons reprendre ici, consiste à analyser
la destinée des différents états du Hilbert dans la limite
ultra-violette: ils doivent alors s'apparenter à des
états de la théorie conforme ultra-violette, que l'on sait
classifier explicitement. Ceci nous permettra de plus une
comparaison directe avec les résultats du \secddvuv.

\noindent $\bullet$ {\it Modèle de Sine--Gordon.}\par\nobreak
L'analyse de Sine--Gordon et de sa limite ultra-violette
n'est qu'un cas particulier de celle qui est effectuée pour
Toda affine complexe dans [\artic04 9,A], mais nous
allons tout de même la reprendre rapidement.

L'action de Sine--Gordon est:
$${\cal S}={1\over 8\pi} \int \d^2 x
\left[(\der_\mu \Phi)^2 + M^2 \cos(\Phi/R)\right]$$
La constante de couplage $M$ fixe, après renormalisation
appropriée, l'échelle de masse du système, et ne nous
intéresse donc pas dans l'UV. Le champ bosonique $\Phi$ appartient {\it a
priori} à la droite réelle.

Le modèle possède les quantités conservées suivantes:

\noindent{$\diamond$} Du fait de la symétrie $\Phi\to\Phi+2\pi R$,
on a un opérateur $T$: $T\Phi T^{-1}=\Phi+2\pi R$ dont les
valeurs propres sont de module $1$. 

On peut donc le diagonaliser,
ce qui correspond à la décomposition suivante du Hilbert ${\cal H}$:
$${\cal H}=\bigoplus_{\theta\in[0,2\pi]} {\cal H}_\theta$$
${\cal H}_\theta$ ($\theta\ne 0$)
est le secteur ``twisté'' de valeur propre $\e{i\theta}$.
${\cal H}_0$ est le secteur non-twisté de
la théorie (le Hilbert de Sine--Gordon
habituel), il correspond à {\it compactifier} le boson $\Phi$:
$\Phi\equiv\Phi+2\pi R$. Nous verrons que tous les modèles
équivalents à Sine--Gordon
ont leur Hilbert inclus dans le ``grand'' Hilbert
${\cal H}$, c'est-à-dire dans des secteurs twistés appropriés.

Dans la limite UV, la symétrie $\Phi\to\Phi+2\pi R$
est étendue en une symétrie $\Phi\to\Phi+{\rm cst}$, d'où une nouvelle
quantité conservée ${\cal N}$, la {\it charge électrique}
(dans le langage des cordes, ${\cal N}/R$ constitue l'impulsion
dans l'espace cible). On fixe la normalisation de ${\cal N}$ par la
relation: $T=\e{2\pi i {\cal N}}$, de sorte que dans un secteur
${\cal H}_\theta$, les valeurs propres $n$ de ${\cal N}$ vérifient:
$n\in {\Bbb Z}+{\theta\over 2\pi}$.
En particulier, dans le secteur non-twisté, la charge électrique
est entière.

\noindent{$\diamond$} Supposons que nous considérons la
théorie sur un espace compactifié de longueur $L$, et
faisons tendre $L$ (plus précisément $ML$, où $M$ est l'échelle
de masse renormalisée) vers l'infini.
Du fait que le potentiel $\cos(\Phi/R)$ a pour
minima $\Phi=2\pi m R$ ($m$ entier), on a nécessairement
$\Phi(x=+\infty)-\Phi(x=-\infty)=2\pi m R$, où $m$ est
la valeur propre entière de la charge ${\cal M}$
associée au courant topologique $\der_x \Phi$.
Par extension, il est naturel (mais non obligatoire) d'imposer,
même pour $ML<\infty$, la condition de quantification:
$\Phi(x+L)=\Phi(x)+2\pi m$ de la charge $\cal M$.
Insistons sur le fait que ceci est un choix de {\it conditions aux
bords}, et non de compactification de $\Phi$. Dans le secteur
non-twisté ${\cal H}_{\rm sG}$, ce sont des conditions aux
bords périodiques pour $\Phi$. $\cal M$ est nommée charge
magnétique (ou, pour les cordistes, nombre d'enroulement).

Dans la limite ultra-violette, on aboutit à la théorie
d'un simple boson libre. Digressons un instant: on appelle aussi
parfois cette théorie
gaz de Coulomb conforme, mais cette appellation est un peu
trompeuse, car le gaz de Coulomb proprement dit, c'est-à-dire
le gaz de particules électriques (ou magnétiques) en
interaction coulombienne n'apparaît
que grâce à la perturbation (les opérateurs de vertex $\e{\pm i\Phi(x)/R}$
créent des sources d'impulsion au point $x$ qui sont les charges électriques,
tandis que les opérateurs ``duaux'' -- voir plus loin --
créent des vortex qui sont les charges magnétiques). Nos charges
magnétique et électrique ne sont pas des vraies charges ponctuelles,
mais des effets de topologie globale de type instantonique.

Revenons au boson libre: la théorie est conforme, elle se scinde en ses
deux chiralités; la solution des équations du mouvement est
$$\Phi(x,t)=\phi^+(x^+)+\phi^-(x^-)$$
avec $x^\pm=t\pm x$, et on a deux courants chiraux $j^\pm(x^\pm)
=\der^\pm \phi^\pm$ qui forment deux copies
de l'algèbre de Kac--Moody $\widehat{{\goth u}(1)}$. Les opérateurs primaires
pour ces deux algèbres sont les opérateurs de vertex
$${\cal O}=\e{i(\alpha^+\phi^+ + \alpha^- \phi^-)}$$
dont les charges $U(1)$ chirales $\alpha^\pm$ se recombinent
pour former les charges électrique $n$ et magnétique $m$ selon la formule:
$$\alpha^\pm=\pm {mR\over 2} + {n\over R}\eqn\alph$$
Les poids conformes sont alors donnés par
$$\Delta^\pm={1\over 2} (\alpha^\pm)^2\eqn\delt$$

Rappelons que dans le secteur non twisté, le réseau des charges
électrique/magnétique est: $(m,n)\in {\Bbb Z}\times {\Bbb Z}$.
On remarque alors 
la dualité électromagnétique qui consiste en la transformation:
$R\to 2/R$ et $m\leftrightarrow n$. Au point self-dual $R=\sqrt{2}$, il
est bien connu que l'on a une symétrie plus grande: l'algèbre
de Kac--Moody $\widehat{{\goth u}(1)}$ est aggrandie en 
$\widehat{{\goth sl}(2)}$, les générateurs supplémentaires étant
les opérateurs de vertex: $m=n=\pm 1$ pour la chiralité droite,
$m=-n=\pm 1$ pour la chiralité gauche. Cette symétrie ${\goth sl}(2)$
est importante pour nous puisqu'elle doit être liée à la limite isotrope
de Sine--Gordon (point de Kosterlitz--Thouless).

Pour mieux comprendre ce point,
discutons des opérateurs perturbants $\e{\pm i\Phi/R}$,
qui sont purement électriques de charge $\pm 1$\foot{Ils
brisent donc la dualité électro-magnétique de la théorie UV.}:
leur dimension vaut $\Delta^\pm={1\over 2R^2}$. Ils sont donc
pertinents pour $R>1/\sqrt{2}$. On identifie le point $R=1/\sqrt{2}$
au {\it point de transition de Kosterlitz--Thouless}: on voit
qu'il est distinct du point self-dual $R=\sqrt{2}$.

Plaçons-nous alors à un rayon double $R'=2R$,
et perturbons avec les mêmes opérateurs $\e{\pm i\Phi/R}=
\e{\pm 2i\Phi/R'}$, qui sont maintenant
de charge électrique $\pm 2$: nous appellerons cette
théorie Sine--Gordon(2). La dimension vaut
toujours $\Delta^\pm={2\over
R'^2}$, de telle sorte que le point de Kosterlitz--Thouless est
au point self-dual $R'=\sqrt{2}$. Cependant, la condition
de quantification de la charge magnétique n'est plus correcte:
en effet, le nouveau potentiel $\cos(2\Phi/R')$ possède des minima
en $m\pi R'$, mais seules les configurations telles que
$\Phi(x+L)-\Phi(x)=2\pi m R'$ sont admises: dans la limite IR,
ceci signifie que l'on n'obtient que les états qui contiennent
{\it un nombre pair} de solitons. Nous reviendrons sur ce point plus
loin.

\noindent $\bullet$ {\it Modèle de Thirring massif.}\par\nobreak
Par souci de complétude, nous introduisons également le
modèle de Thirring massif, bien qu'il ne nous soit pas utile.
Son action vaut:
$${\cal S}=\int \d^2 x \left[i\bar\Psi\dsl\bar\Psi-{g\over 2}
(\bar\Psi \gamma_\mu \Psi)^2 - M\bar\Psi\Psi\right]$$
où $\Psi$ est un fermion de Dirac. Ce modèle
possède une invariance $U(1)$. Dans la limite UV, comme
pour le modèle de Sine--Gordon, le terme de masse tend vers $0$,
et on est ramené au modèle de Thirring non-massif,
caractérisé par sa constante de couplage $g$. Ce modèle possède
une invariance supplémentaire $U(1)_\chi$ chirale, c'est-à-dire
qu'il y a invariance séparée des deux chiralités
$U(1)_+\times U(1)_-$, que l'on
recombine en $(U(1)\times U(1)_\chi)/{\Bbb Z}_2$.

On montre
[\ref\CM{S.~Coleman, {\it Phys. Rev.} D11 (1975), 2088\semi
S.~Mandelstam, {\it Phys. Rev.} D11 (1975), 3026.}]
que le modèle de Thirring massif est équivalent
au modèle de Sine--Gordon; avec les conventions standard,
la correspondance entre constantes de couplage est: $R^2=1+g/\pi$.
Dans l'ultra-violet, si l'on prend des conditions aux
bords spatiales anti-périodiques pour le fermion,
alors les poids conformes primaires (pour Kac-Moody ${\goth
u}(1)\oplus {\goth u}(1)$) de Thirring non-massif sont donnés par
les mêmes formules \alph-\delt, où $m$ est la charge $U(1)$
(donc un entier), mais où $n$ est {\it la moitié} de la
charge $U(1)$ chirale\foot{De manière plus
générale, avec des conditions
aux bords twistées $\Psi(x+L)=-\e{i\theta}\Psi(x)$, $|\theta|<\pi$,
on aurait: $n={p\over 2}+{\theta\over 2\pi}$, $p$ charge $U(1)$ chirale.}. 
Le réseau des charges électrique/magnétique
n'est donc pas celui de Sine--Gordon (non twisté), 
mais plutôt, en tenant compte
du fait que les charges $U(1)$ et $U(1)$ chirale sont congrues
modulo $2$: $(m,n)\in 2{\Bbb Z}\times {\Bbb Z}\cup (2{\Bbb Z}+1)\times
({\Bbb Z}+{1\over 2})$. Le fait que les réseaux de charges
ne coïncident pas est bien connu des cordistes, puisqu'on passe
de l'un à l'autre par une projection GSO. Celle-ci assure
que la fonction de partition du boson libre sur un tore
est invariante modulaire, alors que celle de Thirring non-massif
a des propriétés modulaires caractéristiques d'une théorie 
fermionique\foot{Remarquons
que pour notre théorie massive il y a deux projections GSO possibles
puisqu'il y a deux combinaisons invariantes modulaires:
$PP$ et $AA+AP+PA$; elles correpondent en fait aux
deux choix de rayons $R$ et $R'$ pour la théorie de Sine--Gordon
résultante (avec bien sûr les mêmes opérateurs perturbants
$\e{\pm i\Phi/R}=\e{\pm 2i\Phi/R'}$).}.

Observons que le réseau des charges possède cette fois
une dualité $R\to 1/R$, $m\leftrightarrow 2n$, qui n'a pas
de signification directe pour nous, à part que le point self-dual
est le point bien connu du fermion de Dirac libre.

Enfin, on vérifie que les opérateurs perturbants
$M\bar\Psi_\mp\Psi_\pm$ sont bien les mêmes que ceux
de Sine--Gordon, c'est-à-dire électriques de charge $\pm 1$.
Nous renvoyons le lecteur à [\KMb] pour une analyse plus
minutieuse de la relation entre Sine--Gordon et Thirring massif.

\noindent $\bullet$ {\it Modèle NJL anisotrope.}\par\nobreak
Terminons par le modèle qui nous intéresse le plus directement.
Nous avons déjà donné son action, que nous reproduisons ici:
$${\cal S}=\int \d^2 x\left[
i\psib^a \dsl \psi^a - g j^3_\mu j^3{}^\mu
-f (j^1_\mu j^1{}^\mu+j^2_\mu j^2{}^\mu)\right]$$
où les $\psi^a$ sont deux fermions de Dirac.
On a cette fois une symétrie $U(1)^2$ (un $U(1)$ par fermion),
et une symétrie $U(1)$ chirale qui s'étend en $U(1)^2$ chirale
dans l'UV.

On montre\rem{ref?} que le modèle NJL anisotrope est équivalent
à un boson (ou fermion de Dirac) libre découplé 
et à un modèle de Sine--Gordon. La relation entre
constantes de couplage
dépend là encore de choix de regularisation; avec nos
conventions, le rayon de compactification du boson de Sine--Gordon
est: 
$$R^2={1\over 2(1-g/\pi)}$$

Les poids conformes UV primaires ${\goth u}(1)\oplus{\goth u}(1)
\oplus{\goth u}(1)\oplus{\goth u}(1)$ sont paramétrisés
par les $4$ charges $r^{a\pm}$, $a=1,2$;
du fait du découplage du secteur $U(1)$, on les
reparamétrise ainsi: $r_c^\pm=r_1^\pm+r_2^\pm$,
$r^\pm=r_1^\pm-r_2^\pm$. Alors, les poids sont donnés par:
$$\Delta^\pm={1\over 8}\left(\pm (r^++r^-)R+{(r^+-r^-)\over 2R}\right)^2
+{1\over 4} (r_c^\pm)^2\eqn\cwb$$
En retirant la contribution du secteur $U(1)$,
on retrouve \alph-\delt\ à condition de poser:
$m=r^+ + r^-$ et $n=(r^+-r^-)/4$. Ceci est cohérent avec
le fait que la perturbation est une interaction à $4$ fermions,
et doit avoir une charge électrique $\pm 1$.
Le réseau des charges électriques est cette fois: 
$(m,n)\in 2{\Bbb Z}\times {\Bbb Z}/2\cup (2{\Bbb Z}+1)\times
({\Bbb Z}/2+{1\over 4})$. On a une dualité $R\to 1/2R$, 
$m\leftrightarrow 4n$ dont le point fixe est le point
de transition de Kosterlitz--Thouless, comme pour le modèle
de Sine--Gordon(2). Ceci suggère qu'il existe un lien direct
entre ces deux modèles.

Nous avons représenté sur la figure \res\ les réseaux
des charges pour Sine--Gordon(2) et le modèle NJL anisotrope.
Ceci permet de constater que le second est inclus dans le premier:
les états de Sine--Gordon(2) correspondent exactement
aux états tels que $r$ (et donc $\Delta r$) est pair,
ce qui est évident compte tenu des formules explicites:
$m'=m/2=(r^+ + r^-)/2$, $n'=2n=(r^+ - r^-)/2$ des charges
électrique et magnétique de Sine--Gordon(2).
\fig\res{Réseau des charges électro-magnétiques.
Les abscisses/ordonnées sont les $r^+$/$r^-$.
Les gros cercles forment le réseau de Sine--Gordon(2);
complétés par les petits cercles, ils forment le réseau
du modèle NJL. On a de plus entouré les points
correpondant aux courants ${\goth sl}(2)$ dans le
cas isotrope ($R=1/\sqrt{2}$, $R'=\sqrt{2}$).}{\figdir{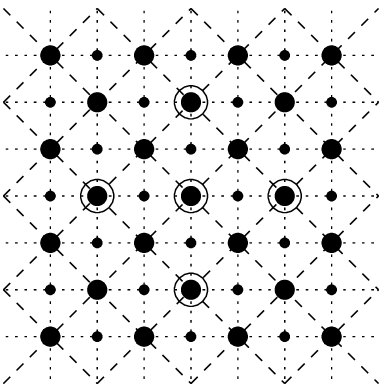}}

\noindent{$\bullet$}
{\it Comparaison avec les résultats de l'Ansatz de Bethe.}\par\nobreak
Prenons maintenant $n=1$ dans la formule \cw\ des poids conformes
donnés par l'Ansatz de Bethe; on trouve:
$$\Delta^\pm={1\over 16(1-\gamma/\pi)}(\pm r+ (1-\gamma/\pi)\Delta r)^2
\eqn\cwc$$
\cwc\ coïncide avec \cwb\ en faisant les identifications normales:
$r=r^++r^-$, $\Delta r=r^+-r^-$, $1-\gamma/\pi=1/(2R^2)$.
De plus, si l'on se restreint au secteur pair (nombre pair
de solitons), on retrouve exactement Sine--Gordon(2) avec
$1-\gamma/\pi=2/R'^2$. C'est cette dernière identification
qu'on retrouve le plus fréquemment dans la littérature.

\noindent{\it Remarques:}\par\nobreak
\item{$\star$} Ces subtiles distinctions entre les différents
modèles équivalents doivent être relativisées.
Dans la limite où la taille $L$
de l'espace compactifié tend vers l'infini,
comme expliqué dans [\artic04.9.1],
tous les secteurs twistés dégénèrent. Il y a donc
dégénérescence dans l'IR de bon nombre d'états
qui ont une destinée différente dans l'UV. La question
non-triviale est de trouver quels sont exactement les différents
états UV qui dégénèrent vers le même état IR;
pour répondre à cette question, il faudrait une analyse
détaillée des NLIE twistées (celle qui a été faite dans 
[\ref\NT{T.~Nassar et O.~Tirkonnen, \pre{hep-th/9707098}.}]
est un premier pas dans ce sens).
\item{$\star$}
Après avoir écrit cette section de ma thèse,
un preprint [\ref\FRT{G.~Feverati,
F.~Ravanini et G.~Tak\'acs, \pre{hep-th/9805117}.}] sur les
NLIE de Sine--Gordon (équation de Destri--De~Vega) est sorti. 
Les auteurs de [\FRT], après avoir
corrigé quelques erreurs sans conséquence dans [\DDVexc] (qui avaient déjà
été corrigées dans [\artic04] pour le cas plus général d'une algèbre
simplement lacée quelconque), ont repris à leur compte un certain
nombre d'idées qui avait déjà été exprimées dans [\artic04]:
\itemitem{$\bullet$} Les racines/trous ``immobiles'' doivent
être considérés comme des artefacts de la contruction par
l'Ansatz de Bethe, car ils conduisent à des formules pour
les poids conformes UV qui ne sont pas interprétables.
\itemitem{$\bullet$} La valeur de $M$ modulo $4$ (de $M_s$ modulo
$2h$ pour les algèbres de rang supérieur, cf remarque de bas de
page numéro $2$ dans [\artic04]) influe sur les propriétés des états que
l'on obtient par l'Ansatz de Bethe. Elle doit être choisie
par des arguments physiques indépendants.
\itemitem{$\bullet$} Les Equations de Destri--De~Vega ne décrivent
pas exactement Sine--Gordon (ou Thirring massif). Ce point
est bien sûr à la base de mes travaux (on part du modèle NJL
déformé). Cependant, du fait
que dans [\FRT], seuls les poids conformes des états
de charge électrique nulle sont calculés explicitement,
il ne leur est pas possible de conclure que l'on décrit
vraiment le modèle NJL déformé. Remarquons que de
toutes façons, les contraintes qu'ils imposent sur l'algèbre
des opérateurs, sont, comme dans [\KMb], trop fortes pour pouvoir
trouver le modèle NJL déformé, qui n'apparaît pas
dans ce cadre du fait du secteur $U(1)$ découplé.
L'existence de ce secteur $U(1)$ peut pourtant être devinée sans
recourir aux BAE nues du modèle NJL, par le raisonnement
suivant: quand on définit l'énergie dans l'approche du cône de
lumière, on définit réellement $\e{iL\epsilon^\pm}$ avec
$E=M(\epsilon^+ +\epsilon^-)$. Quand on prend le logarithme de
$\e{iL\epsilon^\pm}$, il apparaît naturellement une ambiguïté
de $2\pi/L$ dans la définition de $\epsilon^\pm$, qui s'interprète
comme la contribution à l'énergie du secteur $U(1)$. C'est
d'ailleurs cette contribution qui permet d'éviter à la régularisation
sur réseau de l'approche
du cône de lumière le problème du doublement des fermions.

\nref\SOK{A.~Sokal, {\it Soc. Text} 46/47 (1996), 217.}
\newchap\chappcf{Les modèles $\sigma$ 
principal, de Wess--Zumino--Witten et de Kondo.}
Le modèle $\sigma$ principal et sa généralisation, le modèle de
Wess--Zumino--Witten (WZW) sont des modèles de théorie
des champs à deux dimensions que nous allons introduire
(section \secpres),
avant de le résoudre grâce à l'Ansatz de Bethe (section
\secbaepcf). Nous utiliserons la technique des BAE physiques
pour déduire
le spectre des excitations physiques (sections
\secspcf, \secswz), et nous ferons quelques
calculs thermodynamiques grâce aux
TBA (sections \sectbawz, \secmagwz).
Nous généraliserons 
en particulier les résultats obtenus dans
[\ref\ZZ{A.B.~Zamolodchikov et Al.B.~Zamolodchikov,
{\it Nucl.\ Phys.} B379 (1992), 602.}] pour le modèle
de WZW $SU(2)$ niveau $1$\foot{Notons
que le niveau $1$ est un modèle très particulier
puisque la structure solitonique y est ``triviale'',
en un sens à définir -- voir \secswz.} à $SU(N)$ et à un niveau quelconque.

Le modèle de Kondo généralisé est un modèle de matière condensée, dont on
verra dans la section \seckondo\ 
qu'on peut le reformuler comme un problème de théorie
quantique des champs à $2$ dimensions. On pourra
alors lui appliquer l'Ansatz de Bethe. Il existe un lien étroit
entre le modèle de Kondo et la théorie conforme de WZW (point
fixe infra-rouge du modèle de WZW): on verra
que le modèle de Kondo se ramène à placer
une impureté dans une théorie conforme WZW, ou
encore à considérer une théorie WZW {\it à bord}.
C'est pourquoi nous traitons ces différents modèles ensemble.
Nous reviendrons sur une partie du contenu
de [\artic03]: TBA, entropie à température nulle
(\seckontba); mais nous obtiendrons
également un certain nombre de résultats non-publiés,
tels que la structure des excitations de basse énergie
(\seckonphys), ainsi qu'une étude complémentaire
de la thermodynamique de Kondo (\seckonmag). Tous ces
résultats nouveaux confirmeront les hypothèses faites
dans [\artic03] sur la structure des excitations et sur
la nature des différents points fixes infra-rouges.
Enfin nous établirons succintement le lien entre l'approche
de l'Ansatz de Bethe et l'approche de Théorie Conforme
du modèle de Kondo (\seckoncft); nous montrerons
comment la première implique la seconde (et non l'inverse).

\newsec\secpres{Les modèles $\sigma$ principal et WZW}
Le modèle de Wess--Zumino--Witten $U(N)$ [\WI] est
constitué d'un boson $\phi$ vivant dans $U(N)$ et dont l'action est:
$$A={1\over 4 g^2} \int \d^2 x \, \tr(\der_\mu \phi \, \der^\mu \phi^{-1})+f \Gamma\eqn\AWZW$$
où $\Gamma$ est le terme topologique de Wess--Zumino; $f$ est nécessairement
entier. Pour $f=0$, on retrouve l'action du modèle $\sigma$ principal.

L'action \AWZW\ est invariante par $U(N)_L\times U(N)_R$: $\phi\to g_L \phi g_R^{-1}$.
Les équations du mouvement expriment la conservation des courants correspondants:
$$\eqalign{
j^\mu_L&={1\over 2 g^2}\der^\mu \phi\, \phi^{-1}+ {f\over 8\pi} \epsilon^{\mu\nu}
\der_\nu \phi\, \phi^{-1}\cr
j^\mu_R&={1\over 2 g^2}\phi^{-1}\,\der^\mu \phi - {f\over 8\pi} \epsilon^{\mu\nu} \phi^{-1}
\,\der_\nu \phi\cr
}$$
L'action \AWZW\ est renormalisable; le flot de groupe de renormalisation est donné par
la figure \WZWflow.
\fig\WZWflow{Flot de groupe de renormalisation de WZW. Les points fixes 
infra-rouges sont situés sur
les deux demi-droites données par $1/g^2=|f/4\pi|$.}{\figdir{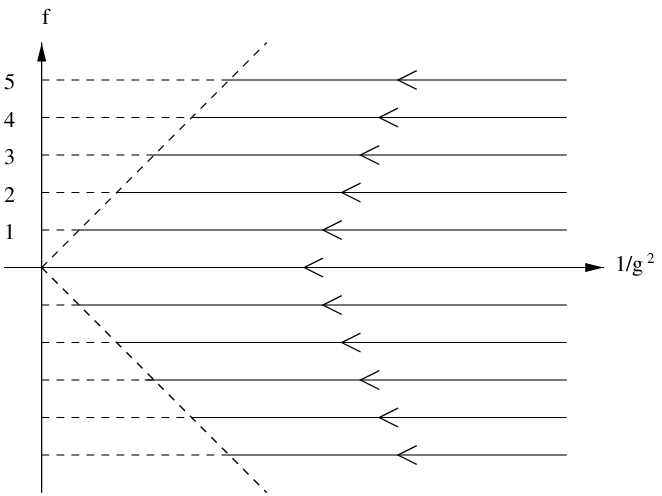}}
Pour tout $f$, la théorie est asymptotiquement libre (dans l'UV); au point
fixe UV, il y a alors $4$ courants conservés $j_L^\pm$, $j_R^\pm$, 
du fait des symétries chirales. Au contraire, 
au point fixe infra-rouge $1/g^2=f/4\pi$
(on suppose dorénavant $f>0$),
on voit que: $\der_\mu j^\mu_L=\der_+ j^-$ 
et $\der_\mu j^\mu_R=\der_- j^+$ ($\der_\pm=\der/\der x^\pm$,
$x^\pm=x^0\pm x^1$), où
$j^\pm$ sont les courants chiraux (c'est-à-dire que
$j^\pm$ ne dépend que de $x^\pm$) définis
par $j^+={1\over g^2} \phi^{-1}\, \der^+ \phi$, $j^-={1\over g^2}
\der^- \phi\,\phi^{-1}$.
Si l'on se débarrasse de la composante $U(1)$,
il reste deux courants chiraux $SU(N)$ qui forment alors deux copies 
$\widehat{{\goth sl}(N)}_f^\pm$ de l'algèbre
de Kac--Moody $\widehat{{\goth sl}(N)}$
au niveau $f$, dont les symétries ${\goth sl}(N)_L$ et ${\goth sl}(N)_R$
sont les sous-algèbres horizontales.

\subsec\secbaepcf{Equations d'Ansatz de Bethe}
Les modèles $\sigma$ principal et WZW sont intégrables, et on peut
donc essayer de leur appliquer l'Ansatz de Bethe.
Formellement, dans tout ce paragraphe,
nous considérerons le modèle $\sigma$ principal
comme le cas particulier $f=0$ du modèle de WZW, et ce bien que
les propriétés physiques du modèle
$\sigma$ principal diffèrent de celles du modèle de WZW,
ce qui nous obligera à les traiter séparément par la suite.

Nous allons rappeler brièvement par quelles
méthodes les Equations d'Ansatz
de Bethe peuvent être obtenues. Il y en a deux:

$\star$ La première méthode consiste
à partir d'Equations d'Ansatz de Bethe nues [\ref\PW{A.M.~Polyakov
et P.B.~Wiegmann, {\it Phys. Lett.} B141 (1984), 223.}]: par
des raisonnements un peu heuristiques, on montre que
le modèle de WZW $U(N)$ niveau $f$ doit
être équivalent à un modèle fermionique invariant 
$U(N)$\foot{Le modèle
que nous résolvons par l'Ansatz de Bethe est donc réellement
une fermionisation du modèle de WZW $U(N)$, qui,
lui-même, se scinde en un
modèle de WZW $SU(N)$ et un secteur
$U(1)$ découplé; cf la discussion du chapitre \chaptba.}
dans lequel les fermions gauches appartiennent à une représentation
symétrique à $f_L$ boîtes, les fermions droits à une représentation
symétrique à $f_R$ boîtes, $f_R-f_L=f$, $f_L, f_R\to\infty$.
On applique alors les techniques exposées aux chapitres
\chapbase-\chapstrucbae, et on obtient le diagramme de BAE nues
donné par la figure \baewz. Dans ce chapitre, les diagrammes
de BAE ne seront dessinés que dans le cas $SU(2)$, pour
éviter des figures trop volumineuses; les dessins $SU(N)$
sont obtenus à partir des dessins $SU(2)$ en plaçant
plusieurs lignes de noeuds.
\fig\baewz{Dans la limite $f_L, f_R\to\infty$,
$f_R-f_L\equiv f$ fixé, ce diagramme de BAE représente
le modèle de WZW $SU(2)$ au niveau $f$.}{\figdir{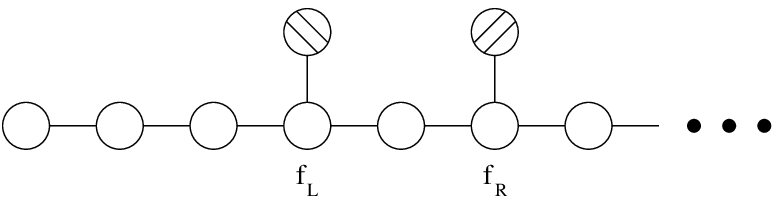}}
L'inconvénient de cette méthode tient au fait que seule la symétrie
$SU(N)_R\equiv SU(N)$ existe tant que $f_L, f_R<\infty$,
la symétrie $SU(N)_L$ étant prétendument restaurée à $f_L, f_R=\infty$.

Il a été objecté à cette méthode [\ref\DDVc{C.~Destri
et H.J.~De~Vega, {\it Phys. Lett.} B201 (1988), 245.}] que l'on n'obtenait
pas réellement le Hilbert complet du modèle de WZW: 
plus précisément, les auteurs de [\DDVc] ont fait
un calcul explicite (dans le cas $f=0$,
mais le raisonnement est sans doute généralisable à $f>0$)
de la transformation de l'intégrale
fonctionnelle du modèle fermionique en celle du modèle
$\sigma$ principal,
en tenant compte des conditions aux bords pour les deux
théories. Ils ont montré que l'on n'obtenait ainsi
que le {\it secteur singlet} pour la symétrie $SU(N)_L$ du Hilbert du
modèle $\sigma$ principal.
Avec les techniques développées au chapitre \chapstrucbae,
nous pouvons à présent résoudre cette difficulté.
Si l'on examine la figure \baewz, on se rend compte que pour
$f_L<\infty$, la symétrie $SU(N)_L$ n'existe pas mais
est remplacée par une symétrie $U_q({\goth sl}(N))$ avec
$q=-\exp(-i\pi/(f_L+2))$. Cependant, comme cela a été expliqué
au \secphyssup, cette symétrie est ``cachée'' au sens où le Hilbert
n'est réellement que le secteur singlet de $U_q({\goth sl}(N))$. Dans
la limite $f\to\infty$, la symétrie de groupe quantique redevient
une symétrie de groupe standard, mais toujours cachée,
et on aboutit bien à la même
conclusion que [\DDVc] (mais pour tout $f$). 
Il est même remarquable que l'on puisse
obtenir cette conclusion par un raisonnement indépendant des BAE.
En fait, toujours en suivant la logique du \secphyssup,
on pourrait modifier les conditions de bord pour obtenir
toutes les ``bonnes'' représentations de $U_q({\goth sl}(N))$,
mais même cela serait tout-à-fait insuffisant, car
le spin par unité de longueur $s/L$, qui est la quantité physique
raisonnable dans la limite thermodynamique $L\to\infty$, resterait
nul\foot{Le problème se ramènerait alors à une question
de non-commutativité des limites $f_L\to\infty$ et $L\to\infty$.}.

Comment résoudre ce problème? La solution
passe par l'observation suivante:
même si les solutions des équations à $f_L=\infty$ obtenues
comme limite des solutions des équations à $f_L<\infty$ sont
de spin nul, rien n'empêche les équations à $f_L=\infty$ d'avoir
plus de solutions! En fait, plusieurs arguments permettent de se
convaincre que les solutions des BAE à $f_L=\infty$ recouvrent
effectivement le Hilbert complet de WZW. Le premier est l'évidente
isosymétrie $SU(N)_L\leftrightarrow SU(N)_R$ qui est restaurée à $f_L=\infty$.
A toute solution de spin $SU(N)_R$ arbitraire, on peut, par retournement
du diagramme de BAE, faire correspondre une solution de spin $SU(N)_L$
arbitraire. Un autre argument est que les BAE à $f_L=\infty$ sont
aussi la limite $f_L\to\infty$ des BAE {\it non-tronquées}
$U_q({\goth sl}(N))$, puisque seuls les noeuds envoyés à l'infini
sont modifiés par la troncation.

$\star$ Une deuxième méthode pour obtenir les BAE consiste à trouver
les matrices $S$ physiques par le bootstrap, et d'écrire alors
des BAE physiques directement. Ceci a été effectivement réalisé
pour le champ chiral principal ($f=0$) et pour le modèle de WZW
au niveau $f=1$ dans le cas $SU(2)$ [\ZZ], et les BAE ainsi
obtenues sont effectivement équivalentes aux BAE nues de la
figure \baewz. 
Le fait que les deux méthodes proposées donnent
le même résultat dans les cas, certes très particuliers, où
elles peuvent toutes deux être appliquées,
est un signe encourageant. Pour $f>1$, il
est difficile de deviner les matrices $S$ physiques par le
bootstrap, et il nous faut faire confiance à la méthode
des BAE nues.

Nous allons maintenant écrire les TBA correspondantes,
et analyser brièvement comment elles permettent
de retrouver le flot de groupe
de renormalisation.

\subsec\sectbawz{TBA, limites ultra-violette et infra-rouge}
Dans la limite d'échelle, et par des méthodes identiques
à celles utilisées dans [\artic03], on obtient les BAE
du modèle de Wess--Zumino--Witten:
$$\rho^r_j
+(C^{-1})^{qr}\star C_{jk}\star \rhot^q_k
= \sin\left({\pi r\over N}\right) {E\over 2T}
\left( \delta_{j0} \e{-\zeta}+\delta_{jf} \e{\zeta} \right)
\eqn\baewzw$$
et donc les TBA:
$$\log(1+(\eta^r_j)^{-1})
-(C^{-1})^{qr}\star C_{jk}\star \log(1+\eta^q_k) 
= \sin\left({\pi r\over N}\right) {E\over 2T}
\left( \delta_{j0} \e{-\zeta}+\delta_{jf} \e{\zeta} \right)
\eqn\tbawz$$
$E$ est une échelle d'énergie engendrée dynamiquement,
caractéristique de la transition du point fixe UV vers
le point fixe IR. Nous verrons au \secswz\ que pour $f>0$,
$E$ s'interprète plus précisément comme l'échelle d'énergie
de l'interaction entre particules gauches et droites.

Nous avons représenté sur la figure \rgbaewz\ les TBA de
WZW $SU(2)$, ainsi que les limites UV et IR.
\fig\rgbaewz{TBA du modèle de WZW $SU(2)$.}{\figdir{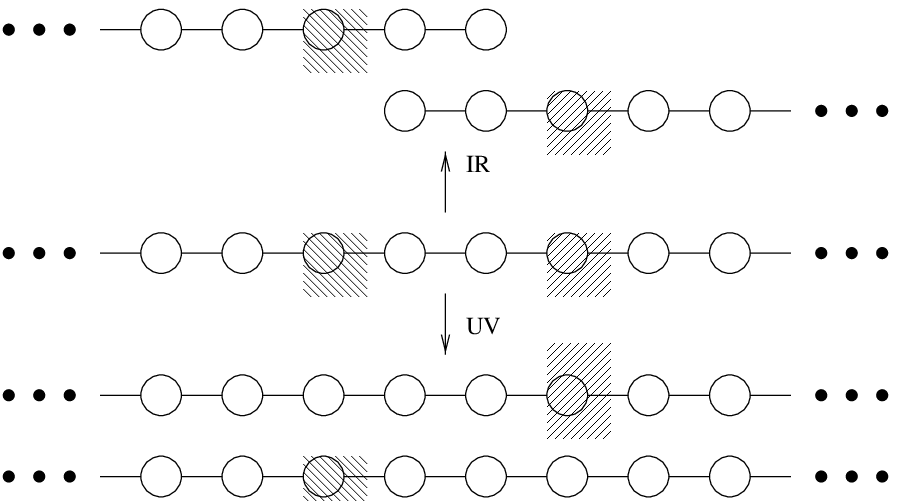}} 
Dans l'ultra-violet, le terme topologique 
de Wess--Zumino ne peut pas jouer de rôle, et la situation
est donc identique quel que soit $f$. 
Il y a alors $4$ diagrammes semi-infinis qui correspondent
à la symétrie $SU(N)_L^+\times SU(N)_R^+
\times SU(N)_L^-\times SU(N)_R^-$.

Cherchons à calculer la charge
centrale ultra-violette. On rencontre alors une difficulté:
la contribution du diagramme ``nu'' est une chaîne de noeuds infinie aux
deux bouts, et celle-ci ne figure pas dans la liste
(figure \dilogval) des sommes dilogarithmiques. En fait,
on peut montrer qu'il n'existe pas de solutions aux équations
\tbalim\ (équations vérifiées par les valeurs limites) pour
le diagramme infini aux deux bouts; le problème provient du fait
que les fonctions chirales $\eta^{r \pm}_j(\zeta)$ tendent vers l'infini
quand $\zeta\to\mp\infty$;
en particulier la valeur $\eta^r_j(\zeta=0)$ diverge quand $m/T\to 0$,
comme je l'ai vérifié par des simulations numériques.
Pour le calcul de la charge centrale, ce n'est pas un problème,
car on voit que la contribution du diagramme est nulle,
de sorte que $c=N^2-1$: c'est la charge centrale d'un boson
libre non-massif vivant dans l'algèbre de Lie ${\goth sl}(N)$.
Par contre, le calcul du terme suivant dans le développement
de haute température de l'énergie
libre n'est pas accessible simplement; on peut
seulement conclure qu'il est d'ordre $T^2/\log(T/m)$, mais
sans pouvoir calculer le coefficient devant ce terme\foot{Une
situation similaire est étudiée dans [\ref\Zamc{Al.B.~Zamolodchikov,
preprint LPS-ENS 335 (1991).}],
pour des TBA plus simples. Malheureusement, il est difficile
de généraliser le raisonnement de [\Zamc] aux TBA plus
compliquées du modèle $\sigma$ principal, afin de pousser plus loin
le développement de $F(T)$, $T\to\infty$.}.

La charge centrale $c=N^2-1$ et les
corrections logarithmiques sont bien sûr une conséquence de la liberté
asymptotique dans l'ultra-violet; quand on fait un développement
perturbatif de l'énergie libre à haute température, c'est-à-dire
de la fonction de partition
sur un cylindre très étroit\foot{En fait, le premier ordre
de la perturbation est exactement égal à la contribution
non-perturbative du mode zéro par rapport à la direction
compactifiée, soit en d'autres termes 
à l'énergie libre du modèle à $1$ dimension.},
on obtient
effectivement comme \eme{0} ordre de la perturbation les bosons
libres vivant sur l'algèbre de Lie, puis une série en puissance de
la constante de couplage renormalisée à l'échelle $T$.
On identifie donc $g(T)\sim 1/\log(T/m)$, ce qui est la
relation attendue. Notons que le développement perturbatif
brise la symétrie $SU(N)_L\times SU(N)_R$ en son sous-groupe
diagonal, la symétrie complète étant donc restaurée au niveau
non-perturbatif.

Passons maintenant à la limite infra-rouge. Pour $f>0$
(voir la section \secspcf\ pour le cas $f=0$), chacune des
deux chiralités reproduit le diagramme que l'on avait
déjà considéré au \secdilog\ (figure \tbaNwz): 
on avait alors calculé la charge
centrale, qui vaut $c=(N^2-1)f/(f+N)$,
ce qui est caractéristique d'une symétrie $\widehat{{\goth sl}(N)}_f$.
De plus, il est remarquable que la forme des TBA dans
l'infra-rouge illustre parfaitement la décomposition du champ
$\phi(x^+,x^-)$ au point fixe IR: en effet, les équations
du mouvement impliquent alors que
$$\phi^{ij}(x^+,x^-)=g^{ia}(x^-) h^{aj}(x^+)\eqn\decwzw$$
où $i$ est un indice de $SU(N)_L$, $j$ est un indice
de $SU(N)_R$ (les symétries $SU(N)_L$ et $SU(N)_R$ sont donc bien
devenues des symétries chirales, comme indiqué plus haut), 
et $a$ s'identifie tout naturellement à un
indice de $U_q({\goth sl}(N))$ avec $q=-\exp(-i\pi/(f+N))$.

Comme pour le modèle NJL anisotrope, dans le modèle de WZW avec
$f>1$, les deux chiralités, au point fixe IR, ne sont pas ``totalement''
découplées au sens où elles restent couplées par les modes zéros
(ainsi la fonction de partition sur un tore, même fermionisée,
ne s'écrit que comme {\it somme}
de produits d'un caractère de $\widehat{{\goth sl}(N)}_f^+$
et d'un caractère de $\widehat{{\goth sl}(N)}_f^-$).
Evidemment, ceci ne peut pas se voir dans les TBA (il faudrait
des ``TBA pour les états excités'' pour voir apparaître ce couplage),
donc on peut dire que le diagramme IR des TBA
associées à chaque chiralité
décrit la théorie conforme de WZW chirale $SU(N)$ niveau $f$.

\subsec\secspcf{Spectre et matrices $S$ du modèle $\sigma$ principal}
Décrivons maintenant la nature des excitations physiques; on doit
séparer les cas $f=0$ et $f>0$, puisque leur physique
diffère sensiblement. 

Le modèle $\sigma$ principal ($f=0$), à la différence
du modèle de WZW, n'exhibe pas
de restauration de la symétrie conforme au point fixe IR:
il s'agit en effet d'une théorie massive,
comme on le voit d'après le second membre $E \sin(\pi r/N)
\delta_{j0} \cosh\zeta$ des TBA. Le spectre de masse est donc:
$m^r= E \sin(\pi r/N)$.

On peut alors obtenir les matrices $S$ des différentes particules,
soit par
le bootstrap [\ref\ORW{E.~Ogievietsky, N.~Reshetikhin et P.~Wiegmann,
{\it Nucl. Phys.} B280 (1987), 45.}], soit par la méthode
des Equations d'Ansatz de Bethe Physiques.

On trouve qu'il y a $N-1$ types de particules:
la \eme{r} particule, de masse $m^r=E \sin(\pi r/N)$,
appartient à la \eme{r} représentation fondamentale de $SU(N)_R$
(tableau de Young à une colonne de $r$ boîtes), et à la
\eme{(N-r)} représentation fondamentale de $SU(N)_L$ (représentation
conjuguée). Pourquoi une telle conjugaison? Rien dans les BAE
ne permet de distinguer cette hypothèse de celle qui est faite
dans [\ORW] et qui consiste à choisir la {\it même} représentation
pour $SU(N)_L$ et $SU(N)_R$: en effet, par une redéfinition
évidente ($g_L\to \bar{g}_L$) de l'action de $SU(N)_L$, on passe
d'une hypothèse à l'autre. Cependant, des arguments généraux
montrent que la première hypothèse est la bonne; par exemple,
on remarque que pour le sous-groupe diagonal $SU(N)\subset SU(N)_L\times
SU(N)_R$, le nombre de boîtes du tableau de Young
d'un état quelconque
est nécessairement congru à $0$ modulo $N$, ce qui est le cas
avec la première hypothèse mais pas avec la seconde.
Ainsi, la dualité gauche/droite $SU(N)_L\leftrightarrow SU(N)_R$
échange particules et anti-particules.

Les matrices $S$ sont données par:
$$S^{qr}(\zeta)=\Sh^{qr}(\zeta) R^{qr}_L(\zeta) R^{qr}_R(\zeta)$$
avec des notations évidentes. Le facteur scalaire qui donne la
diffusion de plus haut poids, obtenu par comparaison entre les BAE
nues et les BAE physiques, vaut, conformément aux méthodes
générales du \secbaephysgen:
$$\Sh^{qr}(\zeta)=
\exp\left[-2i \int_0^{+\infty} \d\omega\, 
{\sin(\omega\zeta)\over \omega}
\left(
2{\sinh|\pi\omega(1-q/N)|\sinh|\pi\omega r/N|\over\sinh|\pi\omega|}
-\delta^{qr}
\right)\right]\eqn\Spcf$$
qui n'est autre que la limite quand $f\to\infty$ de \Ssup\ 
($\zeta=2\pi\lambda/N$).

\subsec\secswz{Les excitations du modèle de WZW}
Du fait de l'absence de mass gap, la notion de particule est
un peu ambiguë dans une théorie non-massive. De plus,
la diffusion de deux particules de même chiralité (allant
donc à la vitesse de la lumière dans la même direction) paraît
dénuée de signification physique.
Cependant, dans le cadre des modèles intégrables, on peut
donner un sens à de tels concepts (voir par exemple l'introduction de
[\ref\FSc{P.~Fendley et H.~Saleur, \pre{hep-th/9310058}.}]).
Ainsi, à partir de l'Ansatz de Bethe, et
par analogie avec le cas massif, il est naturel de considérer
que les noeuds hachurés des BAE physiques correspondent à des
excitations non-massives qui diffusent avec des matrices $S$ que
nous allons maintenant calculer.

D'après le diagramme des BAE physiques,
nous conjecturons la structure suivante des excitations:
tout d'abord, les particules gauches (resp. droites)
de type $r$ ($1\le r\le N-1$) appartiennent
à la \eme{r} représentation fondamentale de $SU(N)_L$
(resp. $SU(N)_R$).
Ensuite, les noeuds compris entre les deux noeuds hachurés ($1\le j\le
f-1$) suggèrent d'introduire une symétrie
$U_q({\goth sl}(N))$, avec $q=-\exp(-i\pi/(f+N))$. 
Si l'on poursuit dans cette logique,
on doit alors effectuer un choix arbitraire:
supposons que les particules gauches de type $r$ (qui sont reliées aux
noeuds $j=1$) appartiennent
à la \eme{r} \rep\ fondamentale de $U_q({\goth sl}(N))$;
les particules droites (qui sont reliées aux noeuds $j=f-1$)
sont alors au contraire dans la \rep\ $r\times (f-1)$ (tableau de Young
rectangulaire). Nous aurions pu
faire le choix inverse; dans tous les cas, les particules
gauche et droite ne jouent pas des rôles symétriques,
mais cette dissymétrie n'est qu'apparente, comme nous
allons le voir plus loin.

Le Hilbert complet de la théorie est finalement construit comme
un espace de Fock tronqué vis-à-vis de la symétrie $U_q({\goth sl}(N))$.
Les particules gauches diffusent entre elles avec une matrice $S$
$$S_{LL}^{qr}(\zeta)=\Sh_{LL}^{qr}(\zeta)
R_L^{qr}(\zeta) R'{}^{qr}(\zeta)$$
où on a utilisé les notations habituelles, et $\zeta=\log(E_1/E_2)
-\alpha^{qr}$
est (à une constante $\alpha^{qr}=\log\sin(\pi q/N)-
\log\sin(\pi r/N)$ près)
le logarithme du rapport des énergies des deux particules.
Il est important de remarquer que la rapidité $\zeta$ ne dépend
d'aucune échelle d'énergie; en particulier dans la limite IR
où les chiralités se découplent, les expressions que
nous allons trouver pour les matrices $S$ doivent rester valables.
On a une formule strictement analogue pour $S_{RR}^{qr}(\zeta)$.
Enfin, la matrice $S$ entre particules de chiralités différentes
est de la forme:
$$S_{LR}^{qr}(\zeta)=\Sh_{LR}^{qr}(\zeta)
R'{}^{q\hat{r}}(\zeta)$$
où $R'{}^{q\hat{r}}$ est la matrice $R$ de $U_q({\goth sl}(N)$
entre les représentations $q\times 1$ et $r\times (f-1)$.
Cette fois $\zeta=\log(E_1 E_2/E^2)-\alpha^{qr}$, donc $\lambda$ dépend
de l'échelle d'énergie engendrée dynamiquement $E$.

Ensuite, la comparaison entre BAE physiques et BAE nues nous
donne de manière standard les facteurs scalaires:
$\Sh_{LL}^{qr}$ est donné par
\Ssup, ce qui était évident puisque dans la limite infra-rouge,
chaque chiralité prise séparément reproduit le modèle fusionné du
\secphysisoN; quant à $\Sh_{LR}^{qr}$, il est donné par:
($q\ge r$)
$$\Sh_{LR}^{qr}(\zeta)=
\exp\left(2i
\int_0^{+\infty} \d\omega\, 
{\sin(\omega\zeta)\over \omega}
{\sinh(\pi\omega(1/N+1))\over\sinh(\pi\omega(f/N+1))}
{\sinh(\pi\omega(1-q/N))\sinh(\pi\omega r/N)\over\sinh(\pi\omega/N)
\sinh(\pi\omega)}
\right)\eqn\SLR$$

Enfin, on peut calculer $\Sh_{RR}^{qr}(\zeta)$, mais
nous ne le donnerons pas explicitement, car $\Sh_{RR}$ n'a pas
de signification physique pour nous: en effet,
après la troncation de groupe quantique,
il est impossible
de faire diffuser deux particules de spin $(f-1)/2$ dans
la représentation de plus haut poids.

Le fait qu'on puisse définir un déphasage
$S_{LL}$ qui ait un sens physique,
tandis que $S_{RR}$ en est dénué peut
paraître surprenant, mais il est dû à la
manière asymétrique dont nous avons traité les particules gauche
et droite. Pour faire disparaître cette asymétrie, il faut
passer à la formulation solitonique des matrices $S$.
Nous ne traiterons en détail que le cas $N=2$.

Tout d'abord, pour $N=2$, on sait (cf Eq. \kinkSb) que
la matrice $S_{LL}$ est exactement donnée par
le produit tensoriel de la matrice 
$$S(\zeta)=R(\zeta) 
{ \Gamma\left(1+i{\zeta\over 2\pi}\right)
  \Gamma\left({1\over 2}-i{\zeta\over 2\pi}\right)
\over
  \Gamma\left(1-i{\zeta\over 2\pi}\right)
  \Gamma\left({1\over 2}+i{\zeta\over 2\pi}\right)}
$$
de $Y({\goth sl}(2))$, et de la matrice $S$ solitonique
\kinkS\ (avec $\lambda=\zeta/\pi$).

Ensuite, on constate que la symétrie des Clebsch--Gordan:
$s\otimes \varphi(s')=\varphi(s\otimes s')$ où $\varphi(s)=f/2-s$,
se répercute au niveau des matrices $S$ \kinkS, puisque
celles-ci sont invariantes par $s\to \varphi(s)$. Il est alors possible
de faire la transformation suivante: plutôt que de considérer
les particules droites comme ayant un spin $(f-1)/2$, on les prend
de spin $1/2$ comme les particules gauches,
en effectuant la transformation $\varphi$
après chaque transition associée à une
particule droite. On calcule sans difficulté
la matrice $S_{RR}$ solitonique dans cette nouvelle base,
et on la trouve égale à $S_{LL}$ (Eq.\ \kinkS), avec
le même préfacteur scalaire 
$\Sh_{LL}$. La symétrie est donc rétablie.

On peut alors effectuer la même transformation à $S_{LR}$
(figure \solibrab).
\fig\solibrab{Etats constitués d'une particule gauche
et d'une particule droite.}{\figdir{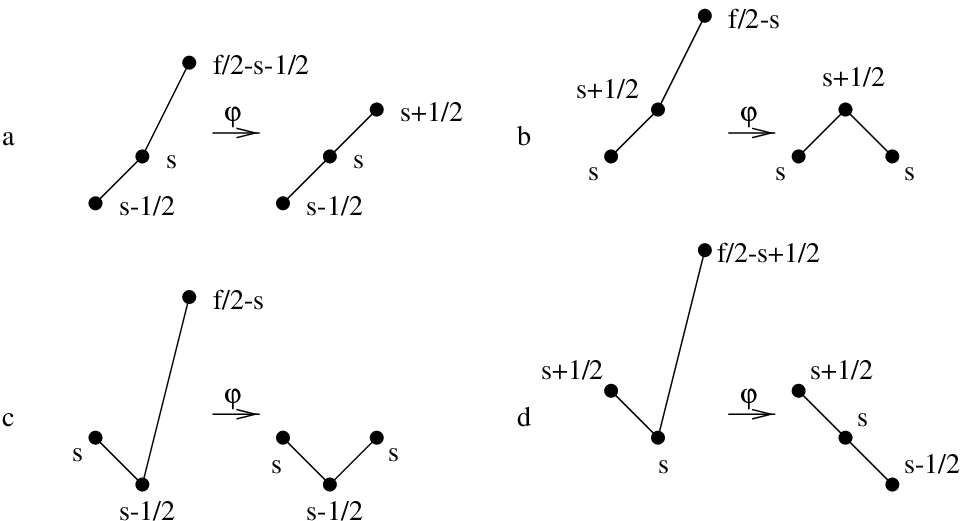}}
Notons que le facteur scalaire \SLR\ s'identifie alors
au déphasage de la diffusion de spin $0$, et non de spin $1$
(on peut par exemple le voir sur la figure \solibrab, diagramme
$b$, $s=0$). Reprenant un instant la
paramétrisation $\zeta=\pi\lambda$,
on trouve le résultat:
$$\eqalign{
S(a\rightarrow a)=S(d\rightarrow d)&=
{\cosh(\tilde{\gamma}(\lambda-i))\over\cosh(\tilde{\gamma}(\lambda+i))}
\Sh_{LR}(\lambda)\cr
S(b\rightarrow c)=S(c\rightarrow b)&={\q{2s+2}^{1/2} \q{2s}^{1/2}\over\q{2s+1}}
{\cosh(\tilde{\gamma}\lambda)\over\cosh(\tilde{\gamma}(\lambda+i))}
\Sh_{LR}(\lambda)\cr
S(b\rightarrow b)&=-{1\over\q{2s+1}}
{\cosh(\tilde{\gamma}(\lambda+i(2s+1)))\over\cosh(\tilde{\gamma}(\lambda+i))}
\Sh_{LR}(\lambda)\cr
S(c\rightarrow c)&={1\over\q{2s+1}}
{\cosh(\tilde{\gamma}(\lambda-i(2s+1)))\over\cosh(\tilde{\gamma}(\lambda+i))}
\Sh_{LR}(\lambda)\cr
}\eqn\kinkSLR$$
avec $\tilde{\gamma}=\pi/(f+2)$.
Ceci s'avère être essentiellement
identique à la matrice $S_{LR}$ conjecturée dans [\FSZ]\foot{Dans
les notations de [\FSZ], $\Sh_{LR}(\zeta)=\tilde{Z}(\zeta)
\cosh(\zeta+i\pi)/(f+2)$.},
ce qui est normal puisque nous avons vu au \secbaephysgen\ que
les matrices $S$, pour $N=2$, se factorisent exactement en matrices $S$
associées aux différentes symétries,
et la matrice $S_{LR}$ ne fait intervenir que la symétrie
$U_q({\goth sl}(2))$ commune aux deux modèles.

Revenons à $N$ quelconque, et discutons enfin du cas $f=1$.
Dans ce cas-là, la structure solitonique est triviale, au sens où,
pour une quelconque diffusion (LL, LR, RR), il n'y a qu'une transition
possible: on peut donc oublier entièrement la structure solitonique.
En particulier, la matrice $S_{LR}$ est une simple phase, donnée
par \SLR\ avec $f=1$.

\subsec\secmagwz{Calculs en champ magnétique}
Nous allons présenter ici des
calculs exacts du modèle de Wess--Zumino--Witten placé
en champ magnétique à température nulle. Ces résultats seront
d'un grand intérêt pour nous puisqu'on verra dans la section
\seckondo\ qu'ils s'appliquent directement au modèle de Kondo.

Expliquons tout d'abord ce qu'il advient des TBA \tbawz\
dans la limite $T\to 0$, à champ magnétique $B$ fixé. 
L'idée générale
est que si l'on pose $\eta^r_j=\e{\epsilon^r_j/T}$, alors
on a $\epsilon^r_j\sim B\gg T$, de sorte que
$\log(1+\eta^r_j)\sim{1\over T}\max(\epsilon^r_j,0)$ et 
$\log(1+(\eta^r_j)^{-1})\sim{1\over T}\max(-\epsilon^r_j,0)$.

Supposons donc que nous avons placé un champ magnétique
$B$ pour la symétrie $SU(N)_R$. Comme il est expliqué dans
[\artic03.3.4], le champ $B$ a $N-1$ composantes, et nous nous
restreindrons au cas où les composantes $b^r$ (dans la
même paramétrisation qu'au [\artic03.3.4]) sont de la forme:
$b^r\propto \sin(\pi r/N)$. Physiquement, il est clair que la présence
de $B$ favorise l'apparition de particules physiques droites
($B$ joue le rôle de potentiel chimique positif pour celles-ci),
mais défavorise au contraire l'apparition d'excitations isotopiques
$SU(N)_R$ qui tendent à réduire le spin $SU(N)_R$ ($B$ joue
le rôle de potentiel chimique négatif pour celles-ci). Quant aux
excitations isotopiques $U_q({\goth sl}(N))$, et aux
particules gauches, elles ne sont 
pas affectées par $B$, et 
ces dernières doivent donc se trouver dans l'état du vide.
Par conséquent, on est conduit aux conjectures suivantes pour les fonctions
$\epsilon^r_j$:
$$\left\{\eqalign{
\epsilon_j^r(\zeta)&>0\qquad j\ne 0,f\cr
\epsilon_0^r(\zeta)&<0\cr
\epsilon_f^r(\zeta)&{>0\quad\zeta<\zeta_0\atop <0\quad\zeta>\zeta_0}\cr
}
\right.$$
On a introduit le paramètre $\zeta_0$, qui est la limite
entre la zone d'énergie dans laquelle des particules droites
sont excitées ($\zeta<\zeta_0$) et la zone d'énergie qui reste
vide de particules ($\zeta>\zeta_0$).

Du fait que le second membre des TBA est proportionnel au
vecteur de Perron--Frobenius de la matrice d'adjacence du
diagramme $A_{N-1}$: $\sin(\pi r/N)$, on peut projeter
toutes les équations sur celui-ci.
On pose donc:
$\epsilon_f^r(\zeta)=\sin(\pi r/N)
(\epsilon^+(\zeta)+\epsilon^-(\zeta))$,
avec $\epsilon^+(\zeta)>0$ et $\epsilon^-(\zeta)<0$ de supports
disjoints, et après quelques manipulations,
les TBA à $T\to 0$ se récrivent:
$$\epsilon^- + K_0\star \epsilon^+=-{E\over 2}\e{\zeta}+b\eqn\tbawzB$$
où $b^r=b(1-\cos(\pi/N))\sin(\pi r/N)$, et
$K_0$ est le noyau de convolution donné en transformée de Fourier
par rapport à $\zeta$\foot{Rappelons que par rapport aux
conventions du chapitre \chapbase, nous avons changé l'échelle des
rapidités; donc de même, en transformée de Fourier, $\kappa=\pi\omega/N$.}:
$$K_0(\omega)={\sinh(\pi\omega/N)\over\cosh(\pi\omega/N)-\cos(\pi/N)}
{\e{f\pi|\omega|/N}\over 2 \sinh(f\pi\omega/N)}\eqn\Kzero$$

Par un décalage $\zeta\to\zeta-\zeta_0$, on peut
supposer que $\epsilon^-$ est non nul sur $[-\infty,0]$ et
que $\epsilon^+$ est non nul sur $[0,+\infty]$. Finalement,
on agit avec l'opérateur différentiel $1-\der/\der\zeta$ sur \tbawzB:
$$u+K\star\epsilon^+=b\eqn\tbawzBb$$
où $u=(1-\der/\der\zeta)\epsilon^-$, $K(\omega)=(1+i\omega) K_0(\omega)$,
et on est ramené
à un classique problème de Wiener--Hopf.
La solution passe par l'introduction de deux fonctions $K^\pm(\omega)$
holomorphes dans le demi-plan $\pm\Im\omega\le 0$, telles que
$K(\omega)=K^+(\omega)/K^-(\omega)$; explicitement,
$$\eqalign{
K^+(\omega)=\alpha{N\over 2\pi^2 f}& 
{\Gamma\left(1+{1+i\omega\over 2N}\right)
\Gamma\left(1+{-1+i\omega\over 2N}\right)
\Gamma\left(1+{i\omega f\over N}\right)
\over\Gamma\left(1+{i\omega\over N}\right)}
\e{-i\omega{f\over N} \log({if\over N{\rm e}}(\omega-i0))}
\e{i\omega{\log2\over N}}\cr
K^-(\omega)^{-1}=\alpha^{-1}&
{\Gamma\left(1-{1+i\omega\over 2N}\right)
\Gamma\left({1-i\omega\over 2N}\right)
\Gamma\left(1-{i\omega f\over N}\right)
\over\Gamma\left(1-{i\omega\over N}\right)}
\e{i\omega{f\over N} \log(-{if\over N{\rm e}}(\omega+i0))}
\e{-i\omega{\log2\over N}}\cr
}$$
Les fonctions $K^+$ et $K^-$ ont été choisies de
telle sorte que lorsque $|\omega|\to\infty$, $K^+(\omega)\sim
|\omega|^{3/2}$ et $K^-(\omega)\sim |\omega|^{1/2}$.
On prend de plus $\alpha=\Gamma(1-1/2N)\Gamma(1/2N)$,
de fa\c con que $K^-(\omega=0)=1$. 
On a alors les solutions de \tbawzBb:
$$\eqalign{
u(\omega)&=i b {K^-(\omega)^{-1}\over\omega+i0}\cr
\epsilon^+(\omega)&=-i b {K^+(\omega)^{-1}\over\omega-i0}\cr}$$
En réintégrant, on obtient
$\epsilon^-$ à partir de $u$, d'où finalement en calculant
l'asymptotique de $\epsilon^-$ 
quand $\zeta\to +\infty$ (qui est directement reliée à 
$K^-(\omega=i)$),
$\zeta_0$ en fonction du champ $b$ (et de $E$):
$$\zeta_0=
\log\left({b\over E}\right)
+\log{2N\over\Gamma(1-1/2N)\Gamma(1/2N)}
+\log \Gamma\left(1+{f\over N}\right)
-{f\over N} \left(\log{f\over N}-1\right)\eqn\zzero$$
Nous verrons la signification physique du paramètre $\zeta_0$
quand nous appliquerons ces formules au modèle de Kondo (section 
\seckonmag).
De fait, nous n'irons pas plus loin dans nos calculs
(bien que la donnée des fonctions $\epsilon^+$ et $\epsilon^-$
permette de calculer les quantités thermodynamiques
physiquement intéressantes du modèle de WZW),
car nous analyserons ces formules dans la limite $N\to\infty$
uniquement (cf section suivante),
et en relation avec le modèle de Kondo.

\noindent{\it Remarque:} après avoir rédigé cette section de ma thèse,
j'ai pris connaissance de l'article [\ref\PT{S.V.~Pokrovsky
et A.M.~Tsvelick, {\it Nucl.\ Phys.} B320 (1989), 696.}],
qui effectue un calcul en champ magnétique
assez similaire (quoique différent)
pour une chaîne de spins dont le point fixe infra-rouge est
la théorie conforme de WZW.

\newsec\seclargNwz{Limite de grand $N$ des TBA}
Avant d'aborder le modèle de Kondo,
nous allons faire quelques remarques concernant la limite de
grand $N$ des équations de TBA $SU(N)$. Celles-ci s'appliquent
à tous les modèles intégrables relativistes invariants $SU(N)$;
nous nous intéresserons tout particulièrement à WZW $SU(N)$ niveau
$f$, pour lequel
nous prendrons la limite $N\to\infty$, $\nu\equiv f/N$ fixé.

Partons des TBA générales $SU(N)$ (cf Eq.\ \tbaN):
$$\log(1+(\eta^r_j)^{-1})
-(C^{-1})^{qr}\star C_{jk}\star \log(1+\eta^q_k) 
=\sin(\pi r/N)(\alpha_j^+ \e{+\zeta}
+\alpha_j^- \e{-\zeta})\eqn\tbaNb$$
Prolongeons analytiquement les fonctions $\eta^r_j(\zeta)$ dans
la bande $|\Im\zeta|\le\pi/N$, et définissons les opérateurs
$\hat{C}^{qr}$ et $\hat{C}_{jk}$ comme dans [\artic03.3.3], 
qui sont tels que $\hat{C}^{qr}\star s=C^{qr}$ et $\hat{C}_{jk}\star s=C_{jk}$.
En appliquant $\hat{C}^{qr}$ à \tbaNb, on obtient:
$$\hat{C}^{qr} \star\log(1+(\eta^r_j)^{-1})
-\hat{C}_{jk}\star\log(1+\eta^q_k)=0\eqn\CC$$
Ces équations ne contiennent plus le second membre, annihilé par
$\hat{C}^{qr}$; elles doivent donc être suppléées par des conditions
aux bords qui fixent le second membre de \tbaNb.

Dans la limite $N\to\infty$, on introduit des indices continus
$x=\pi j/N$ et $y=\pi r/N$ ($0<y<\pi$). 
On voit alors facilement que les opérateurs $\hat{C}^{qr}$
et $\hat{C}_{jk}$, qui étaient des opérateurs de différences
discrètes, deviennent dans cette limite des opérateurs différentiels
du type de Laplace:
$$\eqalign{
\hat{C}^{qr}&\sim -{\pi^2\over N^2} \left({\der^2\over\der y^2}
+{\der^2\over\der\zeta^2}\right)\cr
\hat{C}_{jk}&\sim -{\pi^2\over N^2} \left({\der^2\over\der x^2}
+{\der^2\over\der\zeta^2}\right)\cr
}$$
L'équation \CC\ devient donc une équation aux dérivées partielles
à trois variables pour la fonction $\eta(x,y,\zeta)$:
$$\left({\der^2\over\der y^2}
+{\der^2\over\der\zeta^2}\right)\log(1+\eta^{-1})-
\left({\der^2\over\der x^2}
+{\der^2\over\der\zeta^2}\right)\log(1+\eta)=0\eqn\derpar$$
Cette équation aux dérivées partielles est intégrable
au sens des systèmes intégrables classiques, c'est-à-dire
qu'on peut l'écrire
comme une condition de courbure nulle; malheureusement, il est
difficile d'en écrire des solutions explicites, et encore moins
des solutions qui puissent avoir le bon comportement asymptotique
pour les TBA. Or, toute la physique des différents modèles intégrables
est contenue dans ces conditions aux bords. Signalons aussi
que la validité des équations \derpar\ n'est pas prouvée,
car -- à part pour les valeurs limites $\eta(\zeta=\pm\infty)$ que
l'on sait calculer explicitement et dont on voit
qu'elles ont une limite régulière à $N=\infty$ en
termes des variables $x$ et $y$ -- rien ne nous dit que
les fonctions $\eta$ ne vérifient pas une loi d'échelle
compliquée en fonction de $N$, éventuellement dépendante de $x$,
$y$ et $\zeta$.

On peut tout de même se servir de l'équation \derpar\ dans
la limite $T\to 0$,
pour faire une étude en champ magnétique et à température nulle.
Ainsi, les résultats de [\ref\FKW{V.A.~Fateev, V.A.~Kazakov,
P.B.~Wiegmann, {\it Nucl. Phys.} B424 (1994), 505.}] sur
le modèle $\sigma$ principal à grand $N$ peuvent
être retrouvés de manière élémentaire à partir de \derpar.
De manière similaire, on peut obtenir les
résultats de la section \secmagwz\ directement à $N$ grand:
on part de \derpar, qui se récrit à $T\to 0$:
$$\left({\der^2\over\der y^2}
+{\der^2\over\der\zeta^2}\right)\epsilon^-
+\left({\der^2\over\der x^2}
+{\der^2\over\der\zeta^2}\right)\epsilon^+
=0\eqn\derparb$$
On impose la condition: $\epsilon^-(x,y,\zeta)=\delta(x-\pi\nu)
e^-(y,\zeta)$\foot{Remarquons que $\epsilon^-(x=\nu)$ n'est pas
définie du fait de la fonction $\delta$; ceci provient
à $N$ fini d'une divergence des échelles-types de $\epsilon^-$
et $\epsilon^+$: on a en effet: $\epsilon^-\sim N\epsilon^+$.}.
Ceci ramène le problème à la recherche d'une fonction de Green
(par rapport à $x$ et $\zeta$) qui s'annule en $x=0$. Passant
en double Fourier par rapport à $\zeta$ (variable conjuguée
$\omega$) et $y$ (variable conjuguée -- entière -- $l$), la solution
est:
$$K_0(\omega)={\e{\pi\nu|\omega|}\over \sinh(\pi\nu\omega)} 
{\omega\over\omega^2+l^2}\eqn\Kzerob$$
Finalement, on se restreint au mode de Perron--Frobenius, ce qui
revient à poser $l=1$ dans \Kzerob. On vérifie alors que
\Kzerob\ est bien le noyau limite de \Kzero\ à $N=\infty$.
La suite du calcul est identique à ce que nous avons fait
à $N<\infty$, mais les expressions sont nettement
plus simples; ainsi, $\epsilon^+\equiv\epsilon^+(x=\pi\nu)$ vaut
$$\epsilon^+(\omega)=-i B\pi\nu B {1\over\omega-i0}
\e{i\nu\omega\log(i\nu/{\rm e}(\omega-i0))}
\Gamma(1+i\nu\omega)^{-1}$$
où l'on a redéfini le champ magnétique: $B\equiv \pi b/N$.
Les fonctions $\Gamma$ supplémentaires à $N<\infty$ ne font
que créer un développement régulier en $1/N$ autour
de $N=\infty$. Ainsi, ces formules explicites permettent
d'affirmer que les phénomènes physiques qualitatifs
sont entièrement capturés par la limite $N\to\infty$, et ne
dépendent que du rapport $\nu=f/N$, ce qui constitue
une justification de cette hypothèse, utilisée par
exemple dans [\ref\GP{O.~Parcollet et A.~Georges,
{\it Phys. Rev. Lett.} 79 (1997), 4665 \pre{cond-mat/9707337}\semi
O.~Parcollet, A.~Georges, G.~Kotliar et A.~Sengupta, 
\pre{cond-mat/9711192}.}].
\fig\epsplus{Courbe
de $\epsilon^+(\zeta)/B$ à $N=\infty$, $\nu=1/2$. Notons
la singularité en racine carrée à $\zeta=0$ qui n'existe
qu'à $N=\infty$ (à $N<\infty$ elle est en puissance $3/2$), due à
la divergence des échelles: $\epsilon^-\sim N
\epsilon^+$.}{\figdir{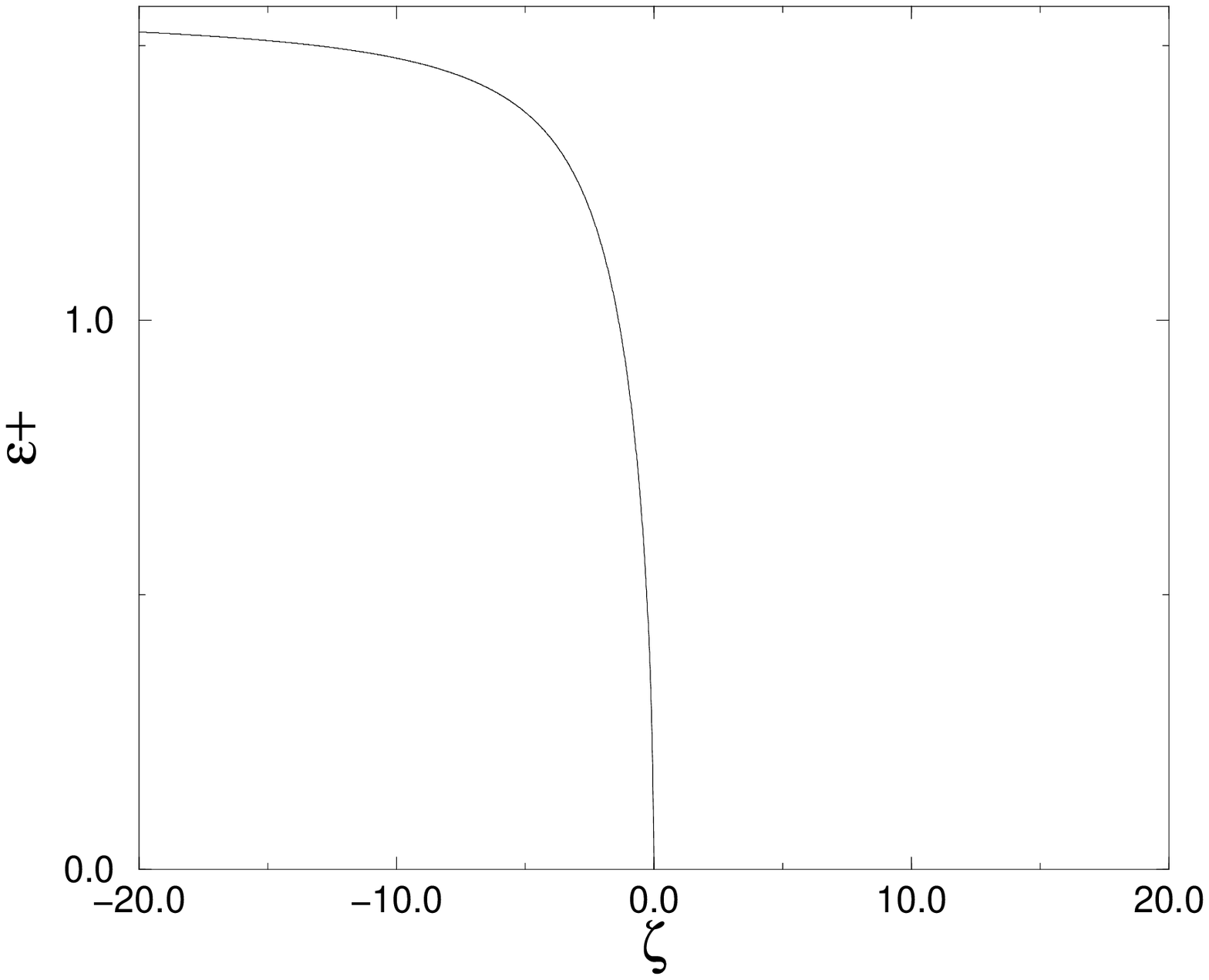}}

\newsec\seckondo{Le modèle de Kondo généralisé}
Le modèle de Kondo décrit l'interaction entre des impuretés
magnétiques en faible concentration dans un métal et
les électrons de conduction. A basse température, on explore
le voisinage du point fixe infra-rouge de la théorie,
qui s'avère être dans une région de couplage fort: la physique
y est non-triviale, et est dominée par un phénomène d'{\it écrantage}
des impuretés par les électrons, de sorte que les impuretés
voient leur spin effectif diminué.
Le lecteur est renvoyé à
[\ref\AFL{N.~Andrei, K.~Furuya et J.H.~Lowenstein,
{\it Rev. Mod. Phys.} 55 (1983), 331.}]
et à l'introduction de [\artic03]
pour plus de détails et pour les références historiques.
Ici, notre objectif est d'effectuer une étude (par l'Ansatz de Bethe)
du modèle de Kondo complémentaire
de celle de [\artic03], et qui utilise les résultats de la
section \secpres\ (en particulier du \secswz)
sur le modèle de WZW.

Dans le modèle de Kondo usuel, on part 
d'impuretés de spin $s$, qui interagissent avec
les électrons (de spin $1/2$) par un couplage
spin-spin antiferromagnétique. Comme les impuretés sont en
faible concentration, on peut supposer qu'il n'y a qu'une impureté.
On se place en coordonnées radiales autour de ladite impureté,
et on ne garde que les modes radiaux. On est alors ramené
à un problème à $1+1$ dimensions, avec la coordonnée spatiale
sur une demi-droite: $x>0$. Plutôt que de se restreindre
à cette demi-droite, on peut identifier les électrons
entrants avec des particules ``droites'' à $x<0$, tandis
que les électrons sortants sont aussi des particules ``droites'' avec
$x>0$. On est donc ramené à un problème avec $-\infty<x<+\infty$,
une seule chiralité d'électrons, et une impureté
localisée en $x=0$. Essentiellement, les électrons arrivent
de $x=-\infty$, interagissent avec l'impureté pour $x\sim 0$,
puis repartent vers $x=+\infty$. Cependant, le modèle de Kondo
reste un modèle de théorie des champs (et non de mécanique
quantique), car l'impureté induit des corrélations non-triviales
entre les électrons successifs qui diffusent sur elle.

Dans le modèle de Kondo multi-canal, il y a plusieurs
bandes de conduction,
de sorte que l'on peut attribuer un nombre
quantique supplémentaire (la ``saveur'')\foot{On peut
également interpréter la saveur
comme le nombre quantique orbital $m$, si l'on tient
compte d'orbitales plus élevées (le nombre de saveurs
est alors égal à $2l+1$).} aux électrons.
On suppose pour simplifier que toutes les saveurs interagissent
identiquement avec l'impureté: s'il y a $f$ saveurs,
le modèle est alors invariant $SU(2)\times SU(f)$.

Enfin, dans le modèle de Kondo généralisé, on remplace
le groupe $SU(2)$, qui provient originellement du spin à $3$ dimensions,
par le groupe $SU(N)$; le modèle résultant est aussi
appelé modèle de Coqblin--Schrieffer. On obtient finalement
un modèle invariant $SU(N)\times SU(f)$, avec des électrons
dans une représentation fondamentale de $SU(N)$\foot{Comme
précédemment, le choix de la représentation
fondamentale ne change rien à la physique (la taille
verticale du tableau de Young ne fait que changer d'un facteur
numérique l'énergie dynamiquement engendrée).},
et une impureté que l'on prendra dans une représentation
de tableau de Young rectangulaire $n\times l$ vis-à-vis de $SU(N)$.

Le Hamiltonien du modèle de Kondo
généralisé est donné au [\artic03.1.1]
(avec une procédure de cutoff précise de fa\c con à permettre
la fusion dynamique), et diagonalisé
grâce à l'Ansatz de Bethe emboîté. Ici, nous nous
contenterons de quelques commentaires généraux qui
éclairent la construction qui est faite.

Retirons provisoirement l'impureté du système. 
Les électrons sont alors des fermions libres $U(Nf)$,
qui constituent la fermionisation
du modèle de WZW $U(Nf)$ niveau $1$. Comme l'interaction
avec l'impureté va briser $U(Nf)$ en $U(1)\times SU(N)\times SU(f)$
(la {\it charge}, le {\it spin} et la {\it saveur}),
nous sommes amenés à considérer les électrons
du point de vue de cette symétrie restreinte. Comme
d'habitude, nous ne nous intéresserons pas à la symétrie
$U(1)$, qui sera reliée à un secteur de charge découplé.
Pour ce qui est de $SU(N)$, on voit que les courants associés
forment une représentation de niveau $f$ de l'algèbre
de Kac--Moody $\widehat{{\goth sl}(N)}$; en effet, 
chaque saveur contribue pour $1$ au
niveau, et il y a $f$ saveurs. De manière analogue,
on a une représentation de niveau $N$ de l'algèbre
$\widehat{{\goth sl}(f)}$. Ceci nous conduit naturellement
à l'hypothèse suivante: le Hilbert des électrons
se décompose naturellement en trois secteurs,
le secteur de charge, le secteur de spin qui est un WZW chiral $SU(N)$
niveau $f$, et le secteur de saveur qui est un WZW chiral $SU(f)$ niveau $f$.
Cette hypothèse est justifiée par le fait qu'il y a un
``plongement conforme'' (les tenseurs énergie-impulsion
des trois secteurs reconstituent le tenseur énergie-impulsion
complet), résumé par l'égalité des charges centrales
$$c=(N^2-1){f\over N+f} + (f^2-1){N\over N+f} + 1 = Nf$$

Bien sûr, cette hypothèse néglige toutes les subtilités reliées
aux ``modes zéros''. De fait, par une analyse plus détaillée
[\ref\ABI{D.~Altschuler, M.~Bauer et C.~Itzykson,
{\it Commun. Math. Phys.} 132 (1990), 349.}], 
on montre que la fonction de partition (sur un tore) des fermions
libres $U(Nf)$ se décompose comme somme de produits
d'un caractère de $\widehat{{\goth u}(1)}$,
d'un caractère de $\widehat{{\goth sl}(N)}_f$ et
d'un caractère de $\widehat{{\goth sl}(f)}_N$, avec
des coefficients déterminés dans [\ABI].
Ces coefficients couplent les différents secteurs,
ce qui est normal puisque la notion de ``WZW chiral'' est quelque
peu ambiguë (cf remarque similaire à la fin du \sectbawz).

Les trois secteurs apparaissent naturellement dans la description
de l'Ansatz de Bethe sous la forme de trois jeux de rapidités
$\lambda_j$, $\omega_\alpha$, $\Lambda_\alpha$ (Eq.\ (\artic03.1.16)).
Si l'on réintroduit maintenant l'impureté,
rien n'est changé dans l'analyse précédente, si ce n'est 
qu'il y a maintenant un secteur
en interaction (qui n'est donc plus conforme):
le secteur de spin, du fait du couplage
spin-spin entre électrons et impureté. On peut donc
se placer dans l'état du vide des deux autres secteurs,
et on aboutit à des BAE effectives pour le secteur
de spin (Eq.\ (\artic03.1.19)), représentées par la figure \baekondo.
\fig\baekondo{BAE nues du modèle de Kondo.}{\figdir{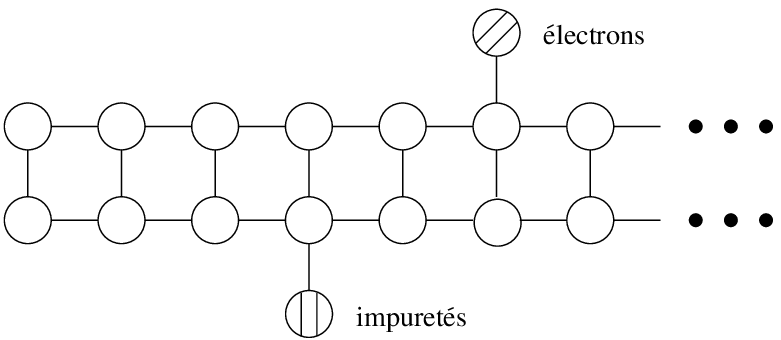}} 
On remarque que
si l'on retire la contribution de l'impureté, ces
équations ont exactement la structure
de celles du modèle de WZW chiral $SU(N)$ niveau $f$, 
ce qui est conforme à la discussion ci-dessus.
En particulier, le fait que l'on aboutisse
à des BAE fusionnées provient de la
{\it fusion dynamique} des électrons
en des composites dans des représentations supérieures
(de largeur du tableau de Young égale à $f$).
Le nombres d'électrons originels
n'étant pas nécessairement multiple de $f$,
on peut supposer que quelques électrons fusionnent
partiellement et assurent des conditions aux bords libres pour
le nombre quantique $U_q({\goth sl}(N))$ (cf section \secphyssup).

La suite de l'article [\artic03] consiste à étudier l'état
fondamental, puis la thermodynamique
du modèle via les TBA:
après être passé à la limite d'échelle, qui, dans le modèle de Kondo,
s'interprète comme l'envoi de la profondeur de la mer de Fermi
à l'infini, on aboutit aux TBA que nous écrirons sous la forme:
$$\log(1+(\eta^r_j)^{-1})
- (C^{-1})^{qr}\star C_{jk}\star \log(1+\eta^q_k) 
= \delta_{jf}
{\sin(\pi r/N)\over\sin(\pi/N)}
{T_K\over 2}e^\zeta\eqn\tbakon$$
et qui s'avèrent être identiques aux TBA de
la théorie conforme WZW $SU(N)$ niveau $f$ (figure \tbaNwz),
du fait que l'impureté
ne contribue pas à l'énergie (à part à travers le champ magnétique).
Cependant, on a tout de même une échelle engendrée dynamiquement:
même si on peut absorber l'échelle $T_K$ 
du second membre par un décalage des rapidités
(cf Eq.\ (\artic03.3.8)),
celle-ci réapparaît dans l'expression de
l'énergie libre (Eq.\ (\artic03.3.10)).
Sans décalage des rapidités, l'échelle $T_K$ {\it (température
de Kondo)} apparaît explicitement dans les TBA \tbakon;
par contre, elle n'apparaît plus dans
l'énergie libre d'une impureté {\it habillée par son interaction avec les
électrons}, obtenue
en soustrayant à l'énergie libre totale
l'énergie libre des électrons en l'absence d'impuretés (et en
divisant par le nombre d'impuretés):
$$F^n_l=-T \int d\zeta \, (\hat{C}^{-1})^{rn}(\zeta)
\log(1+\eta^r_l(\zeta))\eqn\Fkon$$

Plutôt que de continuer dans la même voie que [\artic03] -- 
déduire des TBA les équations de fusion modifiées, puis les
comportements à haute et basse température des quantités
thermodynamiques --
nous allons reprendre les BAE nues
de la figure \baekondo, et montrer leur équivalence, dans
la limite d'échelle, avec des BAE physiques appropriées.
Ceci nous permettra de décrire le Hilbert
de basse énergie, les excitations physiques, etc.

\subsec\seckonphys{BAE physiques et spectre des excitations dans Kondo}
Les BAE physiques du modèle de Kondo contredisent la règle générale
admise jusqu'à présent, selon laquelle le diagramme des TBA
et celui des BAE physiques sont identiques:
en effet, bien que les TBA du modèle de Kondo soient
les mêmes que celles de WZW, les BAE physiques, elles, en
sont distinctes puisque l'impureté y apparaît explicitement
(figure \baekondb).
\fig\baekondb{BAE physiques du modèle de Kondo.}{\figdir{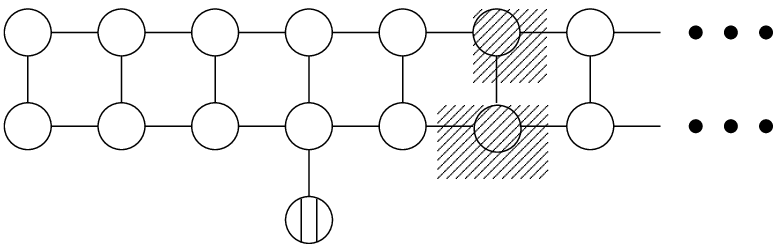}}
Ceci correspond aux équations suivantes, qui ne sont qu'une récriture
des BAE nues:
$$(C^{-1})^{qr}\star C_{jk} \star \rhot^q_k
+ \rho^r_j = \delta_{jf} {T_K\over 4\pi} {\sin(\pi
r/N)\over\sin(\pi/N)} \e{\zeta} + \delta_{jl} D^{\rm imp} (C^{-1})^{rn}\star
s(\zeta)
\eqn\baekonphys$$

On a réintroduit dans \baekonphys\ 
une densité linéique finie d'impuretés $D^{\rm imp}$, pour
éviter d'écrire une équation qui mélange différents ordres de $1/L$
($L$ taille du système). Ceci n'a pas d'importance particulière
puisque l'on suppose toujours que la densité d'impuretés est
suffisamment faible pour qu'elles ne se ``voient'' pas les unes les autres.

Evidemment, les excitations physiques correspondant au noeud $f$ sont
les particules habituelles du modèle de WZW chiral (cf \secswz), c'est-à-dire
des particules ``droites'' de type $1\le r\le N-1$, déterminées
par une rapidité $\zeta$ ($p=E\propto \e{\zeta}$),
et possédant deux nombres quantiques: $SU(N)$ et
$U_q({\goth sl}(N))$ tronqué (on parlera
aussi, suivant [\artic03], de nombre quantique $SU_q(N)$ pour ce dernier), 
avec $q=-\exp(-i\pi/(f+N))$.
Les matrices $S$ de diffusion sont celles qui ont été données
au \secphysisoN\ (Eq.\ \Ssup).

Il reste à se préoccuper des impuretés: chaque impureté {\it habillée}
(comme nous le verrons, cette distinction est importante, car
l'impureté est en général écrantée par les électrons)
doit être considérée comme un ``obstacle'' fixe sur lequel diffusent
les excitations physiques. Dans le cas $N=2$, c'est-à-dire celui
du modèle de Kondo
usuel, les matrices $S$ de diffusion sur l'impureté ont
été trouvées dans [\ref\FEN{P.~Fendley, \pre{cond-mat/9304031}.}];
cependant elles n'ont pas été démontrées par les BAE physiques,
comme nous le faisons maintenant\foot{D'ailleurs,
dans le cas surécranté, le facteur scalaire de la matrice $S$
(Eq.\ \Skonc) diffère de celui proposé dans [\FEN].}.
On distingue trois cas de figures:\goodbreak

\noindent{$\bullet$} $f<l$ {\it (sous-écrantage)}:\par\nobreak
Sur le diagramme des BAE, cela signifie que le noeud associé
aux impuretés est dans la partie droite du diagramme
(par rapport aux noeuds $f$).
Chaque impureté habillée constitue donc un spin $SU(N)$, dans
la représentation $n\times (l-f)$: par rapport à
l'impureté nue qui était un spin $n\times l$,
on voit que toutes les saveurs d'électrons ont écranté
l'impureté de manière à diminuer son spin autant
que possible.

Les excitations physiques diffusent
sur l'impureté avec une matrice $S(\zeta)$ (l'impureté
a par définition une rapidité nulle) qui est proportionnelle
à la matrice $R$ de ${\goth sl}(N)$ dans les représentations
de l'impureté et de l'excitation.
Le facteur de proportionnalité constitue le déphasage de plus
haut poids, qui est donné par:
$$\Sh^{r/{\rm imp}}(\zeta)=\exp\left(
\int_{-\infty}^{+\infty} \d\omega\, 
{\e{i\omega\zeta}\over \omega+i0}
{\sinh(\pi\omega(1-r/N)) \sinh(\pi\omega n/N)
\over \sinh(\pi\omega)\sinh(\pi\omega/N)}
\e{-(l-f)|\pi\omega/N|}\right)\eqn\Skon$$
(où nous avons supposé $r\ge n$; sinon, il faut échanger $r$ et $n$).
Comme nos équations ne nous donnent $\Sh$ qu'à une constante près,
nous avons défini $S_1$ de sorte que $\Sh(\zeta=+\infty)=1$, ce qui
est naturel d'après la liberté asymptotique UV.

Examinons alors le déphasage $\phi={1\over i} \log
\Sh^{r/{\rm imp}}$ à petite énergie $\epsilon\propto \e{\zeta}$:
$$\phi(\epsilon)
\buildrel \epsilon\rightarrow 0 \over =
 2\pi {n(N-r)\over N} + {\alpha\over \log(\epsilon/T_K)} + \cdots$$
On trouve les corrections en $1/\log$ qui
sont le signe de la liberté asymptotique IR.\goodbreak

\noindent{$\bullet$} $f=l$ {\it (écrantage)}:\par\nobreak
Les impuretés sont totalement écrantées, donc leur spin résiduel
est nul. La diffusion sur l'impureté est donc une phase pure,
donnée par \Skon\ avec $f=l$. Explicitement, pour $n=1$, on trouve
$$\Sh^{r/{\rm imp}}(\zeta)=
\e{i\pi r/N}
{\sinh{1\over 2}(\zeta-i\pi r/N)
\over\sinh{1\over 2}(\zeta+i\pi r/N)}\eqn\Skonb$$
Pour $n>1$, on obtient un produit
fini de termes du type de \Skonb.
Le développement à basse énergie est un développement
en puissances entières de $\epsilon$:
$$\phi(\epsilon) \buildrel \epsilon\rightarrow 0 \over =
2\pi {n(N-r)\over N} + \beta {\epsilon\over T_K} + \cdots
$$\goodbreak

\noindent{$\bullet$} $f>l$ {\it (sur-écrantage)}:\par\nobreak
C'est le cas de figure intéressant, où la symétrie de groupe
quantique joue un rôle important; en effet, du fait que le noeud correspondant
aux impuretés est dans la partie gauche du diagramme, on trouve
que l'impureté est un spin $SU_q(N)$ (et non
un spin $SU(N)$; l'impureté est donc {\it totalement écrantée}
au sens où le spin résiduel est nul)
dans la représentation $n\times (f-l)$\foot{Il faut souligner
que nous avons fait un choix différent de celui
de l'article [\artic03], dans lequel l'impureté est considérée
comme ayant une représentation $(N-n)\times l$, mais
ceci est en fait purement conventionnel.}.
La matrice $S$ de diffusion
avec les excitations physiques est donc celle de $U_q({\goth sl}(N))$,
avec un facteur scalaire que l'on déduit des BAE:
$$\Sh^{r/{\rm imp}}(\zeta)=\exp\left(
\int_{-\infty}^{+\infty} \d\omega\, 
{\e{i\omega\zeta}\over \omega+i0}
{\sinh(\pi\omega(1-r/N)) \sinh(\pi\omega n/N)
\over \sinh(\pi\omega)\sinh(\pi\omega/N)}
{\sinh(\pi\omega(l/N+1))\over\sinh(\pi\omega(f/N+1))}
\right)\eqn\Skonc$$
d'où finalement le développement à basse énergie:
$$\phi(\epsilon) \buildrel \epsilon\rightarrow 0 \over =
2\pi {n(N-r)\over N}{l+N\over f+N} 
+ \gamma \left({\epsilon\over T_K}\right)^{N/(f+N)} + \cdots
$$

On trouve la même catégorisation que pour les calculs thermodynamiques
de [\artic03.4.2]. On obtient en particulier l'exposant
critique dans le régime sur-écranté: $\Delta-1=N/(N+f)$, et
on identifie en théorie conforme $\Delta$ avec
la dimension de l'opérateur perturbant ${\cal O}=\vec\jmath_{-1}\cdot
\vec\Phi_{\rm ad}$, où $\vec\jmath$ est le courant
$SU(N)$ et $\vec\Phi_{\rm ad}$ est l'opérateur
primaire de ${\goth sl}(N)_f$ 
associé à la représentation adjointe. On remarque que dans le
développement (\artic03.4.14) apparaissent des termes en $T/T_K$
et en $(T/T_K)^{2(\Delta-1)}$ (et non $(T/T_K)^{\Delta-1}$),
d'où une compétition entre ces deux termes, et une subdivision
supplémentaire entre $f<N$ et $f>N$.
Ceci suggère que lorsqu'on effectue
le développement perturbatif de la fonction de partition
en la constante de
couplage $g$ associée à l'opérateur ${\cal O}$, il ne démarre
qu'en $g^2$; on peut effectivement s'en convaincre
par des arguments de théorie conforme\foot{Je remercie
O.~Parcollet de m'avoir expliqué ce point.}.

Concluons par une remarque sur le comportement dit
de ``non liquide de Fermi'' (Non Fermi Liquid, NFL)
qui est censé être présent dans le modèle
de Kondo sur-écranté. On observe que dans
le régime sur-écranté et seulement dans celui-ci,
l'impureté est un spin de $SU_q(N)$, et donc
il est raisonnable de considérer que le comportement NFL
est directement lié à la symétrie de groupe quantique
sous-jacente. L'impureté est alors le facteur ``déclenchant''
qui met en lumière la symétrie $SU_q(N)$ du
secteur de spin des électrons. Nous définirons
plus précisément ce que l'on entend par ``déclenchant''
au \seckoncft\ (paragraphe théorie conforme
avec bord et groupes quantiques).

\subsec\seckontba{TBA et interprétation
de l'entropie à température nulle}
Nous allons dans cette section revenir sur l'analyse faite
au [\artic03.5] de l'entropie de l'impureté à température nulle,
et la justifier plus en détail.

La base de ce raisonnement est le principe selon
lequel aussi bien
les impuretés du système que les excitations physiques sont
des ``solitons'' en un sens très général, c'est-à-dire 
qu'un tel soliton fait transiter le système d'un ``vide'' initial
vers un ``vide'' final, caractérisé par un certain jeu de nombres
quantiques. Chaque soliton est alors représenté par une
certaine matrice d'adjacence; notre seule hypothèse sera
que ces matrices d'adjacence commutent.

Ce formalisme s'applique naturellement aux impuretés, car elles
n'ont qu'un nombre quantique: $SU(N)$ dans le cas sous-écranté,
$SU_q(N)$ dans le cas sur-écranté (elles ont une
rapidité fixe qui vaut $0$). Donc leurs matrices d'adjacence
sont précisément celles qui sont associées à la théorie de la
représentation de ces deux groupes (quantiques). Nous noterons
$A$ cette matrice d'adjacence.

Considérons alors un état du système caractérisé par des
nombres $N_i$ de solitons de type $i$ correspondant aux
excitations physiques ($i=1\ldots p$), et $N$
impuretés (habillées), et cherchons
quel est le nombre d'états accessibles au système. L'expression
la plus générale que l'on obtient est:
$$\Omega=\sum_{R,R'} d_{R,R'} (A_1^{N_1}\ldots A_p^{N_p}\, A^N)_{RR'}$$ 
où $R$ et $R'$ parcourent tous les ``vides'', 
et les $d_{R,R'}$ sont des coefficients
qui décrivent combien d'états correspondent à une transition
totale du vide $R$ au vide $R'$.

Prenons un exemple concret: supposons que $R$ soit une ``bonne'' représentation
du groupe quantique $U_q({\goth sl}(N))$, de sorte que la structure
solitonique qu'on a introduit ici s'identifie à celle que l'on a
considérée dans la section \secphysisoq. Alors, si $U_q({\goth
sl}(N))$ est une {\it vraie} symétrie de la théorie (comme dans
le modèle NJL anisotrope), on aura: $d_{R,R'}=\delta_{R\emptyset} \dim
R'$. Par contre, si la symétrie est {\it cachée}, alors on aura:
$d_{R,R'}=\delta_{R\emptyset} \delta_{R'\emptyset}$. Cependant,
on verra dans ce qui suit que ces coefficients, qui définissent
dans l'optique solitonique les conditions de bords, n'ont pas
d'effet sur le comportement dominant des quantités thermodynamiques.
Nous n'avons donc pas besoin d'en tenir compte.

Continuons notre raisonnement. Le nombre d'impuretés est
par définition: $N=D^{\rm imp} L$, où la taille du système $L$
est supposée grande, de sorte que $N\gg 1$. De même, si
l'on se place à température finie $T$,
comme les excitations physiques sont non-massives, on a:
$N_i = \alpha_i TL\gg 1$ à petit $T$ (les $\alpha_i$ sont
de coefficients fixes).
Les matrices d'adjacence sont
donc dominées par leur valeur propre de Perron--Frobenius:
$$\Omega \sim \lambda_1^{N_1}\ldots \lambda_p^{N_p} \lambda^N$$
Calculons maintenant l'entropie par unité de longueur $S/L$, à basse
température: ($S=\log\Omega$)
$$S/L\sim D^{\rm imp} \log\lambda + T \sum_{i=1}^p \alpha_i
\log\lambda_i$$
Dans la limite $T\to 0$, la contribution des excitations physiques
disparaît, ce qui est normal (un gaz d'électrons n'a pas d'entropie
à température nulle); par contre subsiste la contribution
$\log\lambda$ par impureté, qui est celle que nous recherchions%
\foot{Remarquons l'intérêt d'avoir supposé, à la différence
de [\artic03], qu'il y a une densité finie d'impuretés:
ceci nous évite d'avoir à retirer explicitement la contribution
des électrons à l'entropie, ce qui serait sinon rendu nécessaire
par le fait que l'on mélange différents ordres de $1/L$.}.

On conclut alors comme dans [\artic03.5]: quand l'impureté
est un spin $SU(N)$ normal (sous-écrantage), $\lambda$ est la dimension
de la représentation de l'impureté, ce qui est conforme au
raisonnement naïf (``l'entropie à température nulle est le
logarithme de la dégénerescence du vide''). Par contre,
quand l'impureté est un spin $SU_q(N)$ (sur-écrantage),
c'est maintenant la dimension quantique qui remplace la dimension
usuelle de la représentation (comme valeur
propre la plus grande de $A$), ce qui lui confère une signification
physique intéressante.

\subsec\seckonmag{Etude en champ magnétique}
Plaçons-nous enfin à température nulle et en présence
d'un champ magnétique: on peut alors résoudre exactement
les TBA. Les informations que nous allons en extraire confirmeront
les résultats obtenus à température non-nulle.

Pour faire les calculs, nous nous placerons à $N=\infty$,
et nous reprendrons les calculs du \seclargNwz. Ceci
n'est absolument pas indispensable, puisqu'on sait
faire les calculs exactement à $N$ fini
(section \secmagwz), mais comme nous l'avons
déjà affirmé au \seclargNwz, les corrections en $1/N$
ne changent pas la physique contenue dans le comportement
dominant à $N\to\infty$, et les formules à $N=\infty$
sont nettement plus claires.

Les TBA du modèle de Kondo généralisé sont les mêmes que
celles du modèle de WZW, et on peut donc reprendre les résultats
obtenus précédemment: la fonction d'énergie
$\epsilon^+=\lim_{T\to 0} T\log(1+\eta^r_f)/\sin(\pi r/N)$ vaut
$$\epsilon^+(\omega)=-i B\pi\nu {1\over\omega-i0}
\e{i\nu\omega\log(i\nu/{\rm e}(\omega-i0))}
\Gamma(1+i\nu\omega)^{-1}$$
en transformée de Fourier $\omega$ de $\zeta-\zeta_0$, où le
décalage $\zeta_0$ est de la forme $\zeta_0=\log(B/B_K)$
avec (cf Eq.\ \zzero)
$$B_K=T_K{\e{\nu(\log\nu-1)}\over\Gamma(1+\nu)}\eqn\BK$$
$B_K$ acquièrera pour nous la signification d'énergie
de transition en champ magnétique: $B_K$ est évidemment
du même ordre que $T_K$, avec un facteur numérique donné
par \BK.

Ensuite, on a besoin de l'expression de $\epsilon^+(x,y,\zeta)$
pour tout $x$ (rappelons que
pour $x\ne \nu$, $\epsilon(x,y,\zeta)>0$). Là encore, le formalisme
d'opérateurs différentiels à $N=\infty$ nous permet
d'exprimer la réponse de manière particulièrement élégante: 
$\epsilon^+(x,y,\zeta)=\sin y\, \epsilon^+(x,\zeta)$ avec
(en transformée de Fourier)
$$\epsilon^+(x,\omega)=\left\{\matrix{
{\displaystyle \sinh(x\omega)\over\displaystyle \sinh(\pi\nu\omega)} 
\epsilon^+(\omega)
\hfill& x\le \pi\nu\cr
\e{-(x-\pi\nu)|\omega|} \epsilon^+(\omega)+B(x-\pi\nu) \delta(\omega)
\hfill& x\ge \pi\nu\cr
}\right.
\eqn\epsB$$

Finalement, on calcule l'énergie libre d'une impureté de
tableau rectangulaire $(xN/\pi)\times (yN/\pi)$
grâce à la formule \Fkon,
soit dans la limite $N\to\infty$ et en Fourier, 
$$F(x,y,\omega)\sim {N^2\over \pi^2}
\sin y {1\over 1+\omega^2} \epsilon(x,\omega)$$
avec, en tenant compte du décalage, $\zeta=-\zeta_0=\log(B_K/B)$.
Le fait que $F\sim N^2$ est normal puisque l'on a affaire
à un tableau de Young avec un nombre de boîtes de l'ordre de $N^2$.
Posons donc $f(x,\omega)\equiv \epsilon^+(x,\omega)/(1+\omega^2)$.
Nous avons représenté sur la figure \fcut\ la structure
analytique de la fonction $f(\omega)\equiv f(x=\pi\nu,\omega)
=\epsilon^+(\omega)/(1+\omega^2)$.
\fig\fcut{Structure analytique de $f(\omega)$. Les zigzags
représentent une coupure, les points des pôles.}{\figdir{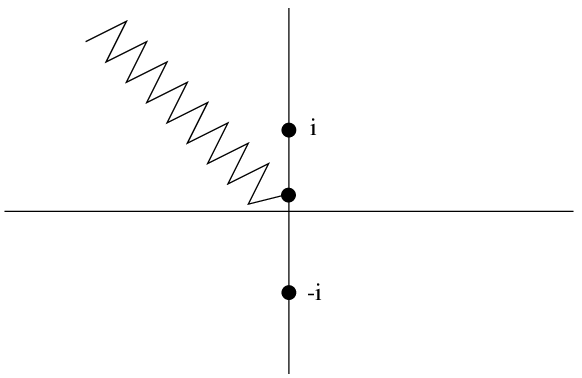}}
Examinons séparément les différents régimes:\goodbreak

\noindent{$\bullet$} {\it Régime écranté.}\par\nobreak

\figg\fscr{L'énergie libre $f/B$ de l'impureté
écrantée en fonction de $\zeta=-\log B$
($B_K\equiv 1$) et de $B$. $\nu=1/2$, $x=\pi\nu$.}%
{\centerline{\epsfbox{\figdir{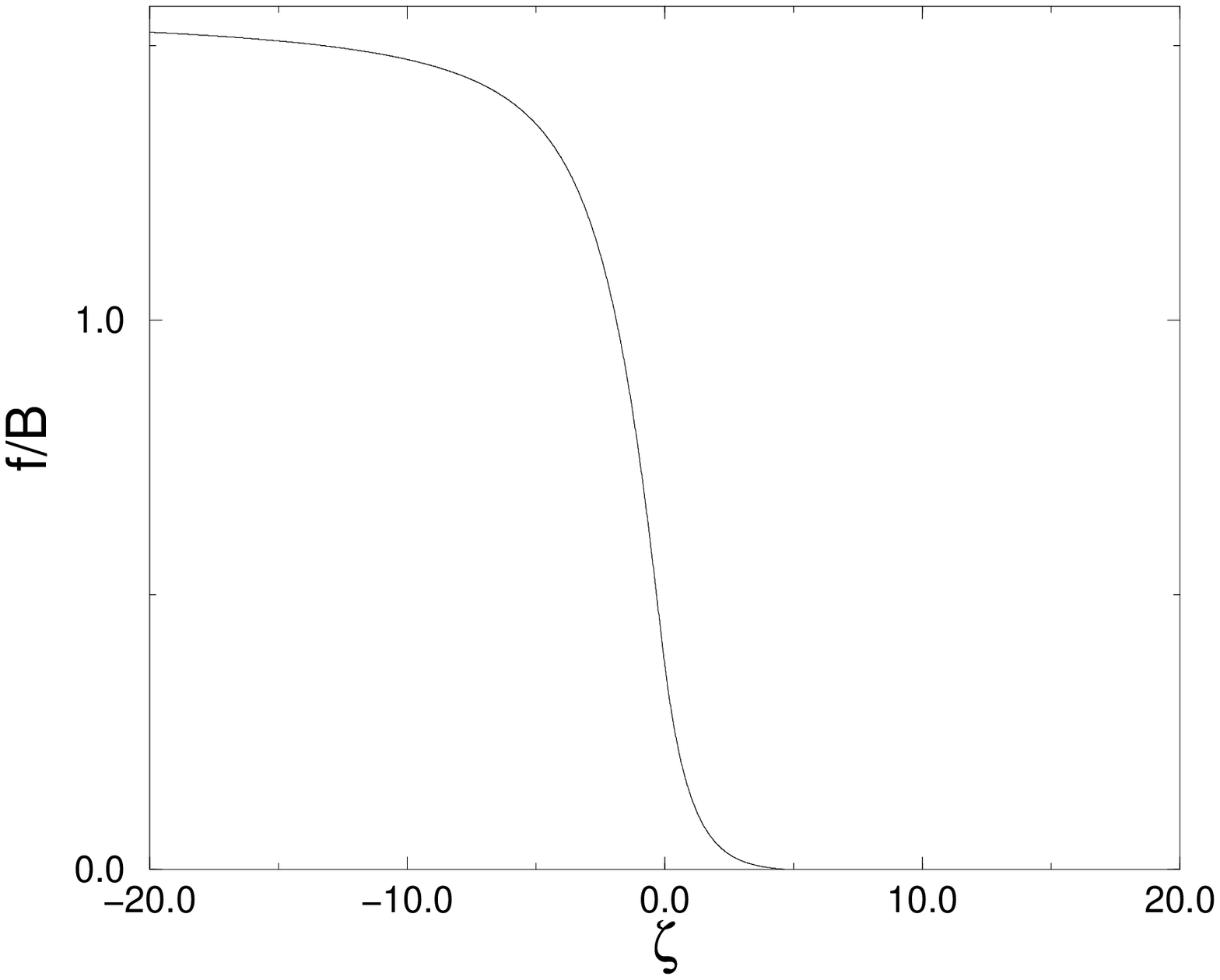}}\epsfbox{\figdir{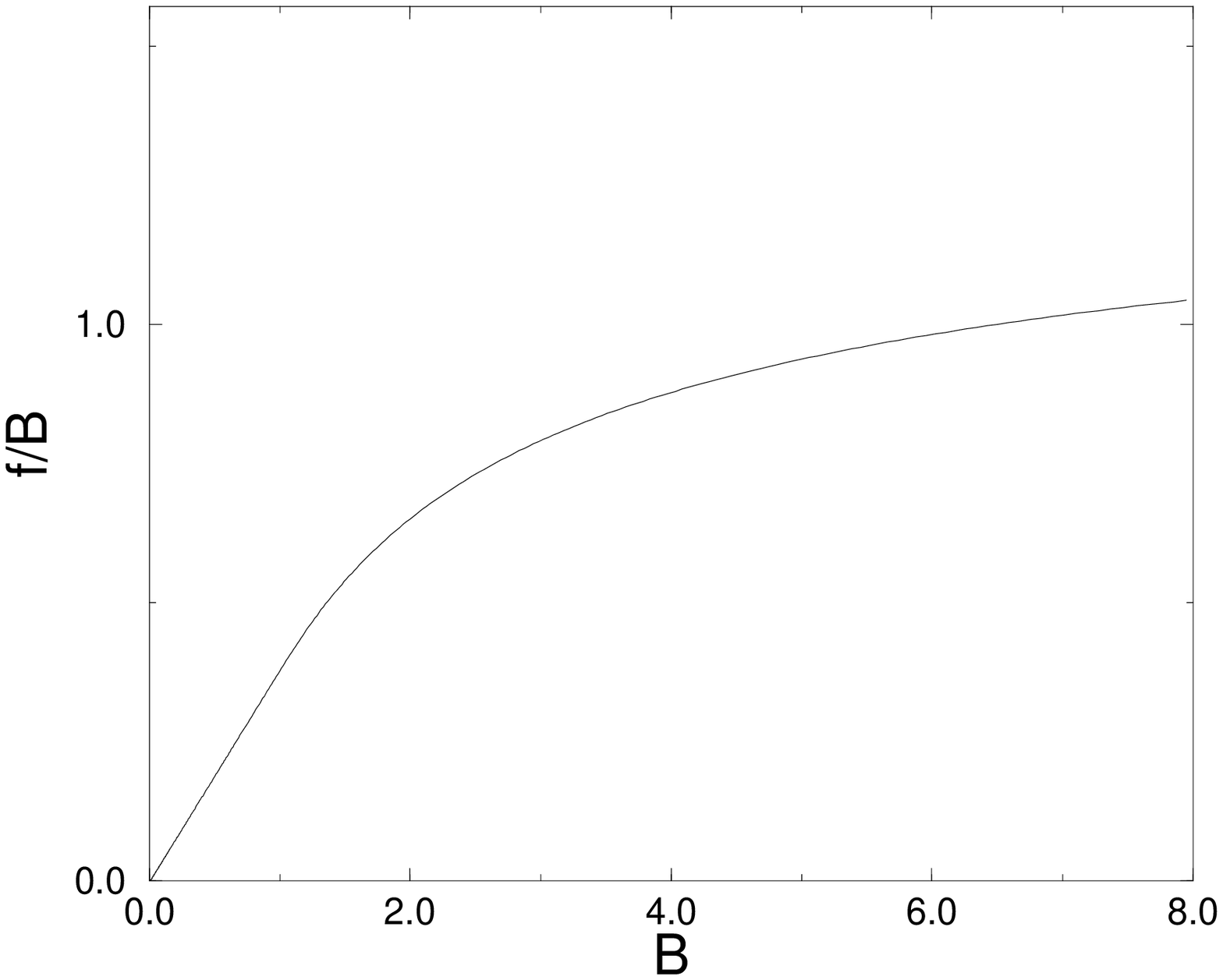}}}}%

Supposons tout d'abord que $B<B_K$ (faible champ magnétique).
Alors, $\zeta>0$, et l'intégrale
$f(\zeta)={1\over 2\pi}\int \d\omega\, \exp(-i\omega\zeta) f(\omega)$
n'attrape que le pôle en $\omega=-i$, d'où la valeur
exacte: $f(\zeta)=-{1\over 2}\epsilon^+(\omega=-i) \e{-\zeta}$. 
En remplaçant
$\zeta$ par sa valeur et en remarquant les compensations qui se produisent
entre $\epsilon^+(\omega=-i)$ et la définition de $B_K$, on obtient:
$$f(B)={B^2\over T_K} {\pi\nu\over 2}\eqn\Becr$$
Cette expression est à comparer avec l'équation (\artic03.4.8)
(dont le domaine de validité est: $T\to 0$, $B\sim T$):
en prenant $T=0$ et $N\to\infty$ dans cette dernière,
on retrouve \Becr\foot{Il faut faire attention que les conventions
de normalisation du second membre des TBA ne sont pas les
mêmes (dans [\artic03], elles ne sont pas adaptées à la
limite de grand $N$). La correspondance est donnée par:
$\sum (B^a)^2=(N/2) B^2$, $T_K=4T_0 \sin(\pi/N)$.}.

Notons que la forme particulièrement simple de \Becr\ n'est valable qu'à
$N=\infty$, puisque les corrections en $1/N$ font apparaître
des pôles en $-ni$ ($n$ entier), donc des puissances (entières)
supérieures de $B$.

Si $B>B_K$ (fort champ magnétique), au contraire, $\zeta<0$ et
on s'intéresse
au demi-plan complexe supérieur de la figure \fcut. Tout d'abord,
on attrape le pôle en $\omega=i0$ qui correspond à la valeur
limite $f(\zeta=-\infty)=\pi\nu B$; ensuite, on doit tenir
compte de la coupure:
on voit qu'on est ramené à calculer l'intégrale
$$B\nu\int_0^{+\infty}\d u\, {\sin(\pi u)\over u}
{\e{u\zeta/\nu} \e{-u(\log u-1)}\over \Gamma(1-u)(1-u^2/\nu^2)}$$
($u\equiv -i\nu\omega$) avec une prescription
pour éviter le pôle en $u=\nu$. Dans
la limite $\zeta\to -\infty$, celui-ci ne nous intéresse pas
car l'intégrale est bien évidemment dominée par
le démarrage de la coupure à $u\sim 0$; en développant,
on obtient:
$$B\pi\nu \int_0^{+\infty} \d u\, \e{u\zeta/\nu} (1-u\log u + O(u))$$
d'où finalement
$$f(B)\buildrel B\to\infty\over =\pi\nu B\left[ 1-{\nu\over\log(B/B_K)}
-{\nu^2\log[\log(B/B_K)]\over \log(B/B_K)^2}
+ O\left({1\over \log(B/B_K)^2}\right)
\right]$$
L'interprétation de ce résultat est claire: le développement
à $B\to\infty$ est celui de couplage faible, autrement dit
la liberté asymptotique.
Le premier terme $\pi\nu B$, soit plus explicitement
$F\sim {N^2\over\pi^2}\sin y\, \pi\nu B$,
n'est autre, comme on le vérifie par
un calcul explicite, que l'énergie
d'un spin libre dans la représentation $f\times (yN/\pi)$
en champ magnétique $B$.
Les deux termes suivants s'identifient
au développement
perturbatif au premier ordre en la constante de couplage
renormalisée (à l'échelle $B$), elle-même
développée à deux boucles.\goodbreak

\noindent{$\bullet$} {\it Régime sur-écranté.}\par\nobreak%
\figg\fover{L'énergie libre $f/B$ de l'impureté
sur-écrantée. $\nu=1/2$, $x=\pi \nu/2$.}%
{\centerline{\epsfbox{\figdir{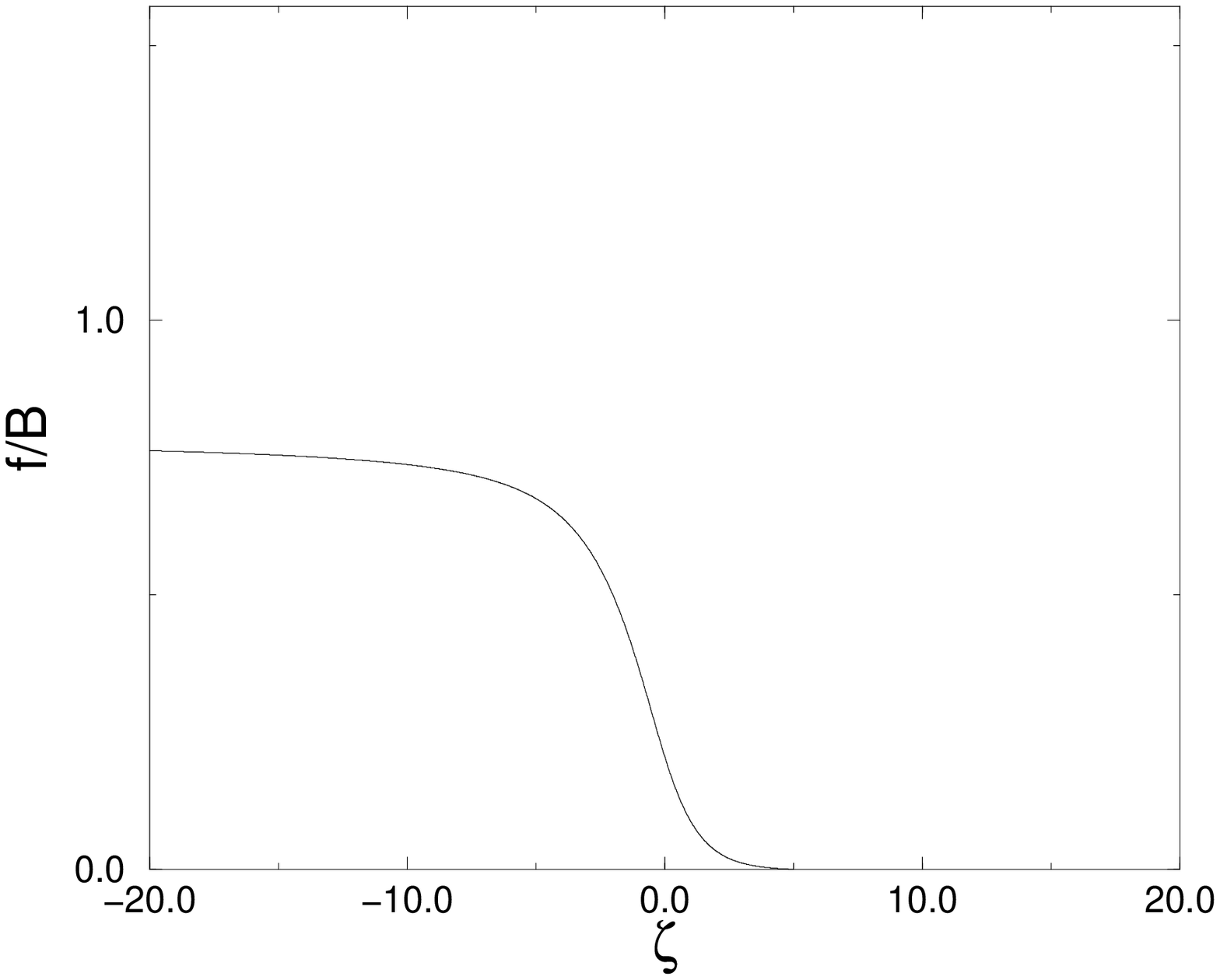}}\epsfbox{\figdir{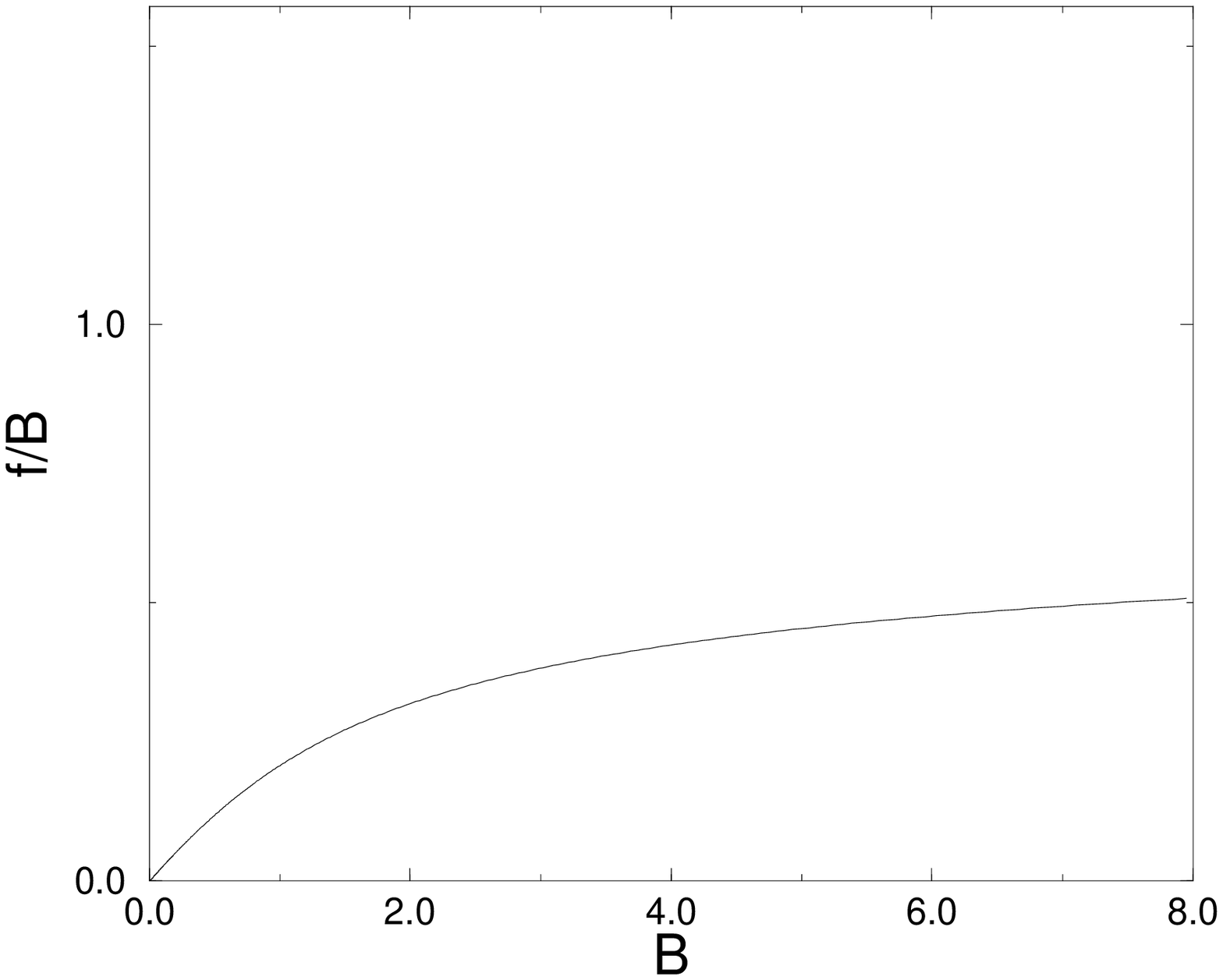}}}}%
\figg\foverb{L'énergie libre $f/B$ de l'impureté
sur-écrantée. $\nu=2$, $x=\pi\nu/2$.}%
{\centerline{\epsfbox{\figdir{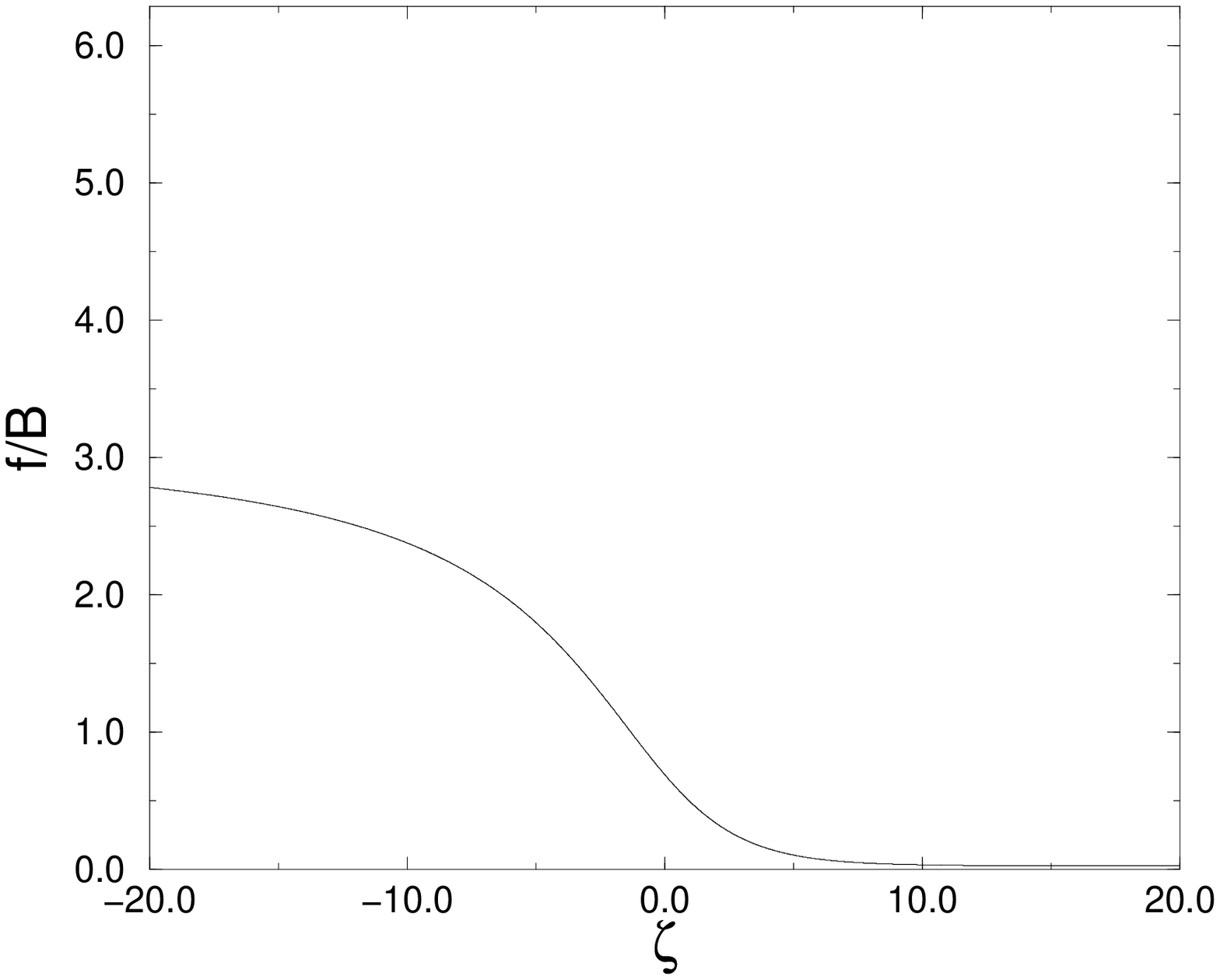}}\epsfbox{\figdir{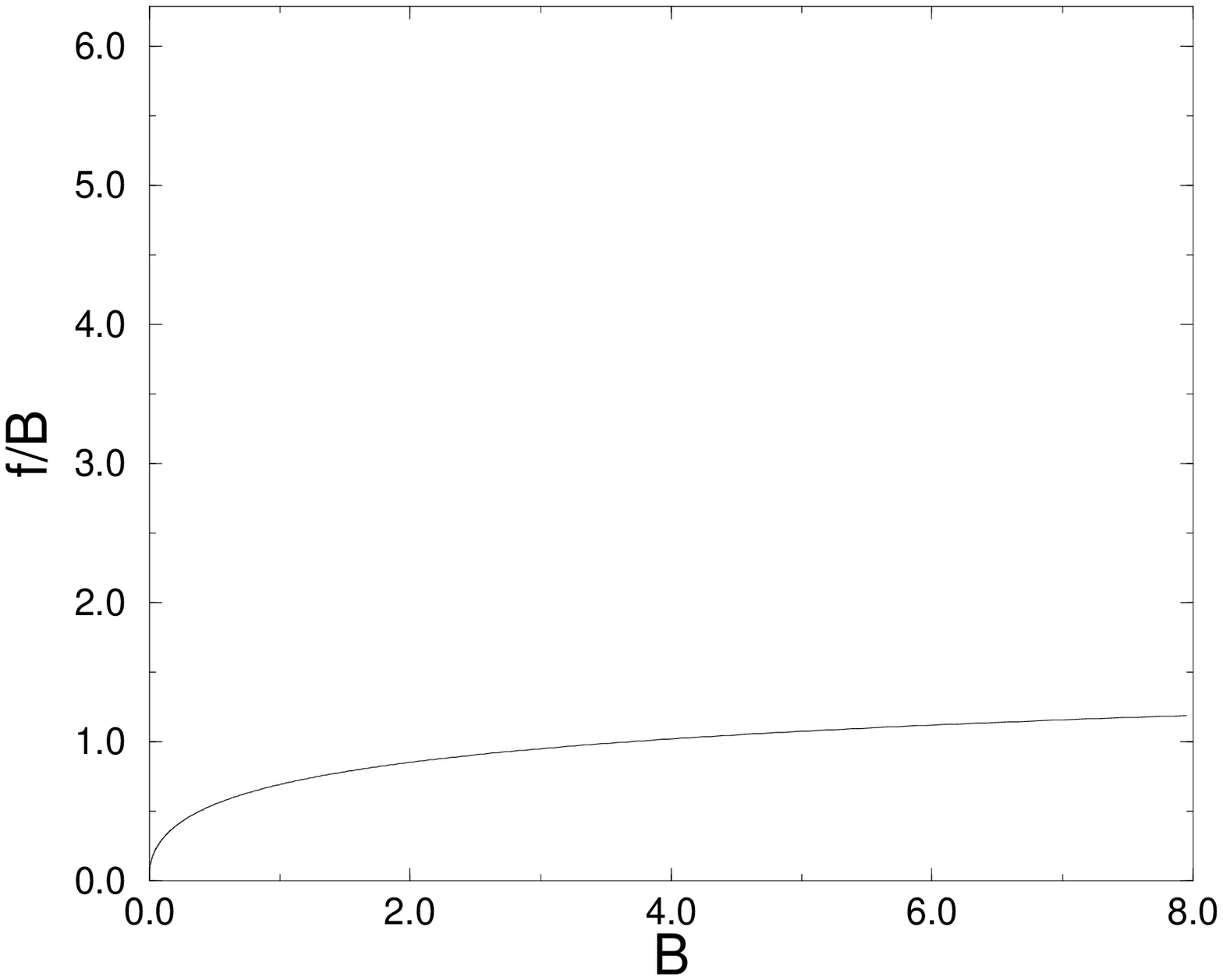}}}}%

On a cette fois: $f(x,\omega)={\sinh(x\omega)\over\sinh(\pi\nu\omega)}
f(\omega)$ avec $x<\pi\nu$.

En faible champ magnétique ($B<B_K$), on a deux types de pôles:
le pôle de $f(\omega)$ à $\omega=-i$ et les pôles
de $1/\sinh(\pi\nu\omega)$ à $\omega=-in/\nu$ ($n$ entier).
Ceci correspond à la compétition entre les deux
opérateurs perturbants au point fixe infra-rouge.
De manière générale,
on a un développement à petit $B$ du type: ($\nu\ne 1$)
$$f(B)/B=c {B\over B_K} + \sum_{n=1}^\infty c_n \left({B\over B_K}\right)
^{n/\nu}$$
et selon que $\nu<1$ ou $\nu>1$, c'est le premier terme
ou le second qui domine. Quand $\nu=1$, on a un pôle double,
d'où en transformée de Fourier un terme en $\zeta\e{-\zeta}$,
soit finalement $f(B)\sim c {B^2\over B_K}\log(B_K/B)$.

Ce résultat confirme que, dans le régime dit sur-écranté,
l'impureté est en fait exactement écrantée (au point
fixe infra-rouge), puisque
son aimantation à petit $B$ est nulle. En termes de groupe
de renormalisation, le point fixe infra-rouge n'est pas le
point de couplage fort (couplage infini), où l'impureté
serait réellement sur-écrantée par les $f$ saveurs
d'électrons; en effet, ce dernier est instable, et
le point fixe IR correspond à un point fixe non-trivial
correpondant à une valeur finie du couplage impuretés/électrons.

Ce résultat montre aussi que les exposants critiques à $B>0$, $T=0$ sont
distincts de ceux à $T>0$, $B\sim T$: ceci s'explique
[\PT] par le fait que, dans la limite $T\to 0$ à $B$ fixé, 
on ``tue'' certains modes (ceux qui sont associés au
nombre quantique $U_q({\goth sl}(N))$)
rendus massifs par le champ magnétique. 

En fort champ magnétique ($B>B_K$), on trouve des résultats identiques
au cas écranté: le comportement dominant est $f(B)\sim x B$,
correspondant à un spin libre $(xN/\pi)\times (yN/\pi)$; puis
on a des corrections logarithmiques caractéristiques de la liberté
asymptotique.\goodbreak

\noindent{$\bullet$} {\it Régime sous-écranté.}\par\nobreak

\figg\funder{L'énergie libre $f/B$ de l'impureté
sous-écrantée. $\nu=1/2$, $x=3\pi\nu$.}%
{\centerline{\epsfbox{\figdir{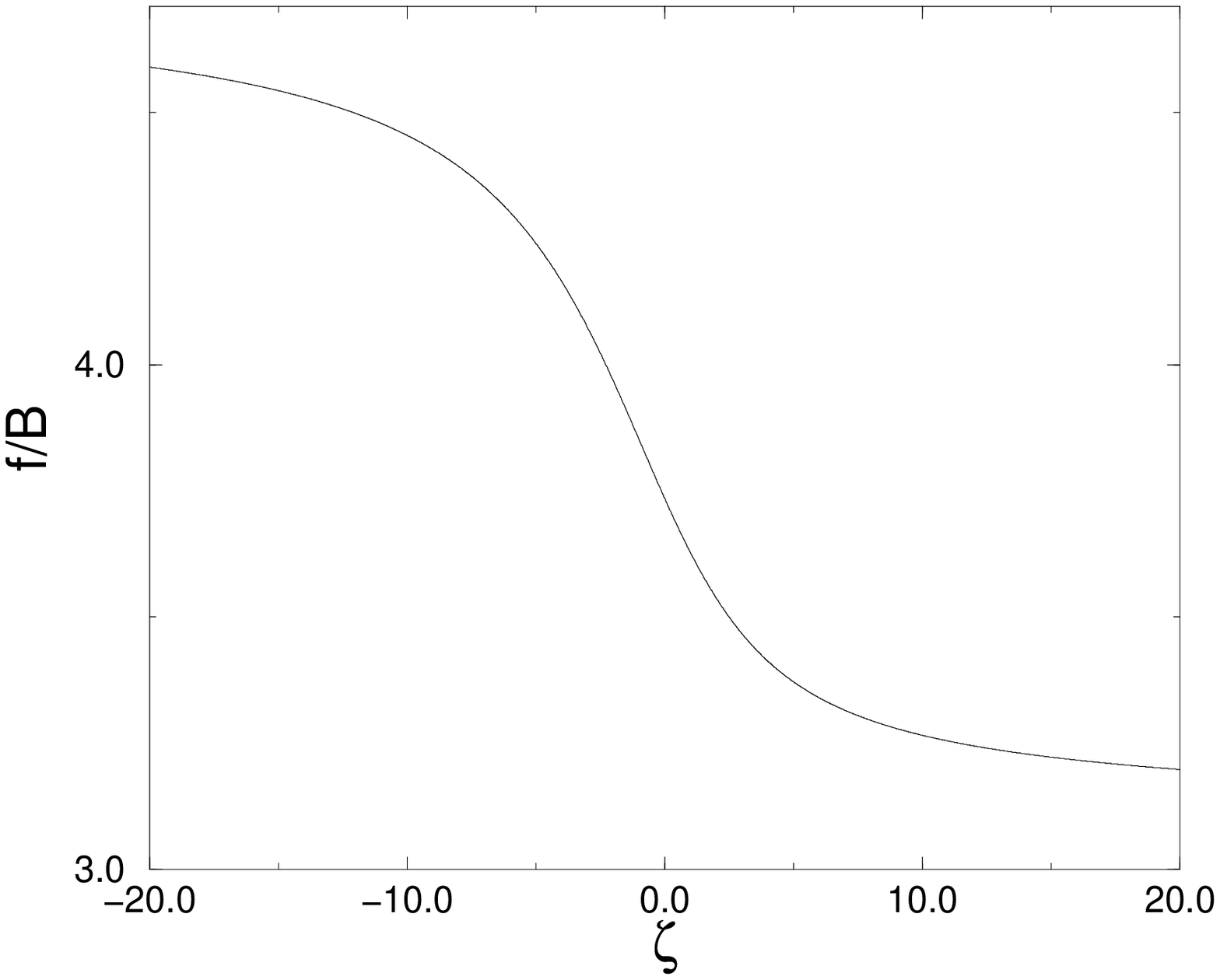}}\epsfbox{\figdir{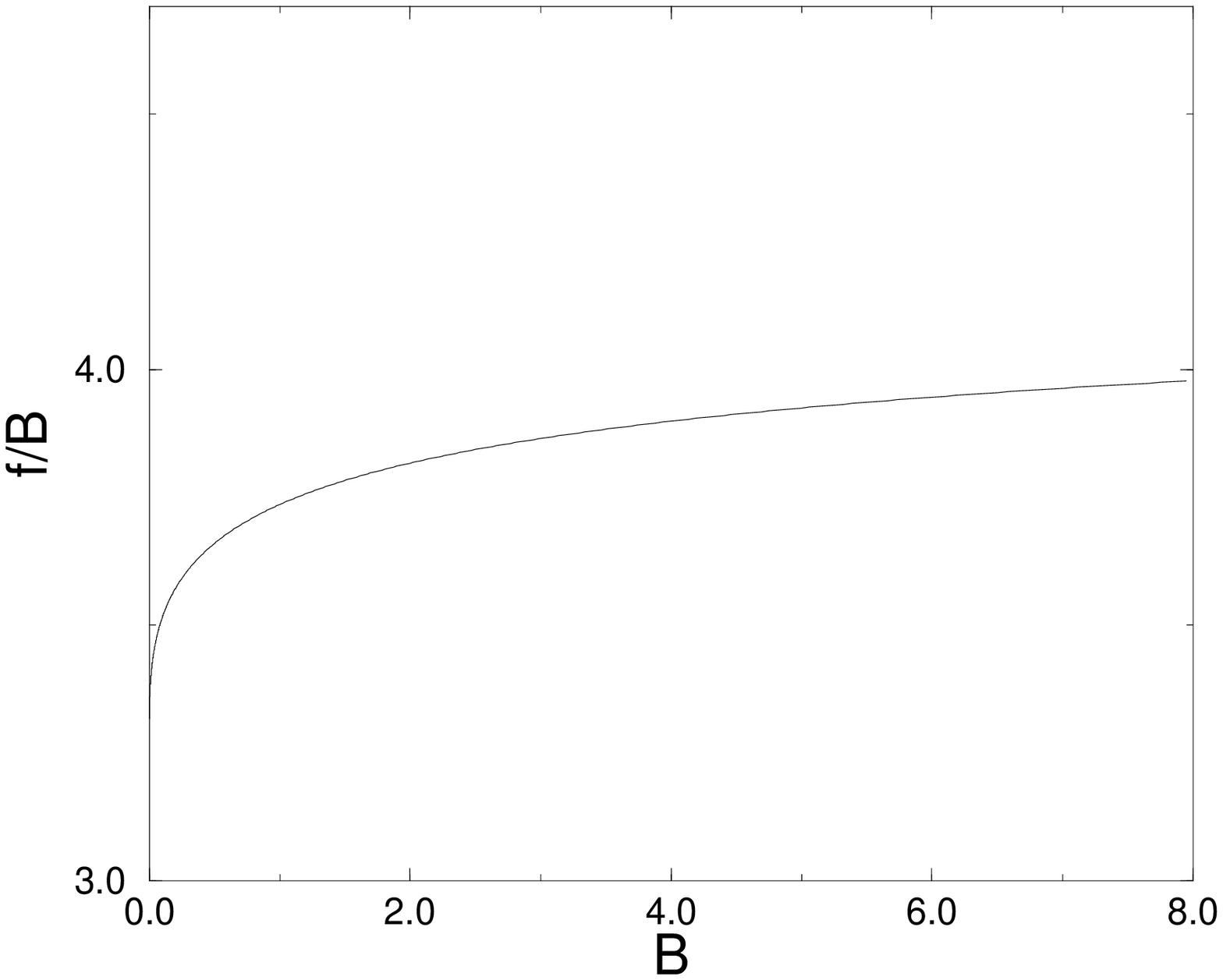}}}}%

L'équation \epsB\ montre que $f(x,\omega)$ a maintenant
à la fois des pôles en $\omega=\pm i0$ (la fonction $\delta(\omega)$
contribue aux deux), et des coupures démarrant en $\omega=\pm i0$.
On trouve facilement, par les mêmes méthodes que précédemment,
que
$$f(B)\sim\left\{\!\!\!\!\matrix{
& (x-\pi\nu)B\hfill& B\to 0\cr
& xB \hfill&B\to\infty\cr}\right.$$
plus les corrections logarithmiques de la liberté asymptotique
(ultra-violette et infra-rouge).

\subsec\seckoncft{Lien avec l'approche de théorie conforme de Kondo}
Nous allons terminer cette thèse par une comparaison des deux
approches principales qui ont été développées pour étudier
le modèle de Kondo: l'approche de l'Ansatz de Bethe et l'approche
de Théorie Conforme avec bord
(pour cette dernière, voir
la revue [\ref\AFF{I.~Affleck, {\it Acta Phys. Polon.} B26 (1995),
1869 \pre{cond-mat/9512099}.}]). Notons tout de suite que
ces deux approches font appel à des mathématiques différentes:
l'une est liée à l'intégrabilité, donc aux groupes quantiques,
tandis que l'autre est liée aux algèbres de Kac--Moody.
Le modèle de Kondo sur-écranté 
fournit un exemple concret de ``dictionnaire''
des correspondances entre ces deux approches et
les objets mathématiques associés; nous allons
donner certaines de ces correspondances. Ce qui suit
se veut être une simple suite de remarques sur
l'approche de théorie conforme de Kondo, dont
certaines mériteraient sans doute d'être plus développées.

\noindent{$\bullet$} {\it Théorie conforme et groupes quantiques.}\par\nobreak
En dehors des modèles intégrables, les groupes quantiques
jouent également un rôle important en théorie conforme. Ceci
peut paraître naturel puisque toutes les théories conformes
connues peuvent également être considérées comme des modèles
intégrables. Cependant, la manière dont les groupes quantiques
apparaissent dans les théories conformes (par exemple
à travers les relations d'échange et de fusion) ne permet
pas de les relier de façon évidente à ceux qui apparaissent
dans les modèles intégrables. Nous allons à présent fournir
quelques indications sur ce lien dans le cas de la théorie
conforme de WZW $SU(N)$ au niveau $f$.

Rappelons qu'une des données fondamentales d'une théorie
conforme est constituée par les {\it règles de fusion}
vérifiées par les opérateurs primaires pour l'algèbre de
symétrie de la théorie. Ainsi, dans WZW $SU(N)$ niveau $f$,
l'algèbre de symétrie en question est constituée de deux
copies (holomorphe et anti-holomorphe, ou gauche et droite)
de (l'algèbre enveloppante de) l'algèbre
de Kac--Moody $SU(N)$ au niveau $f$, et les
opérateurs primaires $\Phi_R$, qui sont indexés par une
représentation $R$ de $SU(N)$ dont le tableau de Young
a moins de $f$ colonnes, satisfont symboliquement:
$$\Phi_R\,\Phi_{R'} \sim A_{RR'}^{R''} \Phi_{R''}$$
où $\sim$ signifie que l'on a gardé les opérateurs primaires
qui apparaissent dans le Développement en Produit d'Opérateurs
de $\Phi_R$ et $\Phi_{R'}$. La matrice $(A_R)_{R'}^{R''}
\equiv A_{RR'}^{R''}$ se trouve être exactement égale
à la matrice de produit tensoriel tronqué de la \rep\ $R$
de $U_q({\goth sl}(N))$
avec $q=-\exp(-i\pi/(f+N))$ [\ref\AGGSb{L.~Alvarez-Gaumé,
C.~G\'omez et G.~Sierra, {\it Phys. Lett.} B220 (1989), 142.}].
Ceci suggère qu'il existe
un groupe quantique sous-jacent qui détermine les règles de fusion
de WZW. Dans l'esprit de la formule \decwzw, on conjecture
que les opérateurs primaires
$\Phi_R$
, qui sont
dans la représentation $R$ pour
les deux sous-algèbres horizontales $SU(N)_L$ et $SU(N)_R$
se décomposent en opérateurs chiraux qui sont eux,
des multiplets de {\it la même} représentation
$R$, mais à la fois pour $SU(N)$ (gauche ou droit selon
la chiralité) et pour le groupe quantique $U_q({\goth sl}(N))$. 
\rem{Plus précisément $\Phi_{aj}(z)$ appartient à la représentation
$R$ de $U_q({\goth sl}(N))$, tandis que $\Phi_{ia}(\bar z)$
appartient à la représentation conjuguée $\bar R$, de
sorte qu'en sommant sur $a$ l'opérateur primaire
complet $\Phi_{ij}(z,\bar z)$ est {\it invariant} par
l'action du groupe quantique: ceci est le signe
que $U_q({\goth sl}(N))$ n'est là encore
qu'une {\it symétrie cachée}, au sens défini précédemment.}%
\rem{fonction de partition et caractères?} Cette hypothèse
est corroborée par la construction explicite des multiplets
de groupe quantique associés aux opérateurs primaires ({\it opérateurs
de vertex écrantés} [\ref\RRR{C.~Ram\'\i rez,
H.~Ruegg et M.~Ruiz--Altaba, {\it Phys. Lett.} B247 (1990),
499.}]) dans un Hilbert un peu élargi, de façon à ce
que la symétrie ne soit plus ``cachée''.

Une fois cette construction faite, il est tentant d'identifier
le groupe quantique $U_q({\goth sl}(N))$ ainsi obtenu avec
celui qui apparaît dans l'Ansatz de Bethe. Rappelons que
ce dernier est la sous-algèbre horizontale d'un groupe
quantique affine $U_q(\widehat{{\goth sl}(N)})$,
et qu'il agit non pas sur les opérateurs de vertex mais
sur les états asymptotiques de la théorie,
qui est donc définie sur un espace non-compactifié.
En particulier, dans les notations de la théorie conforme,
on peut dire que l'action du groupe quantique affine commute
avec les générateurs $L_{-1}$ et $\bar L_{-1}$ de l'algèbre de Virasoro
(quantification ``sur le plan''; voir
[\ref\BMM{D.~Bernard, Z.~Maassarani et P.~Mathieu,
{\it Mod.\ Phys.\ Lett.} A12 (1997), 535 \pre{hep-th/9612217}.}]
pour une remarque similaire sur les Yangiens),
mais pas avec $L_0$ et $\bar L_0$ (qui reconstitueraient
le Hamiltonien si l'on utilisait la quantification
``sur le cylindre''). Heureusement,
on constate que la sous-algèbre horizontale $U_q({\goth sl}(N))$,
a, de par la définition de la gradation (cf remarque sur
gradation et ``spin'' au \secqgaff) un ``spin'' nul, c'est-à-dire
qu'elle commute aussi avec $L_0$ et $\bar L_0$, comme il se doit.
Ceci revient encore à dire que la symétrie de groupe
quantique (non affine) peut subsister sur un espace
compactifié, comme on l'a déjà vu précédemment.
Il n'y a donc pas d'obstacle de principe à identifier
le groupe quantique de la théorie conforme et celui de l'Ansatz
de Bethe.

\noindent{$\bullet$} {\it Théorie conforme avec bord
et groupes quantiques.}\par\nobreak
Pour décrire le modèle de Kondo, nous avons besoin
d'un type de théorie conforme légèrement différent
de ce qui a été considéré ci-dessus:
une théorie conforme avec bord
[\ref\CAR{J.L.~Cardy, {\it Nucl. Phys.} B324 (1989), 581.}].
Considérons donc la théorie de $Nf$ fermions libres
sur un cylindre de rayon $\beta=1/T$ et de taille finie $L$
(au final $L\to\infty$),
les conditions aux bords du cylindre étant libres.
Comme nous l'avons déjà vu, ces conditions aux bord nous permettent
de considérer une théorie équivalente constituée d'une seule
chiralité de fermions; ceci se répercute sur
l'expression de la fonction de partition:
$$Z=\sum_{R,R',r} a_{R,R',r} \chi_R^{\widehat{{\goth sl}(N)}_f}
\chi_{R'}^{\widehat{{\goth sl}(f)}_N}
\chi_{r}^{\widehat{{\goth u}(1)}}\eqn\Zfer$$
(où $R$, $R'$, $r$ parcourent 
les représentations de $\widehat{{\goth sl}(N)}_f$,
$\widehat{{\goth sl}(f)}_N$, $\widehat{{\goth u}(1)}$)
qui, au lieu d'être quadratique en les caractères, est linéaire
en ceux-ci. Les caractères sont évalués en $\e{-\pi\beta/L}$.
Les coefficents $a_{R,R',r}$ ont été calculés
dans [\ABI]: d'après le lien évoqué plus haut entre représentations
de $\widehat{{\goth sl}(N)}_f$ et de $U_q({\goth sl}(N))$,
il est souhaitable de réinterpréter ces coefficients en termes
de groupes quantiques. Pour cela, nous remarquons que le secteur
de spin, lié à l'algèbre $\widehat{{\goth sl}(N)}_f$,
possède un groupe quantique $U_q({\goth sl}(N))$, 
$q=-\e{-i\pi/(f+N)}$, tandis
que le secteur de saveur lié à $\widehat{{\goth sl}(f)}_N$
possède un groupe quantique $U_q({\goth sl}(f))$, avec le même $q$.
Par dualité rang-niveau, ces deux groupes quantiques ont la même
théorie de la représentation, et nous pouvons identifier
leurs représentations: ceci se reflète dans les coefficients
$a_{R,R',r}$ par le fait que $a_{R,R',r}\ne 0$ implique
$R'=R^T$ (tableaux de Young transposés l'un de
l'autre). Ceci est un signe -- mais
pas une démonstration, bien sûr --
que la recombinaison des secteurs de saveur
et de spin, grâce aux
coefficients $a_{R,R',r}$, reconstitue une symétrie de groupe quantique cachée,
alors que si l'on considère le secteur de spin isolément,
la symétrie n'est apparemment pas cachée.

\noindent{$\bullet$} {\it Point fixe infra-rouge de Kondo et hypothèse de
fusion.}\par\nobreak
Continuons sur la voie du paragraphe précédent en 
essayant de faire le lien avec le modèle de Kondo. Rappelons
que dans celui-ci, quand on se ramène de $3$ dimensions
d'espace à une seule, on se retrouve avec un système
de fermions définis sur l'intervalle $[0,L]$ ($L\to\infty$),
et l'impureté placée en $x=0$. On voit donc que pour
incorporer l'impureté dans le formalisme du paragraphe précédent,
il suffit de modifier les conditions aux bords à l'un des bouts
du cylindre. Ces dernières brisent l'invariance conforme du modèle,
mais, aux points fixes ultra-violet et
infra-rouge, elle est restaurée.
Le seul point fixe non-trivial est le point fixe IR de Kondo
sur-écranté, et c'est sur celui-ci que nous allons nous pencher
à présent.

L'impureté sur-écrantée est, on l'a vu, un spin
du groupe quantique $U_q({\goth sl}(N))$; d'après le lien
évoqué plus haut entre opérateurs primaires
et représentations de groupes quantiques,
il est naturel de supposer que l'impureté,
dans l'infra-rouge, va modifier la manière dont les différentes
représentations de $\widehat{{\goth sl}(N)}_f$ apparaissent
dans la fonction de partition. Plus précisément,
conformément à ce qu'on a vu au \secphyssup,
au lieu d'avoir les conditions aux bords caractéristiques
d'une symétrie cachée (spin $U_q({\goth sl}(N))$ total nul,
donc $R'=R^T$ dans la fonction de partition \Zfer),
on va avoir un spin total égal à celui de l'impureté
sur-écrantée (cf figure \bcspin)\foot{C'est en ce
sens que l'impureté est le facteur ``déclenchant'' qui
met en lumière la symétrie cachée $U_q({\goth sl}(N))$.}. 
On aboutit donc
à {\it l'hypothèse de fusion} suivante: au point fixe IR,
la fonction de partition surécrantée est donnée par
$$Z=\sum_{R,R',R'',r} A^R_{R''} a_{R,R',r} 
\chi_{R''}^{\widehat{{\goth sl}(N)}_f}
\chi_{R'}^{\widehat{{\goth sl}(f)}_N}
\chi_{r}^{\widehat{{\goth u}(1)}}\eqn\Zferb$$
où $A$ est la matrice d'adjacence de la \rep\ de l'impureté.
Ceci est exactement l'hypothèse de fusion conjecturée
dans [\ref\AfL{I.~Affleck et A.W.W.~Ludwig,
{\it Nucl. Phys.} B360 (1991), 641; {\it Phys. Rev. Lett.} 67 (1991), 161.}].
On montre qu'elle donne effectivement les résultats
physiques que l'on attendait au point fixe infra-rouge sur-écranté,
et en particulier l'entropie à température nulle.

Terminons sur un commentaire concernant l'hypothèse de fusion:
dans tous les cas que nous avons considérés, la représentation
de groupe quantique de l'impureté habillée (IR) s'avère être
la même, modulo une symétrie de l'algèbre des
représentations de $U_q({\goth sl}(N))$, que la représentation
$SU(N)$ de l'impureté nue (UV). C'est ce qui a permis
de deviner l'hypothèse de fusion appropriée sans avoir
à résoudre explicitement le modèle (par l'Ansatz de Bethe
par exemple). Cependant, cette coincidence des représentations
n'a pas de raison d'être maintenue dans le cas, par exemple,
d'impuretés appartenant à des tableaux de Young
non-rectangulaires. Dans ce cas-là, il paraît
difficile de trouver l'hypothèse de fusion sans
avoir déjà résolu le modèle de façon indépendante.

\newnochap{Conclusion}
Il est à présent temps de tirer quelques conclusions du travail effectué,
et de dégager quelques idées sur les possibilités de développements
ultérieurs.

Comme on a pu le voir, l'Ansatz de Bethe est
un outil puissant, qui permet de sonder la physique non-perturbative
de modèles de physique bidimensionnelle. Les modèles solubles ne
sont bien sûr pas arbitraires, mais on peut espérer que bon
nombre de propriétés que l'on observe chez eux, sont
présents même dans des modèles non-intégrables. Citons également
le fait que l'Ansatz de Bethe diagonalise le Hamiltonien de la théorie
dans une base qui a une interprétation naturelle: dans la limite
thermodynamique $L\to\infty$, c'est celle des états asymptotiques,
composés de (quasi-)particules dont la diffusion est factorisée.
Dans la limite d'échelle relativiste,
on retrouve ainsi la vision intuitive des théories
des champs intégrables.
Enfin, les Equations d'Ansatz de Bethe Thermodynamique et les
Equations Non-Linéaires Intégrales permettent des calculs
explicites de quantités thermodynamiques.

Cependant, il faut aussi constater les limitations de l'Ansatz de
Bethe. Ainsi, le développement perturbatif usuel en diagrammes de
Feynman, qui correpond par exemple au développement de haute
température de l'énergie libre des théories asymptotiquement libres,
est paradoxalement
difficile voire infaisable par l'Ansatz de Bethe; de plus,
l'identification exacte de la constante de couplage renormalisée,
et donc de la fonction $\beta$ exacte est généralement impossible,
à cause des méthodes de régularisation spéciales employées
pour l'Ansatz de Bethe. Il paraît également difficile d'en extraire
les fonctions de corrélation des observables; ceci est d'ailleurs
l'objet de recherches actives dans le domaine des modèles intégrables
(facteurs de forme, etc).

Signalons enfin quelques extensions possibles des travaux contenus
dans cette thèse. Tout d'abord, on peut considérer
des généralisations supersymétriques des 
équations du chapitre \chapstrucbae: on a des familles de
déformations des super-algèbres de Lie, qui donnent lieu
à de nouvelles Equations d'Ansatz de Bethe. Bien que celles-ci
aient déjà été largement
étudiées, il serait intéressant de
classifier de manière générale leurs propriétés, dans l'esprit
de ce qui a été fait au chapitre \chapstrucbae.

Ensuite, les NLIE présentées au \secddv\ (pour le modèle de Toda affine
complexe) doivent pouvoir s'étendre aux modèles fusionnés; ainsi,
le modèle à $19$ vertex admet une NLIE, dont il serait intéressant
de prendre la limite d'échelle relativiste. La structure des NLIE
doit rester plus simple que celle des TBA correspondantes,
mais tout de même plus compliquée que dans le cas non-fusionné:
les NLIE sont-elles alors encore des sortes de ``fonctions de comptage
physiques''? Y a-t-il encore ``resommation des magnons'' associés
à la symétrie $\goth g$?
Il y a là une analyse qu'il reste à effectuer.
\rem{est-ce qu'on peut obtenir DDV par fusion eqs? et pour spin $1$
qu'en est-il? (revoir papier australiens)}

Venons-en finalement au modèle de Kondo généralisé.
L'étude qui en a été faite au \seckondo\ est quelque peu
incomplète. Tout d'abord, les résultats proposés ne
sont pas tous sous une forme qui ait un sens physique
direct: ainsi, il serait intéressant de relier les
matrices $S$ impureté/excitations physiques à la
résistivité, qui est elle observable ``expérimentalement''.
Dans le même ordre d'idées, il faudrait relier de manière
plus précise les résultats de l'Ansatz de Bethe avec
ceux obtenus par des méthodes de théorie conforme:
ceci suppose de développer plus longuement les considérations
du \seckoncft.\rem{lien BA/CFT a approfondir ... aussi CFT perturbe au bord
(extension de: CFT avec bord)}

Ensuite, il serait intéressant de briser la symétrie $SU(f)$,
de façon à considérer des saveurs qui ont des couplages d'intensité
inégale avec l'impureté. Ce problème n'est pas résolu à l'heure
actuelle, car il n'est pas évident, si l'on considère des
constantes de couplage générique pour chaque saveur, de préserver
l'intégrabilité. Cependant, des travaux récents sugg\`erent que
cela est possible, mais cela n'a pas été montré explicitement
au niveau des Equations d'Ansatz de Bethe.

Il reste enfin un cas où le modèle de Kondo généralisé n'a pas été
étudié:
celui où les impuretés appartiennent à des tableaux de Young
non-rectangulaires. 
Le cas non-rectangulaire présente un intérêt particulier; il devrait
permettre de tester sur un cas non-trivial l'idée intuitive qui
se dégage de la solution par l'Ansatz de Bethe (du cas rectangulaire):
l'impureté, au point fixe infra-rouge, et quel que soit le régime,
peut être considérée comme un complexe formé de l'impureté nue
et d'un certain nombre d'électrons (figure \naivspin).
\fig\naivspin{Impureté nue/impureté écrantée.
Sur cette figure, $N=5$, $f=4$. A gauche,
sous-écrantage d'une impureté $3\times 6$;
à droite, sur-écrantage d'une impureté $3\times 3$.
Les boîtes en pointillé sont les boîtes des électrons
qui vont se coller à l'impureté.}{\figdir{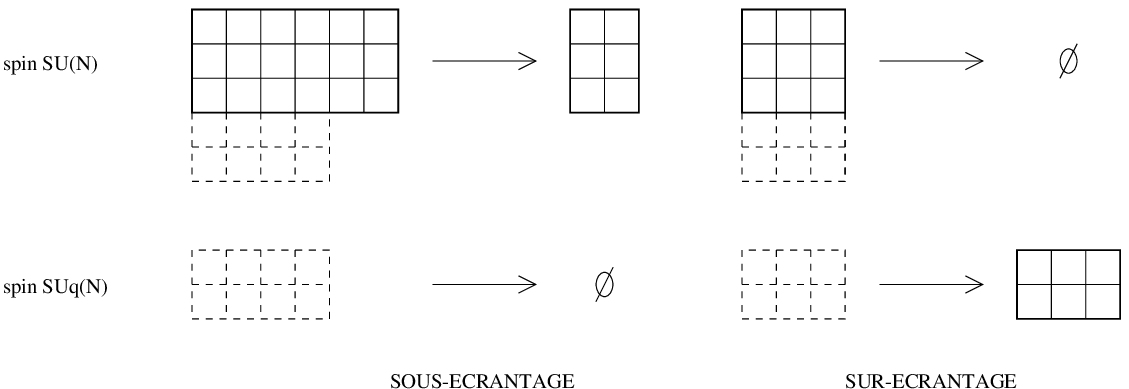}}
Dans le cas rectangulaire sur-écranté, cette vision est
indistinguable\foot{modulo une conjugaison, comme on le voit
sur la figure \naivspin; ce qui est peut-être décelable
dans le signe de certaines expressions.} d'une
autre vision plus na\"ive, qui consiste à dire que le spin $SU(N)$
de l'impureté nue s'est ``transformé'' en le même spin $U_q({\goth sl}(N)$
de l'impureté habillée (et qui conduit à l'hypothèse de fusion).
Le cas non-rectangulaire devrait permettre de trancher.

On peut encore utiliser l'Ansatz de Bethe pour
résoudre le modèle avec impuretés non-rectangulaires,
mais comme l'hypothèse de corde
n'est alors plus valable, il est impossible
d'appliquer le formalisme des BAE continues et des TBA.
Cependant, rien n'empêche de court-circuiter les TBA et d'utiliser
directement les équations de fusion. Pour cela, il est nécessaire
de considérer des équations de fusion pour des tableaux
non nécessairement rectangulaires. Ces équations existent,
on peut même en écrire de toutes sortes: reste
à déterminer lesquelles sont utiles, et comment y incorporer
l'effet des électrons sur l'impureté, ce qui doit
conduire aux équations de fusion modifiées. Il ne
semble pas y avoir de difficulté particulière à cette approche,
et elle paraît donc particulièrement prometteuse.

\newnochap{Appendice}
\write\tfile{\noindent\string\quad\ Article \artic01\string\leaderfill
153\par}
\write\tfile{\noindent\string\quad\ Article \artic02\string\leaderfill
163\par}
\write\tfile{\noindent\string\quad\ Article \artic03\string\leaderfill
193\par}
\write\tfile{\noindent\string\quad\ Article \artic04\string\leaderfill
235\par}
\write\tfile{\noindent\string\quad\ Article \artic05\string\leaderfill
269\par}
Dans cet appendice sont regroupés les cinq articles que j'ai écrits
durant ma thèse. Le premier article [\artic01] a été publié:
sa référence est {\it Nucl.\ Phys.}\ B497 (1997), 725.
Le second article [\artic02], qui est la suite du premier, a été
accepté pour publication dans {\it Commun.\ Math.\ Phys.} Le 
troisième [\artic03], écrit en collaboration avec N.~Andrei, 
a été accepté dans {\it Nucl.\ Phys.}\ B. Le quatrième
[\artic04] a été accepté dans {\it J.\ Phys.}\ A.
Enfin, le cinquième [\artic05], écrit avec V.~Kazakov,
n'a pas encore été soumis.

\ifx\answ\bigans\ifodd\the\pageno{\vfill\eject\ }\fi
\vfill\eject\fi
\pageno=295
\footatend
\newnochap{Références}%
\immediate\closeout\rfile\writestoppt
\baselineskip=14pt
{\frenchspacing%
\parindent=20pt\input refs.tmp\vfill\eject}\nonfrenchspacing
\bye